\documentclass[twocolumn,showpacs,preprintnumbers,amsmath,amssymb,epsfig]{revtex4-2}

\usepackage{graphicx}
\usepackage{dcolumn}
\usepackage{bm}
\usepackage{epsfig}

\usepackage{float}
\usepackage{amsmath}
\usepackage{epsfig,floatflt}
\usepackage{subfigure}
\usepackage[usenames]{color}
\usepackage{rotating}
\usepackage{comment} 
\usepackage{hyperref}

\newcommand{\mii}{Mg\,\textsc{ii}}
\newcommand{\civ}{C\,\textsc{iv}}

\usepackage[T1]{fontenc}
\usepackage{scalerel}
\usepackage{tikz}
\usetikzlibrary{svg.path}

\definecolor{orcidlogocol}{HTML}{A6CE39}
\tikzset{
  orcidlogo/.pic={
    \fill[orcidlogocol] svg{M256,128c0,70.7-57.3,128-128,128C57.3,256,0,198.7,0,128C0,57.3,57.3,0,128,0C198.7,0,256,57.3,256,128z};
    \fill[white] svg{M86.3,186.2H70.9V79.1h15.4v48.4V186.2z}
                 svg{M108.9,79.1h41.6c39.6,0,57,28.3,57,53.6c0,27.5-21.5,53.6-56.8,53.6h-41.8V79.1z M124.3,172.4h24.5c34.9,0,42.9-26.5,42.9-39.7c0-21.5-13.7-39.7-43.7-39.7h-23.7V172.4z}
                 svg{M88.7,56.8c0,5.5-4.5,10.1-10.1,10.1c-5.6,0-10.1-4.6-10.1-10.1c0-5.6,4.5-10.1,10.1-10.1C84.2,46.7,88.7,51.3,88.7,56.8z};
  }
}

\newcommand\orcidicon[1]{\href{https://orcid.org/#1}{\mbox{\scalerel*{
\begin{tikzpicture}[yscale=-1,transform shape]
\pic{orcidlogo};
\end{tikzpicture}
}{|}}}}

\begin{document}


\title{Updated observational constraints on spatially-flat and non-flat $\Lambda$CDM and XCDM cosmological models}

\author{Javier de Cruz P\'erez$^{\orcidicon{0000-0001-8603-5447}}$}%
 \email{jadecruz@ucm.es}
\affiliation{Departamento de F\'isica Te\'orica, Universidad Complutense de Madrid, 28040, Madrid, Spain}%

\author{Chan-Gyung Park$^{\orcidicon{0000-0002-3076-2781}}$}%
 \email{park.chan.gyung@gmail.com (corresponding author)}
\affiliation{Division of Science Education and Institute of Fusion Science, Jeonbuk National University, Jeonju 54896, Republic of Korea}%

\author{Bharat Ratra$^{\orcidicon{0000-0002-7307-0726}}$}%
 \email{ratra@phys.ksu.edu}
\affiliation{Department of Physics, Kansas State University, 116 Cardwell Hall, Manhattan, KS 66506, USA}%

\date{\today}

\begin{abstract}
We study the performance of six $\Lambda$CDM models, with four of them allowing for non-flat spatial hypersurfaces (non-zero current value of the spatial curvature density parameter $\Omega_k$) and three of them allowing for a non-unity value of the lensing consistency parameter $A_L$. We also study a set of six XCDM models where the non-evolving cosmological constant $\Lambda$ dark energy density is replaced by a dynamical dark energy density X-fluid parameterized by a non-evolving equation of state parameter $w$. For the non-flat models we consider two different primordial power spectra, Planck $P(q)$, used by the Planck collaboration, and new $P(q)$, resulting from quantum fluctuations in a not-necessarily-very-slow-roll non-flat inflation model. These models are constrained by and tested against: Planck 2018 CMB temperature and polarization power spectra data (P18); Planck 2018 CMB lensing potential power spectrum data (lensing); and, an updated compilation of baryon acoustic oscillation, type Ia supernova, Hubble parameter [$H(z)$], and growth factor [$f\sigma_8$] data points [collectively denoted by non-CMB (new) data], individually and jointly. P18 data favor $\Omega_k<0$ (closed spatial geometry) for the $\Lambda$CDM and XCDM models and $w<-1$ (phantom-like dynamical dark energy) for the XCDM models while non-CMB (new) data favor $\Omega_k>0$ (open geometry) in the case of the $\Lambda$CDM models and $\Omega_k<0$ (closed geometry) and $w>-1$ (quintessence-like dynamical dark energy) for the XCDM models. When P18 and non-CMB (new) data are jointly analyzed there is weak evidence in favor of open spatial geometry and moderate evidence in favor of quintessence-like dynamical dark energy. On the other hand, regardless of data considered, $A_L>1$ is always favored, with different degrees of evidence, even for P18+lensing+non-CMB (new) data. According to Akaike and deviance information criterion results, $A_L$-varying models are positively favored over the flat $\Lambda$CDM model for P18+lensing+non-CMB (new) data. The XCDM model cosmological parameter constraints obtained from P18 or P18+lensing data and from non-CMB (new) data are incompatible at $> 3\sigma$, ruling out the three $A_L = 1$ XCDM models at $> 3\sigma$. In the nine models not ruled out by $> 3 \sigma$ incompatibilities between parameter values determined from different data sets, for the P18+lensing+non-CMB (new) data set we find little deviation from flat geometry and moderate deviation from a cosmological constant. In all six non-flat models that are not ruled out at $> 3 \sigma$, open geometry is mildly favored (by at most $0.8 \sigma$), and in all three XCDM+$A_L$ models (that are not ruled out at $> 3 \sigma$) quintessence-like dynamical dark energy is moderately favored (by at most $1.6 \sigma)$. In the $A_L = 1$ non-flat $\Lambda$CDM cases, we find for P18+lensing+non-CMB (new) data $\Omega_k = 0.0009 \pm 0.0017$ [$0.0008 \pm 0.0017$] for the Planck [new] $P(q)$ model, favoring open geometry at 0.53$\sigma$ [0.47$\sigma$]. Given these results, the flat $\Lambda$CDM model remains the simplest (largely) observationally-consistent cosmological model. Our cosmological parameter constraints obtained for the flat $\Lambda$CDM model (and other models), when P18+lensing+non-CMB (new) data are considered, are the most restrictive results to date.
\end{abstract}

\maketitle

\maxdeadcycles=200

\section{Introduction}
\label{sec:Introduction} 

The six-parameter spatially-flat $\Lambda$CDM model, \cite{Peebles:1984ge}, built within the framework of general relativity, is the simplest observationally-consistent cosmological model, and is now commonly recognized as the standard model of cosmology. As dominant low-redshift stress-energy building blocks, the model uses a non-evolving cosmological constant $\Lambda$ dark energy density and a pressure-less cold dark matter (CDM) component and assumes flat spatial hypersurfaces. From the perspective of general relativity, the observed currently accelerated cosmological expansion of the Universe is caused by gravity sourced by the currently dominant cosmological constant.

The flat $\Lambda$CDM model is largely observationally consistent, however some recent data hint at potential discrepancies, such as differences in the values of the Hubble constant, $H_0$, and the amplitude of matter fluctuations, $\sigma_8$, measured using different techniques \cite{DiValentino:2020zio, DiValentino:2020vvd, Perivolaropoulos:2021jda, Moresco:2022phi, Abdalla:2022yfr, Hu:2023jqc}, or some anomalies that appear when we compare theoretical predictions of the model, based on best-fit cosmological model parameter values, with actual observations. These potential discrepancies motivate studying extensions of the flat $\Lambda$CDM model.  

Cosmic microwave background (CMB) radiation anisotropy measurements have so far generally provided the most restrictive constraints on cosmological parameters and these data have been used to test extensions of the $\Lambda$CDM model with the aim of studying the issues mentioned above. In particular, the Planck 2018 TT,TE,EE+lowE (hereafter denoted by P18) data set \cite{Planck:2018vyg} has been used to test three single parameter extensions of the standard model: what we call the $\Lambda$CDM Planck $P(q)$ model which makes use of a particular expression for $P(q)$, the primordial power spectrum of matter inhomogeneities in non-flat models that allow for a non-zero $\Omega_k$ spatial curvature density parameter (see Sec.\ \ref{sec:Methods} below for details); the flat XCDM parameterization which assumes a time-evolving dark energy density $\rho_{\text{DE}}\sim a^{-3(1+w)}$ with $a$ being the scale factor and $w$ the constant equation of state parameter of the dynamical dark energy fluid; and, the flat $\Lambda$CDM+$A_L$ model where the amplitude of the gravitational potential power spectrum is rescaled by the phenomenological lensing consistency parameter $A_L$, \cite{Calabreseetal2008}, in such a way that $A_L=1$ corresponds to recovering the theoretically predicted (using the best-fit cosmological parameter values) amount of weak lensing of the CMB anisotropy. When analyzing P18 data, the results for the new non-standard parameter in these three extensions of the flat $\Lambda$CDM model are: $\Omega_k = -0.044^{+0.018}_{-0.015}$ which represents 2.44$\sigma$ evidence in favor of non-flat (closed) spatial hypersurfaces; $w=-1.58^{+0.16}_{-0.35}$ indicating a 3.63$\sigma$ preference for phantom-like behavior of the dynamical dark energy component; and, $A_L=1.180\pm 0.065$ which says $A_L>1$ is preferred over $A_L =1$ at 2.77$\sigma$. Below, using statistical criteria, we will compare the performance of the standard flat $\Lambda$CDM model to these three extensions when it comes to fitting P18 data. It turns out that, to varying degrees of significance, all three extensions are favored over the standard model. 

We showed in \cite{deCruzPerez:2022hfr}, that for P18 data the $\Omega_k$-varying and $A_L$-varying one-parameter $\Lambda$CDM extension models can handle the so-called lensing anomaly, which is related to the amount of weak gravitational lensing in the CMB power spectra. The trajectory of the CMB photons on their way to us are bent due to the gravitational effects produced by the inhomogeneous mass distribution. This effect is commonly referred to as weak gravitational lensing. When we compute the CMB temperature and polarization spectra we must account for this weak lensing effect, getting as a result the lensed CMB spectra. In order to provide a theoretical prediction for the amount of lensing, given a cosmological model, one must assume values for the cosmological parameters, \cite{Lewis:2006fu}. Constraining the flat $\Lambda$CDM model by using P18 CMB data and then using the obtained cosmological parameter values to predict the amount of weak lensing expected in the CMB spectra, one finds a mismatch with the observed CMB power spectra over a small range of multipoles, \cite{Calabreseetal2008, Planck:2018vyg}. Due to the tight constraints provided by Planck CMB data, there seems to be no room in the flat $\Lambda$CDM model to alleviate this anomaly, and consequently alternate models that introduce one or more additional parameters are considered. 

We note that a recent analysis of the updated PR4 Planck data set, \cite{Tristram:2023haj}, results in updated values, $\Omega_k=-0.012\pm 0.010$, $1.2\sigma$ in favor of closed geometry, and $A_L=1.039\pm 0.052$, $0.75\sigma$ in favor of $A_L > 1$. These new measurements are more consistent with the flat $\Lambda$CDM model values and show less evidence in favor of non-flat spatial hypersurfaces and $A_L>1$ than do the P18 data set results, partly because of updated PR4 Planck data and partly because of the different likelihoods used in the new analysis.

In this work, in addition to considering $\Omega_k$ and $A_L$ as additional fitting parameters (see \cite{deCruzPerez:2022hfr} for a detailed study, some results of which we update in the present paper where we use updated data), we also study the possibility of having a time-evolving dark energy density where the equation of state parameter $w$ is constant but allowed to vary from the cosmological constant value of $w = -1$. Here $w$ is the ratio of the pressure to the energy density of the dynamical dark energy X-fluid and we refer to this as the XCDM parametrization or model. Generally, the dark energy fluid can be characterized by the equation of state parameter $w$ and the square of the sound speed $c^2_s \equiv \delta p /\delta \rho$, which is defined in terms of the density and pressure perturbations of the dark energy fluid and is gauge dependent. With $c^2_s$ representing the square of the sound speed in the fluid's rest frame, in the scalar field based dark energy models with a standard kinetic term and no potential energy density term, $c^2_s=1$. Although $c_s^2$ can generally have a wide range of values, it is usually bounded within the range $0 \le c^2_s \le 1$. The reason for this limitation is that values below 0 result in exponentially growing dark energy fluctuations and values above 1 lead to superluminal motion while values in this range lead to oscillatory perturbations (see \cite{Hannestad:2005ak, Fang:2008sn, Weller:2003hw, Ballesteros:2010ks} and references therein for more discussions of the sound speed). We emphasize that the speed of sound squared is arbitrarily set to $c^2_s=1$ in the XCDM parameterization, which is the simplest way to remove the $c^2_s < 0$ instability.
The XCDM parameterization is not a physical model but it has been widely used to analyze data and this is why we study it in detail in this paper. For recent discussions of observational constraints on the XCDM model see \cite{Ooba:2017npx, Ooba:2018dzf, Park:2018bwy, SolaPeracaula:2018wwm, Park:2019emi, Khadka:2020whe, Khadka:2020vlh, Khadka:2020hvb, Cao:2020evz, Cao:2022yvi, Khadka:2022ooh, Cao:2022pdv, Dong:2023jtk} and references therein. In this paper we present results from the most complete analysis of the XCDM models to date.

In physical dynamical dark energy models the dark energy component can be modelled as a dynamical scalar field $\phi$ with its associated potential energy density $V(\phi)$, \cite{Peebles:1987ek, Ratra:1987rm}. For an appropriate $V(\phi)$ the scalar field energy density $\rho_\phi$ evolves slowly over time until it overcomes other contributions to the cosmological energy budget and becomes the dominant component giving rise to the observed currently accelerated cosmological expansion of the Universe. 

In this paper we study twelve cosmological models (six of them are $\Lambda$CDM models, which we previously studied in \cite{deCruzPerez:2022hfr} and present updated results for here, and the other six are XCDM models), namely: the flat $\Lambda$CDM (+$A_L$), the non-flat $\Lambda$CDM Planck $P(q)$ (+$A_L$), the non-flat $\Lambda$CDM new $P(q)$ (+$A_L$), the flat XCDM (+$A_L$), the non-flat XCDM Planck $P(q)$ (+$A_L$), and the non-flat XCDM new $P(q)$ (+$A_L$) model. For the non-flat models we consider two different primordial power spectra, Planck $P(q)$ and new $P(q)$, the details of which are given in Sec.\ \ref{sec:Methods}. For recent discussions of observational constraints on spatial curvature see \cite{Ooba:2017lng, DiValentino:2020hov, Vagnozzi:2020rcz,Vagnozzi:2020dfn, Khadka:2020tlm, Cao:2021ldv, Arjona:2021hmg, Dhawan:2021mel, Khadka:2021vqa, Khadka:2021ukv, Khadka:2021xcc, Cao:2021irf, Khadka:2021sxe, Renzi:2021xii, Geng:2021hqc, Cao:2022wlg, Mukherjee:2022ujw, Glanville:2022xes, Wu:2022fmr, Dahiya:2022avg, Stevens:2022evv, Khadka:2022aeg, Favale:2023lnp, Qi:2023oxv, Zajacek:2023qjm, Cao:2023fpp, Cao:2023daj, Cao:2024vmo, Amendola:2024gkz} and references therein, and see Refs.\ \cite{Baumgartner:2022jdz,Anselmi:2022uvj,Jimenez:2022asc} for recent, more general discussion of non-flat cosmological models. In this paper we present the most restrictive constraints on spatial curvature (in $\Lambda$CDM and XCDM models) to date. We also use the updated data compilation here to constrain the spatially-flat $w_0w_a$CDM parameterization in \cite{Park:2024jns}.

We use combinations of data to place constraints on the cosmological parameters of each model and in particular we want to measure the values of $\Omega_k$, $w$, and $A_L$. We also want to constrain the other six primary parameters that all these models share (the conventional six parameters of the flat $\Lambda$CDM model) as well as constrain the derived parameters $H_0$, $\Omega_m$ (the current value of the non-relativistic matter density parameter), and $\sigma_8$. We are also interested in determining which parameters are measured in a cosmological model independent manner from these data, amongst the twelve models we study.  

The different data sets we use in this work are P18 data, Planck 2018 CMB weak lensing data, and non-CMB data,  consisting of baryon acoustic oscillation (BAO) measurements, type Ia supernova (SNIa) data, Hubble parameter [$H(z)$] data points, and a collection of $f\sigma_8$ growth factor measurements. We consider two different non-CMB data compilations, the non-CMB (old) data set we used in \cite{deCruzPerez:2022hfr}, and an updated non-CMB (new) data set, updated with respect to the non-CMB (old) data set. See Sec.\ \ref{sec:Data} for details. Adding P18 lensing data, and especially adding non-CMB data to the mix, alter the conclusions mentioned above that are based on just P18 data.  

In the following we briefly summarize the main results obtained in this work. Assuming that the data sets we use are correct and that there are no unaccounted systematics, the three XCDM models with $A_L=1$ are ruled out at $> 3\sigma$ due to incompatibilities between P18 data and non-CMB (new) data cosmological parameter constraints. Extending these three models by adding a varying lensing consistency parameter, $A_L$, reduces the incompatibilities between P18 data and non-CMB (new) data constraints, thus allowing for joint analyzes of P18 and non-CMB (new) data in the context of these models. In these models P18 data favor $\Omega_k<0$ (closed geometry), $w<-1$ (phantom-like dynamical dark energy), and $A_L>1$, whereas when Planck CMB lensing data are added to the mix the evidence in favor of $\Omega_k<0$ and $A_L>1$ decreases but that in favor of $w<-1$ is barely affected. When non-CMB (new) data are included in the analysis the conclusions obtained with P18 and P18+lensing data change and the evidence previously favoring $\Omega_k<0$ and $w<-1$ subsides, giving rise to a preference for $\Omega_k>0$ (open geometry) and $w>-1$ (quintessence-like dynamical dark energy). 

Considering only the nine models not ruled out by $> 3 \sigma$ incompatibilities between parameter values determined from different data sets, for the P18+lensing+non-CMB (new) data set we find little deviation from a flat geometry and moderate deviation from a cosmological constant, with the biggest deviations being $\Omega_k = 0.0015 \pm 0.0019$ in the XCDM Planck and new $P(q)+A_L$ models, which favor open geometry and are 0.79$\sigma$ from flat geometry, and $w = -0.958\pm 0.026$ in the XCDM Planck $P(q)+A_L$ model, which favors quintessence-like dark energy and is 1.62$\sigma$ from a  cosmological constant. In all six non-flat models that are not ruled out at $> 3\sigma$, open geometry is mildly favored, and in all three XCDM+$A_L$ models (that are not ruled out at $> 3 \sigma$), quintessence-like dark energy is moderately favored. In the $A_L = 1$ non-flat $\Lambda$CDM cases, we find for P18+lensing+non-CMB (new) data $\Omega_k = 0.0009 \pm 0.0017$ [$0.0008 \pm 0.0017$] for the Planck [new] $P(q)$ model, favoring open geometry at 0.53$\sigma$ [0.47$\sigma$]. Given these results, the flat $\Lambda$CDM model remains the simplest (largely) observationally-consistent cosmological model.   

Our cosmological parameter constraints, when P18+lensing+non-CMB (new) data are considered, are the most restrictive results to date. In particular, for the six primary parameters in the flat $\Lambda$CDM model we get for the current value of the physical baryonic matter density parameter $\Omega_b h^2 = 0.02249\pm 0.00013$, for the current value of the physical cold dark matter density parameter $\Omega_c h^2 = 0.11849 \pm 0.00084$, for the angular size of the sound horizon at recombination $100\theta_{\text{MC}}=1.04109\pm 0.00028$, for the reionization optical depth $\tau = 0.0569\pm 0.0071$, for the primordial scalar-type perturbation power spectral index $n_s = 0.9685\pm 0.0036$, and for the power spectrum amplitude $\ln(10^{10}A_s)=3.046\pm 0.014$, where $h$ is the Hubble constant in units of 100 km s$^{-1}$ Mpc$^{-1}$. Additionally, for the derived parameters, we find $H_0=68.05\pm 0.38$ km s$^{-1}$ Mpc$^{-1}$, $\Omega_m = 0.3059\pm 0.0050$, and $\sigma_8 = 0.8077\pm 0.0057$. Among models with $A_L = 1$, these values show almost model-independent consistency, with differences always below $1\sigma$. However, when we compare these cosmological parameter values with those obtained for the $A_L$-varying models, we observe larger differences. In particular, for the six varying $A_L$ models relative to the flat $\Lambda$CDM model we find the maximum differences for $\sigma_8$, $1.08\sigma$ for models with $w=-1$ and $1.80\sigma$ when comparing to varying-$w$ models, with all other parameters agreeing to better than 1$\sigma$. As in our previous work \cite{deCruzPerez:2022hfr} we once again find that the $A_L$-varying models are the most favored by the most complete data set considered in this work, namely the P18+lensing+non-CMB (new) data set. 

The outline of our paper is as follows. In Sec.\ \ref{sec:Data} we provide details of observational data we use to constrain cosmological parameters in, and to test the performance of, the cosmological models we study. In Sec.\ \ref{sec:Methods} we briefly describe the main features of the models studied, as well as the methods employed for the analyses. Section \ref{sec:Results}, which represents the main part of the article, is dedicated to presenting and commenting in detail on all the results obtained in our analyses. In particular, we discuss the cosmological parameter constraints obtained, compare the performance of the different models under study, and analyze whether there are tensions among the cosmological parameter constraints derived from different data. In Sec.\ \ref{sec:Discussion} we summarize the most significant results obtained in the previous section, and, finally, in Sec.\ \ref{sec:Conclusions} we present our conclusions.

\section{Data}
\label{sec:Data}

The data we use in this work are the Planck cosmic microwave background radiation temperature and polarization anisotropy power spectra and the lensing potential power spectrum, the Pantheon+ type Ia supernovae compilation, and baryon acoustic oscillation, Hubble parameter, and growth rate measurements.

\subsection{Planck 2018 CMB data}

We use the Planck 2018 TT,TE,EE+lowE (P18) CMB temperature and polarization power spectra as well as the Planck lensing potential power spectrum \cite{Planck:2018vyg}. Here TT, TE, and EE denote the temperature-only power spectra at low multipole number $\ell$ ($2 \le \ell \le 29$) and high $\ell$ ($30 \le \ell \le 2508$), TE cross-power spectrum, and $E$-mode polarization power spectrum at high $\ell$ ($30 \le \ell \le 1996$), respectively, while lowE denotes the $E$-mode polarization power spectrum at low $l$ ($2 \le \ell \le 29$). For high-$\ell$ P18 data we use the Planck 2018 baseline Plik likelihood (see Sec.\ 2.2.1 of  \cite{Planck:2018vyg}). We can more restrictively constrain cosmological parameters by adding the power spectrum of the lensing potential measured by Planck \cite{Planck:2018lbu} to the P18 data. In the following we refer to the P18 plus lensing data combination as P18+lensing data. 

\subsection{Non-CMB data}
\label{sec:Non-CMB Data}

In addition to CMB data, we collect and use a number of non-CMB data sets to constrain model parameters. We denote the non-CMB data compilation used in our previous work \cite{deCruzPerez:2022hfr} as non-CMB (old), and the new non-CMB data compilation assembled and used here, with updates described below, as non-CMB (new).

Compared to our earlier work \cite{deCruzPerez:2022hfr}, we replace the 1048 Pantheon SNIa \cite{Pan-STARRS1:2017jku} and the binned DES 3yr SNIa data points \cite{DES:2018paw} with a subset (as discussed below) of the new 1701 Pantheon+ compilation data points \cite{Brout:2022vxf} that also include DES SNIa measurements. We replace the BAO data point $D_A (z = 0.81) / r_d = 10.75 \pm 0.43$ from \cite{DES:2017rfo} with $D_M (z = 0.835) / r_d = 18.92 \pm 0.51$ from \cite{DES:2021wwk}; the distances $D_A$ and $D_M$ are defined below. In the growth rate data we now add the data point $f\sigma_8 (z = 0.013) = 0.46 \pm 0.06$ from \cite{Avila:2021dqv}. In the Hubble parameter data we now include the data point $H(z=0.75) = 98.8 \pm 33.6~\textrm{km} \textrm{s}^{-1} \textrm{Mpc}^{-1}$ from \cite{Borghi:2021rft} and for some of the $H(z)$ data points we now account for a non-diagonal covariance matrix as explained below.

\subsubsection{BAO data}
\label{sec:BAO data}

We use a collection of the latest BAO data points measured at various redshifts. Table \ref{tab:bao} lists the data sets, effective redshifts, observables, and measurement values for the 16 BAO data points we use. All BAO data we use account for all known systematic errors. The six BOSS Galaxy, three eBOSS LRG, three eBOSS Quasar, and two Ly$\alpha$-forest BAO measurements are correlated and their covariance matrices are given below. We do not use the ELGs data from \cite{deMattia:2020fkb,Tamone:2020qrl} because the corresponding posterior distribution is highly non-Gaussian and the full likelihood (not a summary data point) must be used in this case.

The BAO quantities listed in Table \ref{tab:bao} correspond to several distances ($D_V$, $D_M$, and $D_H$) and to the growth rate $f\sigma_8$. The angle-averaged distance is $D_V(z)=\left[ cz D_M^2 (z) / H(z) \right]^{1/3}$, where $H(z)$ is the Hubble parameter at redshift $z$ and $D_M(z)$ is the transverse comoving distance. Using the proper angular diameter distance
\begin{equation}
    D_A (z)=\frac{c}{H_0} \frac{f_k \left[ H_0 \sqrt{|\Omega_k|} \int_0^z \frac{dz'}{H(z')} \right]}{(1+z) \sqrt{|\Omega_k|}},
\end{equation}
where $f_k[x]=\sin x$, $x$, and $\sinh x$ for closed ($k=1$; $\Omega_k <0$), flat ($k=0$; $\Omega_k=0$), and open ($k=-1$, $\Omega_k >0$) Universes, respectively, $D_M(z)$ is expressed as $D_M(z)=(1+z) D_A(z)$. $D_H(z)=c/H(z)$ is the Hubble distance at redshift $z$. In Table \ref{tab:bao} BAO distances are divided by the radius of sound horizon $r_d$ at the drag epoch $z_d$,
\begin{equation}
    r_d=\int_{z_d}^{\infty} \frac{c_s (z)}{H(z)} dz,
\end{equation}
where $c_s(z)$ is the sound speed of the photon-baryon fluid.

\begin{table*}
\caption{BAO measurements.}
\begin{ruledtabular}
\begin{tabular}{lcccc}
  Data Set        & $z_\textrm{eff}$  &  Observable                          &  Measurement        &   Reference    \\[+0mm]
 \hline \\[-2mm]
  6dFGS+SDSS MGS  & $0.122$    & $D_V (r_{d,\textrm{fid}} / r_d)$ [Mpc]   & $539 \pm 17$ [Mpc]     &  \cite{Carter:2018vce} \\[+1mm]
 \hline \\[-2mm]
  BOSS Galaxy     & $0.38$     & $D_M / r_d$                              & $10.274 \pm 0.151$     &  \cite{Gil-Marin:2020bct} \\[+1mm]
                  & $0.38$     & $D_H / r_d$                              & $24.888 \pm 0.582$     &  \cite{Gil-Marin:2020bct} \\[+1mm]
                  & $0.38$     & $f\sigma_8$                              & $0.49729 \pm 0.04508$  &  \cite{Gil-Marin:2020bct} \\[+1mm]
                  & $0.51$     & $D_M / r_d$                              & $13.381 \pm 0.179$     &  \cite{Gil-Marin:2020bct} \\[+1mm]
                  & $0.51$     & $D_H / r_d$                              & $22.429 \pm 0.482$     &  \cite{Gil-Marin:2020bct} \\[+1mm]
                  & $0.51$     & $f\sigma_8$                              & $0.45902 \pm 0.03784$  &  \cite{Gil-Marin:2020bct} \\[+1mm]
 \hline \\[-2mm]
  eBOSS LRG       & $0.698$    & $D_M / r_d$                              & $17.646 \pm 0.302$     & \cite{Gil-Marin:2020bct,Bautista:2020ahg} \\[+1mm]
                  & $0.698$    & $D_H / r_d$                              & $19.770 \pm 0.469$     & \cite{Gil-Marin:2020bct,Bautista:2020ahg} \\[+1mm]
                  & $0.698$    & $f\sigma_8$                              & $0.47300 \pm 0.04429$  & \cite{Gil-Marin:2020bct,Bautista:2020ahg} \\[+1mm]
   \hline \\[-2mm]
  DES Y3           & $0.835$    & $D_M / r_d$                              & $18.92 \pm 0.51$       & \cite{DES:2021wwk} \\[+1mm]
   \hline \\[-2mm]
  eBOSS Quasar    & $1.48$    & $D_M / r_d$                              & $30.21 \pm 0.79$     & \cite{Hou:2020rse,Neveux:2020voa} \\[+1mm]
                  & $1.48$    & $D_H / r_d$                              & $13.23 \pm 0.47$     & \cite{Hou:2020rse,Neveux:2020voa} \\[+1mm]
                  & $1.48$    & $f\sigma_8$                              & $0.462 \pm 0.045$  & \cite{Hou:2020rse,Neveux:2020voa} \\[+1mm]
   \hline \\[-2mm]   
Ly$\alpha$-forest & $2.334$    & $D_M / r_d$                              & $37.5_{-1.1}^{+1.2}$   & \cite{duMasdesBourboux:2020pck} \\[+1mm]
		  & $2.334$    & $D_H / r_d$                              & $8.99_{-0.19}^{+0.20}$ & \cite{duMasdesBourboux:2020pck} \\[+0mm]
\end{tabular}
\\[+1mm]
Note: The sound horizon size of the fiducial model is $r_{d,\textrm{fid}}=147.5~\textrm{Mpc}$ in \cite{Carter:2018vce}.
\end{ruledtabular}
\label{tab:bao}
\end{table*}

\begin{widetext}
The covariance matrix between measurement errors for BOSS Galaxy data is
\begin{equation}
   \mathbf{C}_\textrm{BOSS-Galaxy}=
   \begin{pmatrix}
      0.022897    & -0.02007    & 0.0026481   &  0.013487  & -0.0081402 & 0.0010292  \\
      -0.02007    & 0.33849     & -0.0085213  & -0.016024  &  0.13652   & -0.0038002 \\
      0.0026481   & -0.0085213  & 0.0020319   & 0.001325   & -0.0023012 & 0.000814158 \\
      0.013487    & -0.016024   & 0.001325    & 0.032158   & -0.020091  & 0.0026409 \\
      -0.0081402  & 0.13652     & -0.0023012  & -0.020091  &  0.23192   & -0.0055377 \\
      0.0010292   & -0.0038002  & 0.000814158 & 0.0026409  & -0.0055377 & 0.0014322
   \end{pmatrix} .
\end{equation}
\end{widetext}
The covariance matrix for eBOSS LRG data is
\begin{equation}
   \mathbf{C}_\textrm{eBOSS-LRG}=
   \begin{pmatrix}
       0.09114 & -0.033789 & 0.0024686 \\
       -0.033789 & 0.22009 & -0.0036088 \\
    0.0024686 & -0.0036088 & 0.0019616
\end{pmatrix}.
\end{equation}
The covariance matrix for eBOSS Quasar data is 
\begin{equation}
   \mathbf{C}_\textrm{eBOSS-Quasar}=
  \begin{pmatrix}
    0.6227 & 0.01424 & 0.02257 \\
     0.01424 & 0.2195 & -0.007315 \\
     0.02257 & -0.007315 & 0.002020
\end{pmatrix}.
\end{equation}
The covariance matrix for Ly-$\alpha$ forest data is
\begin{equation}
   \mathbf{C}_{\textrm{Ly}\alpha}=
   \begin{pmatrix}
      1.3225   & -0.1009     \\
      -0.1009  &  0.0380 
   \end{pmatrix} .
\end{equation}

\subsubsection{SNIa data}
\label{sec:SNIa data}

For SNIa data, we use a subset of the new 1701 data point Pantheon+ compilation \cite{Brout:2022vxf}, which is determined by removing all SNIa with $z < 0.01$ since these data points are model dependent due to the peculiar velocities that have to be considered. This cut leaves us with 1590 data points, spanning the redshift range $0.01016 \le z \le 2.26137$. The covariance matrix includes not only the statistical errors but also the systematic ones. The data set, likelihoods, and all necessary information can be found at \url{https://github.com/PantheonPlusSH0ES}. In this work we use the $\chi^2$ of the SNIa data set where the absolute magnitude of the SN is marginalized over, which differs from the $\chi^2$ value obtained from the analysis where the SN absolute magnitude is not marginalized over.

\subsubsection{$H(z)$ data}
\label{sec:H(z) data}

\begin{table}
\caption{$H(z)$ measurements.}
\begin{ruledtabular}
\begin{tabular}{lccc}
  $z$     & $H(z)$                     &    Reference  \\[+0mm]
          & (km s$^{-1}$ Mpc$^{-1}$)   &               \\[+0mm]
  \hline \\[-2mm]
$0.07$    & $69.0 \pm 19.6$    & \cite{Zhang:2012mp}  \\ 
$0.09$    & $69.0 \pm 12.0$    & \cite{Simon:2004tf}  \\
$0.12$    & $68.6 \pm 26.2$    & \cite{Zhang:2012mp}  \\
$0.17$    & $83.0 \pm 8.0$     & \cite{Simon:2004tf}  \\
$0.2$     & $72.9 \pm 29.6$    & \cite{Zhang:2012mp}  \\
$0.27$    & $77.0 \pm 14.0$    & \cite{Simon:2004tf}  \\
$0.28$    & $88.8 \pm 36.6$    & \cite{Zhang:2012mp}  \\
$0.4$     & $95.0 \pm 17.0$    & \cite{Simon:2004tf}  \\
$0.47$    & $89.0 \pm 50.0$    & \cite{Ratsimbazafy:2017vga}  \\
$0.48$    & $97.0 \pm 62.0$    & \cite{Stern:2009ep}  \\
$0.75$    & $98.8 \pm 33.6$    & \cite{Borghi:2021rft}  \\
$0.88$    & $90.0 \pm 40.0$    & \cite{Stern:2009ep}  \\
$0.9$     & $117.0 \pm 23.0$   & \cite{Simon:2004tf}  \\
$1.3$     & $168.0 \pm 17.0$   & \cite{Simon:2004tf}  \\
$1.43$    & $177.0 \pm 18.0$   & \cite{Simon:2004tf}  \\
$1.53$    & $140.0 \pm 14.0$   & \cite{Simon:2004tf}  \\
$1.75$    & $202.0 \pm 40.0$   & \cite{Simon:2004tf}  \\
$0.1791$  & $74.91$            & \cite{Moresco:2020fbm}  \\
$0.1993$  & $74.96$            & \cite{Moresco:2020fbm}  \\
$0.3519$  & $82.78$            & \cite{Moresco:2020fbm}  \\
$0.3802$  & $83.0$             & \cite{Moresco:2020fbm}  \\
$0.4004$  & $76.97$            & \cite{Moresco:2020fbm}  \\
$0.4247$  & $87.08$            & \cite{Moresco:2020fbm}  \\
$0.4497$  & $92.78$            & \cite{Moresco:2020fbm}  \\
$0.4783$  & $80.91$            & \cite{Moresco:2020fbm}  \\
$0.5929$  & $103.8$            & \cite{Moresco:2020fbm}  \\
$0.6797$  & $91.6$             & \cite{Moresco:2020fbm}  \\
$0.7812$  & $104.5$            & \cite{Moresco:2020fbm}  \\
$0.8754$  & $125.1$            & \cite{Moresco:2020fbm}  \\
$1.037$   & $153.7$            & \cite{Moresco:2020fbm}  \\
$1.363$   & $160.0$            & \cite{Moresco:2020fbm}  \\
$1.965$   & $186.5$            & \cite{Moresco:2020fbm}
\end{tabular}
\end{ruledtabular}
\label{tab:hubble}
\end{table}

We use the 32 Hubble parameter measurements provided in Table 1 of \cite{Cao:2023eja}, and listed in Table \ref{tab:hubble} here, that cover the redshift range $0.070 \le z \le 1.965$. This new compilation includes an additional data point, $H(z = 0.75) = 98.8 \pm 33.6~\textrm{km} \textrm{s}^{-1} \textrm{Mpc}^{-1}$, compared to the data set used in our previous paper. We now also account for the correlation between the 15 measurements provided in \cite{Moresco:2012jh,Moresco:2015cya,Moresco:2016mzx}. The corresponding covariance matrix must be computed following the steps in the code at \url{https://gitlab.com/mmoresco/CCcovariance}.

\subsubsection{$f\sigma_8$ data}
\label{sec:fs8}

In addition to the growth rate data included in the BAO data compilation of Table \ref{tab:bao}, we use other growth rate measurements. These $f\sigma_8$ measures are obtained from peculiar velocity analysis \cite{Turnbull:2011ty,Hudson:2012gt,Said:2020epb} or redshift-space distortion analysis \cite{Shi:2017qpr,Simpson:2015yfa,Blake:2013nif,Mohammad:2018mdy,Okumura:2015lvp}, and are listed in Table \ref{tab:growth}, and include an additional data point, $f\sigma_8 (z = 0.013) = 0.46 \pm 0.06$, compared to the data set used in our previous paper.

\begin{table*}
\caption{$f\sigma_8$ measurements.}
\begin{ruledtabular}
\begin{tabular}{lccc}
  Survey          & $z$        & Measurement          &  Reference    \\[+0mm]
 \hline \\[-2mm]
  ALFALFA         & $0.013$    & $0.46 \pm 0.06$      &  \cite{Avila:2021dqv} \\[+1mm]
  IRAS            & $0.02$     & $0.398 \pm 0.065$    &  \cite{Turnbull:2011ty,Hudson:2012gt} \\[+1mm]
  6dFGS+SDSS      & $0.035$    & $0.338 \pm 0.027$    &  \cite{Said:2020epb} \\[+1mm]
  SDSS DR7        & $0.1$      & $0.376 \pm 0.038$    &  \cite{Shi:2017qpr} \\[+1mm]
  eBOSS LRG       & $0.18$     & $0.29 \pm 0.10$      &  \cite{Simpson:2015yfa,Blake:2013nif} \\[+1mm]
  eBOSS LRG       & $0.38$     & $0.44 \pm 0.06$      &  \cite{Simpson:2015yfa,Blake:2013nif} \\[+1mm]
  VIPERS          & $0.6$      & $0.49 \pm 0.12$      &  \cite{Mohammad:2018mdy} \\[+1mm]
  VIPERS          & $0.86$     & $0.46 \pm 0.09$      &  \cite{Mohammad:2018mdy} \\[+1mm]
  FastSound       & $1.36$     & $0.482 \pm 0.116$    &  \cite{Okumura:2015lvp} \\[+1mm]
\end{tabular}
\\[+1mm]
\end{ruledtabular}
\label{tab:growth}
\end{table*}

\section{Methods}\label{sec:Methods}

We use the Markov chain Monte Carlo (MCMC) method to determine the range of $\Lambda$CDM and XCDM model parameters that are favored by the measurements we consider. More precisely, we use the \texttt{CAMB}/\texttt{COSMOMC} program (October 2018 version) \cite{Challinor:1998xk,Lewis:1999bs,Lewis:2002ah} for this purpose. \texttt{CAMB} computes the evolution of spatially inhomogenous perturbations of radiation and matter densities and predicts the power spectra of matter and CMB  anisotropy perturbations as functions of cosmological parameter values, while \texttt{COSMOMC} compares the Planck CMB likelihood data and the non-CMB data sets with these model predictions to find the likelihood distribution of cosmological model parameters. 

When P18 data are used, the cosmological parameters that are measured, that are common across all cosmological models we study, are the current baryon density ($\Omega_b h^2$), the current cold dark matter density ($\Omega_c h^2$), the angular size of the sound horizon at recombination defined in \texttt{CAMB}/\texttt{COSMOMC} ($\theta_{\textrm{MC}}$), the reionization optical depth ($\tau$), and the amplitude ($A_s$) and the spectral index ($n_s$) of the primordial scalar-type perturbation spectrum, with the addition of the current curvature density parameter ($\Omega_k$) for non-flat cosmological models and the dark energy equation of state parameter $w$ for XCDM models. Here $\Omega_b$ and $\Omega_c$ are the current values of the baryon and cold dark matter density parameters, respectively, and $h$ is the Hubble constant in units of 100 km s$^{-1}$ Mpc$^{-1}$. The gravitational lensing consistency parameter $A_L$ is set to unity in some cases, but in some cases $A_L$ is a free parameter to be determined from data. For a detailed definition of the parameters of cosmological models, see the Planck team's paper on cosmological parameter estimation \cite{Planck:2018vyg}. In our paper we consider only the primary parameters that define the cosmological model, and use the Planck team's settings for \texttt{COSMOMC}'s internal calibration or nuisance parameters (e.g., calPlanck). For example, in the flat $\Lambda$CDM model, the primary cosmological parameters are $(\Omega_b h^2$, $\Omega_c h^2$, $100 \theta_{\textrm{MC}}$, $\tau$, $\ln(10^{10} A_s)$, $n_s$). In this paper, when we use P18 data to measure cosmological parameters we assume flat priors for these parameters, non-zero over: $0.005 \le \Omega_b h^2 \le 0.1$, $0.001 \le \Omega_c h^2 \le 0.99$, $0.5 \le 100\theta_\textrm{MC} \le 10$, $0.01 \le \tau \le 0.8$, $0.8 \le n_s \le 1.2$ (for the Planck $P(q)$, see below), and $1.61 \le \ln(10^{10} A_s) \le 3.91$. We also adopt flat priors non-zero over $-0.5 \le \Omega_k \le 0.5$ for non-flat models, $-3.0 \le w \le 0.2$ for XCDM models, $0.8 \le n_s < 1$ for models with the new $P(q)$ (see below), and $0 \le A_L \le 10$ for $A_L$-varying models, except in the case of the P18 data constraints for the non-flat XCDM$+A_L$ models where $0.8 < A_L \le 10$. When estimating parameters from non-CMB data we fix the values of the optical depth $\tau$ and the spectral index $n_s$ to those obtained from P18 data (since non-CMB data are unable to constrain these parameters) and determine the remaining cosmological parameters. When showing the results of the parameter constraints, we also list the three derived parameters $H_0$, $\Omega_m$, and $\sigma_8$, which are obtained from the primary parameters of the cosmological model.

The MCMC chains are considered to have converged when the Gelman and Rubin $R$ statistic, provided by \texttt{COSMOMC}, satisfies the condition $R-1 < 0.01$. For each model and data set, we use the converged MCMC chains to compute the average values, confidence intervals, and likelihood distributions of model parameters. We utilize the \texttt{GetDist} code \cite{Lewis:2019xzd} for this purpose. As for the confidence intervals, we typically use the standard deviation estimated from the MCMC chains. However, for parameters of special interest to us ($H_0$, $\Omega_k$, $w$, and $A_L$) with highly asymmetric likelihood distributions or with bounds, we also compute and record (in parentheses in the tables) a second set of limits (see $\S$2.4 of \cite{Planck:2013pxb}). In the case of two tail limits we compute the second 68\% confidence interval as the one between the two points with equal marginalized probability density such that 68\% of the samples lie within the two points with 16\% of the sample in each tail. The second one tail limit we compute and record is a 95\% confidence limit with 5\% of the samples in the tail. We use these second limits in our discussions of parameters with highly asymmetric likelihood distributions or with bounds. When comparing the sizes of error bars of cosmological parameter values in different models or derived using different data, we always compare the standard deviation error bars. When we compute the increase or decrease in the size of the error bars between two different estimates of a cosmological parameter, we use the expression $[]\% = 100\left(1-\sigma_X/\sigma_Y\right)$ where $\sigma_X$ ($\sigma_Y$) is the error bar of the result obtained with either the smallest (largest) amount of data or with the smallest (largest) number of free parameters.

In our study, we use $\Lambda$CDM and XCDM cosmological models with different initial spectrum of scalar-type perturbations. For the flat tilted $\Lambda$CDM and XCDM models, the primordial scalar-type energy density perturbation power spectrum is given by
\begin{equation}
    P_\delta (k) = A_s \left( \frac{k}{k_0} \right)^{n_s},
\label{eq:powden-flat}
\end{equation}
where $k$ is the wavenumber and $n_s$ and $A_s$ are the spectral index and the amplitude of the spectrum at pivot scale $k_0=0.05~\textrm{Mpc}^{-1}$. This power spectrum arises from quantum fluctuations during an early epoch of power-law inflation with a scalar-field potential energy density of exponential form in a spatially flat cosmological model \cite{Lucchin:1984yf, Ratra:1989uv, Ratra:1989uz}.

For the tilted non-flat models, we use a primordial power spectrum of the form
\begin{equation}
    P_\delta (q) \propto \frac{(q^2-4K)^2}{q(q^2-K)} \left( \frac{k}{k_0} \right)^{n_s -1}, 
\label{eq:powden-nonflat-planck}
\end{equation}
where $q=\sqrt{k^2 + K}$ is the wavenumber in a model with spatial curvature $K = -(H^2_0/c^2)\Omega_k$. This spectrum is a phenomenological version that combines the untilted spectrum of primordial perturbations derived from very-slow-roll inflation in non-flat Universes \cite{RatraPeebles1995, Ratra:2017ezv} and a power-law spectrum to account for tilt. Recently a numerical study of quantum fluctuations in closed inflation models finds that it is possible to get a primordial power spectrum very close to this one in closed models for some initial conditions, \cite{Guth:2024xyz}. We will denote this spectrum as Planck $P(q)$ as it has been widely used in most studies of non-flat cosmological models, including by the Planck team. 

Finally we consider a third power spectrum from a not-necessarily-very-slow-roll non-flat inflation model
\begin{equation}
    P_\delta (q) \propto (q^2-4K)^2 \left|  P_{\mathcal{R}} (A) \right|, 
\label{eq:powden-nonflat-new}
\end{equation}
where $P_{\mathcal{R}} (A)$ is given in Eqs. (14) and (17) of \cite{deCruzPerez:2022hfr}, depending on whether the curvature of the Universe is positive or negative, respectively, \cite{Ratra:2022ksb}.

Equations (\ref{eq:powden-flat}), (\ref{eq:powden-nonflat-planck}), and (\ref{eq:powden-nonflat-new}) correspond to the initial density power spectra $P_\delta$'s that can be derived from the Poisson equation (see Eq.\ 3.43 of Ref.\ \cite{Lesgourgues:2013bra}) but as the \texttt{CAMB} input for the primordial power spectrum we need different forms of power spectra. The function \texttt{ScalarPower(k,ix)} of \texttt{CAMB} \texttt{power\_tilt.f90} computes the input primordial power spectrum (Eq.\ 3.26 of \cite{Lesgourgues:2013bra}), $\mathcal{P}_{\mathcal{R}} (k)=A_s (k/k_0)^{n_s-1}$, which is the appropriate form of the CAMB input primordial power spectrum for the tilted flat and the tilted non-flat model with the Planck $P(q)$. However, for the tilted non-flat model with the new $P(q)$, the primordial power spectrum should be replaced with the one derived from the tilted non-spatially-flat inflation model of \cite{Ratra:2022ksb}. There are slight differences in the notation between \cite{Ratra:2022ksb} and \cite{Lesgourgues:2013bra}. First, the primordial power spectrum in $q$-space $\tilde{\mathcal{P}}_{\mathcal{R}} (q)$, is related to $\mathcal{P}_{\mathcal{R}} (k)$ by $\tilde{\mathcal{P}}_{\mathcal{R}} (q) = \mathcal{P}_{\mathcal{R}} (k) q^2 / (q^2-K)$ (see Eq.\ 3.29 of \cite{Lesgourgues:2013bra}), which is equivalent to $\mathcal{P}_{\mathcal{R}} (A)$ of \cite{Ratra:2022ksb}. Besides, $\mathcal{P}_{\mathcal{R}} (A)=(A+1)^3 P_{\mathcal{R}} (A) / (2\pi^2)$ for the closed model and $\mathcal{P}_{\mathcal{R}} (A)=A^3 P_{\mathcal{R}} (A) / (2\pi^2)$ for the open model (see Eqs.\ B10 and B14 of \cite{Ratra:2022ksb}), where $\nu=q/\sqrt{|K|}=A+1$ and $\nu=A$ for the closed and open models, respectively, and $P_{\mathcal{R}} (A)$ is given in Eqs. (57) and (59) of \cite{Ratra:2022ksb} for closed and open models, respectively. For example, for the closed model, the input primordial spectrum for \texttt{CAMB} becomes $\mathcal{P}_{\mathcal{R}} (k) = (q^2-K)/q^2 \cdot (A+1)^3/(2\pi^2) P_{\mathcal{R}} (A)$. When applying the new primordial spectrum, the amplitude must be normalized so that the amplitude at the pivot scale $k_0$ is $A_s$. For the open model, we can also obtain a similar relation considering the negative sign of curvature $K$ and $\nu=A$.

We constrain the various dark energy models described above by using various combinations of observational data. We focus on how five combinations of data constrain each model: P18 data, P18+lensing data, non-CMB data, P18+non-CMB data, and P18+lensing+non-CMB data. We also investigated differences in cosmological constraints resulting from old and new versions of non-CMB data. We compare the performance of the models by examining how well each model fits different combinations of observations using the AIC and DIC information criteria. 

The Akaike information criterion (AIC) is defined as
\begin{equation}
\label{eq:AIC}    
\textrm{AIC} = \chi^2_{\textrm{min}} + 2n,
\end{equation}
where $\chi^2_{\textrm{min}}$ is the minimum value of $\chi^2$ for the best-fit cosmological parameters and $n$ is the number of independent cosmological parameters. The term $2n$ corresponds to a penalty to the goodness-of-fit for increasing the number of model parameters. The definition of AIC above is valid only for data sets with a large number of data points. All the data combinations we use here have a sufficiently large number of data points to justify using this AIC definition.   

To quantify how much a data set favors the model, we also use the deviance information criterion (DIC) defined as
\begin{equation}\label{eq:DIC}
    \textrm{DIC} = \chi^2(\hat{\theta}) + 2p_D,
\end{equation}
where $p_D= \overline{\chi^2} -  \chi^2(\hat{\theta})$ and the term $2 p_D$ penalizes the goodness-of-fit for increasing the number of model parameters. Here $\overline{\chi^2}$ denotes the average of $\chi^2$'s estimated from the MCMC chains and $\chi^2 (\hat{\theta})$ is the value of $\chi^2$ at the best-fit cosmological parameters $\hat{\theta}$.

We use the differences in the AIC and DIC values of the model under study, computed relative to the tilted flat $\Lambda$CDM model constrained using the same data set. (We also list similarly defined $\chi^2_{\textrm{min}}$ difference values in the Tables.) According to the usual scale, when $-2 \leq \Delta\textrm{AIC},\Delta\textrm{DIC}<0$ there is said to be {\it weak} evidence in favor of the model under study, while when $-6 \leq \Delta\textrm{AIC},\Delta\textrm{DIC} < -2$ there is said to be {\it positive} evidence, when $-10\leq\Delta\textrm{AIC},\Delta\textrm{DIC} < -6$ there is {\it strong} evidence,  and when $\Delta\textrm{AIC},\Delta\textrm{DIC} < -10$ there is {\it very strong} evidence in favor of the model under study relative to the tilted flat $\Lambda$CDM model. This scale also applies if $\Delta\textrm{AIC}$ and $\Delta\textrm{DIC}$ are positive, but then favors the tilted flat $\Lambda$CDM model over the model under study.   

We want to quantitatively compare how consistent the data sets used to constrain cosmological parameters in a cosmological model are with each other, and how the level of consistency varies across models. We used two different statistical estimators to check for consistency between data sets used in a given model.

The first estimator we used to check for consistency is based on the DIC values of the individual data sets, \cite{Joudaki:2016mvz}, and is defined as 
\begin{equation}
\label{eq:Tension_estimator_1}
\mathcal{I}(D_1,D_2) = \textrm{exp}\left(-\frac{\mathcal{G}(D_1,D_2)}{2}\right),
\end{equation}
where $\mathcal{G}(D_1,D_2)=\textrm{DIC}(D_{12})-\textrm{DIC}(D_1)-\textrm{DIC}(D_2)$ and $D_1$ and $D_2$ are the two data sets being compared. Here $\textrm{DIC}(D_1)$ and $\textrm{DIC}(D_2)$ are the DIC values estimated from the MCMC chains when $D_1$ and $D_2$ are used independently to constrain the cosmological parameters of a given model, and $\textrm{DIC}(D_{12})$ is the DIC value that results when $D_1$ and $D_2$ are used together to constrain cosmological parameters of the model. This statistical estimator indicates $\log_{10}\mathcal{I}>0$ when the two data sets used in the cosmological model are consistent with each other. Conversely, $\log_{10}\mathcal{I}<0$ indicates that the two data sets are inconsistent. Applying Jeffreys' scale, we judge the degree of consistency or inconsistency between two data sets as {\it substantial} if $\lvert \log_{10}\mathcal{I} \rvert >0.5$, {\it strong} if $\lvert \log_{10}\mathcal{I} \rvert >1$, and {\it decisive} if $\lvert \log_{10}\mathcal{I} \rvert >2$, \cite{Joudaki:2016mvz}.

The second statistical estimator for determining the consistency between two data sets is the tension probability. The details of this estimator are given in \cite{Handley:2019pqx,Handley:2019wlz,Handley:2019tkm} and in our previous work \cite{deCruzPerez:2022hfr}, and we briefly describe it here. The tension probability is related to the suspiciousness $S_D = R_D / I_D$ that is defined in terms of the Bayes ratio $R_D$ and the information ratio $I_D$. The Bayes ratio $R_D = \mathcal{Z}_{12} / (\mathcal{Z}_1\mathcal{Z}_2 )$ where $\mathcal{Z}_1$, $\mathcal{Z}_2$, and $\mathcal{Z}_{12}$ are the Bayesian evidences estimated from $D_1$, $D_2$, and $D_{12}$, respectively. The Bayesian evidence is defined as  
\begin{equation}
\label{eq:Evidence}
\mathcal{Z}_D = \int \mathcal{L}_D(\theta)\pi(\theta)d\theta,
\end{equation}
where $\mathcal{L}_D(\theta)$ is the likelihood of a model $\theta$ given the data and $\pi(\theta)$ is the prior of the model. We compute the Bayesian evidence by using the method of \cite{Heavens:2017hkr}. The information ratio $I_D$ is defined through $\ln(I_D) = \mathcal{D}_1 + \mathcal{D}_2 - \mathcal{D}_{12}$. Here $\mathcal{D}_D$ is the Kullback-Leibler divergence for data $D$, which is the average over the posterior of the Shannon information 
\begin{equation}
   \mathcal{I}_{S,D}(\theta) = \ln \frac{\mathcal{P}_D(\theta)}{\pi(\theta)},
\end{equation}
where $\mathcal{P}_\mathcal{D}(\theta)$ is the posterior distribution \cite{Shannon:1948zz}. Finally, the tension probability,  \cite{Handley:2019pqx,Handley:2019wlz,Handley:2019tkm},
\begin{equation}
\label{eq:Tension_estimator_2}
p = \int^{\infty}_{d-2\ln(S_D)}\!\!\frac{x^{d/2 -1}e^{-x/2}}{2^{d/2}\Gamma(d/2)}dx,  
\end{equation}
where $d$ is the Bayesian model dimensionality $d = \Tilde{d}_1 + \Tilde{d}_2 - \Tilde{d}_{12}$, where $\Tilde{d}/2 = \langle \mathcal{I}_{S,D}^2\rangle_{\mathcal{P}_D} -  \langle \mathcal{I}_{S,D}\rangle^2_{\mathcal{P}_D}$ and $\Gamma (z)$ is a Gamma function. If $p\lesssim 0.05$ the data sets are in moderate tension whereas if $p\lesssim 0.003$ they are in strong tension. Based on the Gaussian formula, we can convert $p$ into a ``sigma value'', 
\begin{equation}
\label{eq:Tension_estimator_2_sigma}
\sigma = \sqrt{2}\textrm{Erfc}^{-1}(1-p),    
\end{equation}
where $\textrm{Erfc}^{-1}$ is the inverse complementary error function. The value $p=0.05$ ($p=0.003$) corresponds to a 2$\sigma$ (3$\sigma$) Gaussian standard deviation.

\section{Results}\label{sec:Results}

\subsection{$\Lambda$CDM models}
\label{sec:LCDM models}

In this subsection we study how the inclusion of non-CMB (new) data, an updated version of the non-CMB (old) data used in \cite{deCruzPerez:2022hfr}, in different data set combinations we use here, affect the cosmological parameter constraints in the $\Lambda$CDM models and goodness-of-fit results and consistencies between different data subsets. In the last part of this subsection we summarize the updated data constraints on cosmological parameters in the $\Lambda$CDM models.

\subsubsection{Non-CMB (old) vs. non-CMB (new) cosmological parameter constraints}
\label{sec:non-CMB (old) vs. non-CMB (new)}

\begin{figure*}[htbp]
\centering
\mbox{\includegraphics[width=170mm]{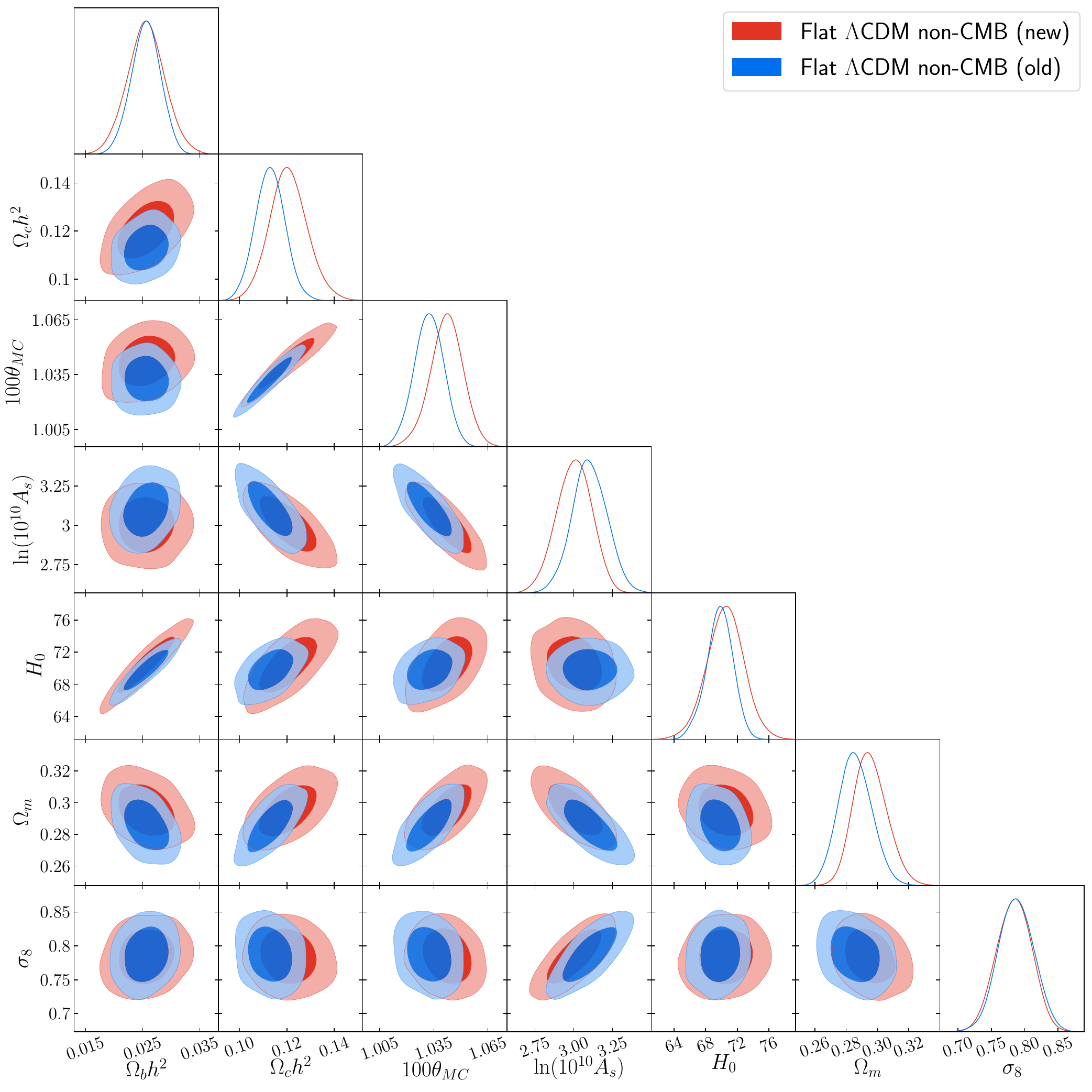}}
\caption{Likelihood distributions of flat $\Lambda$CDM model parameters
         favored by non-CMB data. Non-CMB (old) denotes the compilation of non-CMB data
	 sets used in \cite{deCruzPerez:2022hfr} while non-CMB (new) 
	 denotes the new compilation of non-CMB data sets of Sec.\ \ref{sec:Non-CMB Data}.
}
\label{fig:flat_LCDM_ncmb}
\end{figure*}

Cosmological parameter constraints for the four-parameter flat $\Lambda$CDM model from non-CMB (old) and non-CMB (new) data are in the lower half of Table \ref{tab:results_flat_LCDM} and Fig.\ \ref{fig:flat_LCDM_ncmb}. We observe some differences in the mean parameter values favored by the two data sets, but all are below the 1$\sigma$ level. The differences between the results provided by the two data sets for $\Omega_c{h^2}$, $100\theta_{\textrm{MC}}$, and $\ln(10^{10}A_s)$ are $-0.77\sigma$, $-0.80\sigma$, and $+0.64\sigma$, respectively. Similar levels of differences are observed for the derived parameters, with $\Omega_m$ showing the largest at $-0.64\sigma$. On the other hand, a significant change is observed in the size of the error bars. Compared to the results from the non-CMB (old) data, the results from the non-CMB (new) data increase the error bars of $\Omega_{b}h^2$, $\Omega_{c}h^2$, and $H_0$ by 24\%, 23\%, and 26\%, 
respectively. One major reason for this is the larger error bars associated with updated $H(z)$ data (those for which there is now a non-diagonal covariance matrix).

\begin{figure*}[htbp]
\centering
\mbox{\includegraphics[width=170mm]{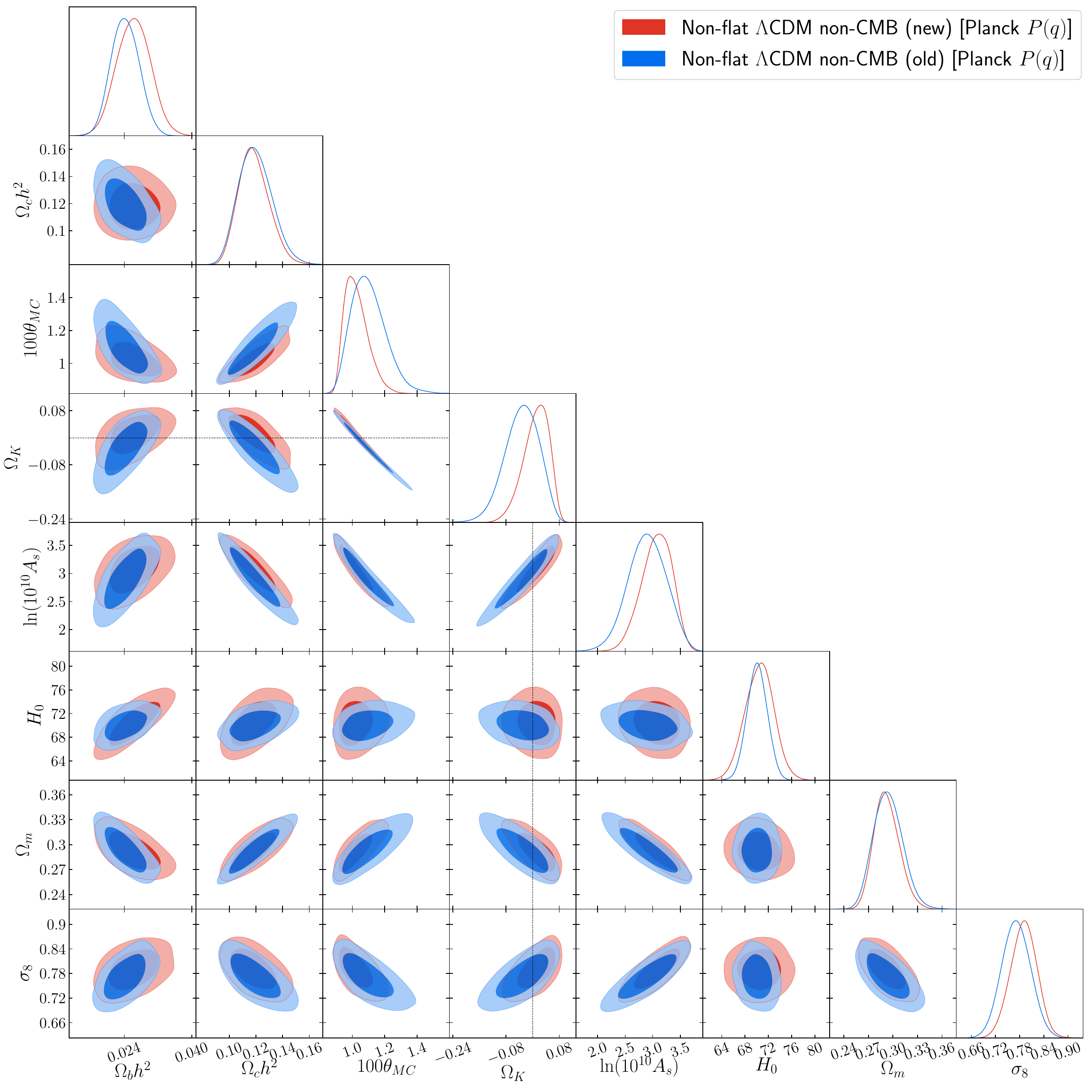}}
\caption{Likelihood distributions of non-flat $\Lambda$CDM model [Planck $P(q)$] parameters
         favored by non-CMB data. Non-CMB (old) denotes the compilation of non-CMB data
	 sets used in \cite{deCruzPerez:2022hfr} while non-CMB (new) 
	 denotes the new compilation of non-CMB data sets of Sec.\ \ref{sec:Non-CMB Data}.
}
\label{fig:Planck_Pq_ncmb}
\end{figure*}

Non-CMB (old) and non-CMB (new) data results obtained for the five-parameter non-flat $\Lambda$CDM Planck $P(q)$ cosmological model are in the lower half of Table \ref{tab:results_Planck_Pq} and in Fig.\ \ref{fig:Planck_Pq_ncmb}. As in the flat $\Lambda$CDM model, the differences in the mean values of the primary parameters remain below 1$\sigma$, with $100\theta_{\textrm{MC}}$ (+0.57$\sigma$) being the largest. The value of the primary spatial curvature parameter is $\Omega_k = -0.032^{+0.056}_{-0.046}$ for non-CMB (old) data which is 0.57$\sigma$ away from flat geometry and $-0.66\sigma$ from the non-CMB (new) value $\Omega_k=0.010^{+0.046}_{-0.030}$ which in turn is 0.33$\sigma$ in favor of an open Universe. For mean derived parameters the largest difference is $-0.39\sigma$ for $\sigma_8$. In contrast to the flat $\Lambda$CDM model, where the error bars of all cosmological parameters obtained from non-CMB (old) data are smaller than those obtained using non-CMB (new) data, we observe a different pattern in the non-flat $\Lambda$CDM Planck $P(q)$ model. This may be due to the degeneracy between parameters and differences in the ability of the two data sets to constrain cosmological parameters. While for $\Omega_{b}h^2$ (+18\%) and $H_0$ (+32\%) there are increases in the error bars when we move from non-CMB (old) to non-CMB (new) data, for the rest of the parameters it is the other way around, with the largest 
decreases affecting $100\theta_{\textrm{MC}}$ ($-51$\%), $\Omega_k$ ($-34$\%), and $\ln(10^{10}A_s)$ ($-26$\%). 

\begin{figure*}[htbp]
\centering
\mbox{\includegraphics[width=170mm]{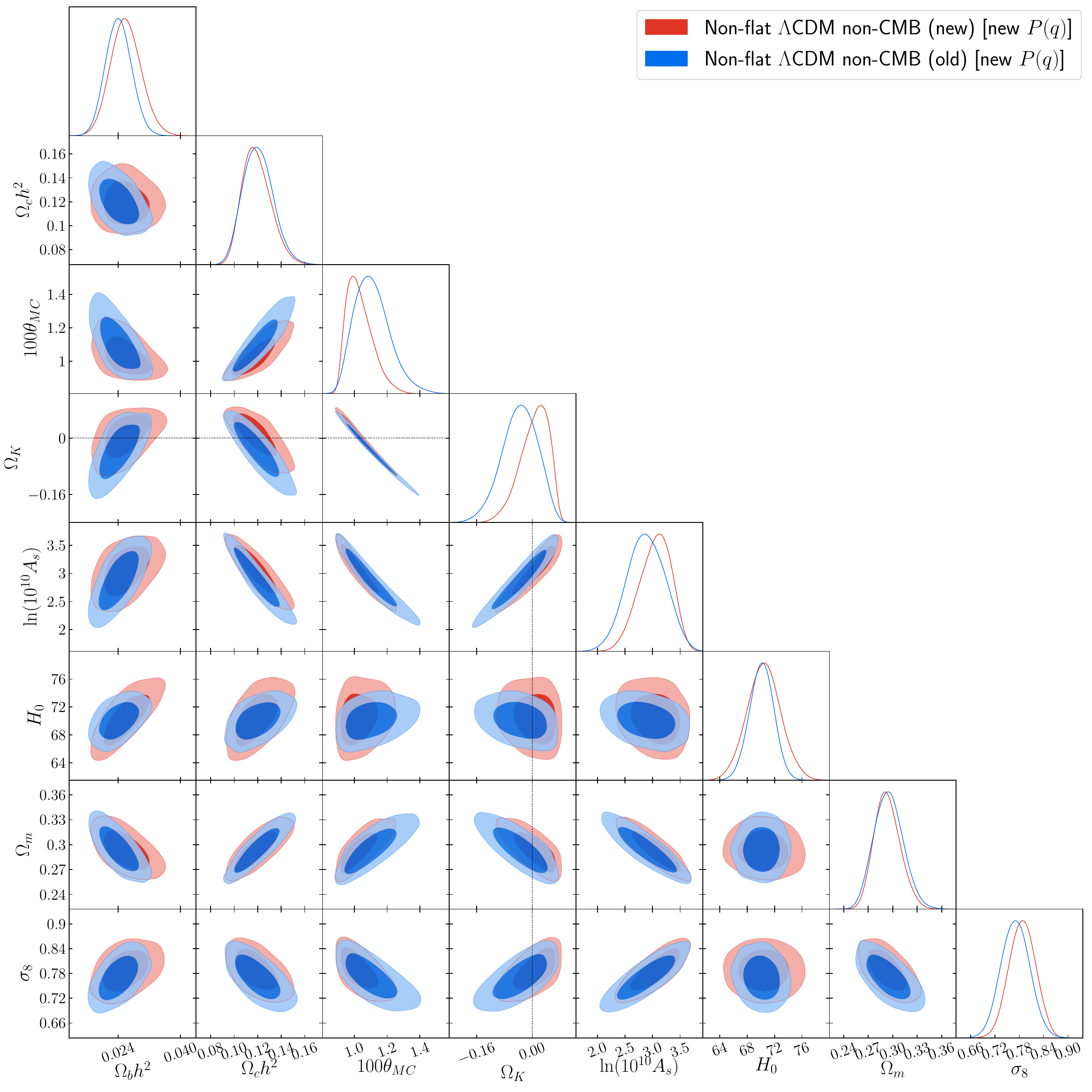}}
\caption{Likelihood distributions of non-flat $\Lambda$CDM model [new $P(q)$] parameters
         favored by non-CMB data. Non-CMB (old) denotes the compilation of non-CMB data
	 sets used in \cite{deCruzPerez:2022hfr} while non-CMB (new) 
	 denotes the new compilation of non-CMB data sets of Sec.\ \ref{sec:Non-CMB Data}.
}
\label{fig:new_Pq_ncmb}
\end{figure*}

When we compare the results obtained with non-CMB (old) and non-CMB (new) data, displayed in Table \ref{tab:results_new_Pq} and in Fig.\ \ref{fig:new_Pq_ncmb}, for the five-parameter non-flat $\Lambda$CDM new $P(q)$ model, we see very similar results to the case of the non-flat $\Lambda$CDM Planck $P(q)$ model. We observe mild differences between the mean values of the primary parameters, with the largest being  +0.57$\sigma$ for $100\theta_{\textrm{MC}}$. As for $\Omega_k$, the non-CMB (old) result $\Omega_k = -0.036^{+0.056}_{-0.047}$ favors non-flat hypersurfaces at  0.64$\sigma$ and differs by $-0.66\sigma$ with the non-CMB (new) value $\Omega_k = 0.006^{+0.051}_{-0.030}$, which is 0.20$\sigma$ away from flat. In regard to changes in the size of the error bars, we observe an increase for $\Omega_{b}h^2$ (+15\%) and for $H_0$ (+28\%), while reductions affect $100\theta_{\textrm{MC}}$ ($-39$\%), $\Omega_k$ ($-27$\%), and $\ln(10^{10}A_s)$ ($-17$\%). 

The non-CMB (old) data and non-CMB (new) data $\Delta \text{DIC}$ and $\Delta \text{AIC}$ values, between the flat $\Lambda$CDM model and the two non-flat $\Lambda$CDM models, in Tables \ref{tab:results_Planck_Pq} and \ref{tab:results_new_Pq}, do not show significant differences. 

\begin{table*}
\caption{Mean and 68\% confidence limits of flat $\Lambda\textrm{CDM}$ model parameters
        constrained by TT,TE,EE+lowE (P18) and P18+lensing, and non-CMB data sets.
        The Hubble constant $H_0$ has units of km s$^{-1}$ Mpc$^{-1}$.
}
\begin{ruledtabular}
\begin{tabular}{lcccc}
\\[-1mm]                         & \multicolumn{4}{c}{Flat $\Lambda$CDM models}        \\[+1mm]
\cline{2-5}\\[-1mm]
  Parameter                      & P18                   & P18+lensing           &  P18+lensing+non-CMB (old) & P18+lensing+non-CMB (new)    \\[+1mm]
 \hline \\[-1mm]
  $\Omega_b h^2$                 & $0.02236 \pm 0.00015$ & $0.02237 \pm 0.00014$ &  $0.02250 \pm 0.00013$  &  $0.02249 \pm 0.00013$  \\[+1mm]
  $\Omega_c h^2$                 & $0.1202 \pm 0.0014$   & $0.1200 \pm 0.0012$   &  $0.11838 \pm 0.00083$  &  $0.11849 \pm 0.00084$  \\[+1mm]
  $100\theta_\textrm{MC}$        & $1.04090 \pm 0.00031$ & $1.04091 \pm 0.00031$ &  $1.04110 \pm 0.00029$  &  $1.04109 \pm 0.00028$  \\[+1mm]
  $\tau$                         & $0.0542 \pm 0.0079$   & $0.0543 \pm 0.0073$   &  $0.0569 \pm 0.0071$    &  $0.0569 \pm 0.0071$  \\[+1mm]
  $n_s$                          & $0.9649 \pm 0.0043$   & $0.9649 \pm 0.0041$   &  $0.9688 \pm 0.0036$    &  $0.9685 \pm 0.0036$  \\[+1mm]
  $\ln(10^{10} A_s)$             & $3.044 \pm 0.016$     & $3.044 \pm 0.014$     &  $3.046 \pm 0.014$      &  $3.046 \pm 0.014$  \\[+1mm]
 \hline \\[-1mm]
  $H_0$                          & $67.28 \pm 0.61$      & $67.34 \pm 0.55$      &  $68.09 \pm 0.38$       &  $68.05 \pm 0.38$  \\[+1mm]
  $\Omega_m$                     & $0.3165 \pm 0.0084$   & $0.3155 \pm 0.0075$   &  $0.3053 \pm 0.0050$    &  $0.3059 \pm 0.0050$  \\[+1mm]
  $\sigma_8$                     & $0.8118 \pm 0.0074$   & $0.8112 \pm 0.0059$   &  $0.8072 \pm 0.0058$    &  $0.8077 \pm 0.0057$  \\[+1mm]
  \hline\\[-1mm]
  $\chi_{\textrm{min}}^2$        & $2765.80$             & $2774.71$             &  $3888.41$              &  $4249.26$            \\[+1mm]
  $\textrm{DIC}$                 & $2817.93$             & $2826.45$             &  $3940.70$              &  $4301.20$            \\[+1mm]
  $\textrm{AIC}$                 & $2819.80$             & $2828.71$             &  $3942.41$              &  $4303.26$             \\[+1mm]
 \hline \hline \\[-1mm]
                                 & \multicolumn{4}{c}{Flat $\Lambda$CDM models}              \\[+1mm]
\cline{2-5}\\[-1mm]
Parameter                        & Non-CMB (old)       &  Non-CMB (new)       &  P18+non-CMB (old)     & P18+non-CMB (new)      \\[+1mm]
 \hline \\[-1mm]
  $\Omega_b h^2$                 & $0.0256 \pm 0.0025$ & $0.0256 \pm 0.0033$  & $0.02250 \pm 0.00012$  & $0.02248 \pm 0.00013$   \\[+1mm]
  $\Omega_c h^2$                 & $0.1129 \pm 0.0062$ & $0.1207 \pm 0.0080$  & $0.11825 \pm 0.00087$  & $0.11839 \pm 0.00087$   \\[+1mm]
  $100\theta_\textrm{MC}$        & $1.0323 \pm 0.0082$ & $1.0421 \pm 0.0091$  & $1.04112 \pm 0.00029$  & $1.04110 \pm 0.00029$   \\[+1mm]
  $\tau$                         & $0.0542$            & $0.0542$             & $0.0548 \pm 0.0076$    & $0.0546 \pm 0.0077$   \\[+1mm]
  $n_s$                          & $0.9649$            & $0.9649$             & $0.9692 \pm 0.0036$    & $0.9689 \pm 0.0036$   \\[+1mm]
  $\ln(10^{10} A_s)$             & $3.10 \pm 0.11$     & $3.00 \pm 0.11$      & $3.041 \pm 0.015$      & $3.041 \pm 0.016$   \\[+1mm]
 \hline \\[-1mm]
  $H_0$                          & $69.8 \pm 1.7$ ($69.8^{+1.8}_{-1.5}$)      & $70.5 \pm 2.3$       & $68.15 \pm 0.39$       & $68.08 \pm 0.39$   \\[+1mm]
  $\Omega_m$                     & $0.286 \pm 0.011$   & $0.296 \pm 0.011$    & $0.3045 \pm 0.0051$    & $0.3054 \pm 0.0051$   \\[+1mm]
  $\sigma_8$                     & $0.787 \pm 0.027$   & $0.784 \pm 0.026$    & $0.8048 \pm 0.0068$    & $0.8052 \pm 0.0067$   \\[+1mm]
    \hline\\[-1mm]
  $\chi_{\textrm{min}}^2$        & $1106.54$           & $1469.93$            & $3879.35$              &  $4240.24$             \\[+1mm]
  $\textrm{DIC}$                 & $1114.45$           & $1478.11$            & $3931.02$              &  $4292.33$              \\[+1mm]
  $\textrm{AIC}$                 & $1114.54$           & $1477.93$            & $3933.35$              &  $4294.24$            \\[+1mm]
\end{tabular}
\\[+1mm]
\end{ruledtabular}
\label{tab:results_flat_LCDM}
\end{table*}

\begin{table*}
\caption{Mean and 68\% confidence limits of non-flat $\Lambda$CDM model [Planck $P(q)$] parameters
        constrained by TT,TE,EE+lowE (P18) and P18+lensing, and non-CMB data sets.
        The Hubble constant $H_0$ has units of km s$^{-1}$ Mpc$^{-1}$.
}
\begin{ruledtabular}
\begin{tabular}{lcccc}
\\[-1mm]                         & \multicolumn{4}{c}{Non-flat $\Lambda$CDM models [Planck $P(q)$]}        \\[+1mm]
\cline{2-5}\\[-1mm]
  Parameter                      & P18                   &  P18+lensing           &  P18+lensing+non-CMB (old)  & P18+lensing+non-CMB (new)   \\[+1mm]
 \hline \\[-1mm]
  $\Omega_b h^2$                 & $0.02260 \pm 0.00017$ & $0.02249 \pm 0.00016$  &  $0.02249 \pm 0.00015$  &  $0.02245 \pm 0.00015$  \\[+1mm]
  $\Omega_c h^2$                 & $0.1181 \pm 0.0015$   & $0.1186 \pm 0.0015$    &  $0.1187 \pm 0.0013$    &  $0.1190 \pm 0.0013$    \\[+1mm]
  $100\theta_\textrm{MC}$        & $1.04116 \pm 0.00032$ & $1.04107 \pm 0.00032$  &  $1.04106 \pm 0.00031$  &  $1.04101 \pm 0.00031$  \\[+1mm]
  $\tau$                         & $0.0483 \pm 0.0083$   & $0.0495 \pm 0.0082$    &  $0.0563 \pm 0.0073$    &  $0.0559 \pm 0.0071$    \\[+1mm]
  $\Omega_k$                     & $-0.043 \pm 0.017$    & $-0.0103 \pm 0.0066$   &  $0.0004 \pm 0.0017$    &  $0.0009 \pm 0.0017$    \\[+1mm]
                                 & ($-0.043^{+0.018}_{-0.015}$)    & ($-0.0103^{+0.0071}_{-0.0060}$)   &     &      \\[+1mm]
  $n_s$                          & $0.9706 \pm 0.0047$   & $0.9687 \pm 0.0046$    &  $0.9681 \pm 0.0044$    &  $0.9672 \pm 0.0043$    \\[+1mm]
  $\ln(10^{10} A_s)$             & $3.027 \pm 0.017$     & $3.030 \pm 0.017$      &  $3.046 \pm 0.014$      &  $3.046 \pm 0.014$      \\[+1mm]
 \hline \\[-1mm]
  $H_0$                          & $54.5 \pm 3.6$ ($54.5^{+3.1}_{-3.9}$)       & $63.7 \pm 2.3$         &  $68.17 \pm 0.55$       &  $68.24 \pm 0.54$       \\[+1mm]
  $\Omega_m$                     & $0.481 \pm 0.062$     & $0.351 \pm 0.024$      &  $0.3051 \pm 0.0053$    &  $0.3053 \pm 0.0051$    \\[+1mm]
  $\sigma_8$                     & $0.775 \pm 0.015$     & $0.796 \pm 0.011$      &  $0.8080 \pm 0.0066$    &  $0.8094 \pm 0.0066$    \\[+1mm]
      \hline\\[-1mm]
  $\chi_{\textrm{min}}^2$        &  $2754.73$            & $2771.53$              &  $3887.99$              &  $4248.74$              \\[+1mm]
  $\Delta\chi_{\textrm{min}}^2$  &  $-11.07$             & $-3.18$                &  $-0.42$                &  $-0.52$              \\[+1mm]
  $\textrm{DIC}$                 &  $2810.59$            & $2826.17$              &  $3942.07$              &  $4302.41$              \\[+1mm]
  $\Delta\textrm{DIC}$           &  $-7.34$              & $-0.28$                &  $+1.37$                &  $+1.21$                 \\[+1mm]
  $\textrm{AIC}$                 &  $2810.73$            & $2827.53$             &  $3943.99$              &  $4304.74$            \\[+1mm]
  $\Delta\textrm{AIC}$           &  $-9.07$             & $-1.18$                &  $+1.58$                &  $+1.48$                \\[+1mm]
 \hline \hline \\[-1mm]
                                 & \multicolumn{4}{c}{Non-flat $\Lambda$CDM models [Planck $P(q)$]}              \\[+1mm]
\cline{2-5}\\[-1mm]
Parameter                        & Non-CMB (old)       &  Non-CMB (new)           &  P18+non-CMB (old)     &  P18+non-CMB (new)      \\[+1mm]
 \hline \\[-1mm]
  $\Omega_b h^2$                 & $0.0242 \pm 0.0033$ & $0.0262 \pm 0.0040$      & $0.02248 \pm 0.00015$   & $0.02244 \pm 0.00015$   \\[+1mm]
  $\Omega_c h^2$                 & $0.120 \pm 0.012$   & $0.118 \pm 0.011$        & $0.1185 \pm 0.0013$     & $0.1189 \pm 0.0013$   \\[+1mm]
  $100\theta_\textrm{MC}$        & $1.10 \pm 0.11$     & $1.025 \pm 0.073$        & $1.04107 \pm 0.00031$   & $1.04102 \pm 0.00031$   \\[+1mm]
  $\tau$                         & $0.0483$            & $0.0483$                 & $0.0543 \pm 0.0077$     & $0.0539 \pm 0.0078$   \\[+1mm]
  $\Omega_k$                     & $-0.032 \pm 0.051$  & $0.010 \pm 0.038$        & $0.0004 \pm 0.0017$     & $0.0009 \pm 0.0017$   \\[+1mm]
                                 & ($-0.032^{+0.056}_{-0.046}$)    & ($0.010^{+0.046}_{-0.030}$)   &     &      \\[+1mm]
  $n_s$                          & $0.9706$            & $0.9706$                 & $0.9687 \pm 0.0043$     & $0.9677 \pm 0.0044$   \\[+1mm]
  $\ln(10^{10} A_s)$             & $2.90 \pm 0.34$     & $3.07 \pm 0.27$          & $3.040 \pm 0.016$       & $3.040 \pm 0.016$   \\[+1mm]
	\hline \\[-1mm]
  $H_0$                          & $70.1 \pm 1.7$      & $70.5 \pm 2.5$           & $68.25 \pm 0.56$        & $68.31\pm 0.56$   \\[+1mm]
  $\Omega_m$                     & $0.294 \pm 0.018$   & $0.292 \pm 0.015$        & $0.3040 \pm 0.0055$     & $0.3043 \pm 0.0054$    \\[+1mm]
  $\sigma_8$                     & $0.771 \pm 0.035$   & $0.790 \pm 0.033$        & $0.8055 \pm 0.0076$     & $0.8070 \pm 0.0077$   \\[+1mm]
      \hline\\[-1mm]
  $\chi_{\textrm{min}}^2$        &  $1106.53$          & $1468.22$                &  $3878.77$              & $4239.58$              \\[+1mm]
  $\Delta\chi_{\textrm{min}}^2$  &  $-0.01$            & $-1.71$                  &  $-0.58$                & $-0.66$              \\[+1mm]
  $\textrm{DIC}$                 &  $1116.92$          & $1479.52$                &  $3933.33$              & $4293.78$              \\[+1mm]
  $\Delta\textrm{DIC}$           &  $+2.47$            & $+1.41$                  &  $+2.31$                & $+1.45$               \\[+1mm]
  $\textrm{AIC}$                 &  $1116.53$          & $1478.22$                &  $3934.77$              & $4295.58$              \\[+1mm]
  $\Delta\textrm{AIC}$           &  $+1.99$            & $+0.29$                  &  $+1.42$                & $+1.34$              \\[+1mm]
\end{tabular}
\\[+1mm]
\begin{flushleft}
\end{flushleft}
\end{ruledtabular}
\label{tab:results_Planck_Pq}
\end{table*}

\begin{table*}
\caption{Mean and 68\% confidence limits of non-flat $\Lambda$CDM model [new $P(q)$] parameters
        constrained by TT,TE,EE+lowE (P18) and P18+lensing, and non-CMB data sets.
        The Hubble constant $H_0$ has units of km s$^{-1}$ Mpc$^{-1}$.
}
\begin{ruledtabular}
\begin{tabular}{lcccc}
\\[-1mm]                         & \multicolumn{4}{c}{Non-flat $\Lambda$CDM models [new $P(q)$]}        \\[+1mm]
\cline{2-5}\\[-1mm]
  Parameter                      & P18                   &  P18+lensing           &  P18+lensing+non-CMB (old)  & P18+lensing+non-CMB (new)    \\[+1mm]
 \hline \\[-1mm]
  $\Omega_b h^2$                 & $0.02255 \pm 0.00017$ & $0.02248 \pm 0.00016$  &  $0.02248 \pm 0.00015$  &  $0.02246 \pm 0.00015$  \\[+1mm]
  $\Omega_c h^2$                 & $0.1188 \pm 0.0015$   & $0.1188 \pm 0.0014$    &  $0.1186 \pm 0.0013$    &  $0.1190 \pm 0.0013$    \\[+1mm]
  $100\theta_\textrm{MC}$        & $1.04109 \pm 0.00032$ & $1.04104 \pm 0.00032$  &  $1.04106 \pm 0.00031$  &  $1.04103 \pm 0.00031$  \\[+1mm]
  $\tau$                         & $0.0525 \pm 0.0083$   & $0.0515 \pm 0.0081$    &  $0.0566 \pm 0.0074$    &  $0.0562 \pm 0.0072$    \\[+1mm]
  $\Omega_k$                     & $-0.033 \pm 0.014$    & $-0.0086 \pm 0.0057$   &  $0.0003 \pm 0.0017$    &  $0.0008 \pm 0.0017$    \\[+1mm]
                                 & ($-0.033^{+0.017}_{-0.011}$)    & ($-0.0086^{+0.0063}_{-0.0050}$)   &     &      \\[+1mm]
  $n_s$                          & $0.9654 \pm 0.0045$   & $0.9661 \pm 0.0043$    &  $0.9679 \pm 0.0042$    &  $0.9671 \pm 0.0041$    \\[+1mm]
  $\ln(10^{10} A_s)$             & $3.039 \pm 0.017$     & $3.035 \pm 0.016$      &  $3.046 \pm 0.014$      &  $3.046 \pm 0.014$      \\[+1mm]
 \hline \\[-1mm]
  $H_0$                          & $56.9 \pm 3.6$        & $64.2 \pm 2.0$         &  $68.13 \pm 0.54$       &  $68.21 \pm 0.54$       \\[+1mm]
  $\Omega_m$                     & $0.444 \pm 0.055$     & $0.345 \pm 0.021$      &  $0.3054 \pm 0.0051$    &  $0.3054 \pm 0.0051$    \\[+1mm]
  $\sigma_8$                     & $0.786 \pm 0.014$     & $0.799 \pm 0.010$      &  $0.8079 \pm 0.0067$    &  $0.8094 \pm 0.0065$    \\[+1mm]
      \hline\\[-1mm]
  $\chi_{\textrm{min}}^2$        & $2757.38$             & $2771.75$              &  $3887.55$              &  $4248.50$              \\[+1mm]
  $\Delta\chi_{\textrm{min}}^2$  & $-8.42$               & $-2.96$                &  $-0.86$                &  $-0.76$              \\[+1mm]
  $\textrm{DIC}$                 & $2811.54$             & $2825.74$              &  $3942.22$              &  $4302.33$              \\[+1mm]
  $\Delta\textrm{DIC}$           & $-6.39$               & $-0.71$                &  $+1.52$                &  $+1.13$               \\[+1mm]
  $\textrm{AIC}$                 & $2813.38$             & $2827.75$              &  $3943.55$              &  $4304.50$              \\[+1mm]
  $\Delta\textrm{AIC}$           & $-6.42$               & $-0.96$                &  $+1.14$                &  $+1.24$              \\[+1mm]
 \hline \hline \\[-1mm]
                                 & \multicolumn{4}{c}{Non-flat $\Lambda$CDM models [new $P(q)$]}              \\[+1mm]
\cline{2-5}\\[-1mm]
Parameter                        & Non-CMB (old)       &  Non-CMB (new)           &  P18+non-CMB (old)     &  P18+non-CMB (new)      \\[+1mm]
 \hline \\[-1mm]
  $\Omega_b h^2$                 & $0.0241 \pm 0.0033$ & $0.0260 \pm 0.0039$      & $0.02249 \pm 0.00015$   & $0.02246 \pm 0.00015$   \\[+1mm]
  $\Omega_c h^2$                 & $0.120 \pm 0.013$   & $0.119 \pm 0.012$        & $0.1184 \pm 0.0013$     & $0.1188 \pm 0.0013$   \\[+1mm]
  $100\theta_\textrm{MC}$        & $1.11 \pm 0.11$     & $1.033 \pm 0.079$        & $1.04108 \pm 0.00031$   & $1.04104 \pm 0.00031$   \\[+1mm]
  $\tau$                         & $0.0525$            & $0.0525$                 & $0.0549 \pm 0.0077$     & $0.0542 \pm 0.0076$   \\[+1mm]
  $\Omega_k$                     & $-0.036 \pm 0.051$  & $0.006 \pm 0.041$        & $0.0003 \pm 0.0017$     & $0.0008 \pm 0.0017$   \\[+1mm]
                                 & ($-0.036^{+0.056}_{-0.047}$)    & ($0.006^{+0.051}_{-0.030}$)   &     &      \\[+1mm]
  $n_s$                          & $0.9654$            & $0.9654$                 & $0.9684 \pm 0.0041$     & $0.9675 \pm 0.0042$   \\[+1mm]
  $\ln(10^{10} A_s)$             & $2.88 \pm 0.34$     & $3.05 \pm 0.29$          & $3.042 \pm 0.016$       & $3.041 \pm 0.015$   \\[+1mm]
\hline \\[-1mm]
  $H_0$                         & $70.1 \pm 1.8$       & $70.4 \pm 2.5$           & $68.21 \pm 0.55$        & $68.29 \pm 0.55$   \\[+1mm]
  $\Omega_m$                    & $0.295 \pm 0.018$    & $0.293 \pm 0.016$        & $0.3043 \pm 0.0054$     & $0.3045 \pm 0.0053$    \\[+1mm]
  $\sigma_8$                    & $0.770 \pm 0.035$    & $0.787 \pm 0.033$        & $0.8057 \pm 0.0074$     & $0.8071 \pm 0.0075$   \\[+1mm]
      \hline\\[-1mm]
  $\chi_{\textrm{min}}^2$       & $1106.49$            & $1468.21$                & $3878.76$               & $4239.45$             \\[+1mm]
  $\Delta\chi_{\textrm{min}}^2$ & $-0.05$              & $-1.72$                  & $-0.59$                 & $-0.79$              \\[+1mm]
  $\textrm{DIC}$                & $1117.31$            & $1480.16$                & $3932.56$               & $4293.50$             \\[+1mm]
  $\Delta\textrm{DIC}$          & $+2.86$              & $+2.05$                  & $+1.54$                 & $+1.17$               \\[+1mm]
  $\textrm{AIC}$                & $1116.49$            & $1478.21$                & $3934.76$               & $4295.45$             \\[+1mm]
  $\Delta\textrm{AIC}$          & $+1.95$              & $+0.28$                  & $+1.41$                 & $+1.21$               \\[+1mm]

\end{tabular}
\\[+1mm]
\begin{flushleft}
\end{flushleft}
\end{ruledtabular}
\label{tab:results_new_Pq}
\end{table*}

\subsubsection{P18+non-CMB (old) vs. P18+non-CMB (new) cosmological parameter constraints}
\label{sec:P18+non-CMB (old) vs. P18+non-CMB (new)}

The aim of this subsubsection is to determine what happens to the changes highlighted in the previous subsubsection when P18 data are added to the mix. 

\begin{figure*}[htbp]
\centering
\mbox{\includegraphics[width=170mm]{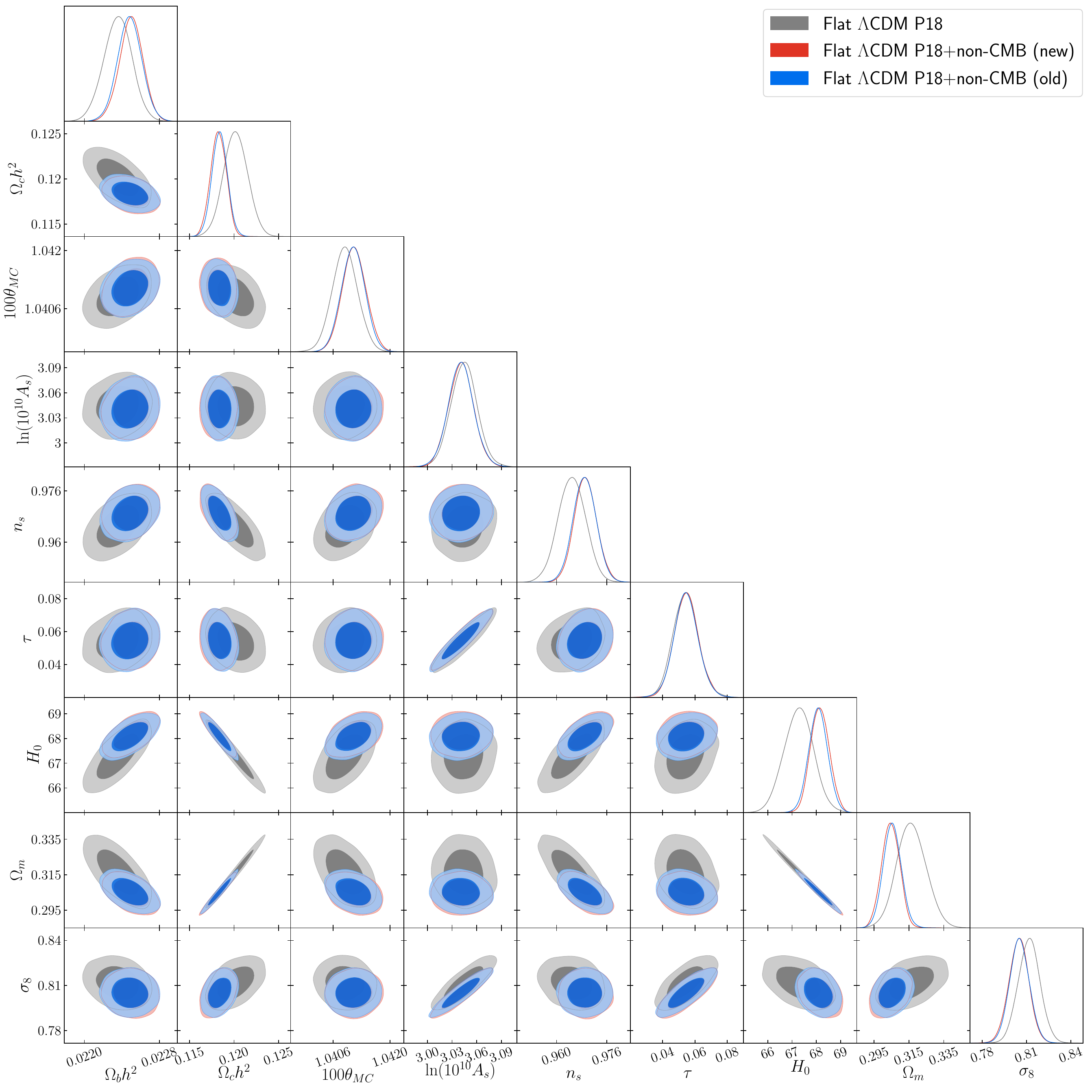}}
\caption{Likelihood distributions of flat $\Lambda$CDM model parameters
         favored by P18 and non-CMB data. Non-CMB (old) denotes the compilation of non-CMB data
	 sets used in \cite{deCruzPerez:2022hfr} while non-CMB (new) 
	 denotes the new compilation of non-CMB data sets of Sec.\ \ref{sec:Non-CMB Data}.
         Likelihood results for P18 data are shown for comparison. 
}
\label{fig:flat_LCDM_p18ncmb}
\end{figure*}

P18+non-CMB (old) and P18+non-CMB (new) cosmological parameter constraints for the six-parameter flat $\Lambda$CDM model are in the lower half of Table \ref{tab:results_flat_LCDM} and in Fig.\ \ref{fig:flat_LCDM_p18ncmb}. Moving from P18 + non-CMB (old) data to P18 + non-CMB (new) data results in smaller changes in the mean parameter values than found in the non-CMB data alone case of the previous subsubsection. For the primary parameters, the most affected mean values are $\Omega_{b}h^2$ and $\Omega_{c}h^2$, whose values differ by +0.11$\sigma$ and $-0.11\sigma$ when compared with the P18+non-CMB (old) data results. Similar conclusions hold for derived parameter means, with the differences in $H_0$ and $\Omega_m$ being +0.13$\sigma$ and $-0.12\sigma$, respectively. No significant changes are observed in the size of the error bars, when we move from P18+non-CMB (old) data to P18+non-CMB (new) data. The largest changes are increases of 7.7\% for $\Omega_b{h^2}$ and 6.3\% for $\ln(10^{10}A_s)$. 

\begin{figure*}[htbp]
\centering
\mbox{\includegraphics[width=170mm]{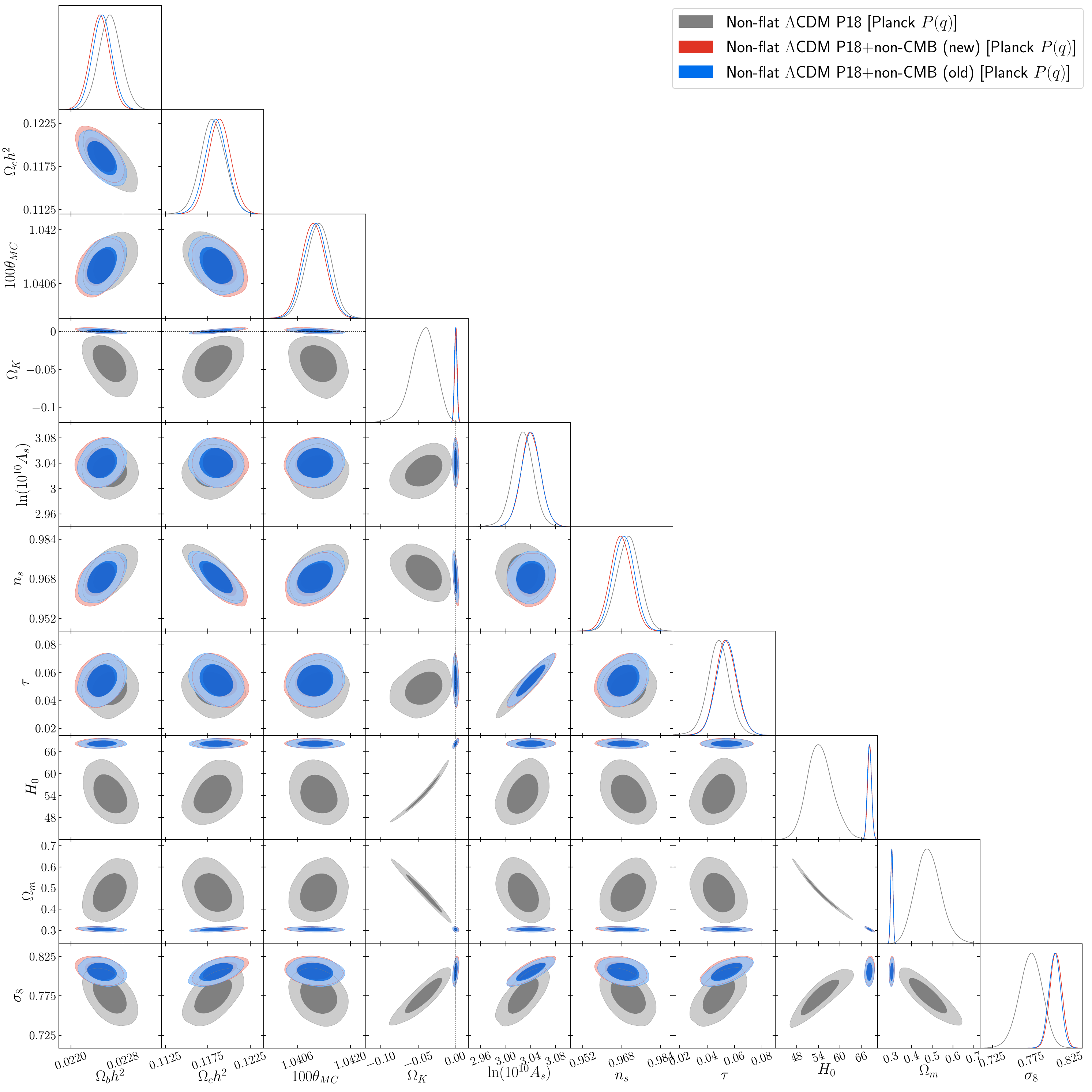}}
\caption{Likelihood distributions of non-flat $\Lambda$CDM model [Planck $P(q)$] parameters
         favored by P18 and non-CMB data. Non-CMB (old) denotes the compilation of non-CMB data
	 sets used in \cite{deCruzPerez:2022hfr} while non-CMB (new) 
	 denotes the new compilation of non-CMB data sets of Sec.\ \ref{sec:Non-CMB Data}.
      Likelihood results for P18 data are shown for comparison. 
}
\label{fig:Planck_Pq_p18ncmb}
\end{figure*}

\begin{figure*}[htbp]
\centering
\mbox{\includegraphics[width=170mm]{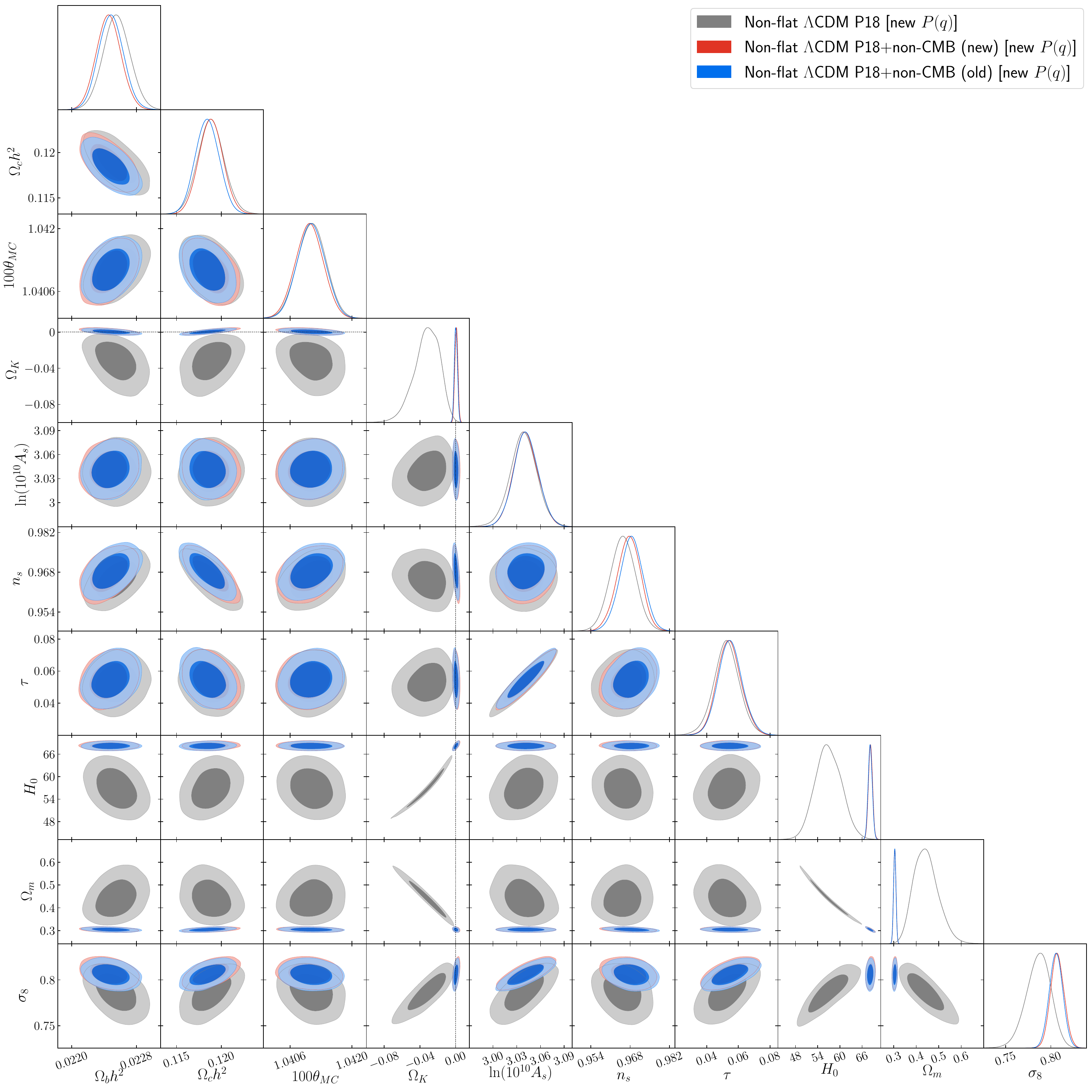}}
\caption{Likelihood distributions of non-flat $\Lambda$CDM model [new $P(q)$] parameters
         favored by P18 and non-CMB data. Non-CMB (old) denotes the compilation of non-CMB data
	 sets used in \cite{deCruzPerez:2022hfr} while non-CMB (new) 
	 denotes the new compilation of non-CMB data sets of Sec.\ \ref{sec:Non-CMB Data}.
     Likelihood results for P18 data are shown for comparison. 
}
\label{fig:new_Pq_p18ncmb}
\end{figure*}

In the non-flat geometry cases there are slightly larger differences between the mean values obtained with P18+non-CMB (old) and P18+non-CMB (new) data. Table \ref{tab:results_Planck_Pq} [\ref{tab:results_new_Pq}] and Fig.\ \ref{fig:Planck_Pq_p18ncmb} [\ref{fig:new_Pq_p18ncmb}] show results for the seven-parameter non-flat $\Lambda$CDM Planck [new] $P(q)$ cosmological model, obtained from P18+non-CMB (old) and P18+non-CMB (new) data. The mean values of $\Omega_b{h^2}$, $\Omega_c{h^2}$, and $n_s$ differ by +0.19$\sigma$ [+0.14$\sigma$], $-0.22\sigma$ [$-0.22\sigma$], and +0.16$\sigma$ [+0.15$\sigma$] respectively. For $\Omega_k$, when  P18+non-CMB (old) data are analyzed we find $\Omega_k =0.0004\pm 0.0017$ [$\Omega_k = 0.0003\pm 0.0017$], a 0.24$\sigma$ [0.18$\sigma$] favoring of open geometry, and differing from the P18+non-CMB (new) value $\Omega_k = 0.0009\pm 0.0017$ [$\Omega_k=0.0008\pm 0.0017$] by $-0.21\sigma$ [$-0.21\sigma$].  We do not find significant changes in the error bars, the largest changes being those for $n_s$ with an increase of 2.3\% [2.4\%], and for $\Omega_m$ with a decrease of $-1.9$\% [$-1.9$\%]. 

\begin{figure*}[htbp]
\centering
\mbox{\includegraphics[width=170mm]{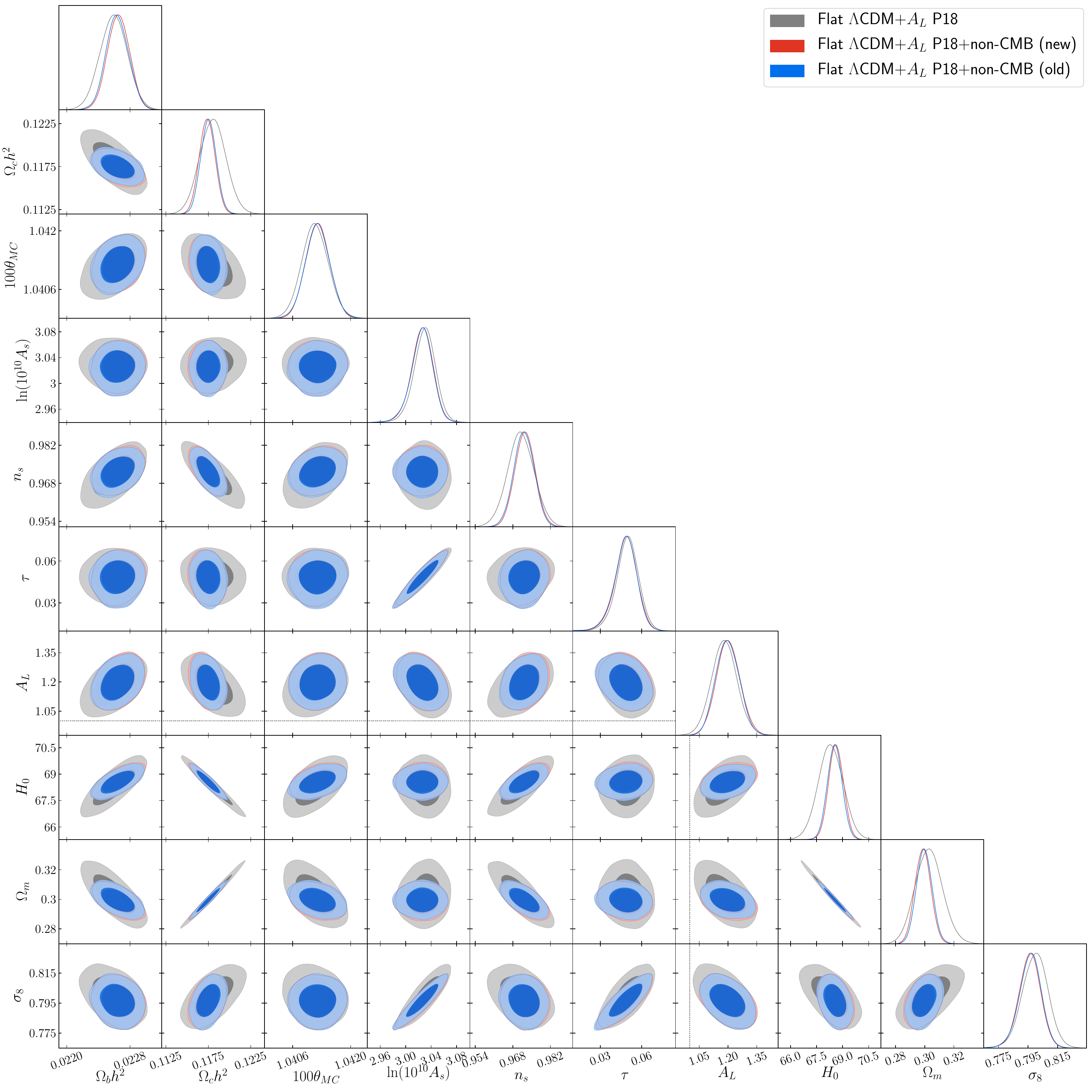}}
\caption{Likelihood distributions of flat $\Lambda$CDM$+A_L$ model parameters favored
      by P18 and non-CMB data. Non-CMB (old) denotes the compilation of non-CMB data
	 sets used in \cite{deCruzPerez:2022hfr} while non-CMB (new) 
	 denotes the new compilation of non-CMB data sets of Sec.\ \ref{sec:Non-CMB Data}.
       Likelihood results for P18 data
      are shown for comparison. 
}
\label{fig:flat_LCDM_p18ncmb_AL}
\end{figure*}

The lensing parameter $A_L$ does not play a significant role at low redshift \cite{deCruzPerez:2022hfr}, but there are small differences in the results obtained with P18+non-CMB (old) and P18+non-CMB (new) data when $A_L$ is allowed to vary. For the flat $\Lambda$CDM+$A_L$ model, comparing results from P18+non-CMB (old) data and from P18+non-CMB (new) data (see Table \ref{tab:results_flat_LCDM_AL} and Fig.\ \ref{fig:flat_LCDM_p18ncmb_AL}) we do not find significant changes in primary parameter values (those of $\Omega_{c}h^2$ differ at $-0.12\sigma$) or in derived parameter values (with changes in $H_0$ and $\Omega_m$ being $+0.12\sigma$ and $-0.12\sigma$ respectively). In addition, the error bars are practically unchanged. The value of the lensing parameter obtained from P18+non-CMB (old) data is $A_L = 1.201\pm 0.061$, differing only +0.04$\sigma$ from $A_L = 1.198\pm 0.060$ obtained from P18-non-CMB (new) data. 

\begin{figure*}[htbp]
\centering
\mbox{\includegraphics[width=170mm]{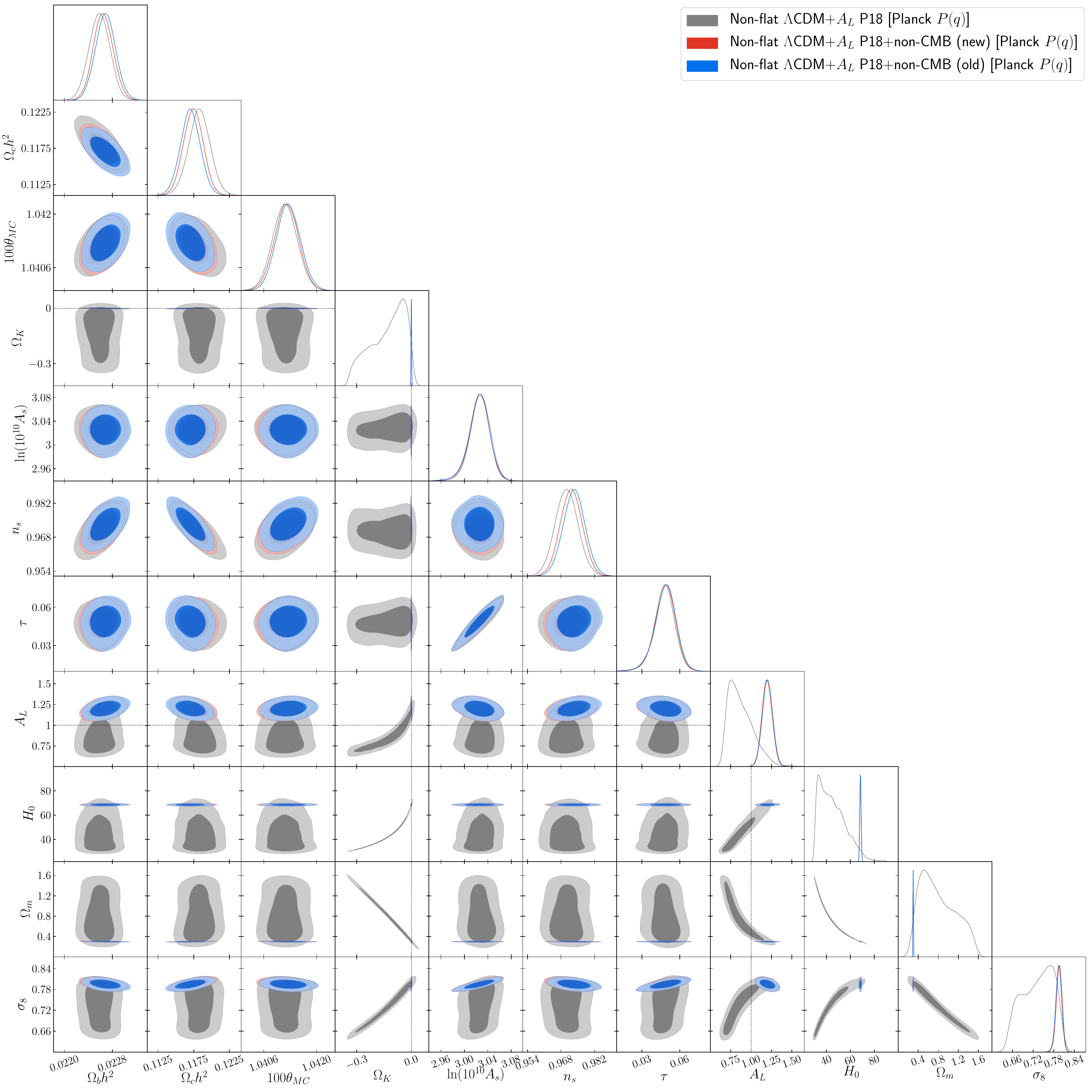}}
\caption{Likelihood distributions of non-flat $\Lambda$CDM$+A_L$ model [Planck $P(q)$]
         parameters favored by P18 and non-CMB data. Non-CMB (old) denotes the compilation of non-CMB data
	 sets used in \cite{deCruzPerez:2022hfr} while non-CMB (new) 
	 denotes the new compilation of non-CMB data sets of Sec.\ \ref{sec:Non-CMB Data}.
          Likelihood results for P18 data are shown for comparison. 
}
\label{fig:Planck_Pq_p18ncmb_AL}
\end{figure*}

\begin{figure*}[htbp]
\centering
\mbox{\includegraphics[width=170mm]{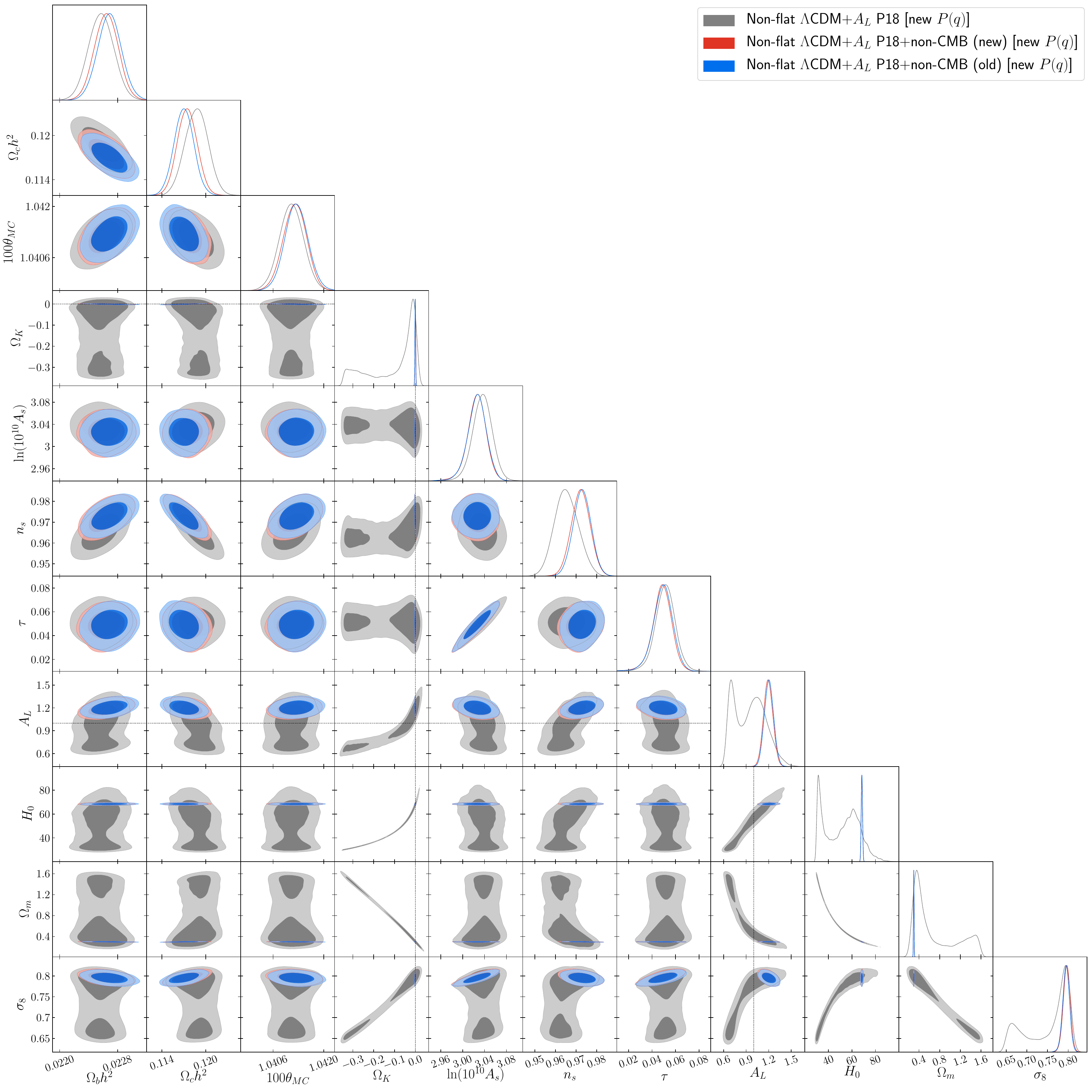}}
\caption{Likelihood distributions of non-flat $\Lambda$CDM$+A_L$ model [new $P(q)$]
         parameters favored by P18 and non-CMB data. Non-CMB (old) denotes the compilation of non-CMB data
	 sets used in \cite{deCruzPerez:2022hfr} while non-CMB (new) 
	 denotes the new compilation of non-CMB data sets of Sec.\ \ref{sec:Non-CMB Data}.
        Likelihood results for P18 data are shown for comparison. 
}
\label{fig:new_Pq_p18ncmb_AL}
\end{figure*}

When the curvature parameter is allowed to vary, the results also do not change much when the non-CMB data set is updated. For the non-flat $\Lambda$CDM+$A_L$ Planck [new] $P(q)$ model (see Table \ref{tab:results_Planck_Pq_AL} [\ref{tab:results_new_Pq_AL}] and Fig.\ \ref{fig:Planck_Pq_p18ncmb_AL} [\ref{fig:new_Pq_p18ncmb_AL}]), for the primary parameters obtained from P18+non-CMB (old) and P18+non-CMB (new) data we find differences in $\Omega_{c}h^2$ and $n_s$ at $-0.20\sigma$ [$-0.26\sigma$] and +0.18$\sigma$ [+0.11$\sigma$] with almost no changes in the size of the error bars. For $\Omega_k$ and $A_L$ the differences are  at $-0.21\sigma$ [$-0.21\sigma$] and +0.080$\sigma$ [+0.12$\sigma$] respectively.

In summary, the combination of P18 CMB data with either non-CMB (old) or non-CMB (new) data give almost identical cosmological parameter results. While it is reassuring that updated non-CMB data do not significantly affect the P18+non-CMB data constraints, this almost certainly is because P18 data have much more weight than current non-CMB data. 

The consistency [see Eqs.\ \eqref{eq:Tension_estimator_1} and \eqref{eq:Tension_estimator_2_sigma}] between the results obtained using P18 data and using either non-CMB (old) data or non-CMB (new) data are quite similar, as can be seen in Table \ref{tab:consistency_LCDM}, 
with a slight exception in the non-flat Planck $P(q)$ model where the discordance between P18 and non-CMB data decreases from $\sigma=3.005$ for non-CMB data (old) to $\sigma=2.704$ for non-CMB (new) data and so there now is a lower level of tension than was found using older data, \cite{DiValentino:2019qzk}. 

No significant differences are observed between the P18+non-CMB (old) and P18+non-CMB (new) data values of $\Delta$AIC and $\Delta$DIC obtained when flat $\Lambda$CDM is compared with the non-flat models (see Tables \ref{tab:results_Planck_Pq} and \ref{tab:results_new_Pq}) and the $A_L$-varying models (Tables \ref{tab:results_flat_LCDM_AL}, \ref{tab:results_Planck_Pq_AL} and \ref{tab:results_new_Pq_AL}).


\begin{turnpage}
\begin{table*}
\caption{Mean and 68\% confidence limits of flat $\Lambda\textrm{CDM}$+$A_L$ model parameters
        constrained by P18, P18+lensing, P18+non-CMB (old), P18+non-CMB (new), P18+lensing+non-CMB (old), and P18+lensing+non-CMB (new) data sets.
        $H_0$ has units of km s$^{-1}$ Mpc$^{-1}$. 
}
\begin{ruledtabular}
\begin{tabular}{lcccccc}
\\[-1mm]                       & \multicolumn{6}{c}{Flat $\Lambda$CDM+$A_L$ models}        \\[+1mm]
\cline{2-7}\\[-1mm]
Parameter                      & P18                   & P18+lensing           &  P18+non-CMB (old)     & P18+non-CMB (new)      &  P18+lensing+non-CMB (old)  & P18+lensing+non-CMB (new) \\[+1mm]
\hline \\[-1mm]
$\Omega_b h^2$                 & $0.02259 \pm 0.00017$ & $0.02251 \pm 0.00017$ & $0.02265 \pm 0.00014$  & $0.02264 \pm 0.00014$  & $0.02258 \pm 0.00014$  & $0.02257 \pm 0.00014$   \\[+1mm]
$\Omega_c h^2$                 & $0.1180 \pm 0.0015$   & $0.1183 \pm 0.0015$   & $0.11736 \pm 0.00092$  & $0.11752 \pm 0.00091$  & $0.11747 \pm 0.00091$  & $0.11759 \pm 0.00090$   \\[+1mm]
$100\theta_\textrm{MC}$        & $1.04114 \pm 0.00032$ & $1.04109 \pm 0.00032$ & $1.04120 \pm 0.00029$  & $1.04120 \pm 0.00029$  & $1.04118 \pm 0.00029$  & $1.04116 \pm 0.00029$   \\[+1mm]
$\tau$                         & $0.0496 \pm 0.0082$   & $0.0487 \pm 0.0087$   & $0.0484 \pm 0.0083$    & $0.0480 \pm 0.0084$    & $0.0476 \pm 0.0085$    & $0.0477 \pm 0.0086$     \\[+1mm]
$n_s$                          & $0.9710 \pm 0.0050$   & $0.9695 \pm 0.0048$   & $0.9726 \pm 0.0038$    & $0.9722 \pm 0.0038$    & $0.9715 \pm 0.0038$    & $0.9711 \pm 0.0038$     \\[+1mm]
$\ln(10^{10} A_s)$             & $3.030 \pm 0.017$     & $3.028 \pm 0.018$     & $3.026 \pm 0.017$      & $3.025 \pm 0.017$      & $3.023 \pm 0.018$      & $3.024 \pm 0.018$       \\[+1mm]
$A_{L}$                        & $1.181 \pm 0.067$     & $1.073 \pm 0.041$     & $1.201 \pm 0.061$      & $1.198 \pm 0.060$      & $1.089 \pm 0.035$      & $1.087 \pm 0.035$       \\[+1mm]
  \hline \\[-1mm]
$H_0$                          & $68.31 \pm 0.71$      & $68.14 \pm 0.69$      & $68.62 \pm 0.43$       & $68.55 \pm 0.42$       & $68.52 \pm 0.42$       & $68.45 \pm 0.42$       \\[+1mm]
$\Omega_m$                     & $0.3029 \pm 0.0093$   & $0.3048 \pm 0.0091$   & $0.2988 \pm 0.0054$    & $0.2997 \pm 0.0054$    & $0.2998 \pm 0.0053$    & $0.3005 \pm 0.0053$    \\[+1mm]
$\sigma_8$                     & $0.7997 \pm 0.0088$   & $0.7996 \pm 0.0089$   & $0.7961 \pm 0.0074$    & $0.7964 \pm 0.0075$    & $0.7955 \pm 0.0075$    & $0.7961 \pm 0.0075$    \\[+1mm]
 \hline \\[-1mm]
$\chi_{\textrm{min}}^2$ (Total)& $2756.12$             & $2771.24$             & $3865.90$              & $4227.27$              & $3881.55$              & $4242.61$              \\[+1mm]
$\Delta\chi_{\textrm{min}}^2$  & $-9.68$               & $-3.47$               & $-13.45$               & $-12.97$               & $-6.86$                & $-6.65$                \\[+1mm]
$\textrm{DIC}$                 & $2812.41$             & $2825.53$             & $3922.11$              & $4283.86$              & $3935.15$              & $4297.19$              \\[+1mm]
$\Delta\textrm{DIC}$           & $-5.52$               & $-0.92$               & $-8.91$                & $-8.47$                & $-5.55$                & $-4.01$                \\[+1mm]
$\textrm{AIC}$                 & $2812.12$             & $2827.24$             & $3921.90$              & $4283.27$              & $3937.55$              & $4298.61$              \\[+1mm]
$\Delta\textrm{AIC}$           & $-7.68$               & $-1.47$               & $-11.45$               & $-10.97$               & $-4.86$                & $-4.65$                 \\[+1mm]
\end{tabular}!
\\[+1mm]
\begin{flushleft}
\end{flushleft}
\end{ruledtabular}
\label{tab:results_flat_LCDM_AL}
\end{table*}
\end{turnpage}

\begin{turnpage}
\begin{table*}
\caption{Mean and 68\% confidence limits of non-flat $\Lambda\textrm{CDM}$+$A_L$ [Planck $P(q)$] model parameters
        constrained by P18, P18+lensing, P18+non-CMB (old), P18+non-CMB (new), P18+lensing+non-CMB (old), and P18+lensing+non-CMB (new) data sets.
        $H_0$ has units of km s$^{-1}$ Mpc$^{-1}$. 
}
\begin{ruledtabular}
\begin{tabular}{lcccccc}
\\[-1mm]                       & \multicolumn{6}{c}{Non-flat $\Lambda$CDM+$A_L$ [Planck $P(q)$] models}        \\[+1mm]
\cline{2-7}\\[-1mm]
Parameter                      & P18                   & P18+lensing             &  P18+non-CMB (old)     & P18+non-CMB (new)   &  P18+lensing+non-CMB (old)  & P18+lensing+non-CMB (new) \\[+1mm]
\hline \\[-1mm]
$\Omega_b h^2$                 & $0.02258 \pm 0.00017$ & $0.02251 \pm 0.00017$   & $0.02268 \pm 0.00017$  & $0.02265 \pm 0.00016$ & $0.02259 \pm 0.00016$  & $0.02255 \pm 0.00015$   \\[+1mm]
$\Omega_c h^2$                 & $0.1183 \pm 0.0015$   & $0.1183 \pm 0.0015$     & $0.1170 \pm 0.0014$    & $0.1174 \pm 0.0014$   & $0.1173 \pm 0.0014$    & $0.1178 \pm 0.0013$     \\[+1mm]
$100\theta_\textrm{MC}$        & $1.04116 \pm 0.00033$ & $1.04110 \pm 0.00032$   & $1.04125 \pm 0.00032$  & $1.04120 \pm 0.00032$ & $1.04118 \pm 0.00032$  & $1.04113 \pm 0.00032$   \\[+1mm]
$\tau$                         & $0.0478 \pm 0.0081$   & $0.0489 \pm 0.0085$     & $0.0485 \pm 0.0087$    & $0.0482 \pm 0.0085$   & $0.0479 \pm 0.0085$    & $0.0480 \pm 0.0085$    \\[+1mm]
$\Omega_k$                     & $-0.130 \pm 0.095$    & $-0.005 \pm 0.027$      & $-0.0006 \pm 0.0017$   & $-0.0001 \pm 0.0017$  & $-0.0002 \pm 0.0017$   & $0.0004 \pm 0.0017$     \\[+1mm]
                               & ($-0.130^{+0.13}_{-0.055}$)    & ($-0.0048^{+0.029}_{-0.0083}$)      &    &   &   &      \\[+1mm]
$n_s$                          & $0.9704 \pm 0.0048$   & $0.9696 \pm 0.0049$     & $0.9735 \pm 0.0046$    & $0.9723 \pm 0.0046$   & $0.9718 \pm 0.0045$    & $0.9705 \pm 0.0044$     \\[+1mm]
$\ln(10^{10} A_s)$             & $3.027 \pm 0.017$     & $3.028 \pm 0.018$       & $3.025 \pm 0.018$      & $3.025 \pm 0.017$     & $3.024 \pm 0.017$      & $3.025 \pm 0.017$      \\[+1mm]
$A_{L}$                        & $0.88 \pm 0.15$       & $1.09 \pm 0.16$         & $1.203 \pm 0.062$      & $1.196 \pm 0.062$     & $1.090 \pm 0.036$      & $1.084 \pm 0.035$      \\[+1mm]
                               & ($0.876^{+0.088}_{-0.19}$)     &  ($1.09^{+0.15}_{-0.17}$)             &      &       &       &      \\[+1mm]
  \hline \\[-1mm]
$H_0$                          & $45 \pm 11$ ($45^{+5}_{-10}$)          & $69 \pm 11$             & $68.48 \pm 0.56$       & $68.52 \pm 0.56$      & $68.49 \pm 0.56$       & $68.53 \pm 0.56$       \\[+1mm]
$\Omega_m$                     & $0.80 \pm 0.35$       & $0.32 \pm 0.11$         & $0.2994 \pm 0.0055$    & $0.2998 \pm 0.0055$   & $0.2998 \pm 0.0055$    & $0.3003 \pm 0.0054$    \\[+1mm]
$\sigma_8$                     & $0.733 \pm 0.045$     & $0.796 \pm 0.016$       & $0.7946 \pm 0.0088$    & $0.7961 \pm 0.0086$   & $0.7952 \pm 0.0085$    & $0.7975 \pm 0.0085$    \\[+1mm]
 \hline \\[-1mm]
$\chi_{\textrm{min}}^2$ (Total)& $2754.99$             & $2771.14$               & $3865.53$              & $4227.07$             & $3881.37$              & $4242.22$              \\[+1mm]
$\Delta\chi_{\textrm{min}}^2$  & $-10.81$              & $-3.57$                 & $-13.82$               & $-13.17$              & $-7.04$                & $-7.04$                \\[+1mm]
$\textrm{DIC}$                 & $2811.63$             & $2827.14$               & $3924.07$              & $4285.58$             & $3936.85$              & $4298.73$              \\[+1mm]
$\Delta\textrm{DIC}$           & $-6.30$               & $0.69$                  & $-6.95$                & $-6.75$               & $-3.85$                & $-2.47$                \\[+1mm]
$\textrm{AIC}$                 & $2812.99$             & $2829.14$               & $3923.53$              & $4285.07$             & $3939.37$              & $4300.22$              \\[+1mm]
$\Delta\textrm{AIC}$           & $-6.81$               & $0.43$                  & $-9.82$                & $-9.17$               & $-3.04$                & $-3.04$                \\[+1mm]
\end{tabular}
\\[+1mm]
\begin{flushleft}
\end{flushleft}
\end{ruledtabular}
\label{tab:results_Planck_Pq_AL}
\end{table*}
\end{turnpage}

\begin{turnpage}
\begin{table*}
\caption{Mean and 68\% confidence limits of non-flat $\Lambda\textrm{CDM}$+$A_L$ [new $P(q)$] model parameters
        constrained by P18, P18+lensing, P18+non-CMB (old), P18+non-CMB (new), P18+lensing+non-CMB (old), and P18+lensing+non-CMB (new) data sets.
        $H_0$ has units of km s$^{-1}$ Mpc$^{-1}$. 
}
\begin{ruledtabular}
\begin{tabular}{lcccccc}
\\[-1mm]                       & \multicolumn{6}{c}{Non-flat $\Lambda$CDM+$A_L$ [new $P(q)$] models}        \\[+1mm]
\cline{2-7}\\[-1mm]
Parameter                      & P18                   & P18+lensing             &  P18+non-CMB (old)     & P18+non-CMB (new)     &  P18+lensing+non-CMB (old)     & P18+lensing+non-CMB (new) \\[+1mm]
\hline \\[-1mm]
$\Omega_b h^2$                 & $0.02257 \pm 0.00017$ & $0.02252 \pm 0.00017$   & $0.02269 \pm 0.00016$  & $0.02265 \pm 0.00016$  & $0.02260 \pm 0.00016$  & $0.02256 \pm 0.00016$     \\[+1mm]
$\Omega_c h^2$                 & $0.1187 \pm 0.0016$   & $0.1183 \pm 0.0015$     & $0.1170 \pm 0.0013$    & $0.1175 \pm 0.0014$    & $0.1174 \pm 0.0013$    & $0.1178 \pm 0.0014$       \\[+1mm]
$100\theta_\textrm{MC}$        & $1.04111 \pm 0.00033$ & $1.04108 \pm 0.00032$   & $1.04125 \pm 0.00032$  & $1.04120 \pm 0.00032$  & $1.04118 \pm 0.00032$  & $1.04113 \pm 0.00032$     \\[+1mm]
$\tau$                         & $0.0512 \pm 0.0086$   & $0.0495 \pm 0.0093$     & $0.0490 \pm 0.0086$    & $0.0484 \pm 0.0085$    & $0.0486 \pm 0.0086$    & $0.0481 \pm 0.0082$       \\[+1mm]
$\Omega_k$                     & $-0.10 \pm 0.11$      & $0.003 \pm 0.016$       & $-0.0006 \pm 0.0017$   & $-0.0001 \pm 0.0017$   & $-0.0002 \pm 0.0017$   & $0.0004 \pm 0.0017$       \\[+1mm]
                               & ($-0.10^{+0.12}_{-0.23}$)    & ($0.003^{+0.019}_{-0.010}$)      &    &   &   &      \\[+1mm]
$n_s$                          & $0.9654 \pm 0.0057$   & $0.9688 \pm 0.0053$     & $0.9730 \pm 0.0043$    & $0.9723 \pm 0.0045$    & $0.9713 \pm 0.0042$    & $0.9707 \pm 0.0046$       \\[+1mm]
$\ln(10^{10} A_s)$             & $3.036 \pm 0.018$     & $3.030 \pm 0.019$       & $3.026 \pm 0.018$      & $3.026 \pm 0.017$      & $3.025 \pm 0.017$      & $3.025 \pm 0.017$         \\[+1mm]
$A_{L}$                        & $0.94 \pm 0.20$       & $1.13 \pm 0.15$         & $1.204 \pm 0.061$      & $1.194 \pm 0.061$      & $1.088 \pm 0.035$      & $1.084 \pm 0.034$         \\[+1mm]
                               & ($0.94^{+0.22}_{-0.30}$)     &  ($1.13^{+0.12}_{-0.17}$)             &      &       &       &      \\[+1mm]
  \hline \\[-1mm]
$H_0$                          & $51 \pm 14$           & $72.0 \pm 9.2$ ($72.0^{+7.7}_{-11}$)         & $68.47 \pm 0.56$       & $68.52 \pm 0.55$       & $68.48 \pm 0.56$       & $68.53 \pm 0.55$          \\[+1mm]
$\Omega_m$                     & $0.70 \pm 0.43$       & $0.287 \pm 0.076$       & $0.2994 \pm 0.0056$    & $0.2999 \pm 0.0054$    & $0.2999 \pm 0.0055$    & $0.3004 \pm 0.0054$       \\[+1mm]
$\sigma_8$                     & $0.752 \pm 0.052$     & $0.801 \pm 0.011$       & $0.7948 \pm 0.0083$    & $0.7964 \pm 0.0084$    & $0.7956 \pm 0.0082$    & $0.7974 \pm 0.0084$       \\[+1mm]
 \hline \\[-1mm]
$\chi_{\textrm{min}}^2$ (Total)& $2756.33$             & $2770.45$               & $3865.41$              & $4227.11$              & $3880.69$              & $4242.01$                 \\[+1mm]
$\Delta\chi_{\textrm{min}}^2$  & $-9.47$               & $-4.26$                 & $-13.94$               & $-13.13$               & $-7.72$                & $-7.25$                \\[+1mm]
$\textrm{DIC}$                 & $2814.83$             & $2827.29$               & $3923.86$              & $4285.29$              & $3937.52$              & $4298.75$                 \\[+1mm]
$\Delta\textrm{DIC}$           & $-3.10$               & $0.84$                  & $-7.16$                & $-7.04$                & $-3.18$                & $-2.45$                   \\[+1mm]
$\textrm{AIC}$                 & $2814.33$             & $2828.45$               & $3923.41$              & $4285.11$              & $3938.69$              & $4300.01$                 \\[+1mm]
$\Delta\textrm{AIC}$           & $-5.47$               & $-0.26$                 & $-9.94$                & $-9.13$                & $-3.72$                & $-3.25$                   \\[+1mm]
\end{tabular}
\\[+1mm]
\begin{flushleft}
\end{flushleft}
\end{ruledtabular}
\label{tab:results_new_Pq_AL}
\end{table*}
\end{turnpage}



\begin{table*}
\caption{Consistency check parameter $\log_{10} \mathcal{I}$ and tension parameters $\sigma$ and $p$ for P18 vs.\ non-CMB data sets and P18+lensing vs.\ non-CMB data sets in the $\Lambda\textrm{CDM}$ models.
}
\begin{ruledtabular}
\begin{tabular}{lcccc}
\\[-1mm]                         & \multicolumn{4}{c}{Flat $\Lambda$CDM model}        \\[+1mm]
\cline{2-5}\\[-1mm]
  Data                           & P18 vs non-CMB (old)  & P18 vs non-CMB (new)  &  P18+lensing vs non-CMB (old) & P18+lensing vs non-CMB (new)    \\[+1mm]
 \hline \\[-1mm]
  $\log_{10} \mathcal{I}$        & $0.296$               & $0.805$               &  $0.029$                &  $0.730$  \\[+1mm]
  $\sigma$                       & $1.749$               & $1.152$               &  $1.747$                &  $1.209$  \\[+1mm]
  $p$ (\%)                       & $8.03$                & $24.9$                &  $8.06$                 &  $22.7$  \\[+1mm]
\hline \\[-1mm]
                                 & \multicolumn{4}{c}{Flat $\Lambda$CDM+$A_L$ model}        \\[+1mm]
\cline{2-5}\\[-1mm]
  Data                           & P18 vs non-CMB (old)  & P18 vs non-CMB (new)  &  P18+lensing vs non-CMB (old) & P18+lensing vs non-CMB (new)    \\[+1mm]
 \hline \\[-1mm]
   $\log_{10} \mathcal{I}$       & $1.033$               & $1.446$               &  $1.033$                &  $1.400$  \\[+1mm]
  $\sigma$                       & $0.835$               & $0.164$               &  $0.774$                &  $0.0872$  \\[+1mm]
  $p$ (\%)                       & $40.4$                & $87.0$                &  $43.9$                 &  $93.1$  \\[+1mm]
\hline \\[-1mm]
                                 & \multicolumn{4}{c}{Non-flat $\Lambda$CDM model [Planck $P(q)$]}        \\[+1mm]
\cline{2-5}\\[-1mm]
  Data                           & P18 vs non-CMB (old)  & P18 vs non-CMB (new)  &  P18+lensing vs non-CMB (old) & P18+lensing vs non-CMB (new)    \\[+1mm]
 \hline \\[-1mm]
   $\log_{10} \mathcal{I}$       & $-1.263$              & $-0.796$              &  $0.297$                &  $0.711$  \\[+1mm]
  $\sigma$                       & $3.005$               & $2.704$               &  $1.837$                &  $1.555$  \\[+1mm]
  $p$ (\%)                       & $0.265$               & $0.687$               &  $6.62$                 &  $12.0$  \\[+1mm]
\hline \\[-1mm]
                                 & \multicolumn{4}{c}{Non-flat $\Lambda$CDM+$A_L$ model [Planck $P(q)$]}        \\[+1mm]
\cline{2-5}\\[-1mm]
  Data                           & P18 vs non-CMB (old)  & P18 vs non-CMB (new)  &  P18+lensing vs non-CMB (old) & P18+lensing vs non-CMB (new)    \\[+1mm]
 \hline \\[-1mm]
   $\log_{10} \mathcal{I}$       & $0.972$               & $1.210$               &  $1.641$                &  $1.719$  \\[+1mm]
  $\sigma$                       & $0.793$               & $0.595$               &  $0.516$                &  $0.241$  \\[+1mm]
  $p$ (\%)                       & $42.8$                & $55.2$                &  $60.6$                 &  $80.9$  \\[+1mm]
\hline \\[-1mm]
                                 & \multicolumn{4}{c}{Non-flat $\Lambda$CDM model [new $P(q)$]}        \\[+1mm]
\cline{2-5}\\[-1mm]
  Data                           & P18 vs non-CMB (old)  & P18 vs non-CMB (new)  &  P18+lensing vs non-CMB (old) & P18+lensing vs non-CMB (new)    \\[+1mm]
 \hline \\[-1mm]
   $\log_{10} \mathcal{I}$       & $-0.806$              & $-0.391$              &  $0.143$                &  $0.775$  \\[+1mm]
  $\sigma$                       & $2.577$               & $2.308$               &  $1.886$                &  $1.544$  \\[+1mm]
  $p$ (\%)                       & $0.996$               & $2.10$                &  $5.93$                 &  $12.3$  \\[+1mm]
\hline \\[-1mm]
                                 & \multicolumn{4}{c}{Non-flat $\Lambda$CDM+$A_L$ model [new $P(q)$]}        \\[+1mm]
\cline{2-5}\\[-1mm]
  Data                           & P18 vs non-CMB (old)  & P18 vs non-CMB (new)  &  P18+lensing vs non-CMB (old) & P18+lensing vs non-CMB (new)    \\[+1mm]
 \hline \\[-1mm]
   $\log_{10} \mathcal{I}$       & $1.798$              & $2.107$                &  $1.500$                &  $1.887$  \\[+1mm]
  $\sigma$                       & $0.402$              & $0.289$                &  $0.573$                &  $0.312$  \\[+1mm]
  $p$ (\%)                       & $68.7$               & $77.2$                 &  $56.7$                 &  $75.5$  \\[+1mm]
\end{tabular}
\\[+1mm]
\end{ruledtabular}
\label{tab:consistency_LCDM}
\end{table*}


\subsubsection{P18+lensing+non-CMB (old) vs. P18+lensing+non-CMB (new) cosmological parameter constraints} 
\label{sec:P18+lensing+non-CMB (old) vs. P18+lensing+non-CMB (new)}

In this subsubsection we check the impact on the cosmological parameter constraints when we move from P18+lensing+non-CMB (old) data to P18+lensing+non-CMB (new) data. 
As we shall see, and as expected, the differences presented in this subsubsection are even smaller than the ones presented in the previous one for P18+non-CMB data. 

\begin{figure*}[htbp]
\centering
\mbox{\includegraphics[width=170mm]{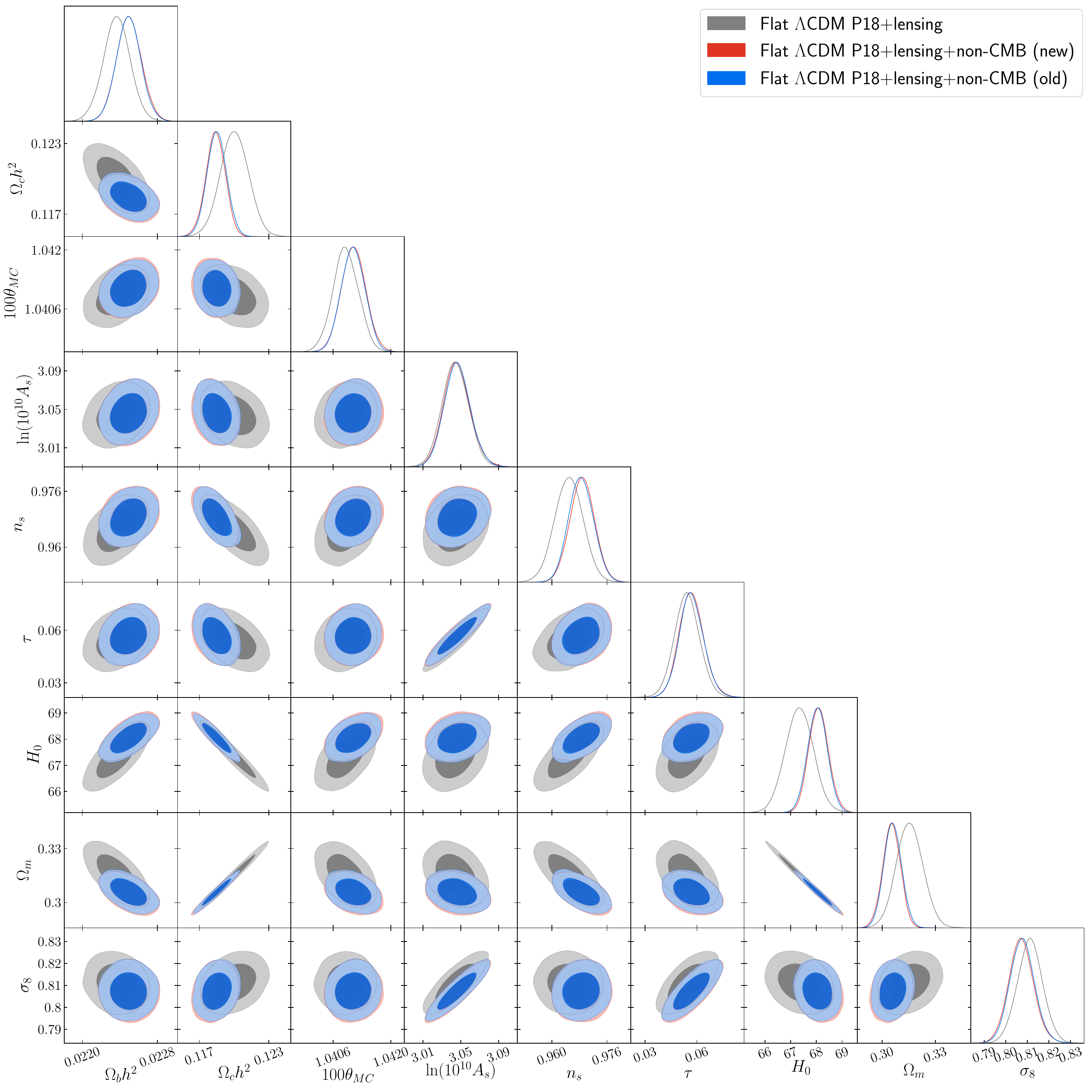}}
\caption{Likelihood distributions of flat $\Lambda$CDM model parameters
         favored by P18+lensing and non-CMB data. Non-CMB (old) denotes the compilation of non-CMB data
	 sets used in \cite{deCruzPerez:2022hfr} while non-CMB (new) 
	 denotes the new compilation of non-CMB data sets of Sec.\ \ref{sec:Non-CMB Data}.
        Likelihood results for P18+lensing data are shown for comparison. 
}
\label{fig:flat_LCDM_p18lenncmb}
\end{figure*}

P18+lensing+non-CMB (old) and P18+lensing+non-CMB (new) cosmological parameter constraints for the six-parameter flat $\Lambda$CDM model are in the upper half of Table \ref{tab:results_flat_LCDM} and Fig.\ \ref{fig:flat_LCDM_p18lenncmb}. For primary parameters the largest differences are for $\Omega_{c}h^2$ and $n_s$, at $-0.093\sigma$ and +0.059$\sigma$ respectively. Derived parameters $H_0$, $\Omega_m$, and $\sigma_8$ differ at +0.074$\sigma$, $-0.085\sigma$, and $-0.061\sigma$. The error bars obtained with P18+lensing+non-CMB (old) data and P18+lensing+non-CMB (new) data are very similar, the largest differences affect primary parameter $100\theta_{\textrm{MC}}$ ($-3.57$\%) and derived parameter $\sigma_8$ ($-1.75$\%). 

\begin{figure*}[htbp]
\centering
\mbox{\includegraphics[width=170mm]{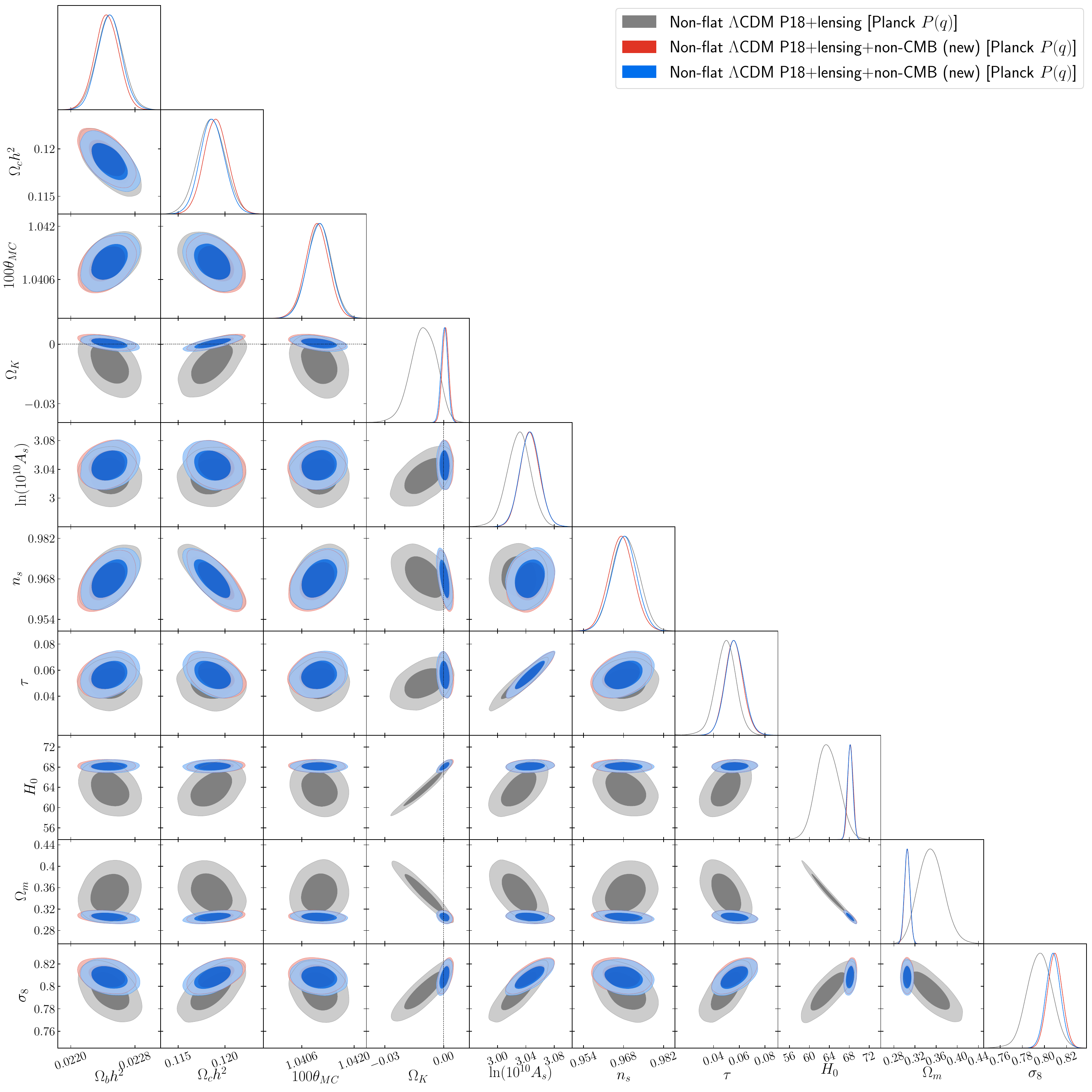}}
\caption{Likelihood distributions of non-flat $\Lambda$CDM model [Planck $P(q)$] parameters
         favored by P18+lensing and non-CMB data. Non-CMB (old) denotes the compilation of non-CMB data
	 sets used in \cite{deCruzPerez:2022hfr} while non-CMB (new) 
	 denotes the new compilation of non-CMB data sets of Sec.\ \ref{sec:Non-CMB Data}.
        Likelihood results for P18+lensing data are shown for comparison. 
}
\label{fig:Planck_Pq_p18lenncmb}
\end{figure*}

\begin{figure*}[htbp]
\centering
\mbox{\includegraphics[width=170mm]{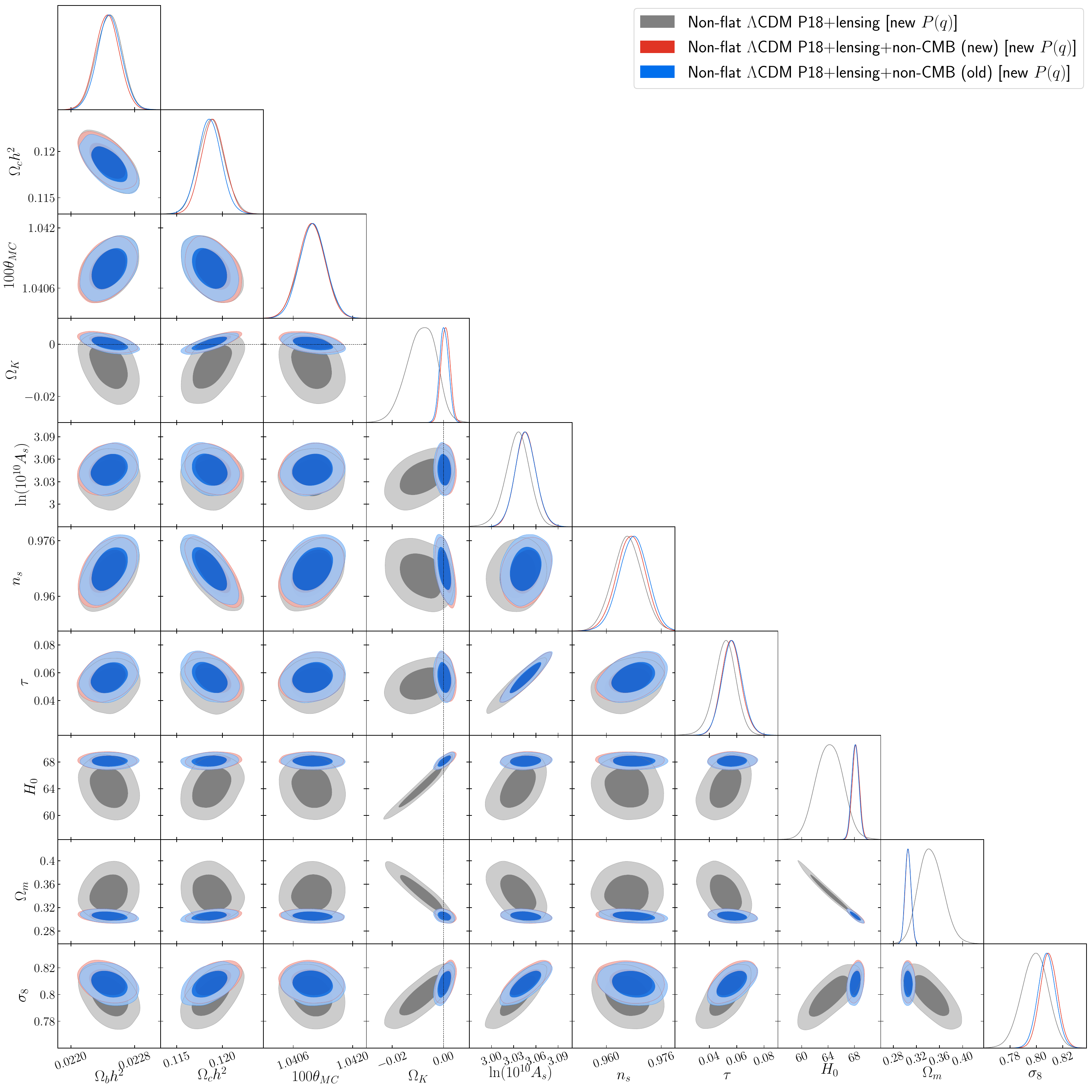}}
\caption{Likelihood distributions of non-flat $\Lambda$CDM model [new $P(q)$] parameters
         favored by P18+lensing and non-CMB data. Non-CMB (old) denotes the compilation of non-CMB data
	 sets used in \cite{deCruzPerez:2022hfr} while non-CMB (new) 
	 denotes the new compilation of non-CMB data sets of Sec.\ \ref{sec:Non-CMB Data}.
        Likelihood results for P18+lensing data are shown for comparison. 
}
\label{fig:new_Pq_p18lenncmb}
\end{figure*}

As for the non-flat spatial geometry models, the results obtained with P18+lensing+non-CMB (old) and P18+lensing+non-CMB (new) data, for the seven-parameter $\Lambda$CDM Planck [new] $P(q)$ model are in the upper half of Table  \ref{tab:results_Planck_Pq} [\ref{tab:results_new_Pq}] and Fig.\ \ref{fig:Planck_Pq_p18lenncmb} [\ref{fig:new_Pq_p18lenncmb}]. The primary parameters $\Omega_b{h^2}$, $\Omega_c{h^2}$, and $n_s$, differ at +0.19$\sigma$ [+0.094$\sigma$], $-0.16\sigma$ [$-0.22\sigma$], and +0.15$\sigma$ [+0.14$\sigma$] respectively. For $\Omega_k$, when P18+lensing+non-CMB (old) data are analyzed we find $\Omega_k = 0.0004\pm 0.0017$ [$0.0003\pm 0.0017$], which is $+0.24\sigma$ [$+0.18\sigma$] away from zero and differing by $-0.21\sigma$ [$-0.21\sigma$] with the P18+lensing+non-CMB (new) value, $\Omega_k = 0.0009\pm 0.0017$ [$0.0008\pm 0.0017$] which is only $+0.53\sigma$ [$+0.47\sigma$] in favor of an open Universe. The error bars do not change much. For the primary parameters $\tau$ and $n_s$ we get a reduction of $-2.82$\% [$-2.78$\%] and $-2.33$\% [$-2.43$\%] whereas for the derived parameters $H_0$ and $\sigma_8$ the error bars are reduced by $-1.85$\% [0.00\%] and 0.00\% [$-3.08$\%], respectively.  

\begin{figure*}[htbp]
\centering
\mbox{\includegraphics[width=170mm]{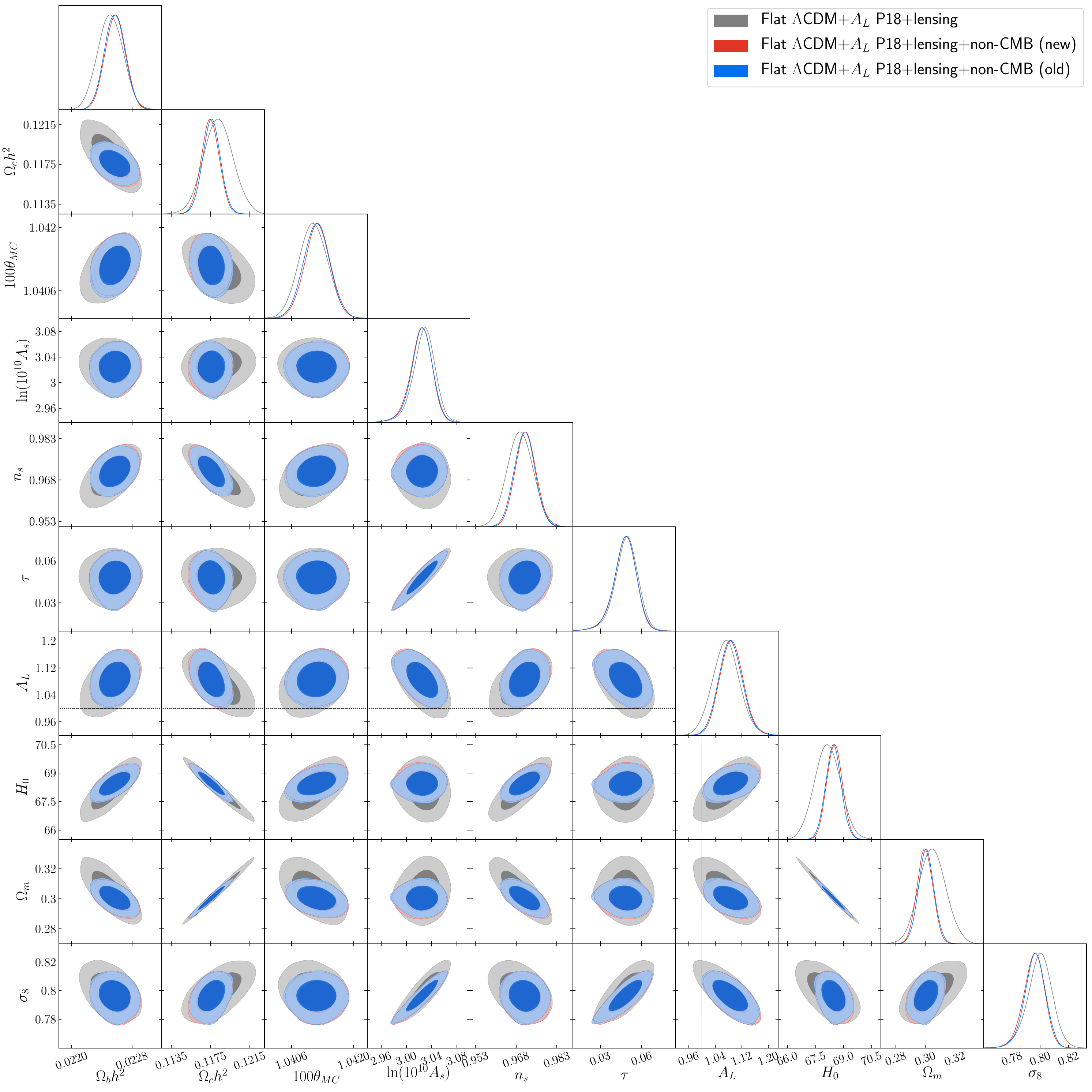}}
\caption{Likelihood distributions of flat $\Lambda$CDM$+A_L$ model parameters
         favored by P18+lensing and non-CMB data. Non-CMB (old) denotes the compilation of non-CMB data
	 sets used in \cite{deCruzPerez:2022hfr} while non-CMB (new) 
	 denotes the new compilation of non-CMB data sets of Sec.\ \ref{sec:Non-CMB Data}.
         Likelihood results for P18+lensing data are shown for comparison. 
}
\label{fig:flat_LCDM_p18lenncmb_AL}
\end{figure*}

Results obtained with P18+lensing+non-CMB (old) and P18+lensing+non-CMB (new) data, provided in Table \ref{tab:results_flat_LCDM_AL} and Fig.\ \ref{fig:flat_LCDM_p18lenncmb_AL}, for the seven-parameter flat $\Lambda$CDM+$A_L$ model, are very similar. The primary parameters $\Omega_{b}h^2$ and $\Omega_{c}h^2$ differ at +0.05$\sigma$ and $-0.09\sigma$ respectively whereas the derived parameter $H_0$ differs at +0.12$\sigma$. When P18+lensing+non-CMB (old) data are analyzed we obtain $A_L = 1.089\pm 0.035$ which shows a small difference of +0.04$\sigma$ with the value $A_L = 1.087\pm 0.035$ obtained after analyzing P18+lensing+non-CMB (new) data. 

\begin{figure*}[htbp]
\centering
\mbox{\includegraphics[width=170mm]{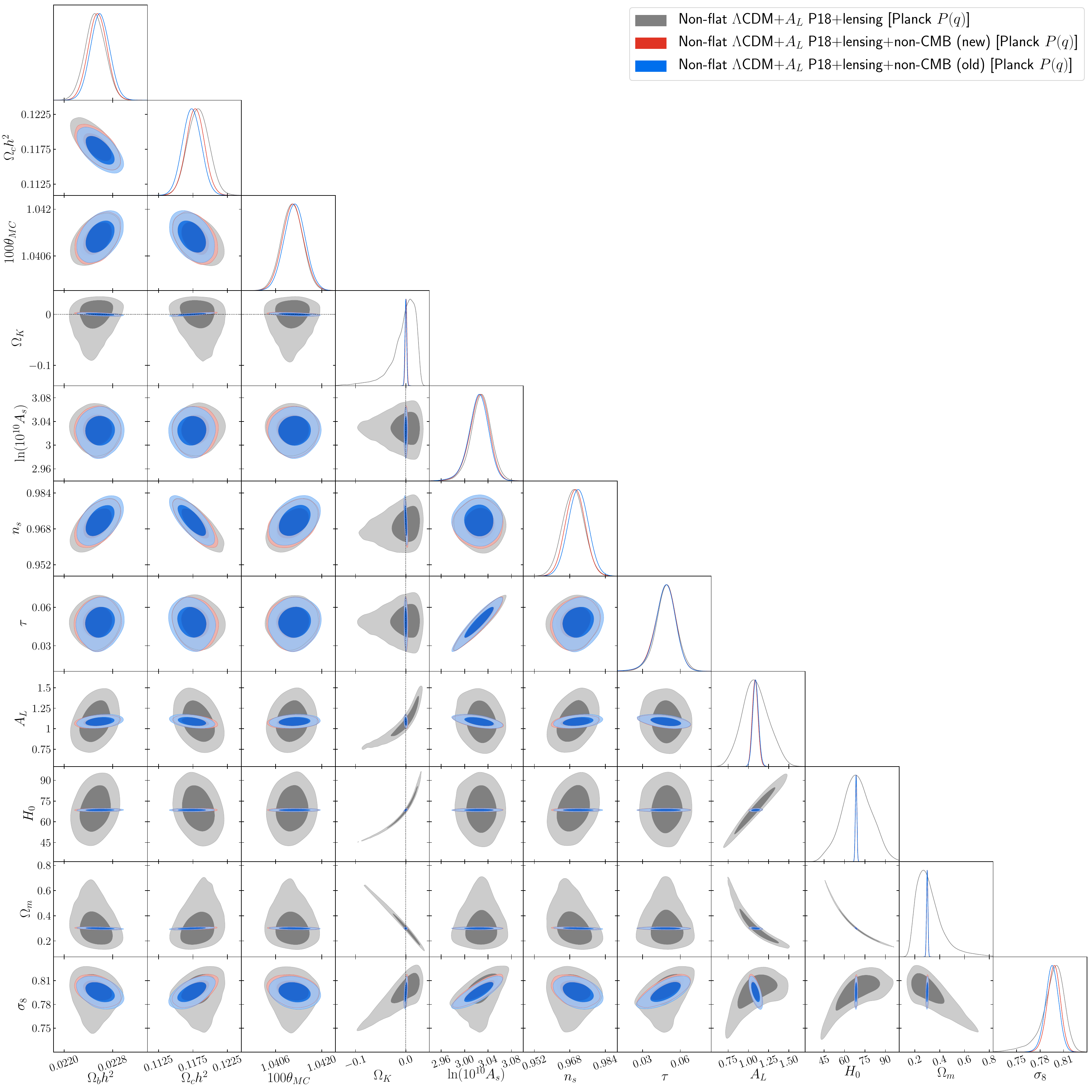}}
\caption{Likelihood distributions of non-flat $\Lambda$CDM$+A_L$ model [Planck $P(q)$]
         parameters favored by P18+lensing and non-CMB data. Non-CMB (old) denotes the compilation of non-CMB data
	 sets used in \cite{deCruzPerez:2022hfr} while non-CMB (new) 
	 denotes the new compilation of non-CMB data sets of Sec.\ \ref{sec:Non-CMB Data}.
         Likelihood results for P18+lensing data are shown for comparison. 
}
\label{fig:Planck_Pq_p18lenncmb_AL}
\end{figure*}

\begin{figure*}[htbp]
\centering
\mbox{\includegraphics[width=170mm]{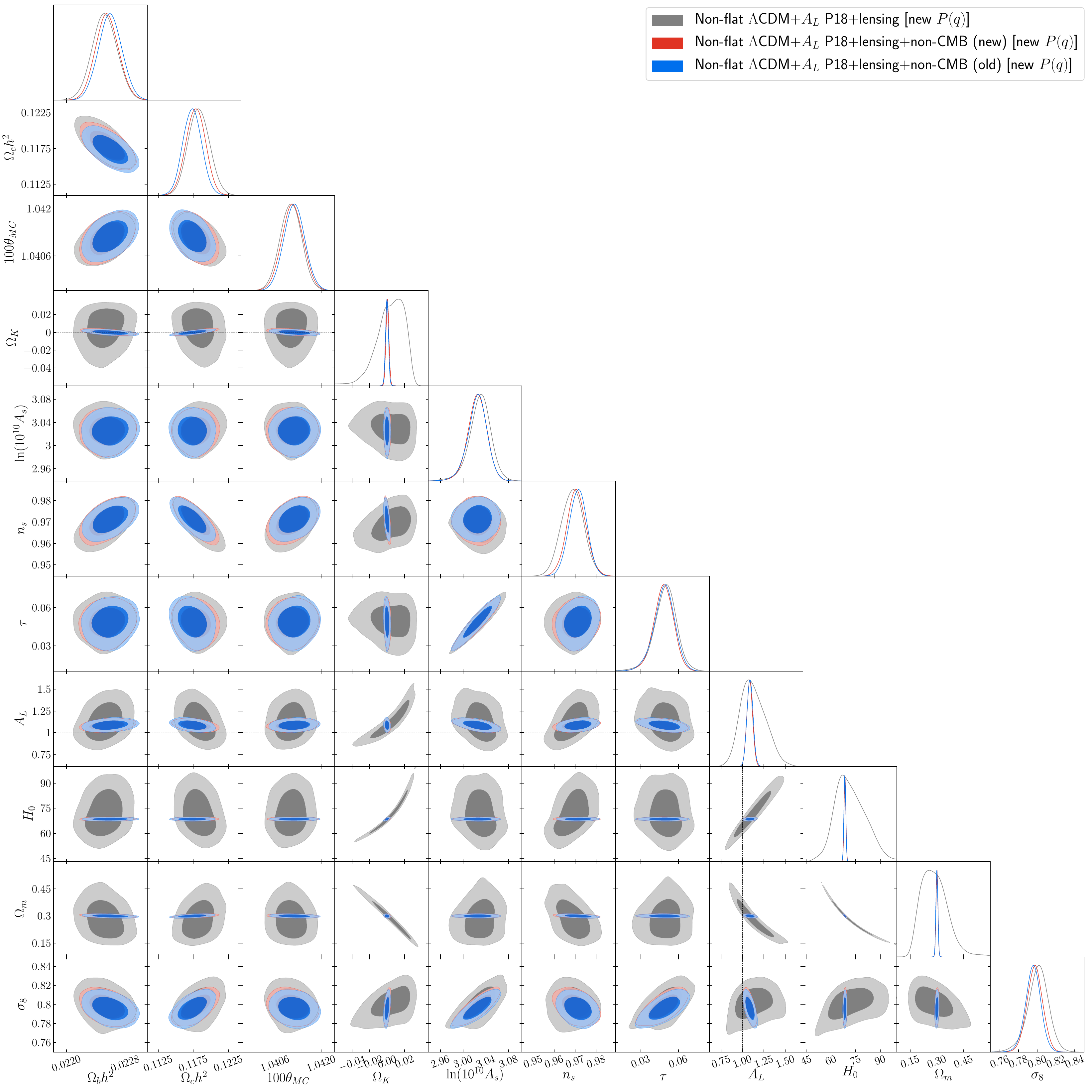}}
\caption{Likelihood distributions of non-flat $\Lambda$CDM$+A_L$ model [new $P(q)$] parameters
         favored by P18+lensing and non-CMB data. Non-CMB (old) denotes the compilation of non-CMB data
	 sets used in \cite{deCruzPerez:2022hfr} while non-CMB (new) 
	 denotes the new compilation of non-CMB data sets of Sec.\ \ref{sec:Non-CMB Data}. Likelihood results for P18+lensing data are shown for comparison. 
}
\label{fig:new_Pq_p18lenncmb_AL}
\end{figure*}

Cosmological parameter constraints for the eight-parameter non-flat $\Lambda$CDM Planck [new] $P(q)$ model from P18+lensing+non-CMB (old) and P18+lensing+non-CMB (new) data are in Table \ref{tab:results_Planck_Pq_AL} [\ref{tab:results_new_Pq_AL}] and in Fig.\ \ref{fig:Planck_Pq_p18lenncmb_AL} [\ref{fig:new_Pq_p18lenncmb_AL}]. Although the differences are greater than in the case of the flat $\Lambda$CDM+$A_L$ model, they are still small. Primary parameters $\Omega_{c}h^2$, $n_s$, and $\Omega_{b}h^2$ differ at $-0.26\sigma$ [$-0.21\sigma$], +0.21$\sigma$ [$+0.10\sigma$], and +0.18$\sigma$ [$+0.18\sigma$], respectively, while derived parameter $\sigma_8$ differs at $-0.19\sigma$ [$-0.15\sigma$]. For $\Omega_k$ and $A_L$ the differences are $-0.25\sigma$ [$-0.25\sigma$] and +0.12$\sigma$ [+0.082$\sigma$].

In summary, the combination of P18+lensing data with either non-CMB (old) or non-CMB (new) data gives almost identical cosmological parameter results, similar to the P18 case of the previous subsubsection, but with even smaller differences. 

Again, similar to the previous subsubsection for P18+non-CMB data, there are no significant differences between the P18+lensing+non-CMB (old) and P18+lensing+non-CMB (new) data values of $\Delta$AIC and $\Delta$DIC obtained when flat $\Lambda$CDM is compared with the non-flat models (see Tables \ref{tab:results_Planck_Pq} and \ref{tab:results_new_Pq}) and the $A_L$-varying models (Tables \ref{tab:results_flat_LCDM_AL}, \ref{tab:results_Planck_Pq_AL} and \ref{tab:results_new_Pq_AL}). 

Contrary to the previous subsubsection for P18 data and non-CMB data, we do not observe a qualitative change between the results obtained using P18+lensing data and using either non-CMB (old) data or non-CMB (new) data regarding the concordance/discordance (see Table \ref{tab:consistency_LCDM}) provided by the two statistical estimators in Eqs.\ \eqref{eq:Tension_estimator_1} and \eqref{eq:Tension_estimator_2_sigma}.



\subsubsection{Summary of updated constraints in $\Lambda$CDM models}\label{subsubsec:summary}

Here we summarize updated results for the $\Lambda$CDM models constraints that follow from the updated non-CMB (new) data we use. The results of cosmological parameter constraints from P18 data and P18+lensing data in the flat and non-flat $\Lambda$CDM ($+A_L$ ) models are described in detail in Sec.\ IV A 1 and 2 of our previous paper \cite{deCruzPerez:2022hfr}.

For numerical values of cosmological parameters and to see how the consideration of different data sets affect the two-dimensional contour plots, see Tables \ref{tab:results_flat_LCDM}---\ref{tab:results_new_Pq_AL} and Figs.\ \ref{fig:flat_LCDM_ncmb}---\ref{fig:new_Pq_p18lenncmb_AL}. 

Within the context of a given cosmological model, it is possible to assess whether the constraints obtained from two different data sets are in tension or not at some level of significance, and when they are in tension they cannot be jointly used to constrain that cosmological model, which is ruled out at that level of significance, if we assume the two data sets are correct. In \cite{deCruzPerez:2022hfr} we utilized the same statistical estimators employed in this work, namely Eqs.\ \eqref{eq:Tension_estimator_1} and \eqref{eq:Tension_estimator_2_sigma}. One of the most important results found there was that for the non-flat $\Lambda$CDM Planck $P(q)$ model P18 data and non-CMB (old) data could not be jointly analyzed since the level of discordance between the two data sets was $\sigma = 3.005$, larger than $3\sigma$, meaning that the non-flat $\Lambda$CDM Planck $P(q)$ model was ruled out at more than 3$\sigma$. On the other hand, the P18 data set and the non-CMB (new) data set have a reduced  discordance level of $\sigma = 2.704$, indicating that although there is still a significant level of tension, it is less than $3\sigma$ and so both data sets can now be jointly analyzed. As can be seen from Table \ref{tab:consistency_LCDM}, the discordance between the P18 and P18+lensing data sets and the non-CMB (new) data set is less than that between the two CMB data sets and the non-CMB (old) data set. A major reason for this is the larger error bars associated with updated $H(z)$ data (those for which there is now a non-diagonal covariance matrix). We return to these points in Sec.\ \ref{sec:Data_Set_Tensions}.

Two other significant results obtained in \cite{deCruzPerez:2022hfr} remain unchanged for the updated non-CMB (new) data. The first is the compatibility of the measured values of the curvature parameter $\Omega_k$, in the non-flat models, with spatially-flat hypersurfaces, when P18 or P18+lensing data are jointly analyzed with non-CMB (new) data. The second is the evidence that the lensing parameter $A_L$ deviates from 1. When P18+lensing+non-CMB (new) data are utilized with the $\Lambda$CDM Planck [new] $P(q)$ model we measure $\Omega_k= 0.0009\pm 0.0017$ [$0.0008\pm 0.0017$], whereas for the $\Lambda$CDM Planck [new] $P(q)$+$A_L$ model we find $\Omega_k= 0.0004\pm 0.0017$ [$0.0004\pm 0.0017$], therefore an open Universe is very mildly favored in all the cases. In regard to the second result, for P18+lensing+non-CMB (new) data, we still detect $\sim 2\sigma$ evidence in favor of $A_L > 1$ for all models in which this phenomenological parameter is allowed to vary and so can be measured. In particular, in the flat $\Lambda$CDM+$A_L$ model we find $A_L = 1.087\pm 0.035$, a deviation of 2.49$\sigma$ from the expected value $A_L=1$. For the non-flat models, the $\Lambda$CDM Planck $P(q)$+$A_L$ model yields $A_L = 1.084\pm 0.035$, while the $\Lambda$CDM new $P(q)$+$A_L$ model yields $A_L=1.084\pm 0.034$. Both results are approximately 2.4$\sigma$ away from 1. When P18+non-CMB (new) data are used with the $\Lambda$CDM Planck [new] $P(q)$ model, the resulting value for $\Omega_k$ is $0.0009\pm 0.0017$ [$0.0008\pm 0.0017$]. For the $\Lambda$CDM Planck [new] $P(q)$+$A_L$ model, $\Omega_k = -0.0001\pm 0.0017$ [$-0.0001\pm 0.0017$]. For P18+non-CMB (new) data we find $\sim 3\sigma$ evidence in favor of $A_L > 1$ for all the models. Specifically, in the case of the flat $\Lambda$CDM+$A_L$ model, $A_L = 1.198\pm 0.060$, indicating a deviation of 3.30$\sigma$ from the theoretically expected value of $A_L=1$. For the non-flat models, the $\Lambda$CDM Planck $P(q)$+$A_L$ model gives $A_L = 1.196\pm 0.062$, while the $\Lambda$CDM new $P(q)$+$A_L$ model yields $A_L=1.194\pm 0.061$ with both values deviating from 1 by 3.2$\sigma$.

The six-parameter tilted flat $\Lambda$CDM model, with $\Omega_k=0$ and $A_L=1$, once again seems to pass all the consistency tests that we have considered. The P18+lensing+non-CMB (new) data set is the largest one that we have analyzed in this work. The primary cosmological parameter constraints obtained from it for this standard model are $\Omega_{b}h^2 = 0.02249\pm 0.00013$, $\Omega_{c}h^2=0.11849\pm 0.00084$, $100\theta_{\text{MC}}=1.04109\pm 0.00028$, $\tau = 0.0569\pm 0.00071$, $n_s=0.9685\pm 0.0036$, and $\ln(10^{10}A_s)=3.046\pm 0.014$. Among the different primary parameters the least well-determined is the reionization optical depth at $8.0\sigma$ and the spectral index $n_s$ deviates from the scale invariant value $n_s = 1$ by 8.75$\sigma$. As we noted in \cite{deCruzPerez:2022hfr} these values are very similar to the corresponding ones in the other models under study, therefore, they are almost model-independent results. As for the derived parameters, for the Hubble constant we get $H_0=68.05\pm 0.38$ km s$^{-1}$ Mpc$^{-1}$, which is in agreement with not only the result obtained from a median statistics analysis $H_0=68\pm 2.8$ km km s$^{-1}$ Mpc$^{-1}$ \cite{Chen:2011ab,Gottetal2001,Calabreseetal2012}, but also with some other local measurements like for instance the flat $\Lambda$CDM model value provided in \cite{Cao:2023eja} $H_0=69.5\pm 2.4$ km s$^{-1}$ Mpc$^{-1}$ from a joint analysis of $H(z)$, BAO, Pantheon+ SNIa, quasar angular size, reverberation-measured \mii\ and \civ\ quasar, and 118 Amati correlation gamma-ray burst data, or the local $H_0=69.8\pm 1.7$ km s$^{-1}$ Mpc$^{-1}$ from TRGB and SNIa data \cite{Freedman:2021ahq}. On the other hand our measured $H_0$ value is still in tension with the local $H_0=73.04\pm 1.04$ km s$^{-1}$ Mpc$^{-1}$ from Cepheids and SNIa data \cite{Riess:2021jrx}, also see \cite{Chen:2024gnu}. As for the other two derived parameters, we get $\Omega_m = 0.3059\pm 0.0050$ and $\sigma_8 = 0.8077\pm 0.0057$, the first of which is in good agreement with the flat $\Lambda$CDM model value of $\Omega_m = 0.313\pm 0.012$ of \cite{Cao:2023eja} (for the same data used to measure $H_0$ listed above). These results have been obtained using the P18+lensing+non-CMB (new) data set, which is possibly the currently largest combination of mutually consistent data sets, and so these results are likely the currently most-restrictive constraints on flat $\Lambda$CDM model parameters.

\subsection{XCDM models} 
\label{sec:XCDM models}

In this subsection we study the XCDM models cosmological parameters constraints, goodness of fit results, and the consistencies between different data set combinations we use. 

Before discussing these we compare our XCDM model results to prior results in the literature for a few cases (where the same data set has previously been used). The Planck team \cite{Planck:2018tab} measure, for the flat XCDM model and P18 data, the dark energy equation of state parameter to be $w=-1.58^{+0.16}_{-0.35}$  (see $\S$17.5 of \cite{Planck:2018tab}), and for P18+lensing data they get $w=-1.57^{+0.16}_{-0.33}$ (see $\S$17.6 of \cite{Planck:2018tab}), both of which are in good agreement with our results shown in Table \ref{tab:results_flat_XCDM}.

\begin{figure*}[htbp]
\centering
\mbox{\includegraphics[width=170mm]{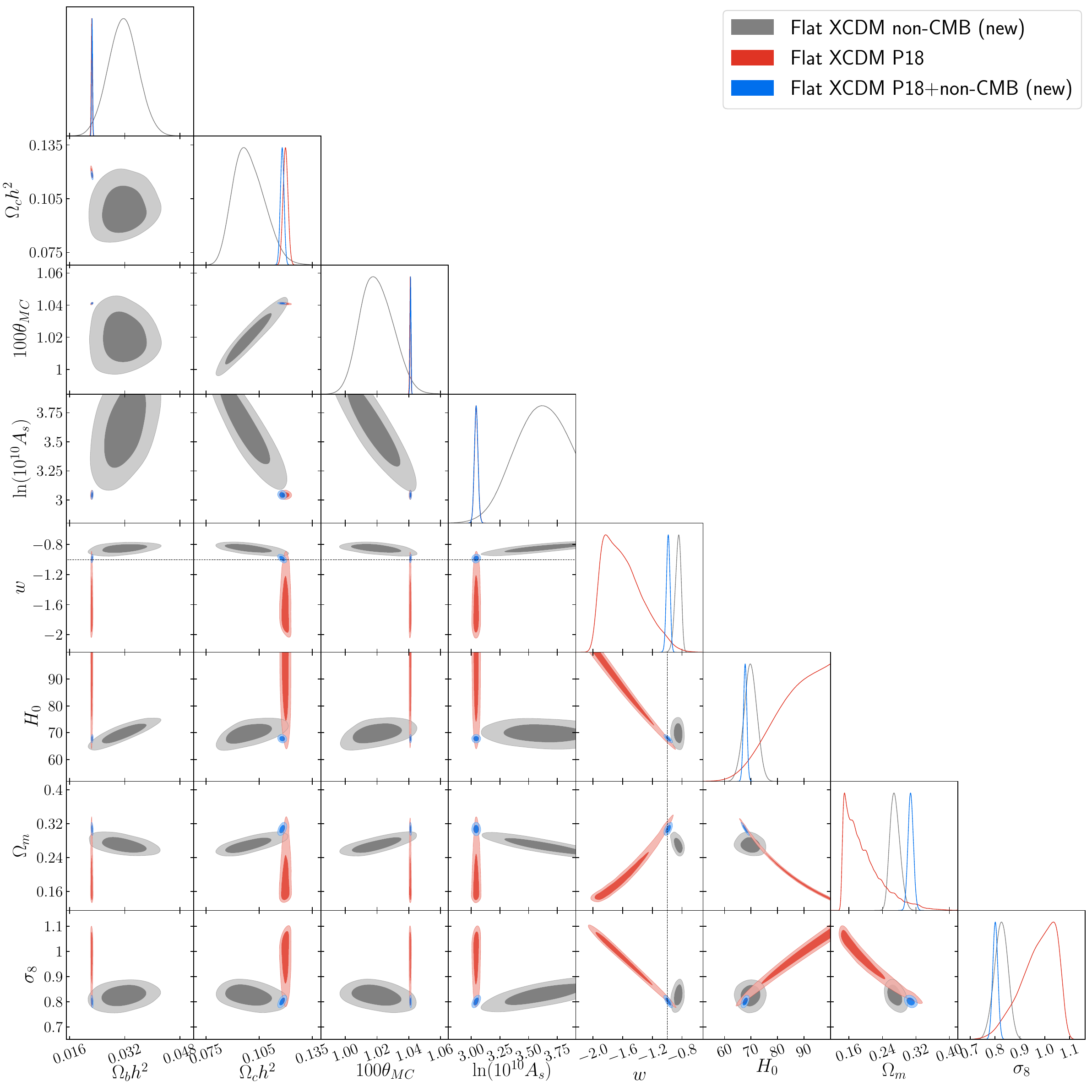}}
        \caption{Likelihood distributions of flat XCDM model 
        parameters favored by non-CMB (new) data.
        Results for P18 and P18+non-CMB (new) data are shown for comparison.
}
\label{fig:flat_XCDM_P18_vs_nonCMB_1}
\end{figure*}

\begin{figure*}[htbp]
\centering
\mbox{\includegraphics[width=170mm]{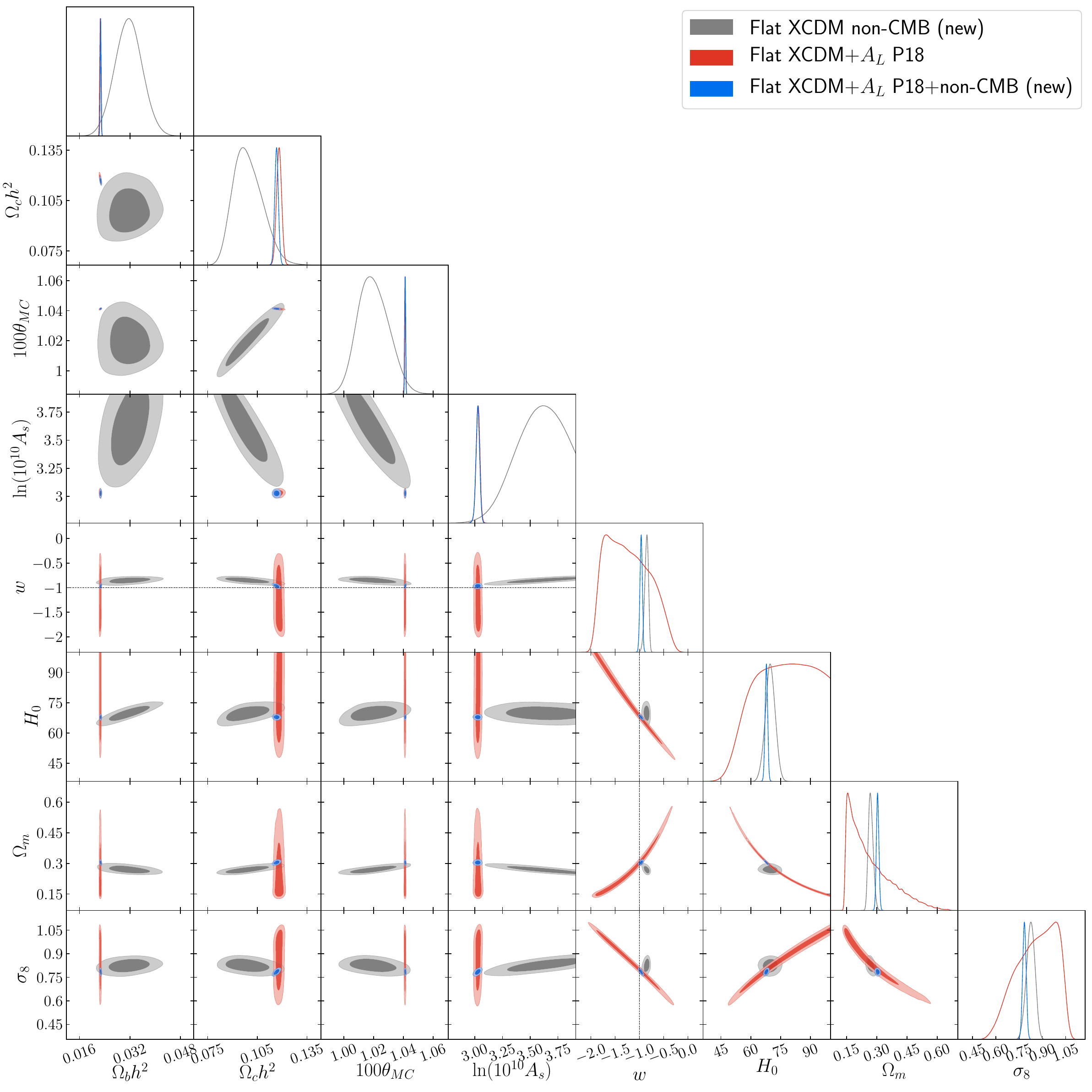}}
        \caption{Likelihood distributions of flat XCDM$+A_L$ model
        parameters favored by non-CMB (new) data. Results for P18 and P18+non-CMB (new) data are shown for comparison.
}
\label{fig:FX_Alens_p18_ncmb}
\end{figure*}

\begin{figure*}[htbp]
\centering
\mbox{\includegraphics[width=170mm]{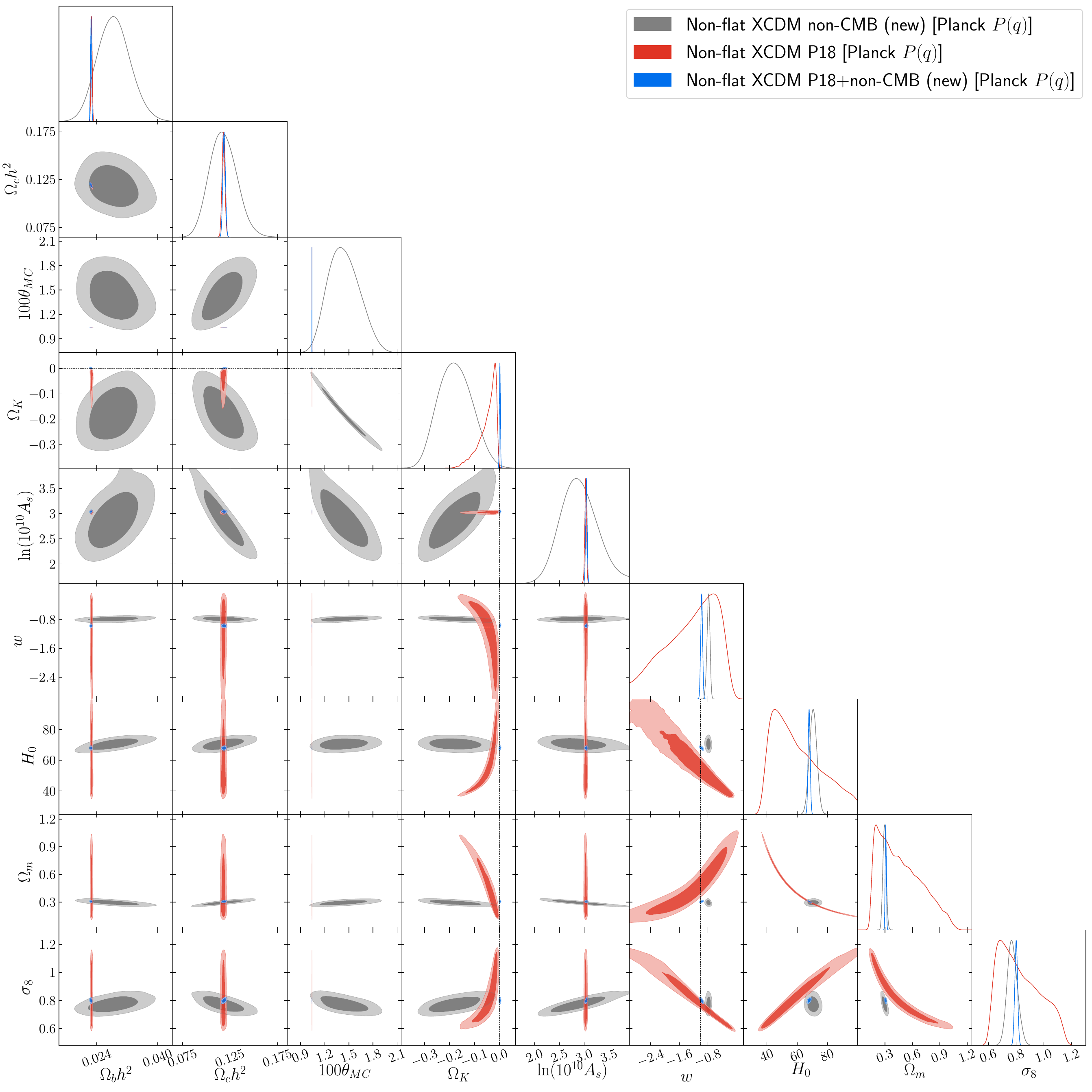}}
	\caption{Likelihood distributions of non-flat XCDM model [Planck $P(q)$]
        parameters favored by non-CMB (new) data.
        Results for P18 and P18+non-CMB (new) data are shown for comparison.
}
\label{fig:XCDM_Planck_Pq_P18_vs_nonCMB_1}
\end{figure*}

\begin{figure*}[htbp]
\centering
\mbox{\includegraphics[width=170mm]{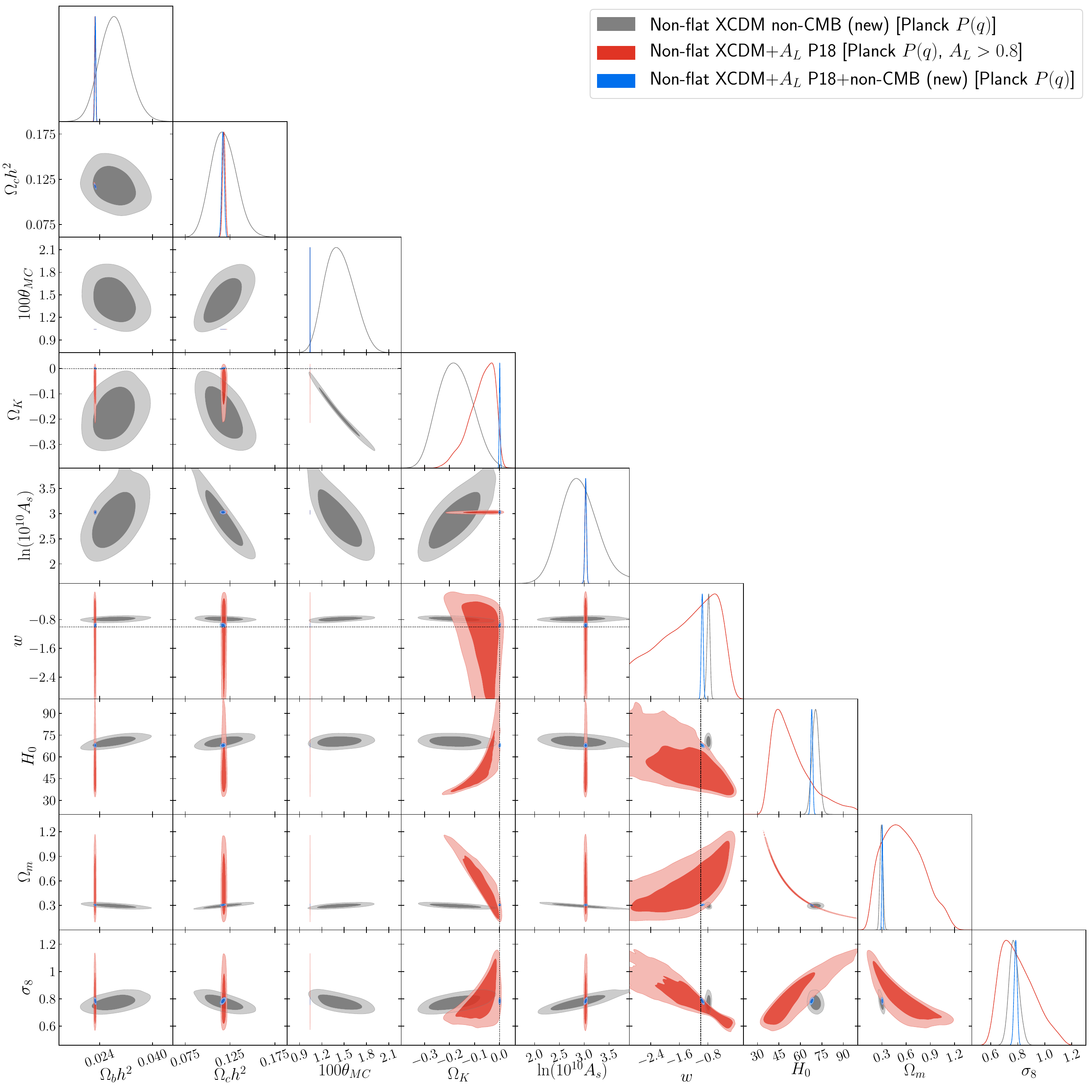}}
        \caption{Likelihood distributions of non-flat XCDM$+A_L$ model [Planck $P(q)$]
        parameters favored by non-CMB (new) data. Results for P18 and P18+non-CMB (new) data are shown for comparison.
}
\label{fig:NX_Alens_ns_p18_ncmb}
\end{figure*}

\begin{figure*}[htbp]
\centering
\mbox{\includegraphics[width=170mm]{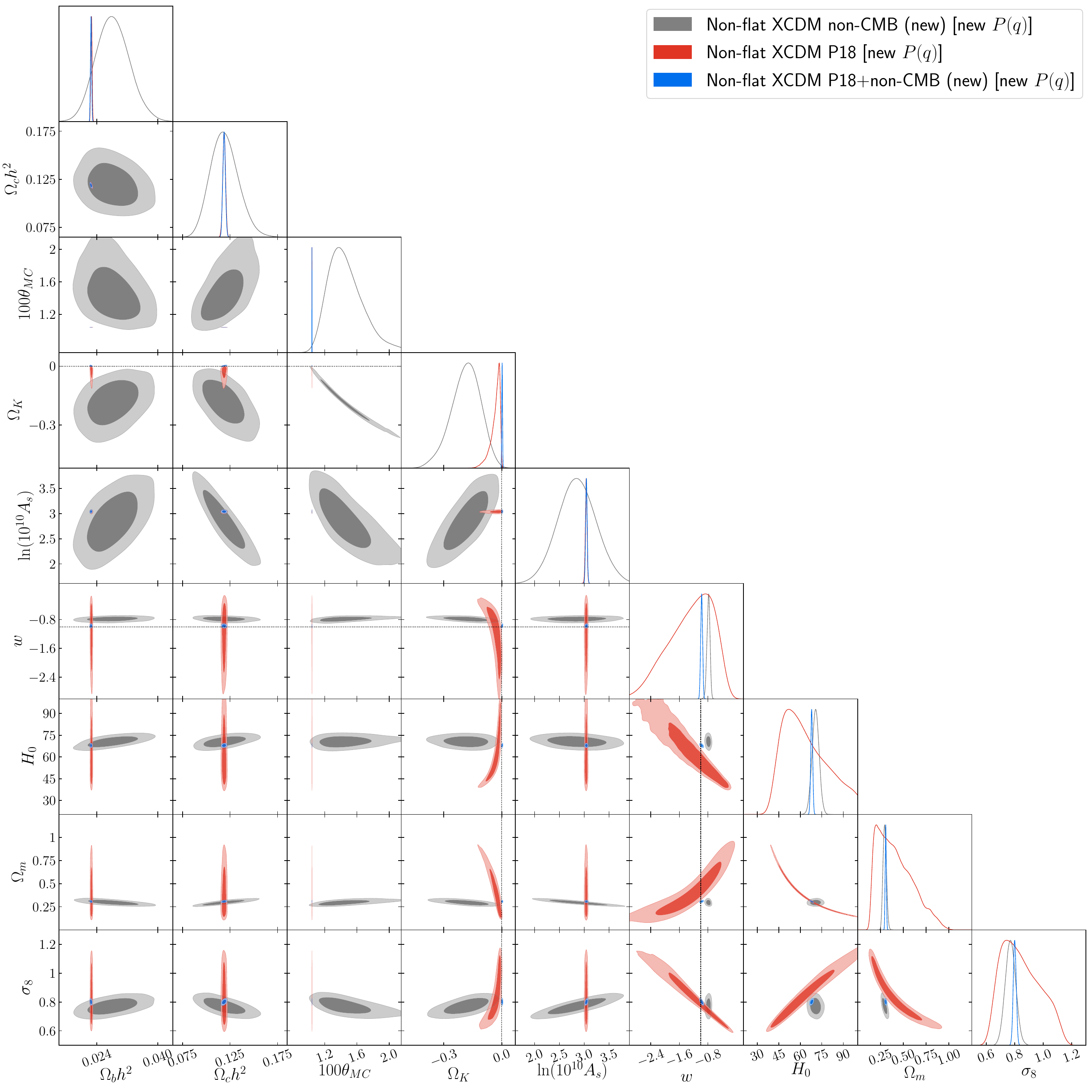}}
        \caption{Likelihood distributions of non-flat XCDM model [new $P(q)$]
        parameters favored by non-CMB data (new).
        Results for P18 and P18+non-CMB (new) data are shown for comparison.
}
\label{fig:XCDM_new_Pq_P18_vs_nonCMB_1}
\end{figure*}

\begin{figure*}[htbp]
\centering
\mbox{\includegraphics[width=170mm]{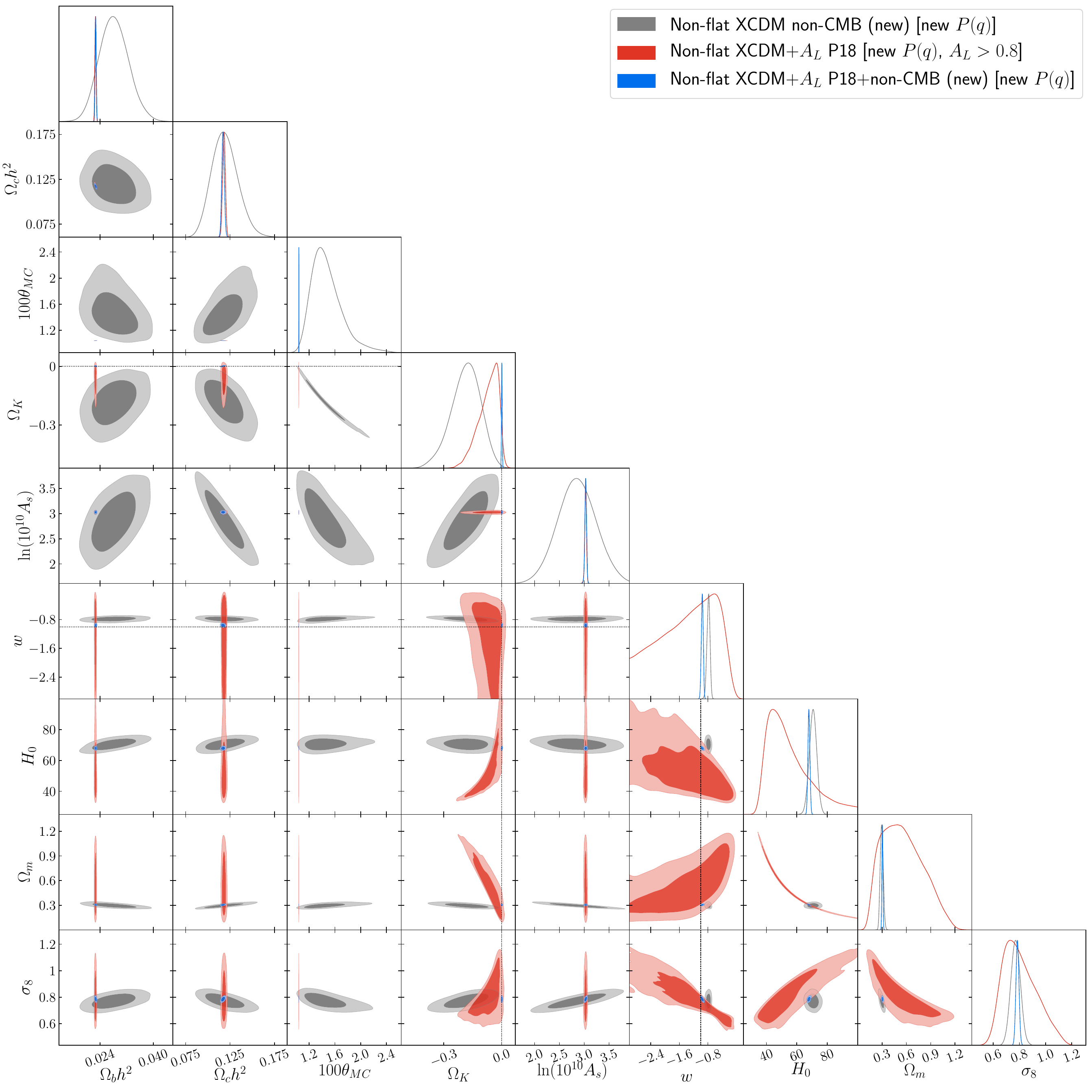}}
        \caption{Likelihood distributions of non-flat XCDM$+A_L$ model [new $P(q)$]
        parameters favored by non-CMB (new) data. Results for P18 and P18+non-CMB (new) data are shown for comparison.
}
\label{fig:TNX_Alens_ns1_p18_ncmb}
\end{figure*}

Interestingly, while Planck CMB data favor a large phantom-like region with $w < -1$, non-CMB data favor a quintessence-like region with $w>-1$, see Figs.\ \ref{fig:flat_XCDM_P18_vs_nonCMB_1}---\ref{fig:TNX_Alens_ns1_p18len_ncmb}. This fact, as we shall see, goes hand in hand with the inconsistencies between results, discussed below, obtained from Planck data and from non-CMB (new) data in the XCDM models. When Planck jointly analyze such data (see $\S$17.21 of \cite{Planck:2018tab} for P18+BAO+SNIa), they measure $w=-1.028\pm 0.033$ favoring phantom-like dynamical dark energy at 0.85$\sigma$, while using our larger non-CMB (new) data compilation we find $w=-0.986 \pm 0.024$ for P18+non-CMB (new) data, disfavoring phantom-like dynamical dark energy at 0.58$\sigma$ in the flat XCDM model.

Another important point is that in all the XCDM cases, non-CMB data better determine $w$ than do P18 or P18+lensing data. From Figs.\ \ref{fig:flat_XCDM_P18_vs_nonCMB_1}---\ref{fig:TNX_Alens_ns1_p18len_ncmb} we see that P18 or P18+lensing data alone are not sensitive to the dark energy equation of state parameter $w$ because in these cases $w$ is strongly degenerate with all other cosmological parameters. From these figures we can also see that this is not the case for non-CMB (new) data. This is because SNIa as well as $H(z)$+BAO data, both included in the non-CMB data compilation we use, have the ability to reasonably restrictively constrain $w$, and the SNIa+$H(z)$+BAO combination constrains $w$ very tightly. For example, the flat XCDM model result from SNIa Pantheon+ data is $w=-0.90^{+0.17}_{-0.12}$, Table V of \cite{Cao:2023eja}, $w=-0.90 \pm 0.14$ from Pantheon+SH0ES SNIa data \cite{Brout:2022vxf}, $w=-0.963 \pm 0.070$ is obtained from analyzing SN+BAO data \cite{Park:2018tgj}, and the $H(z)$+BAO+SNP+ data compilation result is $w=-0.886 \pm 0.053$, Table V of \cite{Cao:2023eja}, the last of which is very consistent with our flat XCDM non-CMB (new) data result of $w=-0.853 \pm 0.039$. 

Additionally, possibly as a consequence of the fact that non-CMB (new) data more restrictively constrain $w$ than do P18 or P18+lensing data, non-CMB (new) data also more restrictively constrain the derived parameters, $H_0$, $\Omega_m$, and $\sigma_8$, than do P18 or P18+lensing data, see Tables \ref{tab:results_flat_XCDM}---\ref{tab:results_new_Pq_XCDM}.


\begin{table*}
	\caption{Mean and 68\% (or 95\%) confidence limits of flat XCDM ($+A_L$) model parameters
        from non-CMB (new), P18, P18+lensing, P18+non-CMB (new), and P18+lensing+non-CMB (new) data.
        $H_0$ has units of km s$^{-1}$ Mpc$^{-1}$.
}
\begin{ruledtabular}
\begin{tabular}{lccccc}
\\[-1mm]                        & \multicolumn{5}{c}{Flat XCDM models}        \\[+1mm]
\cline{2-6}\\[-1mm]
  Parameter                     &  Non-CMB (new)       & P18                    &  P18+lensing           &  P18+non-CMB (new)      & P18+lensing+non-CMB (new)    \\[+1mm]
 \hline \\[-1mm]
  $\Omega_b h^2$                & $0.0316 \pm 0.0043$  & $0.02240 \pm 0.00015$  & $0.02243 \pm 0.00015$  &  $0.02251 \pm 0.00014$  &  $0.02250 \pm 0.00014$  \\[+1mm]
  $\Omega_c h^2$                & $0.0994 \pm 0.0087$  & $0.1200 \pm 0.0014$    & $0.1193 \pm 0.0012$    &  $0.1181 \pm 0.0010$    &  $0.11830 \pm 0.00095$  \\[+1mm]
  $100\theta_\textrm{MC}$       & $1.020 \pm 0.010$    & $1.04094 \pm 0.00032$  & $1.04100 \pm 0.00031$  &  $1.04114 \pm 0.00030$  &  $1.04112 \pm 0.00029$  \\[+1mm]
  $\tau$                        & $0.0537$             & $0.0537 \pm 0.0078$    & $0.0524 \pm 0.0074$    &  $0.0558 \pm 0.0078$    &  $0.0577 \pm 0.0075$    \\[+1mm]
  $n_s$                         & $0.9654$             & $0.9654 \pm 0.0044$    & $0.9667 \pm 0.0041$    &  $0.9696 \pm 0.0039$    &  $0.9690 \pm 0.0038$    \\[+1mm]
  $\ln(10^{10} A_s)$            & $3.57 \pm 0.20$      & $3.043 \pm 0.016$      & $3.038 \pm 0.015$      &  $3.043 \pm 0.016$      &  $3.048 \pm 0.015$      \\[+1mm]
  $w$                           & $-0.853 \pm 0.039$   & $-1.59 \pm 0.26$       & $-1.55 \pm 0.26$       &  $-0.986 \pm 0.024$     &  $-0.990 \pm 0.023$     \\[+1mm]  
                                & ($-0.853^{+0.043}_{-0.033}$)   & ($-1.59^{+0.15}_{-0.34}$)       & ($-1.55^{+0.16}_{-0.35}$)     &   &    \\[+1mm]  
 \hline \\[-1mm]
  $H_0$                         & $69.8 \pm 2.5$       & $86.8 \pm 8.9$ ($>70.2$)        & $86.0 \pm 9.2$ ($>69.6$)        &  $67.78 \pm 0.63$       &  $67.81 \pm 0.63$       \\[+1mm]
  $\Omega_m$                    & $0.270 \pm 0.012$    & $0.197 \pm 0.046$      & $0.200 \pm 0.048$      &  $0.3075 \pm 0.0063$    &  $0.3077 \pm 0.0062$    \\[+1mm]
  $\sigma_8$                    & $0.824 \pm 0.027$    & $0.974 \pm 0.071$      & $0.960 \pm 0.071$     &  $0.801 \pm 0.010$      &  $0.8047 \pm 0.0089$    \\[+1mm]
      \hline\\[-1mm]
  $\chi_{\textrm{min}}^2$       & $1459.18$            & $2761.40$              & $2770.58$              & $4239.85$               &  $4249.05$              \\[+1mm]
  $\Delta\chi_{\textrm{min}}^2$ & $-10.75$             & $-4.40$                & $-4.13$                & $-0.39$                 &  $-0.21$              \\[+1mm]
  $\textrm{DIC}$                & $1468.74$            & $2815.67$              & $2824.21$              & $4294.20$               &  $4303.30$              \\[+1mm]
  $\Delta\textrm{DIC}$          & $-9.37$              & $-2.26$                & $-2.24$                & $+1.87$                 &  $+2.10$                \\[+1mm]
  $\textrm{AIC}$                & $1469.18$            & $2817.40$              & $2826.58$              & $4295.85$               &  $4305.05$              \\[+1mm]
  $\Delta\textrm{AIC}$          & $-8.75$              & $-2.40$                & $-2.13$                & $+1.61$                 &  $+1.79$                \\[+1mm]
  \hline
\\[-1mm]                         & \multicolumn{5}{c}{Flat XCDM$+A_L$ models}        \\[+1mm]
\cline{2-6}\\[-1mm]
  Parameter                      & & P18                   & P18+lensing            & P18+non-CMB (new)       &  P18+lensing+non-CMB (new)   \\[+1mm]
 \hline \\[-1mm]
  $\Omega_b h^2$                 & & $0.02258 \pm 0.00017$ & $0.02250 \pm 0.00017$  & $0.02272 \pm 0.00015$   & $0.02263 \pm 0.00014$    \\[+1mm]
  $\Omega_c h^2$                 & & $0.1181 \pm 0.0015$   & $0.1184 \pm 0.0015$    & $0.1166 \pm 0.0011$     & $0.1168 \pm 0.0011$    \\[+1mm]
  $100\theta_\textrm{MC}$        & & $1.04114 \pm 0.00033$ & $1.04109 \pm 0.00032$  & $1.04130 \pm 0.00030$   & $1.04126 \pm 0.00030$    \\[+1mm]
  $\tau$                         & & $0.0493 \pm 0.0085$   & $0.04908 \pm 0.0084$   & $0.0500 \pm 0.0085$     & $0.0496 \pm 0.0083$      \\[+1mm]
  $n_s$                          & & $0.9706 \pm 0.0049$   & $0.9691 \pm 0.0049$    & $0.9746 \pm 0.0041$     & $0.9733 \pm 0.0040$      \\[+1mm]
  $\ln(10^{10} A_s)$             & & $3.029 \pm 0.018$     & $3.029 \pm 0.018$      & $3.027 \pm 0.017$       & $3.026 \pm 0.017$        \\[+1mm]
  $w$                            & & $-1.23 \pm 0.42$      & $-1.34 \pm 0.37$       & $-0.964 \pm 0.024$      & $-0.968 \pm 0.024$        \\[+1mm]
                                 & & ($-1.23^{+0.31}_{-0.59}$)  & ($-1.34^{+0.26}_{-0.51}$)  &      &         \\[+1mm]
  $A_L$                          & & $1.180 \pm 0.097$     & $1.054 \pm 0.055$      &  $1.222 \pm 0.063$      & $1.101 \pm 0.037$        \\[+1mm]
                                 & & ($1.180^{+0.062}_{-0.10}$) & ($1.054^{+0.039}_{-0.059}$)  &      &         \\[+1mm]
 \hline \\[-1mm]
  $H_0$                          & & $77 \pm 14$ ($77^{+20}_{-10}$)   \ 
                                                           & $80 \pm 12$ ($>58.6$)  & $67.83 \pm 0.63$        & $67.79 \pm 0.63$        \\[+1mm]
  $\Omega_m$                     & & $0.27 \pm 0.11$       & $0.242 \pm 0.083$      & $0.3043 \pm 0.0062$     & $0.3050 \pm 0.0062$    \\[+1mm]
  $\sigma_8$                     & & $0.86 \pm 0.12$       & $0.89 \pm 0.11$        & $0.784 \pm 0.011$       & $0.785 \pm 0.011$       \\[+1mm]
      \hline\\[-1mm]
  $\chi_{\textrm{min}}^2$        & & $2755.89$             & $2770.43$              &  $4224.98$              & $4240.92$                 \\[+1mm]
  $\Delta\chi_{\textrm{min}}^2$  & & $-9.91$               & $-4.28$                &  $-15.26$               & $-8.34$                  \\[+1mm]
  $\textrm{DIC}$                 & & $2813.08$             & $2825.81$              &  $4283.50$              & $4296.89$                \\[+1mm]
  $\Delta\textrm{DIC}$           & & $-4.85$               & $-0.64$                &  $-8.83$                & $-4.31$                  \\[+1mm]
  $\textrm{AIC}$                 & & $2813.89$             & $2828.43$              &  $4282.98$              & $4298.92$                  \\[+1mm]
  $\Delta\textrm{AIC}$           & & $-5.91$               & $-0.28$                &  $-11.26$               & $-4.34$                \\[+1mm]
\end{tabular}
\\[+1mm]
\begin{flushleft}
\end{flushleft}
\end{ruledtabular}
\label{tab:results_flat_XCDM}
\end{table*}



\begin{table*}
\caption{Mean and 68\% (or 95\%) confidence limits of non-flat XCDM ($+A_L$) model [Planck $P(q)$] parameters
        from non-CMB (new), P18, P18+lensing, P18+non-CMB (new), and P18+lensing+non-CMB (new) data.
        $H_0$ has units of km s$^{-1}$ Mpc$^{-1}$.
}
\begin{ruledtabular}
\begin{tabular}{lccccc}
\\[-1mm]                      & \multicolumn{5}{c}{Non-flat XCDM models [Planck $P(q)$]}        \\[+1mm]
\cline{2-6}\\[-1mm]
  Parameter                   &  Non-CMB (new)      & P18                   &  P18+lensing            &  P18+non-CMB (new)       & P18+lensing+non-CMB (new)    \\[+1mm]
 \hline \\[-1mm]
  $\Omega_b h^2$              & $0.0285 \pm 0.0043$ & $0.02260 \pm 0.00017$ & $0.02249 \pm 0.00016$   &  $0.02245 \pm 0.00015$   &  $0.02246 \pm 0.00015$  \\[+1mm]
  $\Omega_c h^2$              & $0.118 \pm 0.014$   & $0.1181 \pm 0.0015$   & $0.1186 \pm 0.0015$     &  $0.1188 \pm 0.0013$     &  $0.1190 \pm 0.0012$  \\[+1mm]
  $100\theta_\textrm{MC}$     & $1.44 \pm 0.19$     & $1.04117 \pm 0.00032$ & $1.04107 \pm 0.00032$   &  $1.04105 \pm 0.00032$   &  $1.04102 \pm 0.00031$  \\[+1mm]
  $\tau$                      & $0.0480$            & $0.0480 \pm 0.0083$   & $0.0495 \pm 0.0082$     &  $0.0548 \pm 0.0076$     &  $0.0573 \pm 0.0074$  \\[+1mm]
  $\Omega_k$                  & $-0.177 \pm 0.067$  & $-0.048 \pm 0.035$    & $-0.011 \pm 0.017$      &  $0.0017 \pm 0.0019$     &  $0.0016 \pm 0.0019$  \\[+1mm]
                              & ($-0.177^{+0.064}_{-0.072}$)  & ($-0.048^{+0.041}_{-0.012}$)    & ($-0.0111^{+0.013}_{-0.00070}$)      &    &    \\[+1mm]
  $n_s$                       & $0.9706$            & $0.9706 \pm 0.0047$   & $0.9687 \pm 0.0046$     &  $0.9678 \pm 0.0044$     &  $0.9674 \pm 0.0042$  \\[+1mm]
  $\ln(10^{10} A_s)$          & $2.89 \pm 0.36$     & $3.027 \pm 0.017$     & $3.030 \pm 0.017$       &  $3.042 \pm 0.015$       &  $3.048 \pm 0.015$  \\[+1mm]
  $w$                         & $-0.786 \pm 0.041$  & $-1.27 \pm 0.71$      & $-1.28 \pm 0.45$        &  $-0.975 \pm 0.026$      &  $-0.980 \pm 0.026$  \\[+1mm]
                              & ($-0.786^{+0.044}_{-0.037}$)   & ($-1.27^{+0.97}_{-0.45}$)     & ($-1.28^{+0.41}_{-0.54}$)     &   &    \\[+1mm]  

 \hline \\[-1mm]
  $H_0$                       & $70.6 \pm 2.4$      & $60 \pm 16$ ($60^{+9}_{-20}$)   & $73 \pm 15$ ($73^{+20}_{-10}$)            &  $67.95 \pm 0.67$        &  $67.95 \pm 0.66$  \\[+1mm]
  $\Omega_m$                  & $0.294 \pm 0.018$   & $0.47 \pm 0.23$       & $0.30 \pm 0.15$         &  $0.3074 \pm 0.0062$     & $0.3078 \pm 0.0062$  \\[+1mm]
  $\sigma_8$                  & $0.774 \pm 0.037$   & $0.83 \pm 0.15$       & $0.87 \pm 0.12$         &  $0.801 \pm 0.010$       &  $0.8049 \pm 0.0088$  \\[+1mm]
      \hline\\[-1mm]
  $\chi_{\textrm{min}}^2$     & $1460.80$           & $2754.91$             & $2770.40$               &  $4238.67$               & $4248.26$             \\[+1mm]
  $\Delta\chi_{\textrm{min}}$ & $-13.29$            & $-10.89$              & $-4.31$                 &  $-1.57$                 & $-1.00$               \\[+1mm]
  $\textrm{DIC}$              & $1468.14$           & $2810.86$             & $2827.00$               &  $4294.75$               & $4303.54$             \\[+1mm]
  $\Delta\textrm{DIC}$        & $-9.97$             & $-7.07$               & $+0.55$                 &  $+2.42$                 & $+2.34$               \\[+1mm]
  $\textrm{AIC}$              & $1468.64$           & $2812.91$             & $2828.40$               &  $4296.67$               & $4306.26$             \\[+1mm]
  $\Delta\textrm{AIC}$        & $-9.29$             & $-6.89$               & $-0.31$                 &  $+2.43$                 & $+3.00$               \\[+1mm]
\hline
\\[-1mm]                         & \multicolumn{5}{c}{Non-flat XCDM$+A_L$ models [Planck $P(q)$]}        \\[+1mm]
\cline{2-6}\\[-1mm]
Parameter                        & & P18 ($A_L > 0.8$)        &   P18+lensing         &  P18+non-CMB (new)     & P18+lensing+non-CMB (new)  \\[+1mm]
 \hline \\[-1mm]
  $\Omega_b h^2$                 & & $0.02260 \pm 0.00017$   & $0.02250 \pm 0.00017$ & $0.02268 \pm 0.00016$  & $0.02259 \pm 0.00016$      \\[+1mm]
  $\Omega_c h^2$                 & & $0.1182 \pm 0.0015$     & $0.1184 \pm 0.0015$   & $0.1171 \pm 0.0014$    & $0.1175 \pm 0.0014$        \\[+1mm]
  $100\theta_\textrm{MC}$        & & $1.04117 \pm 0.00032$   & $1.04107 \pm 0.00032$ & $1.04124 \pm 0.00032$  & $1.04118 \pm 0.00032$      \\[+1mm]
  $\tau$                         & & $0.0479 \pm 0.0081$     & $0.0484 \pm 0.0085$   & $0.0499 \pm 0.0084$    & $0.04923 \pm 0.0082$       \\[+1mm]
  $\Omega_k$                     & & $-0.073 \pm 0.051$      & $-0.012 \pm 0.027$    & $0.0011 \pm 0.0019$    & $0.0015 \pm 0.0019$        \\[+1mm]
                                 & & ($-0.073^{+0.065}_{-0.029}$) & ($-0.012^{+0.027}_{-0.011}$)   &     &         \\[+1mm]

  $n_s$                          & & $0.9706 \pm 0.0048$     & $0.9689 \pm 0.0049$   & $0.9734 \pm 0.0047$    & $0.9717 \pm 0.0046$        \\[+1mm]
  $\ln(10^{10} A_s)$             & & $3.027 \pm 0.017$       & $3.027 \pm 0.018$     & $3.028 \pm 0.017$      & $3.027 \pm 0.017$          \\[+1mm]
  $w$                            & & $-1.36 \pm 0.77$        & $-1.32 \pm 0.58$      & $-0.958 \pm 0.027$     & $-0.958 \pm 0.026$         \\[+1mm]
                                 & & ($-1.36^{+1.1}_{-0.53}$)  & ($-1.32^{+0.71}_{-0.38}$)   &   &          \\[+1mm]

  $A_L$                          & & $0.95 \pm 0.12$ ($< 1.20$)  \ 
                                                            & $1.02 \pm 0.16$       & $1.217 \pm 0.064$      & $1.102 \pm 0.037$          \\[+1mm]
	\hline \\[-1mm]
  $H_0$                          & & $54 \pm 14$ ($54.3^{+5.9}_{-17}$)  \ 
                                                             & $72 \pm 15$           & $67.92 \pm 0.67$       & $67.94 \pm 0.68$           \\[+1mm]
  $\Omega_m$                     & & $0.57 \pm 0.24$         & $0.31 \pm 0.15$       & $0.3045 \pm 0.0062$    & $0.3049 \pm 0.0063$        \\[+1mm]
  $\sigma_8$                     & & $0.80 \pm 0.13$         & $0.86 \pm 0.13$       & $0.784 \pm 0.011$      & $0.785 \pm 0.011$          \\[+1mm]
      \hline\\[-1mm]
  $\chi_{\textrm{min}}^2$        & & $2754.46$               & $2770.28$             & $4224.83$              & $4239.70$                  \\[+1mm]
  $\Delta\chi_{\textrm{min}}^2$  & & $-11.34$                & $-4.43$               & $-15.41$               & $-9.56$                    \\[+1mm]
  $\textrm{DIC}$                 & & $2811.61$               & $2829.13$             & $4285.15$              & $4298.54$                  \\[+1mm]
  $\Delta\textrm{DIC}$           & & $-6.32$                 & $+2.68$               & $-7.18$                & $-2.66$                    \\[+1mm]
  $\textrm{AIC}$                 & & $2814.46$               & $2830.28$             & $4284.83$              & $4299.70$                  \\[+1mm]
  $\Delta\textrm{AIC}$           & & $-5.34$                 & $+1.57$               & $-9.41$                & $-3.56$                    \\[+1mm]
\end{tabular}
\\[+1mm]
\begin{flushleft}
\end{flushleft}
\end{ruledtabular}
\label{tab:results_XCDM_Planck_Pq}
\end{table*}



\begin{table*}
	\caption{Mean and 68\% (or 95\%) confidence limits of non-flat XCDM ($+A_L$) model [new $P(q)$] parameters
        from non-CMB (new), P18, P18+lensing, P18+non-CMB (new), and P18+lensing+non-CMB (new) data.
        $H_0$ has units of km s$^{-1}$ Mpc$^{-1}$.
}
\begin{ruledtabular}
\begin{tabular}{lccccc}
\\[-1mm]                       & \multicolumn{5}{c}{Non-flat XCDM models [new $P(q)$] }        \\[+1mm]
\cline{2-6}\\[-1mm]
  Parameter                    &  Non-CMB (new)      & P18                   &  P18+lensing           &  P18+non-CMB (new)       &  P18+lensing+non-CMB (new)    \\[+1mm]
 \hline \\[-1mm]
  $\Omega_b h^2$               & $0.0282 \pm 0.0044$ & $0.02256 \pm 0.00017$ & $0.02248 \pm 0.00016$  &  $0.02246 \pm 0.00015$   &  $0.02246 \pm 0.00015$  \\[+1mm]
  $\Omega_c h^2$               & $0.119 \pm 0.015$   & $0.1188 \pm 0.0015$   & $0.1188 \pm 0.0014$    &  $0.1188 \pm 0.0013$     &  $0.1190 \pm 0.0013$  \\[+1mm]
  $100\theta_\textrm{MC}$      & $1.47 \pm 0.24$     & $1.04109 \pm 0.00033$ & $1.04104 \pm 0.00032$  &  $1.04104 \pm 0.00031$   &  $1.04102 \pm 0.00032$  \\[+1mm]
  $\tau$                       & $0.0524$            & $0.0524 \pm 0.0082$   & $0.0511 \pm 0.0081$    &  $0.0552 \pm 0.0079$     &  $0.0576 \pm 0.0076$  \\[+1mm]
  $\Omega_k$                   & $-0.186 \pm 0.076$  & $-0.034 \pm 0.025$    & $-0.008 \pm 0.010$     &   $0.0016 \pm 0.0020$    &  $0.0014 \pm 0.0020$   \\[+1mm]
                               & ($-0.186^{+0.083}_{-0.067}$)  & ($-0.0338^{+0.029}_{-0.0086}$)    & ($-0.0080^{+0.0098}_{-0.0023}$)      &    &    \\[+1mm]
  $n_s$                        & $0.9653$            & $0.9653 \pm 0.0044$   & $0.9663 \pm 0.0044$    &   $0.9677 \pm 0.0043$    &  $0.9673 \pm 0.0042$  \\[+1mm]
  $\ln(10^{10} A_s)$           & $2.85 \pm 0.38$     & $3.038 \pm 0.017$     & $3.034 \pm 0.016$      &   $3.043 \pm 0.016$      &  $3.049 \pm 0.015$  \\[+1mm]
  $w$                          & $-0.785 \pm 0.042$  & $-1.27 \pm 0.61$      & $-1.27 \pm 0.41$       &  $-0.976 \pm 0.026$      &  $-0.982 \pm 0.026$   \\[+1mm]
                               & ($-0.785^{+0.045}_{-0.038}$)   & ($-1.27^{+0.79}_{-0.44}$)       & ($-1.27^{+0.40}_{-0.49}$)     &   &    \\[+1mm]  
 \hline \\[-1mm]
  $H_0$                        & $70.6 \pm 2.5$      & $63 \pm 15$ ($63^{+10}_{-20}$)          & $74 \pm 14$            &  $67.94 \pm 0.66$        &  $67.96 \pm 0.66$  \\[+1mm]
  $\Omega_m$                   & $0.296 \pm 0.018$   & $0.41 \pm 0.19$       & $0.29 \pm 0.12$        &  $0.3074 \pm 0.0062$     &  $0.3077 \pm 0.0061$  \\[+1mm]
  $\sigma_8$                   & $0.771 \pm 0.036$   & $0.85 \pm 0.13$       & $0.87 \pm 0.11$        & $0.801 \pm 0.010$        &  $0.8056 \pm 0.0089$  \\[+1mm]
      \hline\\[-1mm]
  $\chi_{\textrm{min}}^2$      & $1459.51$           & $2757.86$             & $2770.57$              & $4238.57$                & $4247.96$                      \\[+1mm]
  $\Delta\chi_{\textrm{min}}^2$& $-13.27$            & $-7.94$               & $-4.14$                & $-1.67$                  & $-1.30$              \\[+1mm]
  $\textrm{DIC}$               & $1468.73$           & $2811.78$             & $2826.60$              & $4294.90$                & $4304.26$                      \\[+1mm]
  $\Delta\textrm{DIC}$         & $-9.38$             & $-6.15$               & $+0.15$                & $+2.57$                  & $+3.06$               \\[+1mm]
  $\textrm{AIC}$               & $1468.66$           & $2815.86$             & $2828.57$              & $4296.57$                & $4305.96$                      \\[+1mm]
  $\Delta\textrm{AIC}$         & $-9.27$             & $-3.94$               & $-0.14$                & $+2.33$                  & $+2.70$               \\[+1mm]
  \hline
\\[-1mm]                         & \multicolumn{5}{c}{Non-flat XCDM$+A_L$ models [new $P(q)$]}              \\[+1mm]
\cline{2-6}\\[-1mm]
Parameter                        & & P18 ($A_L > 0.8$)       & P18+lensing           &  P18+non-CMB (new)     & P18+lensing+non-CMB (new) \\[+1mm]
 \hline \\[-1mm]
  $\Omega_b h^2$                 & & $0.02260 \pm 0.00017$   & $0.02250 \pm 0.00017$ & $0.02268 \pm 0.00017$  & $0.02258 \pm 0.00016$     \\[+1mm]
  $\Omega_c h^2$                 & & $0.1182 \pm 0.0015$     & $0.1185 \pm 0.0016$   & $0.1171 \pm 0.0014$    & $0.1175 \pm 0.0014$       \\[+1mm]
  $100\theta_\textrm{MC}$        & & $1.04116 \pm 0.00033$   & $1.04106 \pm 0.00032$ & $1.04124 \pm 0.00032$  & $1.04117 \pm 0.00032$     \\[+1mm]
  $\tau$                         & & $0.0478 \pm 0.0083$     & $0.0503 \pm 0.0086$   & $0.0500 \pm 0.0084$    & $0.04936 \pm 0.0082$      \\[+1mm]
  $\Omega_k$                     & & $-0.072 \pm 0.051$      & $-0.003 \pm 0.018$   & $0.0011 \pm 0.0019$    & $0.0015 \pm 0.0019$       \\[+1mm]
                                 & & ($-0.072^{+0.065}_{-0.030}$)  & ($-0.003^{+0.018}_{-0.011}$)   &    &        \\[+1mm]
  $n_s$                          & & $0.9706 \pm 0.0048$     & $0.9676 \pm 0.0055$   & $0.9731 \pm 0.0046$    & $0.9717 \pm 0.0046$       \\[+1mm]
  $\ln(10^{10} A_s)$             & & $3.027 \pm 0.018$       & $3.032 \pm 0.018$     & $3.028 \pm 0.017$      & $3.027 \pm 0.017$         \\[+1mm]
  $w$                            & & $-1.39 \pm 0.77$        & $-1.18 \pm 0.48$      & $-0.959 \pm 0.026$     & $-0.959 \pm 0.027$       \\[+1mm]
                                 & & ($-1.39^{+1.1}_{-0.54}$)  & ($-1.18^{+0.54}_{-0.37}$)  &      &         \\[+1mm]
  $A_L$                          & & $0.95 \pm 0.13$         & $1.07 \pm 0.14$       & $1.213 \pm 0.064$      & $1.101 \pm 0.038$         \\[+1mm]
                                 & & ($<1.19$)              & ($1.07^{+0.12}_{-0.16}$)    &    &          \\[+1mm]
	\hline \\[-1mm]
  $H_0$                          & & $55 \pm 14$ ($54.5^{+5.7}_{-17}$)   \ 
                                                             & $74 \pm 15$           & $67.92 \pm 0.65$       & $67.95 \pm 0.66$         \\[+1mm]
  $\Omega_m$                     & & $0.56 \pm 0.24$         & $0.30 \pm 0.13$       & $0.3046 \pm 0.0061$    & $0.3047 \pm 0.0062$      \\[+1mm]
  $\sigma_8$                     & & $0.80 \pm 0.12$         & $0.84 \pm 0.12$       & $0.785 \pm 0.011$      & $0.786 \pm 0.011$        \\[+1mm]
      \hline\\[-1mm]
  $\chi_{\textrm{min}}^2$        & & $2754.40$               & $2770.27$             & $4224.52$              &  $4239.76$                \\[+1mm]
  $\Delta\chi_{\textrm{min}}^2$  & & $-11.40$                & $-4.44$               & $-15.72$               &  $-9.50$                  \\[+1mm]
  $\textrm{DIC}$                 & & $2811.71$               & $2828.10$             & $4284.84$              &  $4298.27$                \\[+1mm]
  $\Delta\textrm{DIC}$           & & $-6.22$                 & $+1.65$               & $-7.49$                &  $-2.93$                  \\[+1mm]
  $\textrm{AIC}$                 & & $2814.40$               & $2830.27$             & $4284.52$              &  $4299.76$                \\[+1mm]
  $\Delta\textrm{AIC}$           & & $-5.40$                 & $+1.56$               & $-9.72$                &  $-3.50$                  \\[+1mm]
\end{tabular}
\\[+1mm]
\begin{flushleft}
\end{flushleft}
\end{ruledtabular}
\label{tab:results_new_Pq_XCDM}
\end{table*}



\subsubsection{Non-CMB (new) cosmological constraints}
\label{sec:XCDM_non-CMB (new)}

Here we summarize the XCDM models cosmological parameter constraints from the non-CMB (new) data set. The non-CMB (new) data constraints on the XCDM$+A_L$ models are identical to those on the XCDM models so we do not distinguish between these models in this subsubsection.

The results for the five-parameter flat XCDM model obtained with non-CMB (new) data are in Table \ref{tab:results_flat_XCDM} and Figs.\ \ref{fig:flat_XCDM_P18_vs_nonCMB_1} and \ref{fig:FX_Alens_p18_ncmb}. For the equation of state parameter we obtain $w=-0.853^{+0.043}_{-0.033}$ which is 4.45$\sigma$ away from the cosmological constant $w=-1$ value and favors quintessence-like evolution. For earlier indications that non-CMB data favor quintessence-like, $w > -1$, evolution in flat and non-flat XCDM models, see \cite{Park:2018tgj, Cao:2021cix, Cao:2022ugh, Cao:2023eja, Dong:2023jtk, VanRaamsdonk:2023ion}. Regarding the derived parameters, we obtain $H_0=69.8\pm 2.5$ km s$^{-1}$ Mpc$^{-1}$ and $\Omega_m = 0.270\pm 0.012$, in good agreement with the flat XCDM model values of $H_0=69.05\pm 2.4$ km s$^{-1}$ Mpc$^{-1}$ and $\Omega_m = 0.292\pm 0.016$ from a joint analysis of $H(z)$, BAO, Pantheon+ SNIa, quasar angular size, reverberation-measured \mii\ and \civ\ quasar, and 118 Amati correlation gamma-ray burst data, \cite{Cao:2023eja} Table VII. 

Table \ref{tab:results_XCDM_Planck_Pq} [\ref{tab:results_new_Pq_XCDM}] and Figs.\ \ref{fig:XCDM_Planck_Pq_P18_vs_nonCMB_1} and \ref{fig:NX_Alens_ns_p18_ncmb} [\ref{fig:XCDM_new_Pq_P18_vs_nonCMB_1} and \ref{fig:TNX_Alens_ns1_p18_ncmb}] show results for the six-parameter non-flat XCDM Planck [new] $P(q)$ cosmological model from non-CMB (new) data. We obtain $\Omega_k = -0.177^{+0.064}_{-0.072}$ [$-0.186^{+0.083}_{-0.067}$] and $w=-0.786^{+0.044}_{-0.037}$ [$-0.785^{+0.045}_{-0.038}$]. The first result indicates that closed non-flat hypersufaces are favored by 2.77$\sigma$ [2.24$\sigma$] and the second one favors quintessence-like evolution over a cosmological constant by 5.78$\sigma$ [5.66$\sigma$]. As for the derived parameters, as in the flat XCDM model, we find agreement with the $H_0=69.26\pm 2.45$ km s$^{-1}$ Mpc$^{-1}$ and $\Omega_m = 0.296\pm 0.020$ values of \cite{Cao:2023eja} from a joint analysis of the data sets listed at the end of the previous paragraph. We obtain $H_0=70.6\pm 2.4$ km s$^{-1}$ Mpc$^{-1}$ [$70.6\pm 2.5$ km s$^{-1}$ Mpc$^{-1}$] and $\Omega_m = 0.294\pm 0.018$ [$0.296\pm 0.018$].


\subsubsection{P18 data cosmological constraints}
\label{sec:XCDM P18 }

In the case of the flat XCDM cosmological model with seven primary parameters (see Table \ref{tab:results_flat_XCDM} and Fig.\ \ref{fig:flat_XCDM_P18_vs_nonCMB_1}), from P18 data alone we obtain $\Omega_m = 0.197\pm 0.046$, which differs by $-1.94\sigma$ from the flat XCDM model value $\Omega_m = 0.292\pm 0.016$ from a joint analysis of $H(z)$, BAO, Pantheon+ SNIa, quasar angular size, reverberation-measured \mii\ and \civ\ quasar, and 118 Amati correlation gamma-ray burst data, \cite{Cao:2023eja} Table VII. The error bars associated to the $H_0$ parameter cannot be determined and the best estimation possible is $H_0> 70.2$ km s$^{-1}$ Mpc$^{-1}$ (95\% confidence limit).

The improvement in the fit with respect to the flat $\Lambda$CDM cosmological model with $w=-1$ is positive according to the DIC and AIC statistical criteria, see Table \ref{tab:results_flat_XCDM}. This is reflected in the P18 data value for the X-fluid equation of state parameter $w=-1.59^{+0.15}_{-0.34}$, a 3.93$\sigma$ deviation from $w=-1$. This difference in the equation of state parameter is not accompanied by equally significant changes in the other primary parameters when compared to the values obtained in the flat $\Lambda$CDM model, with the largest being $-0.19\sigma$ for $\Omega_b{h^2}$. On the other hand, there are significant differences in derived parameters when compared to the flat $\Lambda$CDM values. In particular, for $\Omega_m$, and $\sigma_8$ we find differences of $2.56\sigma$ and $-2.27\sigma$, respectively. As for the error bars, those associated with the primary parameters barely change (the largest change is an increase of 3.1\% for 100$\theta_{\textrm{MC}}$). However, the error bars of the derived parameters are significantly affected, with increases of 82\% and 90\% for $\Omega_m$ and $\sigma_8$, respectively.

Using Table \ref{tab:results_flat_XCDM} we can compare P18 data results for the seven-parameter flat XCDM model (upper half) and for the eight-parameter flat XCDM+$A_L$ model (lower half and Fig.\ \ref{fig:FX_Alens_p18_ncmb}). There are changes in the values of the primary parameters but none above 1$\sigma$. In particular, the values of $\Omega_{b}h^2$, $\Omega_{c}h^2$, and $n_s$ differ by $-0.79\sigma$, $+0.93\sigma$, and $-0.79\sigma$, respectively. For the equation of state parameter in the flat XCDM model, we find $w=-1.59^{+0.15}_{-0.34}$ (deviating by $3.93\sigma$ from $w=-1$). However, in the flat XCDM+$A_L$ model we find $w=-1.23^{+0.31}_{-0.59}$ ($0.74\sigma$ away from $w=-1$), resulting in a difference of $-0.59\sigma$ between the flat XCDM and flat XCDM+$A_L$ values from P18 data. In both cases there is a preference for phantom-like behavior. In the flat XCDM+$A_L$ model, we find $A_L = 1.180^{+0.062}_{-0.10}$ which is 1.8$\sigma$ away from $A_L = 1$. As for the derived parameters, the matter density parameters $\Omega_m$ differ by $-0.61\sigma$, while the values of $\sigma_8$ differ by $0.82\sigma$. When comparing the flat XCDM and flat XCDM+$A_L$ results, we observe an increase of 38\% in the size of the error bars of the dark energy equation of state parameters. For the rest of the primary parameters the largest increase is 12\% for $\Omega_{b}h^2$.
As for the derived parameters $H_0$, $\Omega_m$, and $\sigma_8$, the corresponding increases are 36\%, 58\%, and 41\% respectively.

As expected, the simultaneous consideration of both $w$ and $\Omega_k$ (see Tables \ref{tab:results_XCDM_Planck_Pq} and \ref{tab:results_new_Pq_XCDM} and Figs.\ \ref{fig:XCDM_Planck_Pq_P18_vs_nonCMB_1} and \ref{fig:XCDM_new_Pq_P18_vs_nonCMB_1}) makes the already existing degeneracies even bigger. In particular, in the non-flat XCDM cases the values of $w$ and $\Omega_m$ increase and the value of $H_0$ decreases with respect to the flat XCDM model values. Therefore, from the results obtained, it is clear that P18 data alone cannot break the degeneracies between $H_0$, $\Omega_{m}$, $\Omega_k$, and $w$. For the non-flat XCDM Planck $P(q)$ and the non-flat XCDM new $P(q)$ models, we find $H_0=60^{+9}_{-20}$ km s$^{-1}$ Mpc$^{-1}$ and $63^{+10}_{-20}$ km s$^{-1}$ Mpc$^{-1}$, respectively, whereas for the matter parameter we get $\Omega_m =0.47\pm 0.23$ and $0.41\pm 0.19$, respectively, which are in agreement within 1$\sigma$ with the values obtained in \cite{Cao:2023eja}, $H_0=69.26\pm 2.45$ km s$^{-1}$ Mpc$^{-1}$ and $\Omega_m =0.296\pm 0.020$ .

Comparing the results of the seven-parameter non-flat $\Lambda$CDM Planck $P(q)$ model (see Table \ref{tab:results_Planck_Pq}) and the results for the eight-parameter non-flat XCDM Planck $P(q)$ model (see Table \ref{tab:results_XCDM_Planck_Pq}) we observe that there is almost no difference in the values of the six primary cosmological parameters in common with the flat $\Lambda$CDM model, with the largest difference being +0.026$\sigma$ for $\tau$. Regarding the curvature parameter, for the $\Lambda$CDM Planck $P(q)$ model we obtain $\Omega_k = -0.043^{+0.018}_{-0.015}$ while for the XCDM Planck $P(q)$ case the corresponding value is $\Omega_k = -0.048^{+0.041}_{-0.012}$, indicating a difference of $+0.11\sigma$. Both models show a clear preference for closed geometry, deviating from flat by 2.39$\sigma$ and 1.17$\sigma$, respectively. The $\Omega_k$ error bars are 51\% larger in the XCDM Planck $P(q)$ model compared to the $\Lambda$CDM Planck $P(q)$ model value. For the equation of state parameter in the XCDM Planck $P(q)$ case we obtain $w=-1.27^{+0.97}_{-0.45}$ which indicates a preference for phantom-like behavior at 0.28$\sigma$. As for the derived parameters $H_0$, $\Omega_m$, and $\sigma_8$, the values obtained for both models differ at $-0.27\sigma$, 0.05$\sigma$, and $-0.36\sigma$ and they have error bars +75.9\%, +73.0\%, and +90.0\% larger in the XCDM Planck $P(q)$ case compared to the $\Lambda$CDM Planck $P(q)$ model. 

If we look at Tables \ref{tab:results_new_Pq} and \ref{tab:results_new_Pq_XCDM} we can compare the results obtained with P18 data for the seven-parameter non-flat $\Lambda$CDM new $P(q)$ model and the eight-parameter non-flat XCDM new $P(q)$ model. As we noted in the Planck $P(q)$ case, there are also no significant differences in the values of the primary cosmological parameters in common with the flat $\Lambda$CDM model in the new $P(q)$ case, the largest being $-0.05\sigma$ for $\Omega_{b}h^2$. As for the curvature parameter, the $\Lambda$CDM new $P(q)$ model yields $\Omega_k=-0.033^{+0.017}_{-0.011}$ (1.94$\sigma$), while the XCDM new $P(q)$ model gives $\Omega_k=-0.0338^{+0.029}_{-0.0086}$ (1.17$\sigma$). Again, the results indicate a preference for closed geometry. The equation of state parameter value in the XCDM new $P(q)$ model obtained from an analysis of P18 data is $w=-1.27^{+0.79}_{-0.44}$ which differs from the cosmological constant by 0.34$\sigma$. Regarding the derived parameters, the differences in $H_0$, $\Omega_m$, and $\sigma_8$ are $-0.30\sigma$, +0.17$\sigma$, and $-0.49\sigma$, and the error bars for the $w\neq -1$ model are larger by 76.0\%, 71.1\%, and 89.2\%, respectively.  

Comparing P18 data results for the eight-parameter non-flat XCDM Planck [new] $P(q)$ model and the nine-parameter non-flat XCDM Planck (new) $P(q)+A_L$ model (see Tables \ref{tab:results_XCDM_Planck_Pq} and \ref{tab:results_new_Pq_XCDM} and Figs.\ \ref{fig:XCDM_Planck_Pq_P18_vs_nonCMB_1} --- \ref{fig:TNX_Alens_ns1_p18_ncmb}), we find similar values of the primary parameters in common with the flat $\Lambda$CDM standard model, with the largest difference affecting $\Omega_{c}h^2$ [$\tau$] by $-0.047\sigma$ [$0.39\sigma$]. While for the non-flat XCDM Planck [new] $P(q)$ model we get $\Omega_k = -0.048^{+0.041}_{-0.012}$ [$-0.0338^{+0.029}_{-0.0086}$], which is $1.17\sigma$ [$1.17\sigma$] away from flat hypersurfaces, for the non-flat XCDM Planck [new] $P(q)+A_L$ model we obtain $\Omega_k = -0.073^{+0.065}_{-0.029}$ [$-0.072^{+0.065}_{-0.030}$], favoring a closed geometry at a significance level of 1.12$\sigma$ [1.11$\sigma$]. The difference between the {$P(q)$ and $P(q)+A_L$ model} values is $0.38\sigma$ [$0.58\sigma$]. For the dark energy equation of state parameter, in the non-flat XCDM Planck [new] $P(q)$ model we find $w=-1.27^{+0.97}_{-0.45}$ [$-1.27^{+0.79}_{-0.44}$] deviating from $w=-1$ by $0.28\sigma$ [$0.34\sigma$] while for the non-flat XCDM Planck [new] $P(q)+A_L$ model we get $w=-1.36^{+1.1}_{-0.53}$ (0.33$\sigma$) [$-1.39^{+1.1}_{-0.54}$ (0.35$\sigma$)] with $P(q)$ and $P(q)+A_L$ model values differing by 0.076$\sigma$ [0.10$\sigma$]. For the non-flat XCDM Planck [new] $P(q)+A_L$ model, P18 data are only able to provide a 95\% upper bound of $<1.20$ [$<1.19$] on the lensing consistency parameter $A_L$. The differences observed in the values of the derived parameters are also not significant, in particular for $H_0$, $\Omega_m$, and $\sigma_8$ we find $0.27\sigma$ [$0.38\sigma$], $-0.30\sigma$ [$-0.49\sigma$], and $0.15\sigma$ [$0.28\sigma$]. Due to degeneracies, in some cases we find smaller error bars when $A_L$ is allowed to vary (compared to the $A_L = 1$ case). For $\Omega_k$ and $w$, they increase by +31.37\% [+50.98\%] and +7.79\% [20.78\%], respectively, while for the derived parameters, $H_0$, $\Omega_m$, and $\sigma_8$, the corresponding error bar changes are $-14.29$\% [$-7.14$\%], $+4.17$\% [$+20.83$\%] and $-15.38$\% [$-8.33$\%].

\begin{figure*}[htbp]
\centering
\mbox{\includegraphics[width=170mm]{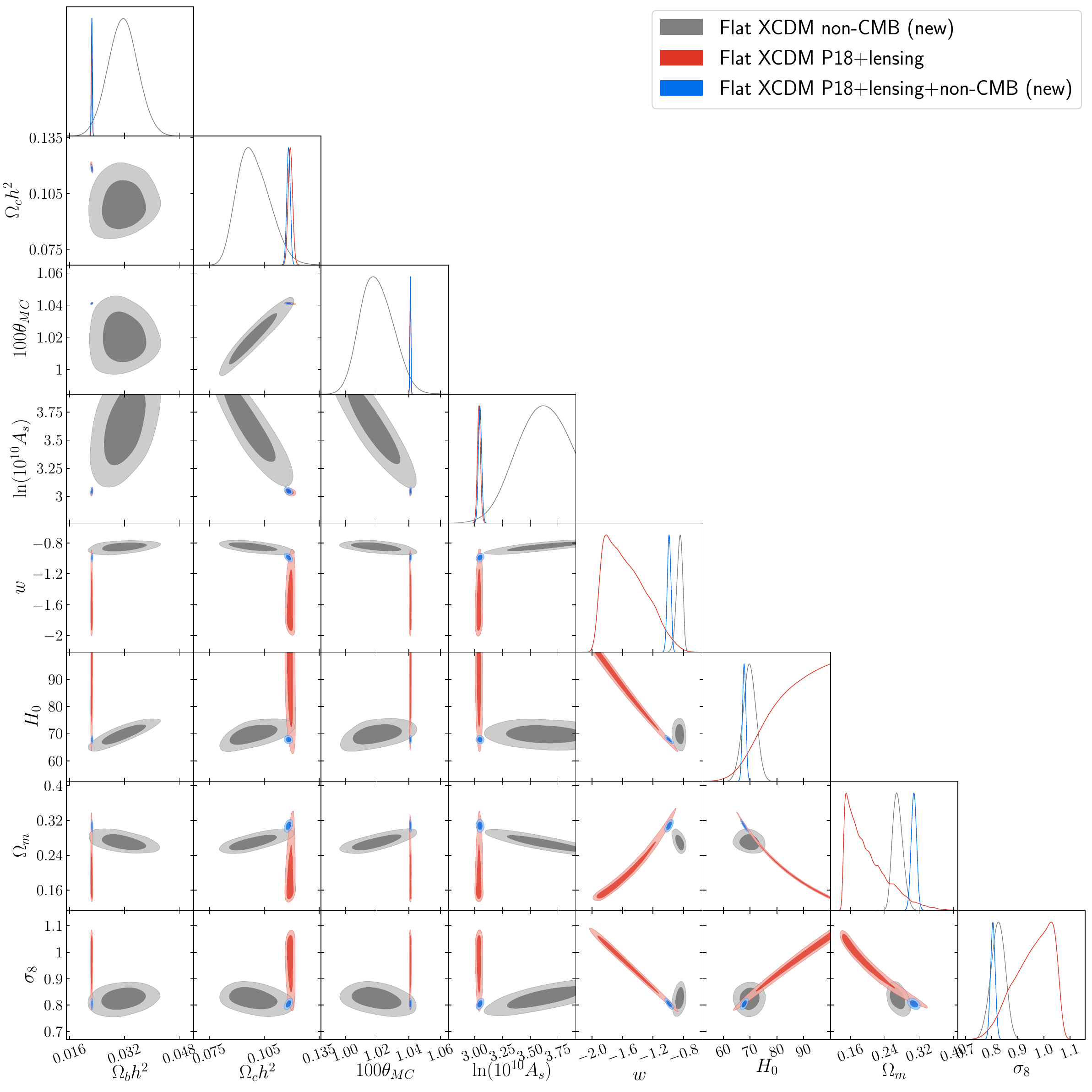}}
        \caption{Likelihood distributions of flat XCDM model 
        parameters favored by non-CMB (new) data.
        Results for P18+lensing and P18+lensing+non-CMB (new) data are shown for comparison.
}
\label{fig:flat_XCDM_P18_vs_nonCMB_2}
\end{figure*}
\begin{figure*}[htbp]
\centering
\mbox{\includegraphics[width=170mm]{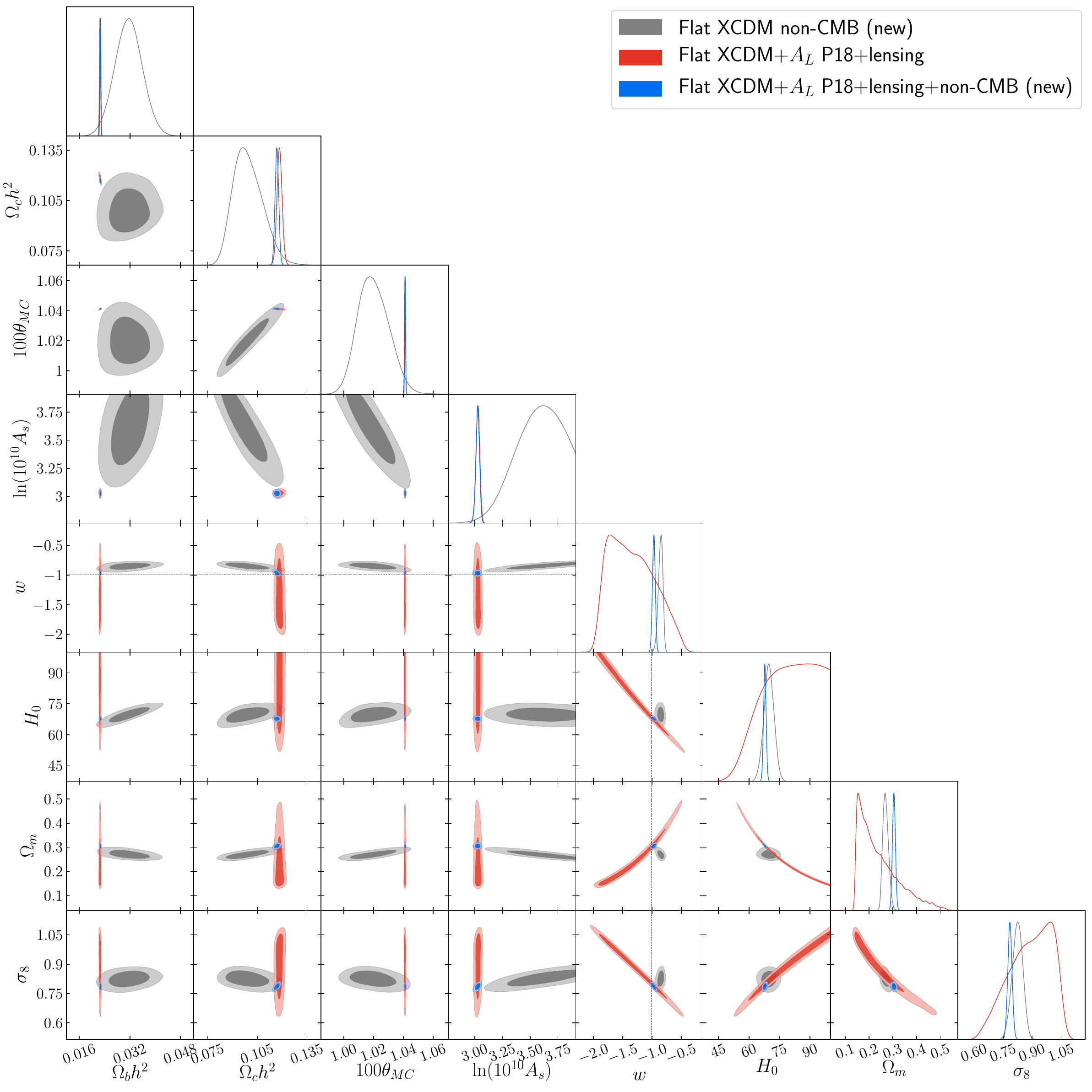}}
        \caption{Likelihood distributions of flat XCDM$+A_L$ model
        parameters favored by non-CMB (new) data. Results for P18+lensing and P18+lensing+non-CMB (new) data are shown for comparison.
}
\label{fig:FX_Alens_p18len_ncmb}
\end{figure*}
\begin{figure*}[htbp]
\centering
\mbox{\includegraphics[width=170mm]{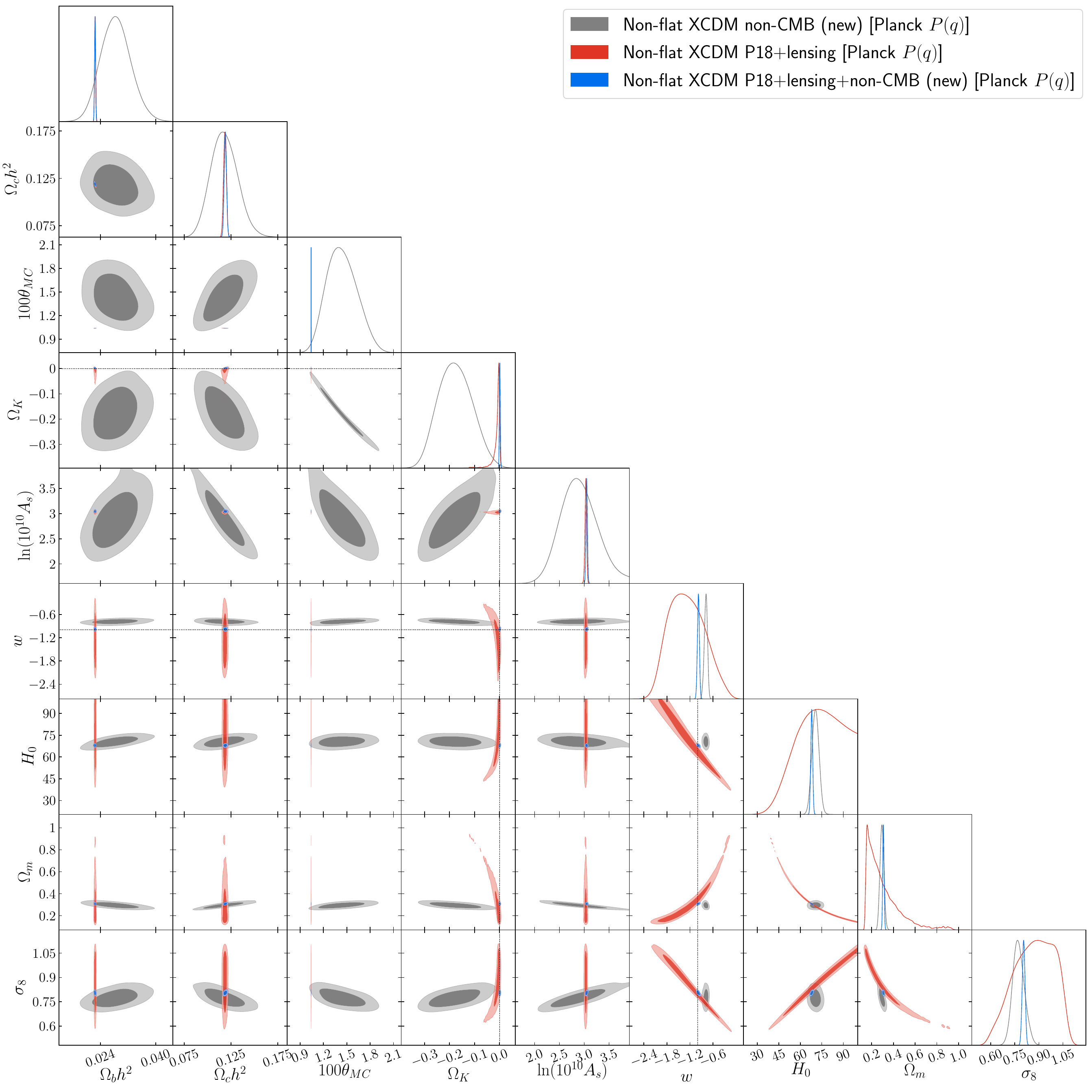}}
	\caption{Likelihood distributions of non-flat XCDM model [Planck $P(q)$]
        parameters favored by non-CMB (new) data.
        Results for P18+lensing and P18+lensing+non-CMB (new) data are shown for comparison.
}
\label{fig:XCDM_Planck_Pq_P18_vs_nonCMB_2}
\end{figure*}
\begin{figure*}[htbp]
\centering
\mbox{\includegraphics[width=170mm]{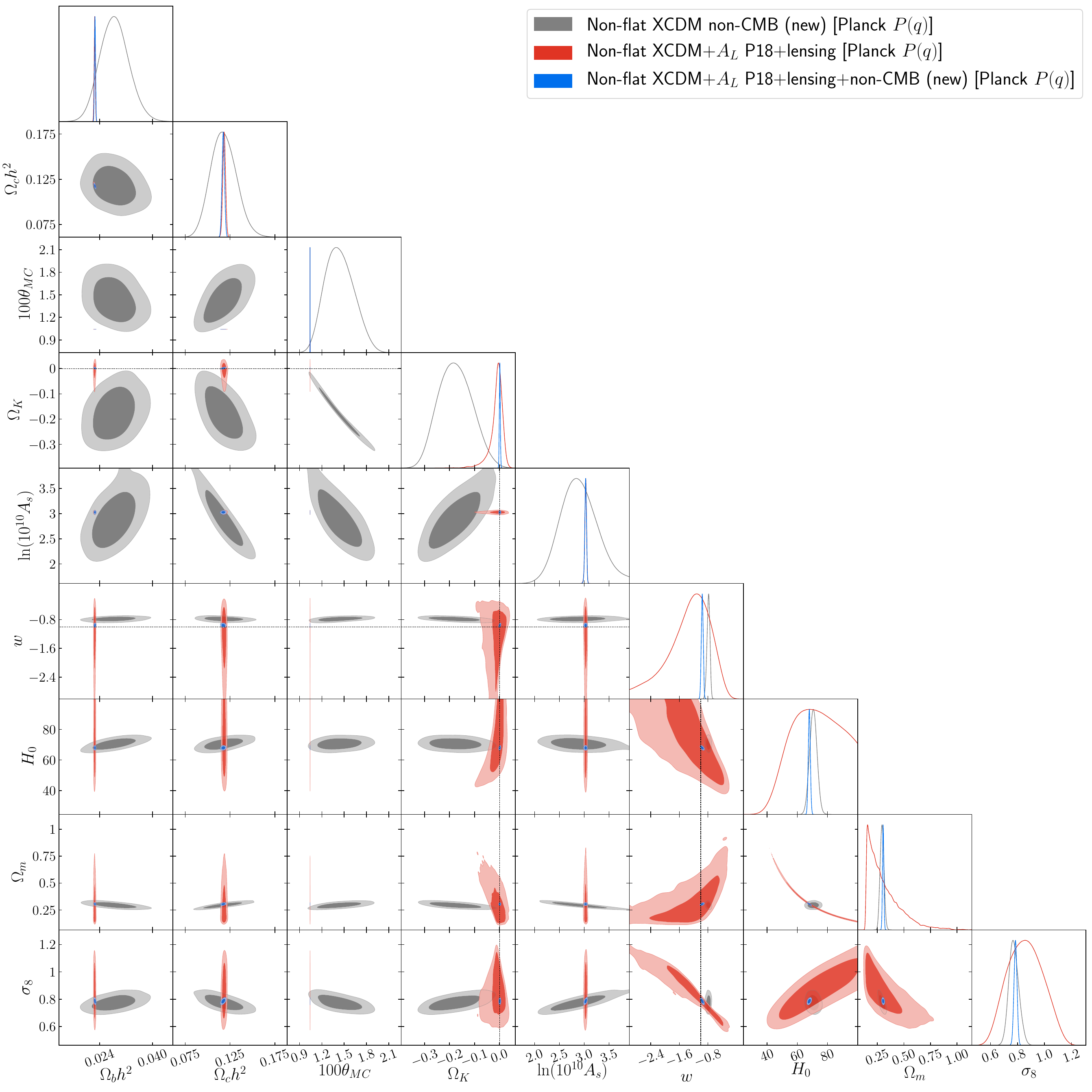}}
        \caption{Likelihood distributions of non-flat XCDM$+A_L$ model [Planck $P(q)$]
        parameters favored by non-CMB (new) data. Results for P18+lensing and P18+lensing+non-CMB (new) data are shown for comparison.
}
\label{fig:NX_Alens_ns_p18len_ncmb}
\end{figure*}
\begin{figure*}[htbp]
\centering
\mbox{\includegraphics[width=170mm]{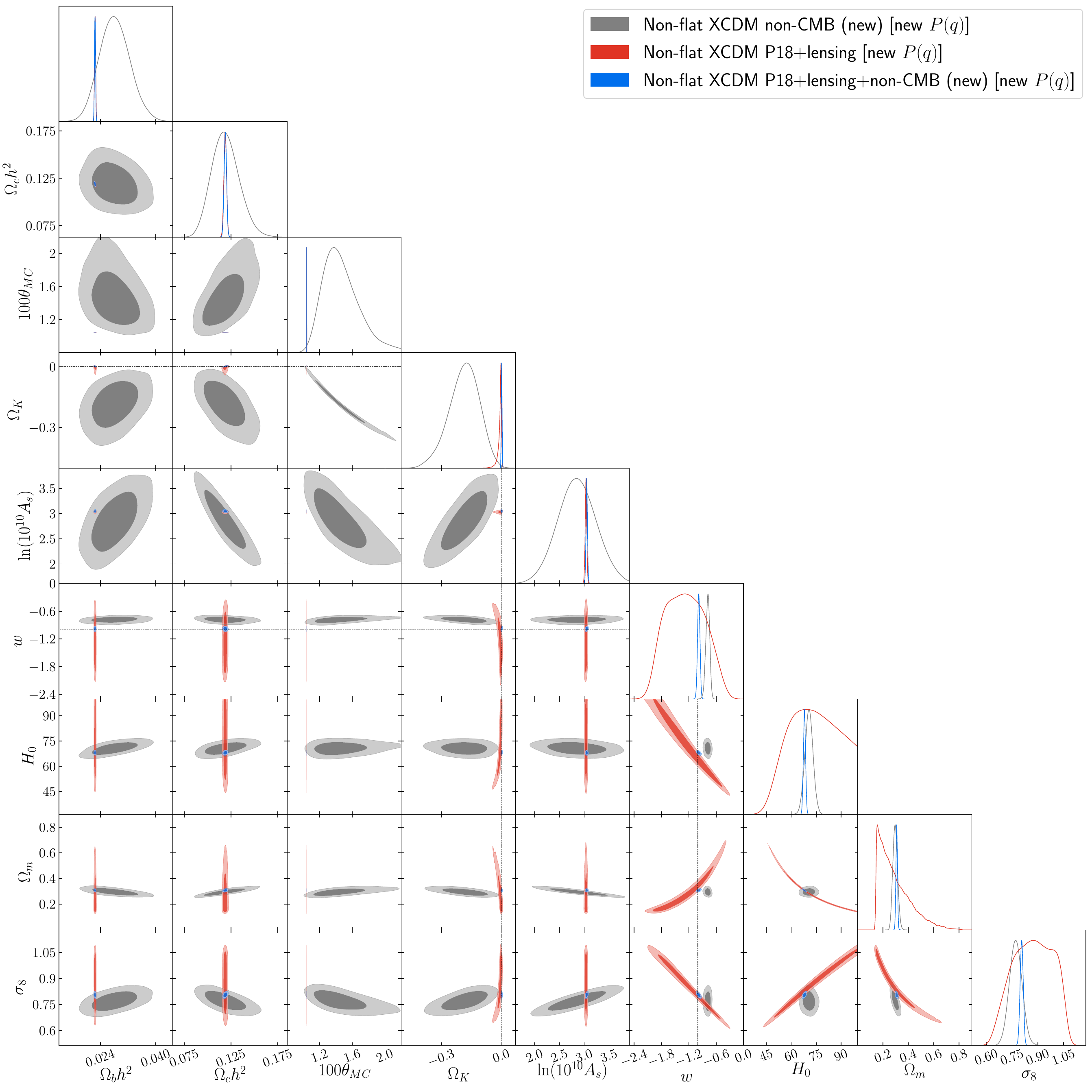}}
        \caption{Likelihood distributions of non-flat XCDM model [new $P(q)$]
        parameters favored by non-CMB (new) data.
        Results for P18+lensing and P18+lensing+non-CMB (new) data are shown for comparison.
}
\label{fig:XCDM_new_Pq_P18_vs_nonCMB_2}
\end{figure*}
\begin{figure*}[htbp]
\centering
\mbox{\includegraphics[width=170mm]{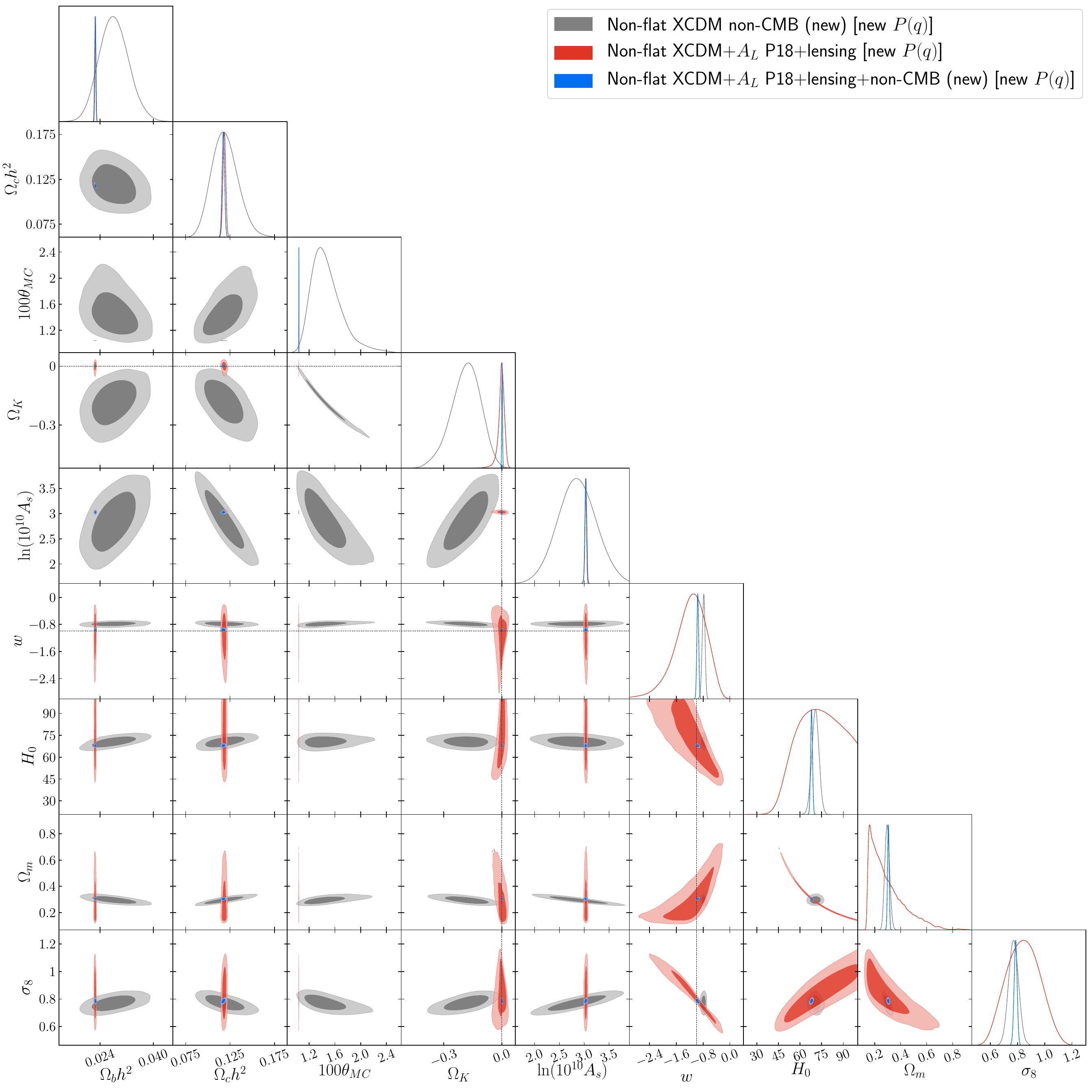}}
        \caption{Likelihood distributions of non-flat XCDM$+A_L$ model [new $P(q)$]
        parameters favored by non-CMB (new) data. Results for P18+lensing and P18+lensing+non-CMB (new) data are shown for comparison.
}
\label{fig:TNX_Alens_ns1_p18len_ncmb}
\end{figure*}

\subsubsection{P18+lensing cosmological constraints}
\label{sec:XCDM P18+lensing }

Comparing the seven-parameter flat XCDM model primary cosmological parameter constraints for P18 and P18+lensing data, shown in Table \ref{tab:results_flat_XCDM} and in Figs.\ \ref{fig:flat_XCDM_P18_vs_nonCMB_1} and \ref{fig:flat_XCDM_P18_vs_nonCMB_2}, we observe only minor differences, with the largest occurring in $\Omega_{c}h^2$ (+0.38$\sigma$) and $n_s$ ($-0.22\sigma$). When P18+lensing data are analyzed, the resulting cosmological parameter error bars are similar to those for P18 data but slightly smaller, with the largest decrease being $-16$\% for $\Omega_{c}h^2$. 
For the equation of state parameter, we obtain $w=-1.55^{+0.16}_{-0.35}$ using P18+lensing data, which is 3.44$\sigma$ in favor of phantom-like behavior and differs by only 0.11$\sigma$ from the P18 data value ($w=-1.59^{+0.15}_{-0.34}$), which is 3.93$\sigma$ away from $w = -1$ of the cosmological constant. For the derived parameters, $H_0$ cannot be properly constrained and the 95\% limit is $H_0>69.6$ km s$^{-1}$ Mpc$^{-1}$, whereas for the other two we have $\Omega_m = 0.200\pm 0.048$ and $\sigma_8 = 0.960\pm 0.071$, which differ by $-0.045\sigma$ and 0.14$\sigma$ from the corresponding values obtained with P18 data. Interestingly, for $\Omega_m$ we find that the error bars obtained using P18 data are smaller than those obtained  using P18+lensing data, by 4.17\%. This could mean that adding lensing data to P18 data still does not break the large degeneracies between cosmological parameters.  

The results in Table \ref{tab:results_flat_XCDM} (and Figs.\ \ref{fig:FX_Alens_p18_ncmb} and \ref{fig:FX_Alens_p18len_ncmb}) allow us to compare P18 and P18+lensing data constraints on the eight-parameter flat XCDM+$A_L$ model. No significant differences in the values of the primary parameters are seen when lensing data are added to the mix, with the largest changes appearing in the $\Omega_{b} h^2$ (0.33$\sigma$) and $n_s$ (0.22$\sigma$) mean values. When P18 data are considered we obtain $w=-1.23^{+0.31}_{-0.59}$ whereas when we use P18+lensing data we get $w=-1.34^{+0.26}_{-0.51}$, both values favoring phantom-like behavior at 0.74$\sigma$ and 1.31$\sigma$. The difference between the two result is 0.17$\sigma$, and the size of the $w$ error bars decreases by $-13.51$\% when moving from P18 data to P18+lensing data. For the lensing consistency parameter $A_L$, from P18 data we obtain $A_L = 1.180^{+0.062}_{-0.10}$ which deviates from the expected value $A_L =1$ by 1.80$\sigma$, while from P18+lensing data we obtain $A_L = 1.054^{+0.039}_{-0.059}$ which prefers $A_L>1$ at 0.92$\sigma$. We observe a shrinkage in the $A_L$ error bars of $-76.36$\% when lensing data are included, and the two mean $A_L$ values differ at 1.17$\sigma$. In regard to the derived parameters, $\Omega_m$ and $\sigma_8$, the mean value results differ by 0.20$\sigma$ and $-0.18\sigma$ and the error bars decrease by $-32.53$\% and $-9.09$\% when lensing data are also included.  

From the results shown in Table \ref{tab:results_flat_XCDM} we can compare P18+lensing data constraints on the seven-parameter flat XCDM (upper half of the table and Fig.\ \ref{fig:flat_XCDM_P18_vs_nonCMB_2}) and the eight-parameter flat XCDM+$A_L$ (lower half of the table and Fig.\ \ref{fig:FX_Alens_p18len_ncmb}) cosmological models. For the primary parameters the differences between the two sets of results are less than 1$\sigma$. The larger changes affect $\Omega_{c}h^2$ and $n_s$ which differ at 0.47$\sigma$ and $-0.57\sigma$ respectively, whereas for the dark energy equation of state parameter $w$ the difference between the two results is $-0.39\sigma$. As expected, in the case of the flat XCDM+$A_L$ model we find larger error bars for the primary parameters, than those found in the flat XCDM model, with the largest increases of 20\%, 17\%, and 30\% corresponding to the error bars of $\Omega_{c}h^2$, $\ln(10^{10}A_s)$, and $w$. For the derived parameters, the $\Omega_m$ and $\sigma_8$ values differ by $-0.44\sigma$ and 0.53$\sigma$ with increases in the size of the error bars of 42\% and 35\% when moving from the flat XCDM model to the flat XCDM+$A_L$ model. 
 
Tables \ref{tab:results_XCDM_Planck_Pq} and \ref{tab:results_new_Pq_XCDM} and Figs.\ \ref{fig:XCDM_Planck_Pq_P18_vs_nonCMB_2} and \ref{fig:XCDM_new_Pq_P18_vs_nonCMB_2} list and show results for the non-flat XCDM Planck $P(q)$ and the non-flat XCDM new $P(q)$ models when P18 and P18+lensing data are used in the analyses. Regarding the primary parameters in common with the flat $\Lambda$CDM model, we do not find significant changes (the differences are less than 1$\sigma$) when we move from P18 data to P18+lensing data, with the largest difference for the XCDM Planck [new] $P(q)$ case being 0.47$\sigma$ [0.34$\sigma$] for $\Omega_{b}h^2$. As for the error bars, we find a general reduction in size when lensing data are added to the mix. In the XCDM Planck $P(q)$ case we find a change of $-6.25$\% in the $\Omega_{b}h^2$ error bar value, while in the XCDM new $P(q)$ model we obtain changes of $-6.25$\% for both $\Omega_{b}h^2$ and $\ln(10^{10}A_s)$. On the other hand, there are non-negligible changes in $\Omega_k$ and $w$ when lensing data are included in the analysis. For the XCDM Planck [new] $P(q)$ model the value $\Omega_k=-0.0111^{+0.013}_{-0.00070}$ [$-0.0080^{+0.0098}_{-0.0023}$] indicates a $0.85\sigma$ [$0.82\sigma$] preference for a closed Universe and a mild difference with the P18 data value, $\Omega_k=-0.048^{+0.041}_{-0.012}$ [$-0.0338^{+0.029}_{-0.0086}$], of $-0.90\sigma$ [$-0.89\sigma$]. The $\Omega_k$ error bars obtained with P18 data are approximately a factor of 2.1 (2.5) larger than those obtained with P18+lensing data. Interestingly, for both of the non-flat XCDM models, the central value of $w$ is barely affected by the addition of lensing data, however there is a reduction in the size of the $w$ error bars (of 50\% for the XCDM Planck $P(q)$ and 38\% for the XCDM new $P(q)$ cases) which increases the evidence favoring phantom behavior. In particular, for the XCDM Planck [new] $P(q)$ case, from P18+lensing data, we get $w=-1.28^{+0.41}_{-0.54}$ [$-1.27^{+0.40}_{-0.49}$] which deviates from $w=-1$ by 0.68$\sigma$ [0.68$\sigma$]. As for the derived parameters, in the case of the XCDM Planck (new) $P(q)$ model, the differences between the P18+lensing data and P18 data values, for $H_0$, $\Omega_m$, and $\sigma_8$ are $-0.97\sigma$ [$-0.64\sigma$], +0.62$\sigma$ [0.53$\sigma$], and $-0.21\sigma$ [$-0.12\sigma$], respectively. The largest reduction in the error bars affects $\Omega_m$, with a decrease of 53\% [58\%]. 

Results in the lower half of Table \ref{tab:results_XCDM_Planck_Pq} [\ref{tab:results_new_Pq_XCDM}] (also see Figs.\ \ref{fig:NX_Alens_ns_p18_ncmb} and \ref{fig:NX_Alens_ns_p18len_ncmb} [\ref{fig:TNX_Alens_ns1_p18_ncmb} and \ref{fig:TNX_Alens_ns1_p18len_ncmb}]) allows us to compare P18 data and P18+lensing data constraints on the nine-parameter non-flat XCDM Planck [new] $P(q)+A_L$ model. The differences in the mean values of the primary parameters are less than 1$\sigma$, with the largest differences occurring in $\Omega_{b}h^2$ (0.42$\sigma$ [0.42$\sigma$]), $100\theta_{\text{MC}}$ (0.22$\sigma$ [0.22$\sigma$]), and $n_s$ (0.22$\sigma$ [0.41$\sigma$]). The primary parameter error bars also do not significantly differ, with the largest differences being $+4.71$\% [$-6.25$\%] for $\ln(10^{10}A_s)$ [$\Omega_{c}h^2$]. For P18 data we obtain $\Omega_k=-0.073^{+0.065}_{-0.029}$ [$-0.072^{+0.065}_{-0.030}$] (favoring closed spatial hypersurfaces by 1.12$\sigma$ [1.11$\sigma$]). When P18+lensing data are used we find $\Omega_k=-0.012^{+0.027}_{-0.011}$ [$-0.003^{+0.018}_{-0.011}$] (in favor of closed geometry by 0.44$\sigma$ [0.17$\sigma$]), with the P18 and P18+lensing values differing by $-0.93\sigma$ [$-1.05\sigma$]. The $\Omega_k$ error bars obtained with P18 data are about 1.9 [2.8] times larger than those obtained with P18+lensing data. In regard to the dark energy equation of state parameter, from P18 data we find $w=-1.36^{+1.1}_{-0.53}$ [$-1.39^{+1.1}_{-0.54}$], whereas when P18+lensing data are used we get $w=-1.32^{+0.71}_{-0.38}$ [$-1.18^{+0.54}_{-0.37}$], with the two results differing by $-0.034\sigma$ [$-0.17\sigma$]. Both values favor phantom-like behavior at 0.33$\sigma$ [0.35$\sigma$] and 0.45$\sigma$ [0.42$\sigma$], respectively, with a shrinkage in the error bars of $-32.76$\% [$-60.42$\%]. P18 data cannot properly constrain the lensing consistency parameter $A_L$ and only provide a 95\% upper bound of $A_L<1.20$ [$<1.19$]. On the other hand, when P18+lensing data are used we obtain $A_L = 1.02\pm 0.16$ [$ 1.07^{+0.12}_{-0.16}$], which favors $A_L > 1$ values by only 0.13$\sigma$ [0.44$\sigma$]. When we look at the values of the derived parameters $H_0$, $\Omega_m$, and $\sigma_8$, we observe differences at $-1.10\sigma$ [$-1.29\sigma$], 0.92$\sigma$ [0.95$\sigma$], and $-0.33\sigma$ [0.24$\sigma$], respectively. The most significant reduction in the size of the error bars is for $\Omega_m$ with a decrease of $-60$\% [$-85$\%]. 

Comparing the P18+lensing data results for the eight-parameter non-flat XCDM Planck $P(q)$ model and the nine-parameter non-flat XCDM Planck $P(q)+A_L$ model, see Table \ref{tab:results_XCDM_Planck_Pq} and Figs.\ \ref{fig:XCDM_Planck_Pq_P18_vs_nonCMB_2} and \ref{fig:NX_Alens_ns_p18len_ncmb}, we see that there are no significant differences in the mean values of the primary cosmological parameters. For $\Omega_{c}h^2$ and $\ln(10^{10} A_s )$ we find values differing at 0.094$\sigma$ and 0.12$\sigma$, the dark energy equation of state parameter $w$ values differ by 0.054$\sigma$, and for the curvature parameter $\Omega_k$ the difference is 0.031$\sigma$. In regard to the increase in the size of the error bars of the primary parameters for the non-flat XCDM Planck $P(q)+A_L$ model, the largest is 37\% for $\Omega_k$, followed by the error bars of $w$ (22\%) and $n_s$ (6.5\%). As for the derived parameters $H_0$, $\Omega_m$, and $\sigma_8$, we observe minimal differences in the mean values, 0.047$\sigma$, $-0.047\sigma$, and 0.057$\sigma$, respectively, with an increase in error bar size only for $\sigma_8$ (8.3\%). 

When we compare P18+lensing data results for the eight-parameter non-flat XCDM new $P(q)$ (upper half of Table \ref{tab:results_new_Pq_XCDM} and Fig.\ \ref{fig:XCDM_new_Pq_P18_vs_nonCMB_2}) and the nine-parameter non-flat XCDM new $P(q)+A_L$ (lower half of Table \ref{tab:results_new_Pq_XCDM} and Fig.\ \ref{fig:TNX_Alens_ns1_p18len_ncmb}) cosmological models we are lead to very similar conclusions to the ones obtained from the comparison of P18+lensing data results for the non-flat XCDM Planck $P(q)$ and XCDM Planck $P(q)+A_L$ models. For the primary parameters in common with the flat $\Lambda$CDM model, the largest mean values differences are $-0.086\sigma$, $0.14\sigma$, and $-0.18\sigma$ for $\Omega_{b}h^2$, $\Omega_{c}h^2$, and $n_s$, respectively. For the curvature parameter $\Omega_k$ the values differ at $-0.34\sigma$ and for the equation of state parameter of dark energy they differ at $-0.17\sigma$. As expected, some increases in the size of the error bars are observed when the $A_L$ parameter is allowed to vary, in particular for $n_s$, $\ln(10^{10}A_s)$, and $w$ the enlargement of the error bars is 20\%, 11\%, and 15\%, respectively. As for the derived parameters $\Omega_m$ and $\sigma_8$, we find differences of $-0.057\sigma$ and 0.18$\sigma$ and the error bars increase by 8.3\% and 9.1\%.

\begin{figure*}[htbp]
\centering
\mbox{\includegraphics[width=170mm]{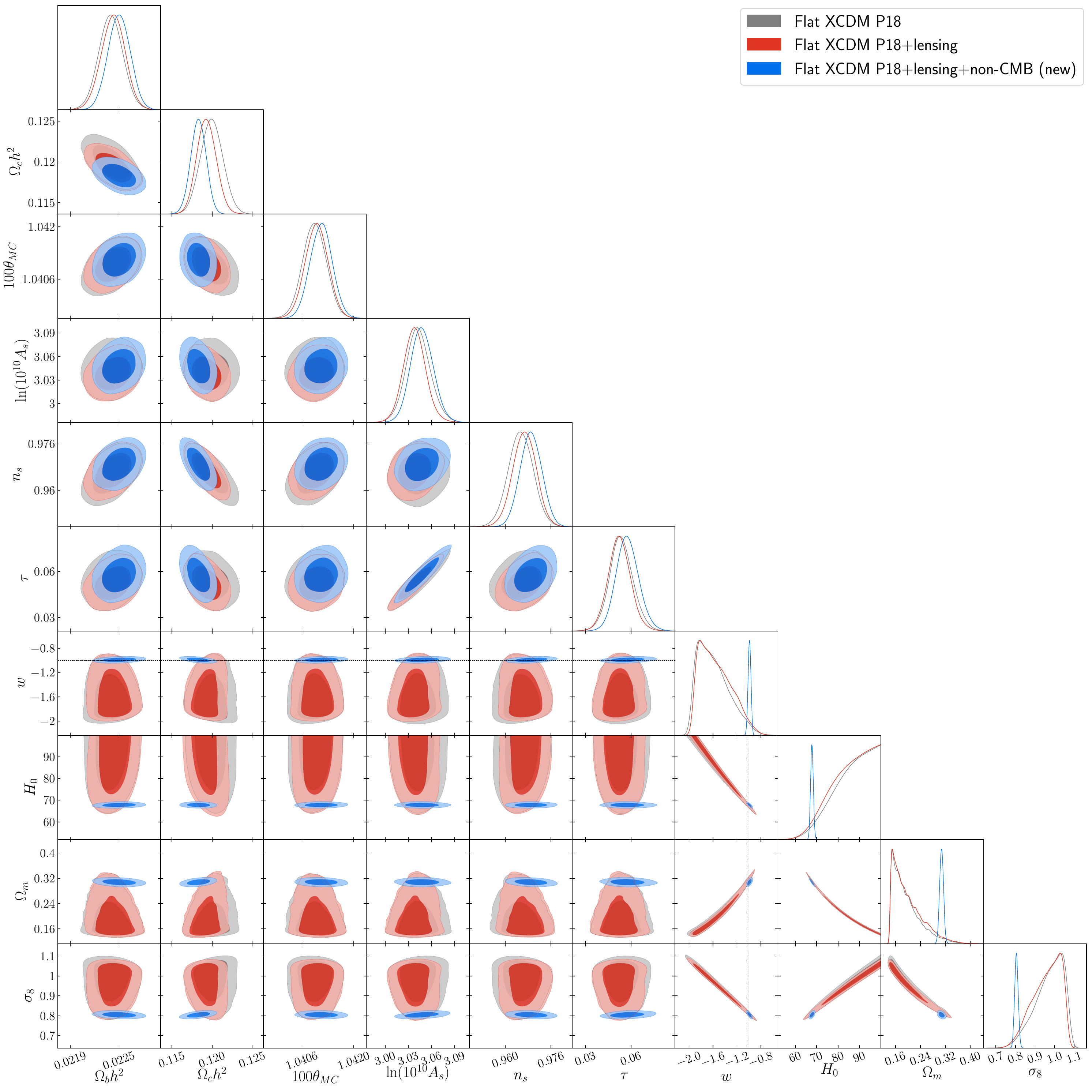}}
\caption{Likelihood distributions of flat XCDM model parameters
         favored by P18, P18+lensing, and P18+lensing+non-CMB (new) data sets. 
}
\label{fig:flat_XCDM}
\end{figure*}
\begin{figure*}[htbp]
\centering
\mbox{\includegraphics[width=170mm]{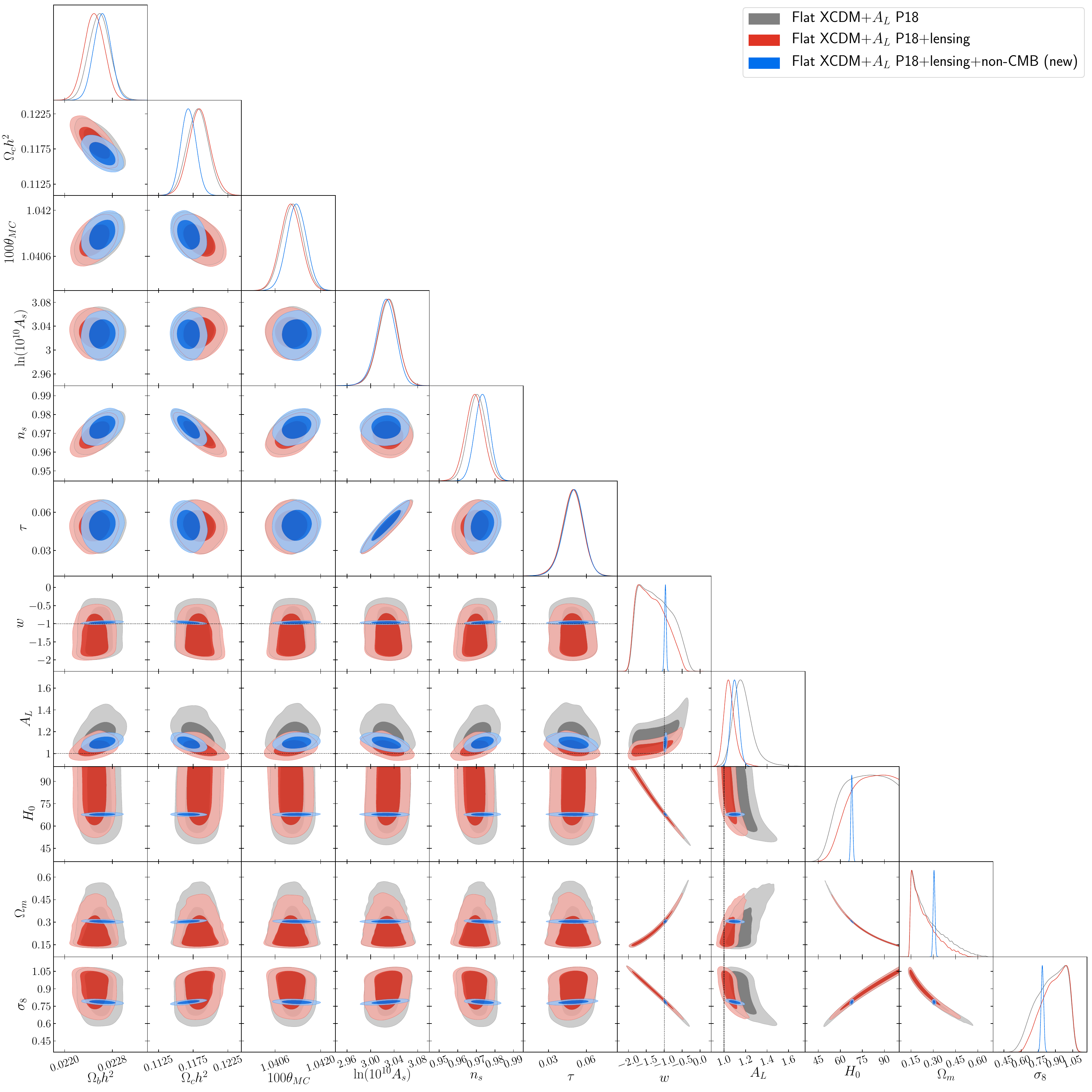}}
        \caption{Likelihood distributions of flat XCDM$+A_L$ model
        parameters favored by P18, P18+lensing, and P18+lensing+non-CMB (new) data sets.
}
\label{fig:FX_Alens_p18lenncmb_v2}
\end{figure*}
\begin{figure*}[htbp]
\centering
\mbox{\includegraphics[width=170mm]{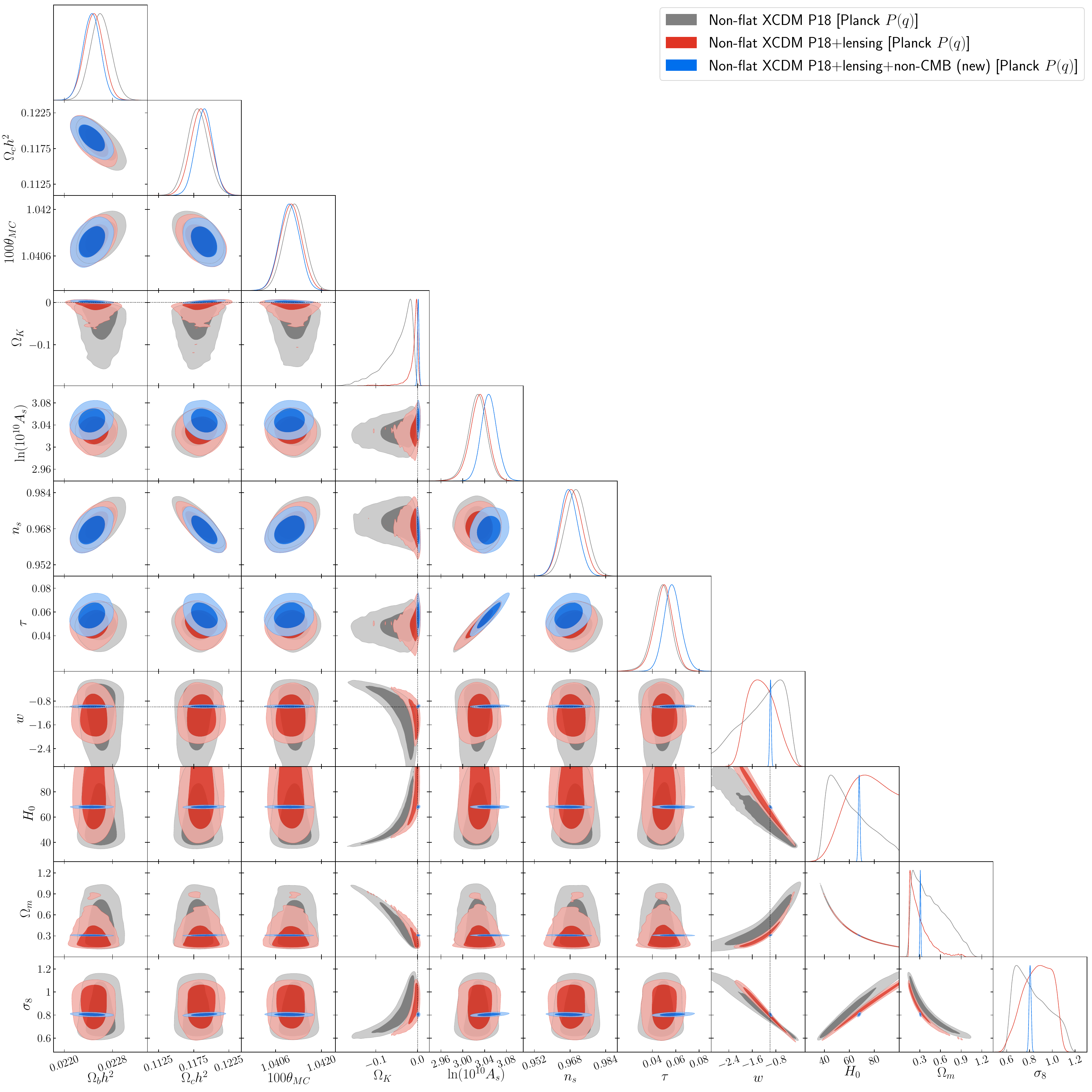}}
	\caption{Likelihood distributions of non-flat XCDM model
	[Planck $P(q)$] parameters favored by P18, P18+lensing, and
         P18+lensing+non-CMB (new) data sets. 
}
\label{fig:XCDM_Planck_Pq}
\end{figure*}
\begin{figure*}[htbp]
\centering
\mbox{\includegraphics[width=170mm]{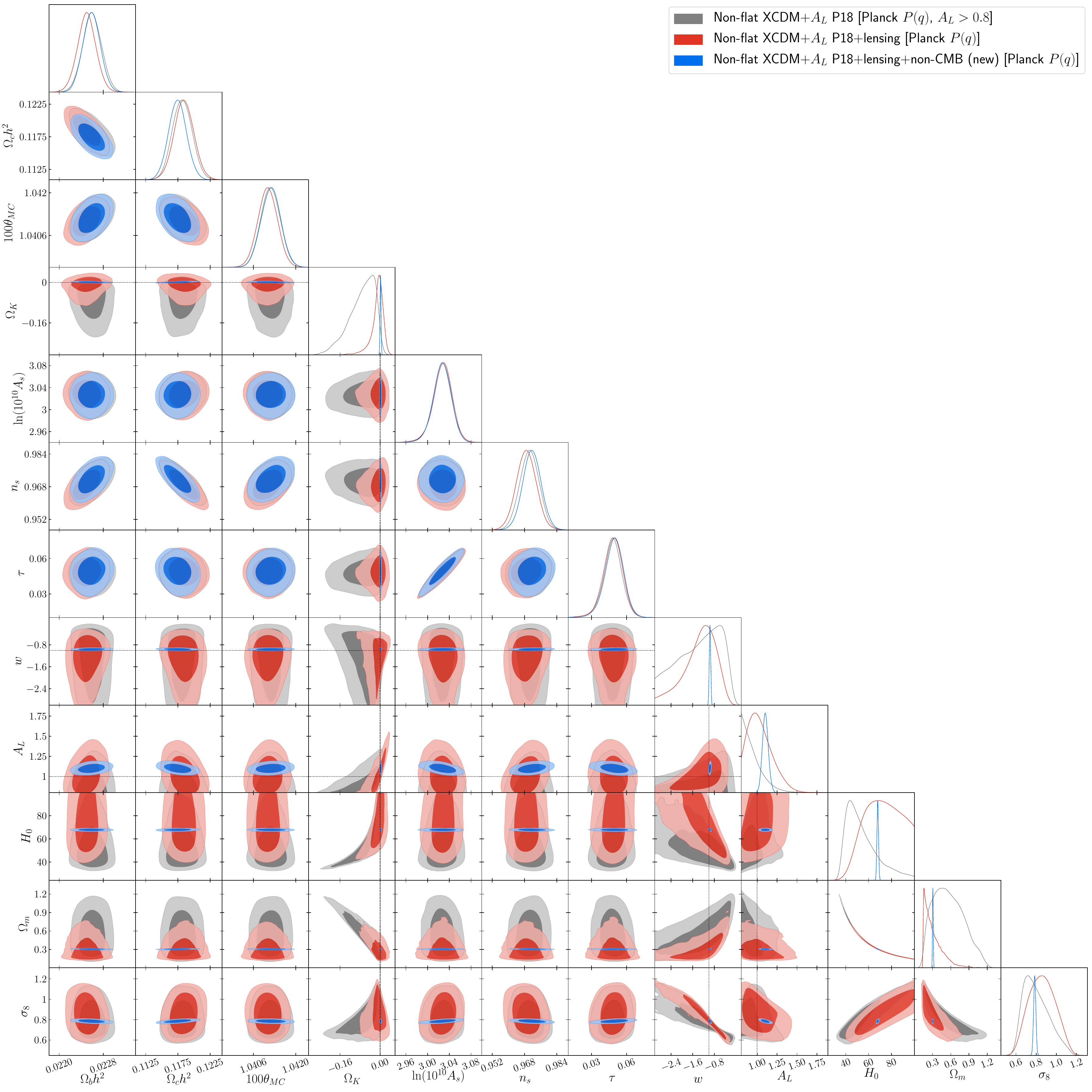}}
        \caption{Likelihood distributions of non-flat XCDM$+A_L$ [Planck $P(q)$] model
        parameters favored by P18, P18+lensing, and P18+lensing+non-CMB (new) data sets. Note that in the model for the P18 data set, a prior of $A_L > 0.8$ is applied.
}
\label{fig:NX_Alens_ns_p18lenncmb_v2}
\end{figure*}
\begin{figure*}[htbp]
\centering
\mbox{\includegraphics[width=170mm]{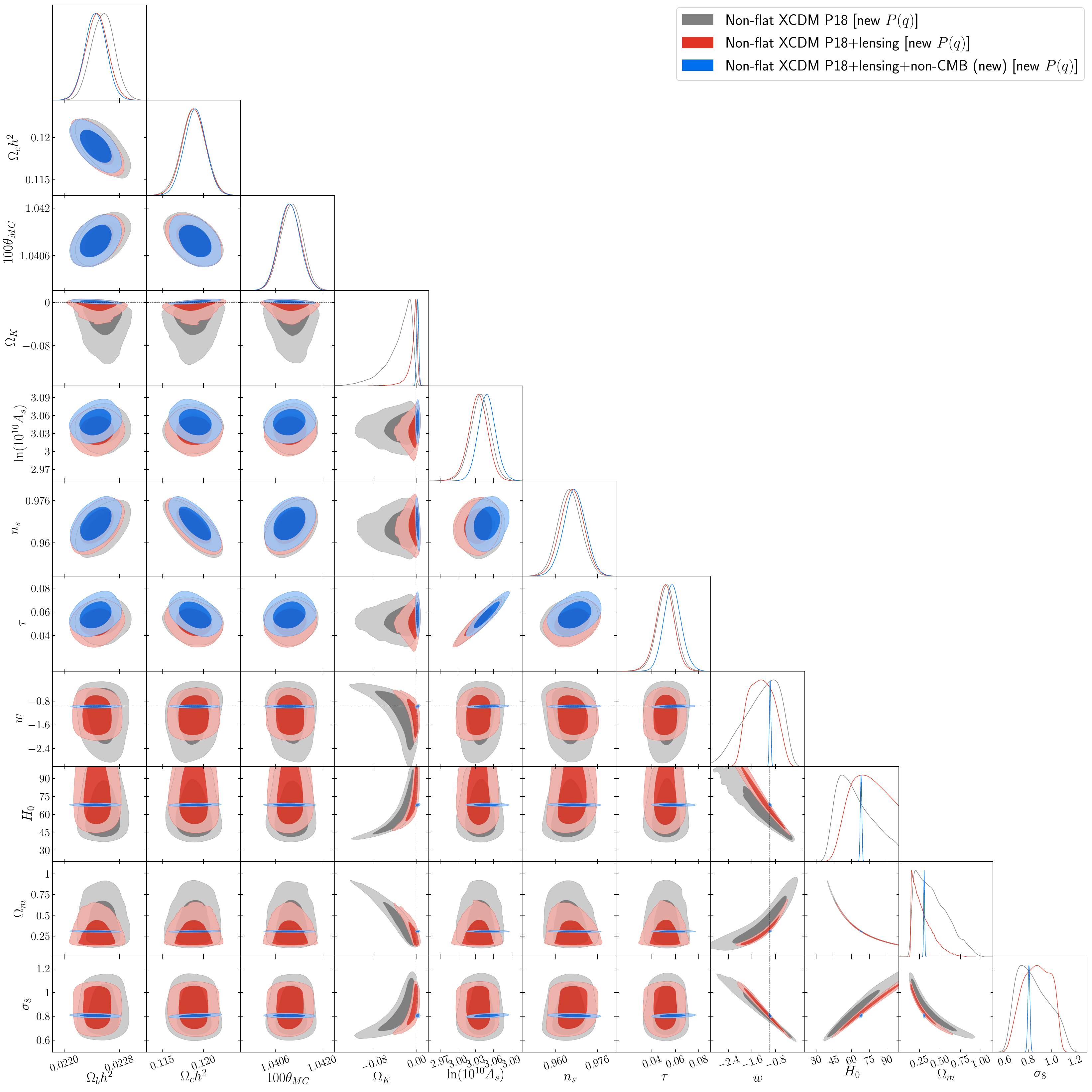}}
	\caption{Likelihood distributions of non-flat XCDM model
	[new $P(q)$] parameters favored by
	P18, P18+lensing, and P18+lensing+non-CMB (new) data sets. 
}
\label{fig:XCDM_new_Pq}
\end{figure*}
\begin{figure*}[htbp]
\centering
\mbox{\includegraphics[width=170mm]{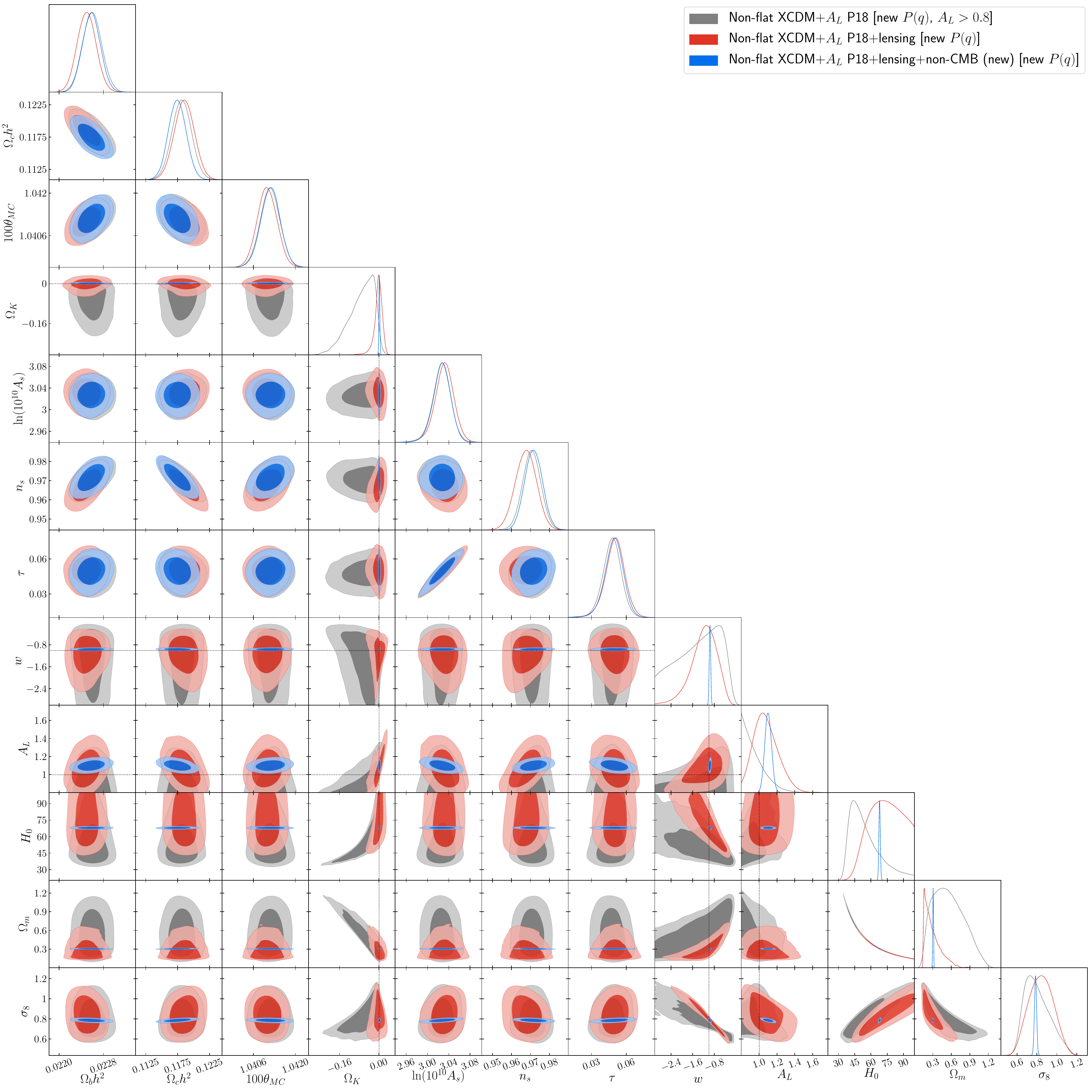}}
        \caption{Likelihood distributions of non-flat XCDM$+A_L$ [new $P(q)$] model
                parameters favored by P18, P18+lensing, and P18+lensing+non-CMB (new) data sets. Note that in the model for the P18 data set, a prior of $A_L > 0.8$ is applied.
}
\label{fig:TNX_Alens_ns1_p18lenncmb_v2}
\end{figure*}


\subsubsection{P18+lensing+non-CMB (new) cosmological constraints}
\label{sec:XCDM P18+lensing+non-CMB (new)}

Seven-parameter flat XCDM model cosmological parameter constraints obtained from analyses of P18+lensing and P18+lensing+non-CMB (new) data are listed in Table \ref{tab:results_flat_XCDM} and shown in Figs.\ \ref{fig:flat_XCDM_P18_vs_nonCMB_2} and \ref{fig:flat_XCDM}. Both sets of primary parameter values are similar, with largest differences for $\Omega_{c}h^2$ ($+0.65\sigma$), $\tau$ ($-0.50\sigma$), and $\ln(10^{10}A_s)$ ($-0.47\sigma$). The error bars decrease when non-CMB (new) data are added to the mix, with the $-26$\% decrease for $\Omega_{c}h^2$ being the largest. The equation of state parameter value from P18+lensing data, $w=-1.55\pm0.26$, is $-2.15\sigma$ away from the P18+lensing+non-CMB (new) data value, $w=-0.990\pm0.023$, with the error bars of the P18+lensing+non-CMB (new) data value being 11 times smaller than the P18+lensing data value. The P18+lensing data value favors phantom-like behavior at 2.12$\sigma$ significance while the P18+lensing+non-CMB (new) data value signifies 0.43$\sigma$ support for quintessence-like dynamical dark energy. As for the derived parameters $H_0$, $\Omega_m$, and $\sigma_8$, we find some non-negligible differences at 1.97$\sigma$, $-2.23\sigma$, and 2.17$\sigma$ respectively. Major reductions in the size of the error bars are observed for these parameters when we move from P18+lensing data to P18+lensing+non-CMB (new) data, in particular the error bars are 14.6, 7.7, and 8.0 times smaller, respectively. 

If we examine the lower half of Table \ref{tab:results_flat_XCDM} and Figs.\ \ref{fig:FX_Alens_p18len_ncmb} and \ref{fig:FX_Alens_p18lenncmb_v2} we can compare the results obtained for the eight-parameter flat XCDM+$A_L$ cosmological model when analyzing P18+lensing and P18+lensing+non-CMB (new) data. Again, both sets of primary parameter values are similar. 
In particular, the $\Omega_{b}h^2$, $\Omega_{c}h^2$, $100\theta_{\text{MC}}$, and $n_s$ values differ by $-0.59\sigma$, $+0.86\sigma$, $-0.39\sigma$, and $-0.66\sigma$, respectively. The size of the error bars is also affected and, as expected,  including the non-CMB (new) data results in smaller error bars, particularly for the parameters $\Omega_{b}h^2$ ($-21.43$\%), $\Omega_{c}h^2$ ($-36.36$\%), and $n_s$ ($-22.50$\%). When P18+lensing data are used we obtain $w=-1.34^{+0.26}_{-0.51}$, which is 1.31$\sigma$ away from $w=-1$, while from P18+lensing+non-CMB (new) data we obtain $w=-0.968\pm 0.024$ which deviates from $w=-1$ by 1.33$\sigma$. The two data sets favor different behaviors of the dark energy equation of state parameter (phantom-like vs.\ quintessence-like) and the two values differ at $-1.42\sigma$ with the $w$ error bars becoming 15 times smaller when moving from P18+lensing to P18+lensing+non-CMB (new) data. Regarding the lensing consistency parameter $A_L$, with P18+lensing data we find $A_L = 1.054^{+0.039}_{-0.059}$ (0.92$\sigma$ away from $A_L = 1$), whereas with P18+lensing+non-CMB (new) data we get $A_L = 1.101\pm 0.037$ (2.73$\sigma$ in favor of $A_L > 1$), where the two values differ by  $-0.87\sigma$. In this case, the reduction in the size of the error bars is $-48.65$\%. The values of the derived parameters $\Omega_m$ and $\sigma_8$ differ by $-0.76\sigma$ and 0.95$\sigma$, respectively, with the error bars obtained with P18+lensing+non-CMB (new) data being 13 and 10 times smaller than the ones obtained with P18+lensing data. 

Looking at Table \ref{tab:results_flat_XCDM} we can compare the results obtained from P18+lensing+non-CMB (new) data for the seven-parameter flat XCDM model (upper half of the table and Figs.\ \ref{fig:flat_XCDM_P18_vs_nonCMB_2} and \ref{fig:flat_XCDM}) and the eight-parameter flat XCDM+$A_L$ model (lower half of the table and Figs.\ \ref{fig:FX_Alens_p18len_ncmb} and \ref{fig:FX_Alens_p18lenncmb_v2}). We observe some differences in the values of the primary parameter, in particular for $\Omega_{c}h^2$, $\tau$, $n_s$, and $\ln(10^{10}A_s)$ the differences are at 1.03$\sigma$, 0.72$\sigma$, $-0.78\sigma$, and $-0.97\sigma$, respectively. As for the equation of state parameter of dark energy the difference between the two values is $-0.66\sigma$. There are no significant increases in the error bars when moving from the flat XCDM model to the flat XCDM+$A_L$ model, the largest being those of $\tau$ (10\%) and $\ln(10^{10}A_s)$ (12\%), whereas for $w$ the error bars increase is 4.2\%. In regard to the derived parameters $H_0$, $\Omega_m$, and $\sigma_8$ the mean values differ at 0.022$\sigma$, $0.31\sigma$, and 1.39$\sigma$, while only the error bars of $\sigma_8$ increase (by 19\%). 

Comparing the eight-parameter non-flat XCDM Planck [new] $P(q)$ primary cosmological parameter constraints for P18+lensing data and for P18+lensing+non-CMB (new) data, listed in Table \ref{tab:results_XCDM_Planck_Pq} [\ref{tab:results_new_Pq_XCDM}] and shown in Figs.\ \ref{fig:XCDM_Planck_Pq_P18_vs_nonCMB_2} and \ref{fig:XCDM_Planck_Pq} [\ref{fig:XCDM_new_Pq_P18_vs_nonCMB_2} and \ref{fig:XCDM_new_Pq}], we observe small differences, all less than 1$\sigma$, with those for $\tau$ and $\ln(10^{10}A_s)$ being the largest at $-0.71\sigma$ [$-0.59\sigma$] and $-0.79\sigma$ [$-0.68\sigma$] respectively. The curvature parameter value obtained using P18+lensing data is $\Omega_k = -0.011\pm 0.017$ [$-0.008\pm 0.010$] a 0.65$\sigma$ [0.80$\sigma$] evidence in favor of closed geometry, whereas for P18+lensing+non-CMB (new) data we get $\Omega_k = 0.0016\pm 0.0019$ [$0.0014\pm 0.0020$] a 0.84$\sigma$ [0.70$\sigma$] preference for open geometry. The  P18+lensing data and the P18+lensing+non-CMB (new) data $\Omega_k$ values differ by $-0.74\sigma$ [$-0.92\sigma$]. It is interesting that when non-CMB (new) data are added to the mix, the preference changes from a closed Universe to an open one; we follow up on this point in the next paragraph. Regarding the dark energy equation of state parameter value, from P18+lensing data we get $w=-1.28\pm 0.45$ [$-1.27\pm 0.41$] which differs by $-0.66\sigma$ [$-0.70\sigma$] from the result obtained with P18+lensing+non-CMB (new) data, $w=-0.980\pm 0.026$ [$-0.982\pm 0.026$]. While the P18+lensing data result is 0.62$\sigma$ [0.66$\sigma$] in favor of  phantom-like behavior, the P18+lensing+non-CMB (new) data result prefers quintessence-like behavior at 0.73$\sigma$ [0.69$\sigma$]. The P18+lensing and P18+lensing+non-CMB (new) data values of the derived parameters $H_0$, $\Omega_m$, and $\sigma_8$ differ at 0.34$\sigma$ [0.43$\sigma$], $-0.052\sigma$ [$-0.15\sigma$], and 0.54$\sigma$ [0.58$\sigma$], respectively, with error bars 23 [21], 25 [20], and 14 [12] times smaller when P18+lensing+non-CMB (new) data are employed.  

From Table \ref{tab:results_XCDM_Planck_Pq} [\ref{tab:results_new_Pq_XCDM}], we see that for the non-flat XCDM Planck [new] $P(q)$ model non-CMB (new) data favor closed geometry at 2.77$\sigma$ [2.24$\sigma$], P18 data favor closed geometry at 1.17$\sigma$ [1.17$\sigma$], P18+lensing data favor closed geometry at 0.85$\sigma$ [0.82$\sigma$], while joint P18+non-CMB (new) data favor open geometry at 0.89$\sigma$ [0.80$\sigma$], and joint P18+lensing+non-CMB (new) data favor open geometry at 0.84$\sigma$ [0.70$\sigma$]. As in the non-flat $\Lambda$CDM models, \cite{deCruzPerez:2022hfr}, this is likely due to the $\Omega_m - \Omega_k - H_0$ degeneracy and the fact that P18 data favor a smaller $H_0$ and a larger $\Omega_m$ than do non-CMB (new) data, see Table \ref{tab:results_XCDM_Planck_Pq} [\ref{tab:results_new_Pq_XCDM}]. This point is also relevant for the non-flat XCDM$+A_L$ models discussed below. 

We can compare the cosmological constraints obtained for the nine-parameter non-flat XCDM Planck [new] $P(q)+A_L$ model from P18+lensing data and P18+lensing+non-CMB (new) data by looking at the lower half of Table \ref{tab:results_XCDM_Planck_Pq} [Table \ref{tab:results_new_Pq_XCDM}] and Figs.\ \ref{fig:NX_Alens_ns_p18len_ncmb} and \ref{fig:NX_Alens_ns_p18lenncmb_v2} [\ref{fig:TNX_Alens_ns1_p18len_ncmb} and \ref{fig:TNX_Alens_ns1_p18lenncmb_v2}]. We do not observe significant differences in the values of the primary parameters, with $\Omega_{b}h^2$, $\Omega_{c}h^2$, and $n_s$ showing the largest differences at $-0.39\sigma$ [$-0.34\sigma$], 0.44$\sigma$ [0.47$\sigma$], and $-0.42\sigma$ [$-0.57\sigma$], respectively. As for the error bars, the most affected parameters are again $\Omega_{b}h^2$, $\Omega_{c}h^2$, and $n_s$ with decreases of $-6.25$\% [$-6.25$\%], $-7.14$\% [$-14.29$\%], and $-6.52$\% [$-19.57$\%]. For the curvature parameter when P18+lensing data are utilized we obtain $\Omega_k = -0.012^{+0.027}_{-0.011}$ [$-0.003^{+0.018}_{-0.011}$] whereas when P18+lensing+non-CMB (new) data are considered we get $\Omega_k = 0.0015\pm 0.0019$ [$0.0015\pm 0.0019$] with the two values differing at $-0.50\sigma$ [$-0.25\sigma$]. While the first result shows a 0.44$\sigma$ [0.17$\sigma$] preference for a closed Universe, the second one is 0.79$\sigma$ [0.79$\sigma$] in favor of an open Universe. Furthermore, the error bars from P18+lensing+non-CMB (new) data are 14 [9] times smaller than the ones obtained with P18+lensing data. In regard to the dark energy equation of state parameter, we get $w=-1.32^{+0.71}_{-0.38}$ (a $0.45\sigma$ phantom-like deviation from $w=-1$) [$w=-1.18^{+0.54}_{-0.37}$ $(0.33\sigma)$] for P18+lensing data, and $w=-0.958\pm 0.026$ (a 1.62$\sigma$ quintessence-like deviation from $w = -1$) [$w=-0.959\pm 0.027$ $(1.52\sigma)$] for P18+lensing+non-CMB (new) data, with P18+lensing+non-CMB (new) data error bars 22 [18] times smaller than the P18+lensing error bars. The two values differ at $-0.51\sigma$ [$-0.43\sigma$]. For the third non-standard-model parameter, $A_L$, with P18+lensing data we get $A_L = 1.02\pm 0.16$ [$1.07^{+0.12}_{-0.16}$] and when we analyze P18+lensing+non-CMB (new) data we find $A_L = 1.102\pm 0.037$ [$1.101\pm 0.038$]. The differences between the two values is $-0.50\sigma$ [$-0.25\sigma$]. Both results show a preference for $A_L > 1$ with a significance of 0.13$\sigma$ [0.44$\sigma$] and 2.76$\sigma$ [2.66$\sigma$], respectively, with error bars 4 [4] times smaller in the second case. The differences in the values of the derived parameters are not significant, where for $H_0$, $\Omega_m$, and $\sigma_8$ we obtain 0.27$\sigma$ [0.43$\sigma$], 0.034$\sigma$ [$-0.036\sigma$], and 0.57$\sigma$ [0.45$\sigma$], respectively, with error bars obtained from P18+lensing+non-CMB (new) being 22 [21], 24 [21], and 12 [11] times smaller, respectively. 

Results obtained from P18+lensing+non-CMB (new) data for the eight-parameter non-flat XCDM Planck $P(q)$ model and the nine-parameter non-flat XCDM Planck $P(q)+A_L$ model are listed in Table \ref{tab:results_XCDM_Planck_Pq} and shown in Figs.\ \ref{fig:XCDM_Planck_Pq_P18_vs_nonCMB_2}, \ref{fig:XCDM_Planck_Pq}, \ref{fig:NX_Alens_ns_p18len_ncmb}, and \ref{fig:NX_Alens_ns_p18lenncmb_v2}. Comparing the results, we find that the values of primary parameters $\Omega_{c}h^2$, $\tau$, $n_s$, and $\ln(10^{10}A_s)$ differ at 0.81$\sigma$, 0.73$\sigma$, $-0.69\sigma$, and 0.93$\sigma$, respectively. Looking at the values of the curvature parameter $\Omega_k$ we observe a difference of 0.037$\sigma$ whereas if we compare the two values of the dark energy equation of state parameter $w$ the difference is $-0.60\sigma$. No significant increase in the size of the error bars for the primary parameters is observed, with the largest error bars increases being associated with $\Omega_{c}h^2$ (14\%), $\tau$ (10\%), and $\ln(10^{10}A_s)$ (12\%). For the derived parameters $H_0$, $\Omega_m$, and $\sigma_8$, the differences between the mean values are at 0.011$\sigma$, 0.33$\sigma$, and 1.41$\sigma$ and the increase in the size of the error bars of $\sigma_8$ is 20\%.   

Comparing the P18+lensing+non-CMB (new) data constraints obtained for the eight-parameter non-flat XCDM new $P(q)$ (upper half of Table \ref{tab:results_new_Pq_XCDM} and Figs.\ \ref{fig:XCDM_new_Pq_P18_vs_nonCMB_2} and \ref{fig:XCDM_new_Pq}) and the nine-parameter non-flat new $P(q)+A_L$ model (lower half of Table \ref{tab:results_new_Pq_XCDM} and Figs.\ \ref{fig:TNX_Alens_ns1_p18len_ncmb} and \ref{fig:TNX_Alens_ns1_p18lenncmb_v2}), we find very similar results to the ones obtained for the non-flat XCDM Planck $P(q)$ ($+A_L$) models from these data. The values of the primary parameters $\Omega_{c}h^2$, $\tau$, $n_s$, and $\ln(10^{10}A_s)$ differ by 0.79$\sigma$, 0.74$\sigma$, $-0.71\sigma$, and 0.97$\sigma$, respectively. For the curvature parameter $\Omega_k$, there is a difference of $-0.036\sigma$ between the two values, and for the equation of state parameter of dark energy $w$ the results differ at $-0.61\sigma$. We do not observe significant changes in the size of the error bars of the primary parameters. In particular, for $n_s$ and $\ln(10^{10}A_s)$, the error bars increase by 8.7\% and 12\%, respectively, while that for the curvature parameter $\Omega_k$ decrease by $-5.3$\%. For the derived parameters $H_0$, $\Omega_m$, and $\sigma_8$, the differences between the two values are 0.011$\sigma$, 0.34$\sigma$, and 1.39$\sigma$, respectively. The corresponding increases in the size of the error bars are 0.0\%, 1.61\%, and 19\%.

\subsubsection{Comparing P18, P18+lensing, and P18+lensing+non-CMB (new) data cosmological constraints}
\label{sec:XCDM comparison}

Cosmological parameter contour plots are very useful for understanding the level of correlation between the different variables considered in the analysis and for detecting inconsistencies between cosmological parameter constraints obtained either for different cosmological models or for the same model but from different data sets. In this subsubsection we discuss the changes observed in the contour plots as we include more data in the analysis, namely when we compare P18, P18+lensing, and P18+lensing+non-CMB (new) data contours. We first discuss the $A_L = 1$ models and then comment on the $A_L$-varying cases.
 
For the seven-parameter flat XCDM $A_L = 1$ cosmological model, as seen in Fig.\ \ref{fig:flat_XCDM}, there are significant overlaps of contours. When we compare P18 (grey) and P18+lensing (red) data contours we see significant overlaps of the 1$\sigma$ contours for all parameters. On the other hand, when we compare grey P18 or red P18+lensing data with blue P18+lensing+non-CMB (new) data contours, we observe a number of panels in which the two sets of 1$\sigma$ contours do not overlap: these are panels labeled with either one or both of the primary parameter $w$ and the derived parameters $H_0$, $\Omega_m$, and $\sigma_8$. This is consistent with comments in Sec.\ \ref{sec:XCDM P18+lensing+non-CMB (new)} that noted that non-CMB (new) data favored quintessence-like dark energy evolution while P18 data favored phantom-like dark energy evolution. 

For the eight-parameter non-flat XCDM Planck $P(q)$ and new $P(q)$ $A_L = 1$ cosmological models (see Figs.\ \ref{fig:XCDM_Planck_Pq} and \ref{fig:XCDM_new_Pq}), we find non-overlapping $1\sigma$ regions, where two separate $1\sigma$ regions either do not overlap or the overlap is infinitesimally small, even when we compare grey P18 and red P18+lensing contours. For the Planck $P(q)$ model in Fig.\ \ref{fig:XCDM_Planck_Pq} these are in the $w-\Omega_k$, $\sigma_8-\Omega_k$, $H_0-w$, $\Omega_m-w$, $\sigma_8-H_0$, and $\sigma_8-\Omega_m$ panels, while in the new $P(q)$ model in Fig.\ \ref{fig:XCDM_new_Pq} these are in the $\Omega_k-H_0$, $\Omega_m-w$, $\sigma_8 -H_0$, and $\sigma_8-\Omega_m$ panels.
When we compare the grey P18 and the blue P18+lensing+non-CMB (new) data contours in the non-flat XCDM models we find less overlap, now with some cases where even the 2$\sigma$ contours do not overlap. For the Planck and the new $P(q)$ models in Figs.\ \ref{fig:XCDM_Planck_Pq} and \ref{fig:XCDM_new_Pq} these are in the $w-\Omega_k$, $H_0-\Omega_k$, $\Omega_m-\Omega_k$, $\sigma_8-\Omega_k$, $H_0-w$, $\Omega_m-w$, $\sigma_8-H_0$, and $\sigma_8-\Omega_m$ panels. These differences are likely caused by non-CMB (new) data favoring slower quintessence-like dark energy evolution while P18 and P18+lensing data favor more rapid phantom-like evolution.

When $A_L$ is allowed to vary and no longer fixed to unity we see, from Figs.\ \ref{fig:FX_Alens_p18lenncmb_v2}, \ref{fig:NX_Alens_ns_p18lenncmb_v2}, and \ref{fig:TNX_Alens_ns1_p18lenncmb_v2} for the flat and non-flat XCDM models, the differences between the grey P18 data contours and the blue P18+lensing+non-CMB (new) data contours subside, with no cases with non-overlapping 2$\sigma$ contours, but with some non-overlapping 1$\sigma$ contours, for the non-flat XCDM models, including some new ones in $A_L$ panels. For the XCDM Planck and new $P(q)+A_L$ models these non-overlapping contours are in all panels that contain $\Omega_k$ and in the $A_L-\sigma_8$, $A_L-H_0$, and $A_L-w$ panels.
If we now compare grey P18 and red P18+lensing contour plots for the three flat and non-flat XCDM$+A_L$ models there are no panels where the contour plots do not overlap at 1$\sigma$.

The results commented on in this subsubsection, as we shall see, largely agree with those we discuss in Sections \ref{sec:Model_Selection} and \ref{sec:Data_Set_Tensions}. Namely, P18 and P18+lensing data results seem to be in tension with non-CMB (new) data results in the context of the XCDM model with $A_L=1$. As stated previously when $A_L$ is allowed to vary we no longer see significantly non-overlapping contours and this translates into a better performance when it comes to fitting those data sets that include non-CMB (new) data (see results in Table \ref{tab:chi2_lcdm} for the $\Lambda$CDM models and Table \ref{tab:chi2_xcdm} for the XCDM models and discussion in Section \ref{sec:Data_Set_Tensions} below) and also into reduced tension between pairs of data sets (see Table \ref{tab:consistency_LCDM} for $\Lambda$CDM models and Table \ref{tab:consistency_XCDM} for XCDM models and discussion in Section \ref{sec:Data_Set_Tensions} below). This may be indicating that in order to jointly analyze CMB and non-CMB data, in the context of the XCDM models, the lensing parameter $A_L$ should be considered as a free parameter.

\subsubsection{Comparing P18 data and non-CMB (new) data cosmological constraints}
\label{sec:XCDM_P18_vs_nonCMB}

In this subsubsection we study the mutual consistency of cosmological parameter constraints derived from P18 and non-CMB (new) data in XCDM models. If these constraints are mutually consistent in an XCDM model then P18 and non-CMB (new) data can be jointly used to constrain cosmological parameters in that XCDM model. If they are not mutually consistent in an XCDM model this implies the model is inconsistent with at least one of these data sets and so can be rejected at some level of confidence (if one assumes that both sets of data are correct). The results presented here are complemented by those provided in Sec.\ \ref{sec:Data_Set_Tensions} where we use the two statistical estimators to assess the level of tension between P18 and non-CMB (new) data in the context of a given XCDM model. 

Since non-CMB (new) data are unable to constrain $\tau$ and $n_s$, in the non-CMB (new) data analyses we set their values to those obtained in the corresponding P18 data analysis. Also, non-CMB (new) data are practically insensitive to variations in the lensing consistency parameter $A_L$, therefore when we compare P18 and non-CMB (new) cosmological constraints for the XCDM+$A_L$ models, the constraints from non-CMB (new) data are those for the corresponding models with $A_L = 1$.

P18 and non-CMB (new) data results obtained for the seven-parameter and the five-parameter flat XCDM cosmological models are listed in Table \ref{tab:results_flat_XCDM} and shown in Fig.\ \ref{fig:flat_XCDM_P18_vs_nonCMB_1}. We observe significant differences between the two sets of cosmological parameter constraints not only in the primary parameters but also in the derived parameters. In regard to the primary parameters we find that the two values for $\Omega_{b}h^2$, $\Omega_{c}h^2$, 100$\theta_{\text{MC}}$, and $\ln(10^{10}A_s)$ disagree by $-2.14\sigma$, 2.34$\sigma$, 2.09$\sigma$, and $-2.63\sigma$, respectively. With non-CMB (new) data we obtain $w=-0.853^{+0.043}_{-0.033}$, which indicates a 4.45$\sigma$ preference for quintessence-like behavior whereas with P18 data we get $w=-1.59^{+0.15}_{-0.34}$ which indicates preference for phantom behavior 3.93$\sigma$ away from the cosmological constant. The values of $w$ estimated from P18 and non-CMB (new) data differ by $-4.80\sigma$. In the flat XCDM model P18 data and non-CMB (new) data values of all five primary parameters are mutually inconsistent at greater than 2$\sigma$ confidence. As for the derived parameters, we find for $\Omega_m$ and $\sigma_8$ differences of $-1.54\sigma$ and 1.97$\sigma$, respectively, between the P18 data and non-CMB (new) data values. These results might mean that in the context of the flat XCDM model P18 data and non-CMB (new) data should not be analyzed together, and that the flat XCDM model might be inconsistent with at least one of the two data sets (under the assumption that both data sets are correct). We will return to this issue in Sec.\ \ref{sec:Data_Set_Tensions}.

We can compare the results obtained for the eight-parameter flat XCDM+$A_L$ model and the five-parameter flat XCDM model, when P18 and non-CMB (new) data are employed, respectively, if we look at Table \ref{tab:results_flat_XCDM} and Fig.\ \ref{fig:FX_Alens_p18_ncmb}. We find significant differences in the values of the primary parameters, which exceed 2$\sigma$ confidence for four of the five primary parameters. In particular, for $\Omega_{b}h^2$, $\Omega_{c}h^2$, $100\theta_{\text{MC}}$, and $\ln(10^{10}A_s)$ we observe differences of $-2.10\sigma$, $2.12\sigma$,  $2.11\sigma$, and $-2.69\sigma$, respectively. As for the equation of state parameter of dark energy, for the flat XCDM+$A_L$ model constrained with P18 data we obtain $w=-1.23^{+0.31}_{-0.59}$, indicating preference for phantom-like behavior at 0.74$\sigma$. Compared to the $w$ value for non-CMB (new) data in the flat XCDM model, the two values differ at $-1.21\sigma$ confidence, significantly smaller than the corresponding $-4.80\sigma$ in the flat XCDM model where $A_L$ is not allowed to vary and is set to unity. When we compare the values of the derived parameters we find less severe discrepancies than in the flat XCDM case, with none of them even reaching the 1$\sigma$ level. The two values of $H_0$ and $\sigma_8$ disagree by $0.70\sigma$ and $0.29\sigma$, respectively. To determine whether P18 and non-CMB (new) data can be jointly analyzed, and whether the flat XCDM+$A_L$ model is inconsistent with at least one of these data sets, we will make use of the statistical estimators in Sec.\ \ref{sec:Data_Set_Tensions}. 

The primary cosmological parameters values of the eight-parameter and six-parameter non-flat XCDM Planck [new] $P(q)$ models from P18 data and from non-CMB (new) data are listed in Table \ref{tab:results_XCDM_Planck_Pq} [\ref{tab:results_new_Pq_XCDM}] and shown in Fig.\ \ref{fig:XCDM_Planck_Pq_P18_vs_nonCMB_1} [\ref{fig:XCDM_new_Pq_P18_vs_nonCMB_1}]. Here we observe smaller differences than those in the flat XCDM model, with two [none] of the six primary parameters disagreeing at 2$\sigma$ confidence. The largest differences are for $\Omega_{b}h^2$ and 100$\theta_{\text{MC}}$ with disagreements at $-1.37\sigma$ [$-1.28\sigma$] and $-2.10\sigma$ [$-1.79\sigma$], respectively, as well as for $\Omega_k$ as discussed next. When analyzing P18 data we obtain $\Omega_k=-0.048^{+0.041}_{-0.012}$ [$-0.0338^{+0.029}_{-0.0086}$] for the curvature parameter, indicating a 1.17$\sigma$ [1.17$\sigma$] preference for closed geometry and a 1.98$\sigma$ [1.82$\sigma$] disagreement with the value obtained from non-CMB (new) data $\Omega_k=-0.177^{+0.064}_{-0.072}$ [$-0.186^{+0.083}_{-0.067}$], which in turn is 2.77$\sigma$ [2.24$\sigma$] in favor of closed hypersurfaces. As for the equation of state parameter, using P18 data we obtain $w=-1.27^{+0.97}_{-0.45}$ [$-1.27^{+0.79}_{-0.44}$], while using non-CMB (new) data we get $w=-0.786^{+0.044}_{-0.037}$ [$-0.785^{+0.045}_{-0.038}$]. The P18 value indicates a 0.28$\sigma$ [0.34$\sigma$] preference for phantom-like behavior while the non-CMB (new) value represents a 5.78$\sigma$ [5.66$\sigma$] preference for quintessence-like behavior, and the P18 and non-CMB (new) values differ at $-0.50\sigma$ [$-0.61\sigma$]. Regarding the derived parameters, for $H_0$, $\Omega_m$, and $\sigma_8$ we observe differences of $-1.14\sigma$ [$-0.74\sigma$], 0.76$\sigma$ [0.60$\sigma$], and 0.36$\sigma$ [0.59$\sigma$], almost all below 1$\sigma$. 

P18 and non-CMB (new) data results obtained for the nine-parameter non-flat XCDM Planck [new] $P(q)+A_L$ model and the six-parameter non-flat XCDM Planck [new] $P(q)$ model can be found in Table \ref{tab:results_XCDM_Planck_Pq} [\ref{tab:results_new_Pq_XCDM}] and seen in Fig.\ \ref{fig:NX_Alens_ns_p18_ncmb} [\ref{fig:TNX_Alens_ns1_p18_ncmb}]. Non-negligible differences are observed when we look at the values of some primary parameters with one [none] of the six primary parameters disagreeing at 2$\sigma$ significance. For $\Omega_{b}h^2$, $100\theta_{\text{MC}}$, and $\ln(10^{10}A_s)$, we see differences at $-1.37\sigma$ [$-1.27\sigma$], $-2.10\sigma$ [$-1.79\sigma$], and $0.38\sigma$ [$0.47\sigma$], respectively. For the non-flat XCDM Planck [new] $P(q)+A_L$ model, when P18 data is considered, we obtain $\Omega_k = -0.073^{+0.065}_{-0.029}$ [$-0.072^{+0.065}_{-0.030}$], which is $1.12\sigma$ [$1.11\sigma$] away from flat. For the non-flat XCDM Planck [new] $P(q)$ model, using non-CMB (new) data, we find $\Omega_k = -0.177^{+0.064}_{-0.072}$ [$-0.186^{+0.083}_{-0.067}$], indicating a preference for a closed Universe at $2.77\sigma$ [$2.24\sigma$]. The two values disagree at $1.48\sigma$ [$1.29\sigma$]. In regard to the equation of state parameter of dark energy for the XCDM Planck [new] $P(q)+A_L$ model $w=-1.36^{+1.1}_{-0.53}$ [$-1.39^{+1.1}_{-0.54}$] from P18 data, with this value favoring phantom-like behavior at $0.33\sigma$ [$0.35\sigma$], whereas from non-CMB data in the XCDM Planck [new] $P(q)$ model we obtain $w=-0.786^{+0.044}_{-0.037}$ [$-0.785^{+0.045}_{-0.038}$], indicating a preference for quintessence at $5.78\sigma$ [$5.66\sigma$], with a discrepancy of $-0.52\sigma$ [$-0.55\sigma$] between the two values. 
As for the derived parameters $H_0$, $\Omega_m$, and $\sigma_8$, we find that the two values disagree at $-2.56\sigma$ [$-2.55\sigma$], $1.15\sigma$ [$1.10\sigma$], and $0.19\sigma$ [$0.23\sigma$], respectively.

\subsubsection{Comparing P18+lensing data and non-CMB (new) data cosmological constraints}
\label{sec:XCDM_P18_lensing_vs_nonCMB}

In the previous subsubsection we showed that in the XCDM models the differences between P18 data and non-CMB (new) data cosmological parameter constraints are not negligible.  
We now determine whether the same is true when we compare P18+lensing data and non-CMB (new) data XCDM models cosmological constraints.

Cosmological parameter constraints for the seven-parameter and the five-parameter flat XCDM model obtained from P18+lensing data and non-CMB (new) data are listed in Table \ref{tab:results_flat_XCDM} and shown in Fig.\ \ref{fig:flat_XCDM_P18_vs_nonCMB_2}. As in the previous subsubsection for the P18 and non-CMB (new) data primary parameter values, there are significant differences between the two sets of cosmological constraints here, with all five primary parameter values differing at more than 2$\sigma$ significance. In particular, $\Omega_{b}h^2$, $\Omega_{c}h^2$, 100$\theta_{\text{MC}}$, and $\ln(10^{10} A_s)$ values differ at $-2.13\sigma$, 2.27$\sigma$, 2.10$\sigma$, and $-2.65\sigma$. From P18+lensing data, for the dark energy equation of state parameter we get $w=-1.55^{+0.16}_{-0.35}$ showing a 3.44$\sigma$ preference for phantom-like behavior and a difference of $-4.27\sigma$ with the non-CMB (new) data result $w=-0.853^{+0.043}_{-0.033}$ which favors quintessence-like behavior at 4.45$\sigma$. Differences in the derived parameters are not as severe as those found for the primary cosmological parameter values but they are non-negligible. For $\Omega_m$ and $\sigma_8$ we obtain differences of $-1.41\sigma$ and $1.79\sigma$. These results probably mean that P18+lensing and non-CMB (new) data should not be jointly analyzed in the context of the flat XCDM model, and that the flat XCDM model probably is inconsistent with at least one of the two data sets (under the assumption that both data sets are correct), but this needs to be confirmed with the help of the statistical estimators.  

The primary and derived parameter values of the eight-parameter flat XCDM+$A_L$ model and the five-parameter flat XCDM model obtained with P18+lensing and non-CMB (new) data, respectively, are listed in Table \ref{tab:results_flat_XCDM} and shown in Fig.\ \ref{fig:FX_Alens_p18len_ncmb}. When we look at the values of the primary parameters, we observe significant differences, with four of the five primary parameter values differing at more than 2$\sigma$. For $\Omega_{b}h^2$, $\Omega_{c}h^2$, $100\theta_{\text{MC}}$, and $\ln(10^{10}A_s)$ the differences between the two values are $-2.11\sigma$, $2.15\sigma$, $2.11\sigma$, and $-2.69\sigma$, respectively. When the non-CMB (new) data are analyzed in the flat XCDM model we obtain $w=-0.853^{+0.043}_{-0.033}$ which indicates a 4.45$\sigma$ preference for quintessence-like behavior. In contrast, when we consider P18+lensing data in the context of the flat XCDM+$A_L$ model we get $w=-1.34^{+0.26}_{-0.51}$ which indicates a $1.31\sigma$ deviation from $w=-1$. The two values differ by $-1.86\sigma$, significantly smaller than the corresponding $-4.27\sigma$ value in the flat XCDM model discussed in the previous paragraph. As in the previous case less significant differences are found when we look at the values of the derived parameters. In particular for $\Omega_m$ and $\sigma_8$ the two results disagree at $-0.33\sigma$ and $0.58\sigma$. As in the previous case with $A_L=1$ these results may indicate that P18+lensing and non-CMB (new) data should not be jointly analyzed in the flat XCDM+$A_L$ model but further tests are needed before we definitively conclude this. 

The results presented in Table \ref{tab:results_XCDM_Planck_Pq} [\ref{tab:results_new_Pq_XCDM}] and shown in Fig.\ \ref{fig:XCDM_Planck_Pq_P18_vs_nonCMB_2} [\ref{fig:XCDM_new_Pq_P18_vs_nonCMB_2}] allow us to compare the eight-parameter and the six-parameter non-flat XCDM Planck [new] $P(q)$ models primary cosmological parameter constraints obtained from P18+lensing data and non-CMB (new) data. In this case two [one] of the six primary parameters differ at more than 2$\sigma$ significance. For $\Omega_{b}h^2$ and 100$\theta_{\text{MC}}$ the differences are at $-1.40\sigma$ [$-1.30\sigma$] and $-2.10\sigma$ [$-1.79\sigma$], respectively, which are very similar to the results found for the case of P18 and non-CMB (new) data. Using P18+lensing data we obtain for the curvature parameter $\Omega_k =-0.0111^{+0.013}_{-0.00070}$ [$-0.0080^{+0.0098}_{-0.0023}$] which is 0.85$\sigma$ [0.82$\sigma$] away from flat and differs at 2.59$\sigma$ [2.14$\sigma$] with the non-CMB (new) data value $\Omega_k=-0.177^{+0.064}_{-0.072}$ [$-0.186^{+0.083}_{-0.067}$] that also favors closed geometry but now at 2.77$\sigma$ [2.24$\sigma$]. For the dark energy equation of state parameter, P18+lensing data give $w=-1.28^{+0.41}_{-0.54}$ [$-1.27^{+0.40}_{-0.49}$] representing a phantom-like deviation of 0.68$\sigma$ [0.68$\sigma$] from the cosmological constant value of $w=-1$ and differing at $-1.20\sigma$ [$-1.21\sigma$] from the non-CMB (new) data value $w=-0.786^{+0.044}_{-0.037}$ [$-0.785^{+0.045}_{-0.038}$] that favors quintessence-like behavior at 5.78$\sigma$ [5.66$\sigma$]. As for the derived parameters we find mild differences, in particular for $H_0$, $\Omega_m$, and $\sigma_8$ these are 0.23$\sigma$ [0.24$\sigma$], $0.040\sigma$ [$-0.049\sigma$], and 0.76$\sigma$ [0.86$\sigma$].  

P18+lensing and non-CMB (new) data results for the nine-parameter non-flat XCDM Planck [new] $P(q)+A_L$ and the six-parameter non-flat XCDM Planck [new] $P(q)$ model can be seen in Table \ref{tab:results_XCDM_Planck_Pq} [\ref{tab:results_new_Pq_XCDM}] and in Fig.\ \ref{fig:NX_Alens_ns_p18len_ncmb} [\ref{fig:TNX_Alens_ns1_p18len_ncmb}]. We observe some non-negligible differences between the values of the primary parameters, with two [one] of the six differing at more than 2$\sigma$ significance. For those parameters common to the flat $\Lambda$CDM model the ones that show larger differences are $\Omega_{b}h^2$ and $100\theta_{\text{MC}}$ with disagreements at $-1.39\sigma$ [$-1.29\sigma$] and $-2.10\sigma$ [$-1.79\sigma$] respectively. With non-CMB (new) data and the non-flat XCDM Planck [new] $P(q)$ model, we find $\Omega_k = -0.177^{+0.064}_{-0.072}$ [$-0.186^{+0.083}_{-0.067}$] which favors closed spatial hypersurfaces at 2.77$\sigma$ [2.24$\sigma$] whereas when we analyze P18+lensing data within the context of the non-flat XCDM Planck [new] $P(q)+A_L$ model we get $\Omega_k = -0.012^{+0.027}_{-0.011}$ [$-0.003^{+0.018}_{-0.011}$] which shows a preference for a closed Universe at 0.44$\sigma$ [0.17$\sigma$]. The two values differ at 2.54$\sigma$ [2.19$\sigma$] significance. In regard to the equation of state parameter of dark energy, when non-CMB (new) data are utilized we find $w=-0.786^{+0.044}_{-0.037}$ [$-0.785^{+0.045}_{-0.038}$] and when P18+lensing data are considered we find $w=-1.32^{+0.71}_{-0.38}$ [$-1.18^{+0.54}_{-0.37}$], with the differences between the two values being $-0.75\sigma$ [$-0.73\sigma$]. In the first case we observe evidence of 5.78$\sigma$ [5.66$\sigma$] in favor of quintessence-like behavior while in the second case the result shows a preference for phantom-like behavior at 0.45$\sigma$ [0.33$\sigma$] significance. 
As for the derived parameters we find less significant differences than for the primary parameters, in particular for $H_0$, $\Omega_m$, and $\sigma_8$ the values disagree at $0.092\sigma$  [$0.24\sigma]$, $0.11\sigma$ [$0.03\sigma$], and $0.64\sigma$ [$0.55\sigma]$, respectively.

\subsection{Model Selection}\label{sec:Model_Selection} 

In this subsection we discuss how (relatively) well each data set is fit by each of the twelve models we study. In particular we consider the following combinations of data: non-CMB (new), P18, P18+lensing, P18+non-CMB (new), and P18+lensing+non-CMB (new). We also consider three pairs of $\Lambda$CDM models and three pairs of XCDM models, with one of each pair having $A_L = 1$ and the other with varying $A_L$, with one of the three being spatially flat and the other two allowing for non-zero spatial curvature. As we have mentioned previously, the lensing consistency parameter $A_L$ does not impact the non-CMB data analysis and so when we comment on the results obtained with non-CMB (new) data we do not mention the $A_L$-varying models. We focus on the results obtained for the DIC, Eq.\ (\ref{eq:DIC}), since it is considered to be more reliable than the AIC, Eq.\ (\ref{eq:AIC}). The different levels of significance for the $\Delta$DIC values are defined below Eq.\ (\ref{eq:DIC}) in Sec.\ \ref{sec:Methods}.  

\textbf{Non-CMB (new).} When the equation of state parameter of dark energy is fixed to $w=-1$, the flat $\Lambda$CDM model is weakly [positively] favored over the non-flat Planck [new] $P(q)$ model. On the other hand, when $w$ is allowed to vary we find that the flat XCDM model is strongly favored over the flat $\Lambda$CDM model. Simultaneous variations of $w$ and the curvature parameter $\Omega_k$ are favored by non-CMB (new) data since both the non-flat XCDM Planck $P(q)$ and new $P(q)$ models are strongly favored compared to the flat $\Lambda$CDM model, with the XCDM Planck $P(q)$ case very weakly favored over the XCDM new $P(q)$ case, while both non-flat XCDM models are very weakly favored over the flat XCDM model. Finally, the $\Lambda$CDM Planck [new] $P(q)$ model is very strongly [very strongly] disfavored when compared to the XCDM Planck [new] $P(q)$ model. These results indicate that allowing $w$ to be a freely varying parameter is significantly more important than allowing $\Omega_k$ to be a freely varying parameter in improving the fit to non-CMB (new) data relative to the performance of the flat $\Lambda$CDM model. 

\textbf{P18.}
For $A_L=1$ and $w=-1$, the non-flat $\Lambda$CDM Planck $P(q)$ and new $P(q)$ models are strongly favored over the flat $\Lambda$CDM model, with the Planck $P(q)$ model weakly favored over the new $P(q)$ one. However, when $A_L=1$ and $w$ is allowed to vary, the non-flat XCDM Planck $P(q)$ and new $P(q)$ models are positively favored over the flat XCDM model, with the Planck $P(q)$ model weakly favored over the new $P(q)$ one. When the lensing consistency parameter $A_L$ is allowed to vary, while keeping $w=-1$, the flat $\Lambda$CDM+$A_L$ model is positively favored over the flat $\Lambda$CDM model while the $\Lambda$CDM Planck [new] $P(q)+A_L$ model is weakly [positively] disfavored compared to the $\Lambda$CDM Planck [new] $P(q)$ model. When $w$ is allowed to vary while keeping $A_L=1$, the flat XCDM model is positively favored over the flat $\Lambda$CDM model whereas the non-flat XCDM Planck [new] $P(q)$ model is weakly [weakly] disfavored compared to the $\Lambda$CDM Planck [new] $P(q)$ model. We now consider the results obtained when both $w$ and $A_L$ are allowed to vary. Compared to the flat $\Lambda$CDM model, we find that the flat XCDM+$A_L$ model is positively favored, while the non-flat XCDM Planck and new $P(q)+A_L$ models are strongly favored. The flat XCDM+$A_L$ model is positively favored over the flat XCDM model. The non-flat XCDM Planck [new] $P(q)+A_L$ model is weakly disfavored [weakly favored] compared to the non-flat XCDM Planck [new] $P(q)$ model and very weakly [positively] favored compared to the $\Lambda$CDM Planck [new] $P(q)+A_L$ model. Additionally, the non-flat $\Lambda$CDM Planck [new] $P(q)+A_L$ model is weakly favored [positively disfavored] over the flat $\Lambda$CDM$+A_L$ model and the non-flat XCDM Planck and new $P(q)+A_L$ models are weakly favored over the flat XCDM$+A_L$ model. As we found previously in the $\Lambda$CDM case, \cite{deCruzPerez:2022hfr}, we find for XCDM here that allowing $A_L$ to vary weakens the support for the non-flat models over the corresponding varying-$A_L$ flat model, with the additional caveat that unlike in the $A_L = 1$ $\Lambda$CDM case where both non-flat models are strongly favored over the flat model, in the XCDM case the non-flat models are only positively favored over the flat model.

\textbf{P18+lensing.}
When we analyze the results obtained from P18+lensing data we do not find the level of evidence found when lensing data is not included in the mix. If we set $w=-1$ and $A_L=1$ we see that the non-flat $\Lambda$CDM Planck $P(q)$ and new $P(q)$ models are weakly favored relative to the flat $\Lambda$CDM model. For $w=-1$ and varying $A_L$, the flat $\Lambda$CDM+$A_L$ model is weakly favored over the flat $\Lambda$CDM model, and the non-flat $\Lambda$CDM Planck [new] $P(q)+A_L$ model is weakly [weakly] disfavored when compared with the $\Lambda$CDM Planck [new] $P(q)$ model. When $w$ is allowed to vary but $A_L=1$, the flat XCDM model is positively favored over the flat $\Lambda$CDM model. However, on the other hand, the non-flat XCDM Planck $P(q)$ and new $P(q)$ models are both weakly disfavored when compared to the non-flat $\Lambda$CDM Planck $P(q)$ and new $P(q)$ models, respectively. When allowing simultaneous variation of $w$ and $A_L$ we find that the flat XCDM+$A_L$ model is weakly favored over the flat $\Lambda$CDM model, while the XCDM Planck [new] $P(q)+A_L$ model is positively [weakly] disfavored compared to the standard model. Finally, the XCDM Planck [new] $P(q)+A_L$ model is positively [positively] disfavored when compared with the $\Lambda$CDM Planck [new] $P(q)$ model and is positively [weakly] disfavored over the XCDM Planck [new] $P(q)$ model. 

\textbf{P18+non-CMB (new).} We now consider how the P18 data analysis results change when we include non-CMB (new) data in the analysis. For $w=-1$ and $A_L=1$ we find that the non-flat $\Lambda$CDM Planck and new $P(q)$ models are weakly disfavored when compared to the flat $\Lambda$CDM model. Varying $A_L$ while holding $w=-1$, we observe that the flat $\Lambda$CDM+$A_L$ model is strongly favored over the flat $\Lambda$CDM model, and both the non-flat $\Lambda$CDM Planck and new $P(q)+A_L$ models are strongly favored over the $\Lambda$CDM Planck and new $P(q)$ models, respectively. Considering the complementary case, when $w$ varies and $A_L=1$ we see that the flat XCDM model is weakly disfavored when compared to the flat $\Lambda$CDM model, whereas the $\Lambda$CDM Planck [new] $P(q)$ model is weakly [weakly] favored over the XCDM Planck [new] $P(q)$ model. When $w$ and $A_L$ are both varied, we observe that the three XCDM+$A_L$ models, namely the flat XCDM+$A_L$ one and the non-flat XCDM Planck and new $P(q)+A_L$ ones, are strongly favored over the corresponding three $\Lambda$CDM models, the flat one and the non-flat Planck and new $P(q)$ ones, respectively. On the other hand the flat $\Lambda$CDM+$A_L$ model is very strongly favored over the flat XCDM model, and the two non-flat $\Lambda$CDM+$A_L$ models are strongly favored over the corresponding non-flat XCDM models. The flat XCDM+$A_L$ model is very strongly favored over the flat XCDM model, the XCDM Planck $P(q)+A_L$ model is on the verge of being very strongly preferred over the XCDM Planck $P(q)$ model, and the XCDM new $P(q)+A_L$ model is very strongly favored over the XCDM new $P(q)$ model. Finally we observe that the XCDM Planck and new $P(q)+A_L$ models are weakly preferred over the $\Lambda$CDM Planck and new $P(q)+A_L$ models, respectively. When $A_L$ is allowed to vary, P18+non-CMB (new) data typically strongly favor $A_L > 1$. 

\textbf{P18+lensing+non-CMB (new).} Finally, we now analyze the results obtained for the most complete data set employed in this work. We note again that when lensing data are included in the analysis such data compilations tend to more poorly distinguish between the models. The flat $\Lambda$CDM model is weakly favored over the $\Lambda$CDM Planck and new $P(q)$ models. When considering the case where $w=-1$ but $A_L$ is freely varied in the analysis, we find that the flat $\Lambda$CDM+$A_L$ model is positively favored over the flat $\Lambda$CDM model. A similar level of support is found for the $\Lambda$CDM Planck and new $P(q)+A_L$ models compared to the $\Lambda$CDM Planck and new $P(q)$ models, respectively. For $A_L=1$ but varying $w$, we observe that the flat XCDM model is positively disfavored in comparison to the flat $\Lambda$CDM model. Additionally, the XCDM Planck [new] $P(q)$ model is weakly [weakly] disfavored compared to the $\Lambda$CDM Planck [new] $P(q)$ model. When both $w$ and $A_L$ are free parameters, we find that the flat XCDM+$A_L$ model is positively preferred over the flat $\Lambda$CDM model and strongly favored over the flat XCDM model, while the XCDM Planck [new] $P(q)+A_L$ model is positively [positively and almost strongly] favored over the XCDM Planck [new] $P(q)$ model. We also note that when we compare the XCDM Planck [new] $P(q)+A_L$ model with the $\Lambda$CDM Planck [new] $P(q)+A_L$ model, the former is only weakly [weakly] preferred over the latter. Compared to our previous analysis with the non-CMB (old) data, \cite{deCruzPerez:2022hfr}, with the non-CMB (new) data here, in the flat and two non-flat $\Lambda$CDM models the varying $A_L$ option is no longer as positively favored over the $A_L = 1$ option, however it is positively or strongly favored over the $A_L = 1$ option in the flat and two non-flat XCDM models. 

\begin{table*}
\caption{Consistency check parameter $\log_{10} \mathcal{I}$ and tension parameters $\sigma$ and $p$ for P18 vs.\ non-CMB (new) data sets and P18+lensing vs.\ non-CMB (new) data sets in the XCDM ($+A_L$) models.
}
\begin{ruledtabular}
\begin{tabular}{lcccc}
\\[-1mm]                         & \multicolumn{2}{c}{Flat XCDM model}  &  \multicolumn{2}{c}{Flat XCDM+$A_L$ model} \\[+1mm]
\cline{2-3}\cline{4-5}\\[-1mm]
   Data                          &  P18 vs non-CMB  & P18+lensing vs non-CMB   &  P18 vs non-CMB   & P18+lensing vs non-CMB  \\[+1mm]
 \hline \\[-1mm]
  $\log_{10} \mathcal{I}$        &   $-2.125$     &  $-2.247$       &  $-0.364$  &  $-0.506$  \\[+1mm]
  $\sigma$                       &   $3.448$      &  $3.555$        &  $2.095$   &  $2.378$   \\[+1mm]
  $p$ (\%)                       &   $0.056$      &  $0.039$        &  $3.619$   &  $1.742$   \\[+1mm]
\hline \\[-1mm]
                                 & \multicolumn{2}{c}{Non-flat XCDM model [Planck $P(q)$]}   & \multicolumn{2}{c}{Non-flat XCDM+$A_L$ model [Planck $P(q)$]}     \\[+1mm]
\cline{2-3}\cline{4-5}\\[-1mm]
  Data                           &   P18 vs non-CMB     & P18+lensing vs non-CMB   &   P18 vs non-CMB        & P18+lensing vs non-CMB  \\[+1mm]
 \hline \\[-1mm]
  $\log_{10} \mathcal{I}$        &   $-3.421$     &  $-1.824$       &  $-1.173$  &  $-0.275$  \\[+1mm]
  $\sigma$                       &   $4.294$      &  $3.396$        &  $2.611$   &  $2.167$   \\[+1mm]
  $p$ (\%)                       &   $0.003$      &  $0.069$        &  $0.902$   &  $3.026$   \\[+1mm]
\hline \\[-1mm]
                                 & \multicolumn{2}{c}{Non-flat XCDM model [new $P(q)$]}    & \multicolumn{2}{c}{Non-flat XCDM+$A_L$ model [new $P(q)$]}     \\[+1mm]
\cline{2-3}\cline{4-5}\\[-1mm]
  Data                           &   P18 vs non-CMB    & P18+lensing vs non-CMB    &   P18 vs non-CMB        & P18+lensing vs non-CMB    \\[+1mm]
 \hline \\[-1mm]
  $\log_{10} \mathcal{I}$        &   $-3.125$     &  $-1.942$       &  $-0.957$  &  $-0.312$ \\[+1mm]
  $\sigma$                       &   $3.960$      &  $3.164$        &  $2.662$   &  $2.256$ \\[+1mm]
  $p$ (\%)                       &   $0.007$      &  $0.155$        &  $0.778$   &  $2.409$ \\[+1mm]
\end{tabular}
\\[+1mm]
\end{ruledtabular}
\label{tab:consistency_XCDM}
\end{table*}


\subsection{Data Set Tensions}
\label{sec:Data_Set_Tensions}

After analyzing in detail the cosmological parameter constraints for each of the twelve models under study, obtained for the different data sets considered in this work, and after comparing how (relatively) well each data set is fit by each model, in this subsection we test whether, in the context of a given cosmological model, there is concordance (discordance) between the results obtained for pairs of some of the data sets and also whether this data set consistency (inconsistency) is model independent. To accomplish this we use the two statistical estimators, presented in Sec.\ \ref{sec:Methods}, whose expressions are provided in Eqs.\ (\ref{eq:Tension_estimator_1}), (\ref{eq:Tension_estimator_2}), and (\ref{eq:Tension_estimator_2_sigma}) respectively. See there for discussions of the different levels of significance of the estimator values. The values for the two statistical estimators, $\log_{10}\mathcal{I}$ and $p (\sigma)$, are provided in Table \ref{tab:consistency_LCDM} for the $\Lambda$CDM models and in Table \ref{tab:consistency_XCDM} for the XCDM models, for the P18 and non-CMB and for the P18+lensing and non-CMB data sets comparisons. For the $\Lambda$CDM models in Table \ref{tab:consistency_LCDM} we provide both non-CMB (old) and non-CMB (new) data results, and as mentioned earlier non-CMB (new) data are more consistent with P18 and P18+lensing data than non-CMB (old) data are. For the XCDM models in Table \ref{tab:consistency_XCDM} we provide only non-CMB (new) data results. In the following discussion we focus only on the results obtained with the non-CMB (new) data set.  

{\bf P18 vs. non-CMB (new).} In the flat $\Lambda$CDM model P18 data and non-CMB (new) data are not inconsistent. For the first statistical estimator we get $\log_{10}\mathcal{I} = 0.805$ which indicates substantial consistency whereas for the second one we find $p = 24.9\%$ ($\sigma = 1.152$) which indicates neither significant concordance nor significant discordance. For the flat $\Lambda$CDM+$A_L$ model we obtain similar results with both estimators, $\log_{10}\mathcal{I}= 1.446$ and $p = 87\%$ ($\sigma = 0.164$), pointing to a strong consistency between the results obtained with the P18 and non-CMB (new) data sets.

When we consider varying $\Omega_k$ but keep $A_L=1$ and $w=-1$ we observe non-negligible tensions between the two sets of cosmological parameter constraints. For the non-flat $\Lambda$CDM Planck $P(q)$ model we get $\log_{10}\mathcal{I} = -0.796$ and $p = 0.687\%$ ($\sigma = 2.704$) indicating a substantial inconsistency between P18 data and non-CMB (new) data results, but less than $3\sigma$ and also less than the $\sigma = 3.005$ found when comparing the P18 and non-CMB (old) data results, \cite{deCruzPerez:2022hfr}. On the other hand, for the non-flat $\Lambda$CDM new $P(q)$ model we have $\log_{10}\mathcal{I} = -0.391$ and $p = 2.10\%$ ($\sigma = 2.308$), with the first estimator indicating neither consistency nor inconsistency while the second estimator points to a moderate inconsistency, which means that the tension between the two data sets is reduced in the new $P(q)$ case relative to the Planck $P(q)$ case. In view of these results the degree of discordance is not significant enough to prevent P18 data and non-CMB (new) data being used together in the $\Lambda$CDM Planck and new $P(q)$ models. 

When $w = -1$ but $\Omega_k$ and $A_L$ are allowed to vary, there is agreement between the results obtained with P18 and non-CMB (new) data. While for the $\Lambda$CDM Planck $P(q)+A_L$ model we find $\log_{10}\mathcal{I} = 1.210$ and $p = 55.2\%$ ($\sigma = 0.595$), for the $\Lambda$CDM new $P(q)+A_L$ model we get $\log_{10}\mathcal{I} = 2.107$ and $p = 77.2\%$ ($\sigma = 0.289$). In the first case the estimators indicate substantial consistency between the pair of data sets whereas in the second case the level of concordance is closer to being decisive. 

When $A_L = 1$ but the dark energy equation of state parameter $w$ varies we find some very significant tensions between P18 and non-CMB (new) data in the context of the XCDM models. For the flat XCDM model we get $\log_{10}\mathcal{I} = -2.125$ and $p = 0.056\%$ ($\sigma = 3.448$), indicating a decisive degree of discordance for this model. The simultaneous variation of $\Omega_k$ and $w$ increases the disagreement between the two data sets. For the XCDM Planck $P(q)$ model we obtain $\log_{10}\mathcal{I} = -3.421$ and $p = 0.003\%$ ($\sigma = 4.294$) and similarly for the XCDM new $P(q)$ model we get $\log_{10}\mathcal{I} = -3.125$ and $p = 0.007\%$ ($\sigma = 3.960$). In both cases the two statistical estimators point to a decisive degree of inconsistency, which makes the joint consideration of P18 and non-CMB (new) data inadvisable in the context of the non-flat XCDM models with $A_L = 1$. When $A_L = 1$ all three XCDM models, flat and non-flat, are inconsistent at $ > 3\sigma$ with either P18 data or non-CMB data or both (if these data sets are correct).

Allowing for a variation in the lensing consistency parameter $A_L$ results in more mutually consistent P18 and non-CMB (new) data results in the context of the XCDM models. In the flat XCDM+$A_L$ model we obtain $\log_{10}\mathcal{I} = -0.364$ and $p = 3.619\%$ ($\sigma = 2.095$) which indicates that the degree of discordance is at most moderate. For the XCDM Planck $P(q)+A_L$ model we find $\log_{10}\mathcal{I} = -1.173$ and $p = 0.902\%$ ($\sigma = 2.611$) whereas for the XCDM new $P(q)+A_L$ model we obtain $\log_{10}\mathcal{I} = -0.957$ and $p = 0.778\%$ ($\sigma = 2.662$). Both statistical estimators still show an almost strong discordance between P18 and non-CMB (new) data results, but the degree of inconsistency is probably not high enough to prevent us from considering these two data sets jointly in our analysis in the context of the non-flat XCDM+$A_L$ models.  

In summary, P18 and non-CMB (new) data are largely consistent in the six flat and non-flat $\Lambda$CDM($+A_L$) models and in the three flat and non-flat XCDM$+A_L$ models, but are inconsistent at $> 3\sigma$ in the three flat and non-flat XCDM $A_L=1$ models, thus ruling out these three models at $> 3\sigma$, if the two data sets are correct. We see next that very similar results hold in the P18+lensing and non-CMB (new) data sets comparison.

{\bf P18+lensing vs. non-CMB (new).} When looking at the results obtained for the flat $\Lambda$CDM model, including lensing data with P18 data in the analysis results in conclusions similar to the case when just P18 data results are compared to the non-CMB (new) data results. From the P18+lensing and non-CMB (new) data comparison, for the first statistical estimator we find $\log_{10}\mathcal{I} = 0.730$ and for the second one we obtain $p = 22.7\%$ ($\sigma = 1.209$). In the first case we find a substantial degree of concordance between the P18+lensing and non-CMB (new) data results whereas in the second case neither significant consistency nor significant inconsistency can be claimed. 

Compared to the previous case, where we compare P18 and non-CMB (new) data results, for $A_L=1$ and $w=-1$ but the curvature parameter $\Omega_k$ is allowed to vary, we find less significant disagreement between the P18+lensing and non-CMB (new) data results. For the non-flat $\Lambda$CDM Planck $P(q)$ model we get $\log_{10}\mathcal{I} = 0.711$ in addition to $p = 12\%$ ($\sigma = 1.555$) and very similarly for the non-flat $\Lambda$CDM new $P(q)$ model we get $\log_{10}\mathcal{I} = 0.755$ for the first estimator and $p = 12.3\%$ ($\sigma = 1.544$) for the second one. So, the first statistical estimator, in both cases, points to substantial consistency between the results obtained with P18+lensing and non-CMB (new) data while the second one indicates neither significant consistency nor significant inconsistency between the two data sets. This means that P18+lensing and non-CMB (new) data can be jointly analyzed in the context of the non-flat $\Lambda$CDM models. 

As expected, in light of the results obtained in the previous P18 vs.\ non-CMB (new) data case, the simultaneous variation of $\Omega_k$ and $A_L$ in the analysis enhances the consistency between results obtained from P18+lensing data and from non-CMB (new) data. In the case of the non-flat $\Lambda$CDM Planck $P(q)+A_L$ model we get $\log_{10}\mathcal{I} = 1.719$ and for the second estimator $p = 80.9\%$ ($\sigma = 0.241$) while for the non-flat $\Lambda$CDM new $P(q)+A_L$ model we have $\log_{10}\mathcal{I} = 1.887$ as well as $p = 75.7\%$ ($\sigma = 0.312$), so again we find very similar results for the two cosmological models. In the context of the non-flat $\Lambda$CDM+$A_L$ models the degree of concordance between P18+lensing and non-CMB (new) data results is strong. 

Like in the P18 vs.\ non-CMB (new) data comparison analysis, when we look at the cases that allow the equation of state parameter of dark energy $w$ to vary, we find some non-negligible disagreements between results obtained with P18+lensing and non-CMB (new) data. For the flat XCDM model we obtain for the two statistical estimators $\log_{10}\mathcal{I} = -2.247$ and $p = 0.039\%$ ($\sigma = 3.555$) respectively. These results indicate a decisive degree of discordance between the P18+lensing and non-CMB (new) data sets in the context of the flat XCDM model and rule out the flat XCDM model at $> 3\sigma$, if both data sets are correct.

As for the cases where $w$ and $\Omega_k$ are allowed to simultaneously vary, we observe a slight reduction in the disagreement between P18+lensing and non-CMB (new) data results, compared to the $\Omega_k=0$ case, although they are probably still too large to allow a joint analysis of the two data sets in the context of the non-flat XCDM models. For the non-flat XCDM Planck $P(q)$ model the first and second statistical estimators give $\log_{10}\mathcal{I} = -1.824$ and $p = 0.069\%$ ($\sigma = 3.396$) whereas for the non-flat XCDM new $P(q)$ model we get $\log_{10}\mathcal{I} = -1.942$ and $p = 0.155\%$ ($\sigma = 3.164$). Therefore, for both models there is still an almost decisive degree of inconsistency between the results obtained with P18+lensing data and with non-CMB (new) data, with both non-flat XCDM models ruled out at $> 3\sigma$. 

As expected, when we also allow $A_L$ to vary the disagreements between the results obtained with the two data sets subside. For the flat XCDM+$A_L$ model we find $\log_{10}\mathcal{I} = -0.506$ which indicates a substantial degree of inconsistency and for the second estimator we  have $p = 1.742\%$ ($\sigma = 2.378$) which goes a step further and seems to point to a strong inconsistency, but $ < 3\sigma$, between the P18+lensing and non-CMB (new) data results. In the context of the flat XCDM+$A_L$ model it is probably safe to jointly analyze P18+lensing and non-CMB (new) data. 

In the case of the non-flat XCDM Planck $P(q)+A_L$ model for the two statistical estimators we have $\log_{10}\mathcal{I} = -0.275$ and $p = 3.026\%$ ($\sigma = 2.167$) whereas for the non-flat XCDM new $P(q)+A_L$ model we obtain $\log_{10}\mathcal{I} = -0.312$ and $p = 2.409\%$ ($\sigma = 2.256$). While the first estimator for both models do not indicate a significant degree of discordance, the second one points to a strong disagreement between the two sets of results. However, the level of disagreement does not seem to be enough to disallow the joint consideration of P18+lensing and non-CMB (new) data within the context of the non-flat XCDM+$A_L$ cosmological models.   

In our previous work \cite{deCruzPerez:2022hfr}, we statistically confirmed that in the tilted non-flat $\Lambda$CDM models with $A_L=1$ there is a tension between P18 data and non-CMB data results that rules out the non-flat models at more than $2.5\sigma$ significance (just over $3.0\sigma$ for the Planck $P(q)$ model and just under $2.6\sigma$ for the new $P(q)$ model). Here, from the results of the non-flat models with $A_L=1$ or varying $A_L$ (in Tables \ref{tab:consistency_LCDM} and \ref{tab:consistency_XCDM}), except for the non-flat $\Lambda$CDM$+A_L$ new $P(q)$ model, we find larger values of $\log_{10} \mathcal{I}$ and $p$ for the P18+lensing vs.\ non-CMB (new) data comparison than for the P18 vs.\ non-CMB (new) data comparison, which suggests that the addition of lensing data to P18+non-CMB (new) data reduces the tension between P18 and non-CMB (new) data. However, this seems to contradict our finding that adding lensing data to P18+non-CMB (new) data does not improve the model fit. For example, in the non-flat $\Lambda$CDM models with $A_L=1$, adding lensing data to the P18+non-CMB (new) data only slightly decreases the DIC values. Furthermore, all $A_L$-varying $\Lambda$CDM models constrained with P18+lensing+non-CMB (new) data are positively disfavored compared to the corresponding models constrained with P18+non-CMB (new) data, and similar behavior is seen for all XCDM models except the non-flat XCDM Planck $P(q)$ model with $A_L=1$. This is likely due to the fact that lensing data slightly shifts the values of the cosmological parameters towards the region of parameter space that is preferred by non-CMB (new) data, and thus appears to be more consistent with non-CMB (new) data. Consequently, when we compare P18+lensing with non-CMB (new) data we observe less disagreement between the two results than when we compare P18 data and non-CMB (new) data results.

In summary, P18+lensing and non-CMB (new) data are largely consistent in the six flat and non-flat $\Lambda$CDM($+A_L$) models and in the three flat and non-flat XCDM$+A_L$ models, but are inconsistent at $> 3\sigma$ in the three flat and non-flat XCDM $A_L=1$ models, thus ruling out these three models at $> 3\sigma$, if the two data sets are correct. From Figs.\ \ref{fig:flat_XCDM_P18_vs_nonCMB_1}, \ref{fig:XCDM_Planck_Pq_P18_vs_nonCMB_1}, and \ref{fig:XCDM_new_Pq_P18_vs_nonCMB_1}, we can compare P18 and non-CMB (new) data contours for the XCDM models with $A_L =1$. We see that in some panels the 3$\sigma$ contours are the first to overlap, in agreement with our discussion in this subsection about inconsistencies between these two data sets within the context of these models. The effect of considering a varying $A_L$ parameter can be seen in Figs.\ \ref{fig:FX_Alens_p18_ncmb}, \ref{fig:NX_Alens_ns_p18_ncmb}, and \ref{fig:TNX_Alens_ns1_p18_ncmb}. The inclusion of the lensing consistency parameter brings the contours closer together thus reducing the tension between P18 and non-CMB (new) cosmological parameter constraints and this agrees with the information in Table \ref{tab:consistency_XCDM} where this tension reduction can be appreciated in a more quantitative way. In Sec.\ \ref{sec:XCDM comparison} we compare P18, P18+lensing, and P18+lensing+non-CMB (new) contours obtained for the XCDM models for fixed and varying $A_L$. We have some at most 2$\sigma$ non-overlapping regions (for the non-flat XCDM models with $A_L =1$) and this was not unexpected since when we jointly analyze either P18 or P18+lensing data with non-CMB (new) data we are forcing the cosmological parameters to have values between P18 (or P18+lensing) data values and non-CMB (new) data values. In Sec.\ \ref{sec:Model_Selection} we compare the performance of the different models tested against the different combinations of data sets. The results provided in Tables \ref{tab:chi2_lcdm} and \ref{tab:chi2_xcdm} clearly show that when non-CMB (new) data are included in the analysis it is possible to improve the performance of the flat $\Lambda$CDM and flat XCDM models by allowing $A_L$ to vary and this goes hand-in-hand with the fact that the inclusion of a varying lensing consistency parameter in the analysis can increase the consistency between P18 and non-CMB (new) data sets in the context of the XCDM models. This again highlights the importance of considering varying $A_L$ in the analysis since in the $\Lambda$CDM and the XCDM models it helps to reduce the tension between CMB and non-CMB data results. These results are very similar to those found above for the P18 data set and the non-CMB (new) data sets comparison.

\begin{table*}
\caption{Individual and total $\chi^2$ values for the best-fit flat and non-flat $\Lambda\textrm{CDM}$ inflation models. Deviance information criterion (DIC) and Akaike information criterion (AIC) are also listed.  }
{\scriptsize
\begin{ruledtabular}
\begin{tabular}{lcccccccccccccc}
    Data sets   & $\chi_{\textrm{plik}}^2$  & $\chi_{\textrm{lowl}}^2$  & $\chi_{\textrm{simall}}^2$  & $\chi_{\textrm{lensing}}^2$ &  $\chi_{\textrm{prior}}^2$  &  $\chi_{\textrm{SN}}^2$  & $\chi_{\textrm{BAO}}^2$  &  $\chi_{H(z)}^2$   &  $\chi_{f\sigma_8}^2$ &  $\chi^2_{\textrm{total}}$      & $\Delta\chi^2$  & DIC & $\Delta\textrm{DIC}$  & $\Delta\textrm{AIC}$ \\[+0mm]
 \hline \\[-2mm]
 \multicolumn{15}{c}{Flat $\Lambda\textrm{CDM}$ model} \\
  \hline \\[-2mm]
   Non-CMB (new)             &          &       &         &       &      & 1416.49 & 26.43 & 14.57 & 12.44 & 1469.93  &         & 1478.11 &         & \\[+1mm]
   P18                       & 2344.71  & 23.39 & 396.05  &       & 1.66 &         &       &       &       & 2765.80  &         & 2817.93 &         & \\[+1mm]
   P18+lensing               & 2344.66  & 23.39 & 396.06  & 8.79  & 1.82 &         &       &       &       & 2774.71  &         & 2826.45 &         & \\[+1mm]
   P18+non-CMB (new)         & 2347.28  & 22.53 & 396.08  &       & 1.83 & 1414.42 & 25.11 & 14.97 & 18.03 & 4240.24  &         & 4292.33 &         & \\[+1mm]
   P18+lensing+non-CMB (new) & 2346.87  & 22.63 & 396.33  & 8.89  & 1.65 & 1414.32 & 25.03 & 14.96 & 18.57 & 4249.26  &         & 4301.20 &         & \\[+1mm]
 \hline\\[-2mm]
    \multicolumn{15}{c}{Flat $\Lambda$\textrm{CDM}$+A_L$ model}  \\
  \hline \\[-2mm]
   P18                       & 2337.23  & 21.92 & 395.66  &       & 1.31 &         &       &       &       & 2756.12  & $-9.68$ & 2812.41 & $-5.52$ & $-7.68$ \\[+1mm]
   P18+lensing               & 2341.62  & 22.29 & 395.68  & 9.94  & 1.71 &         &       &       &       & 2771.24  & $-3.47$ & 2825.53 & $-0.92$ & $-1.47$ \\[+1mm]
   P18+non-CMB (new)         & 2337.37  & 21.80 & 395.70  &       & 1.36 & 1415.51 & 25.39 & 14.90 & 15.23 & 4227.27  &$-12.97$ & 4283.86 & $-8.47$ & $-10.97$ \\[+1mm]
   P18+lensing+non-CMB (new) & 2342.15  & 22.01 & 395.69  & 9.80  & 1.84 & 1415.26 & 25.47 & 14.93 & 15.46 & 4242.61  & $-6.65$ & 4297.19 & $-4.01$ & $-4.65$ \\[+1mm]
   \hline \\[-2mm]
    \multicolumn{15}{c}{Non-flat $\Lambda\textrm{CDM}$ model [Planck $P(q)$]}  \\
  \hline \\[-2mm]
   Non-CMB (new)             &          &       &         &       &      & 1414.52 & 26.98 & 14.57 & 12.15 & 1468.22  & $-1.71$  & 1479.52 &  +1.41   & +0.29    \\[+1mm]
   P18                       & 2336.45  & 21.29 & 395.60  &       & 1.38 &         &       &       &       & 2754.73  & $-11.07$ & 2810.59 & $-7.34$ & $-9.07$ \\[+1mm]
   P18+lensing               & 2342.29  & 21.86 & 395.66  & 10.09 & 1.63 &         &       &       &       & 2771.53  & $-3.18$  & 2826.17 & $-0.28$ & $-1.18$ \\[+1mm]
   P18+non-CMB (new)         & 2345.93  & 23.07 & 396.21  &       & 1.89 & 1414.50 & 24.80 & 14.84 & 18.32 & 4239.58  & $-0.66$  & 4293.78 &  +1.45   & +1.34   \\[+1mm]
   P18+lensing+non-CMB (new) & 2345.79  & 23.29 & 395.95  &  8.99 & 1.97 & 1414.27 & 24.90 & 14.84 & 18.75 & 4248.74  & $-0.52$  & 4302.41 &  +1.21   & +1.48   \\[+1mm]
 \hline\\[-2mm]
    \multicolumn{15}{c}{Non-flat $\Lambda$\textrm{CDM}$+A_L$ model [Planck $P(q)$]}  \\
  \hline \\[-2mm]
   P18                       & 2336.57  & 21.51 & 395.61  &       & 1.29 &         &       &       &       & 2754.99  & $-10.81$ & 2811.63 & $-6.30$ & $-6.81$  \\[+1mm]
   P18+lensing               & 2341.32  & 22.55 & 395.71  & 9.44  & 2.12 &         &       &       &       & 2771.14  & $-3.57$  & 2827.14 & $+0.69$  & $+0.43$   \\[+1mm]
   P18+non-CMB (new)         & 2337.36  & 21.67 & 395.77  &       & 1.34 & 1415.46 & 25.50 & 14.95 & 15.02 & 4227.07  & $-13.17$ & 4285.58 & $-6.75$ & $-9.17$ \\[+1mm]
   P18+lensing+non-CMB (new) & 2341.49  & 22.19 & 395.70  & 9.78  & 1.82 & 1415.38 & 25.47 & 14.88 & 15.51 & 4242.22  & $-7.04$  & 4298.73 & $-2.47$ & $-3.04$  \\[+1mm]
   \hline \\[-2mm]
    \multicolumn{15}{c}{Non-flat $\Lambda\textrm{CDM}$ model [new $P(q)$]}  \\
  \hline \\[-2mm]
   Non-CMB (new)             &          &       &         &       &      & 1414.72 & 26.79 & 14.55 & 12.15 & 1468.21  & $-1.72$  & 1480.16 & +2.05   &  +0.28    \\[+1mm]
   P18                       & 2338.26  & 21.42 & 396.28  &       & 1.42 &         &       &       &       & 2757.38  & $-8.42$  & 2811.54 & $-6.39$ & $-6.42$  \\[+1mm]
   P18+lensing               & 2342.99  & 21.18 & 395.90  & 9.92  & 1.76 &         &       &       &       & 2771.75  & $-2.96$  & 2825.74 & $-0.71$ & $-0.96$  \\[+1mm]
   P18+non-CMB (new)         & 2346.77  & 22.78 & 395.77  &       & 1.81 & 1414.47 & 25.00 & 14.85 & 17.98 & 4239.45  & $-0.79$  & 4293.50 & +1.17   & +1.21    \\[+1mm]
   P18+lensing+non-CMB (new) & 2345.86  & 22.98 & 396.27  & 8.94  & 1.73 & 1414.35 & 24.79 & 14.86 & 18.73 & 4248.50  & $-0.76$  & 4302.33 & +1.13   & +1.24    \\[+1mm]
 \hline\\[-2mm]
    \multicolumn{15}{c}{Non-flat $\Lambda$\textrm{CDM}$+A_L$ model [new $P(q)$]}  \\
  \hline \\[-2mm]                
   P18                       & 2337.56  & 21.31 & 395.93  &       & 1.52 &         &       &       &       & 2756.33  & $-9.47$  & 2814.83 & $-3.10$ & $-5.47$  \\[+1mm]
   P18+lensing               & 2341.21  & 22.62 & 395.75  & 9.49  & 1.37 &         &       &       &       & 2770.45  & $-4.26$  & 2827.29 & $+0.84$  & $-0.26$  \\[+1mm]
   P18+non-CMB (new)         & 2336.94  & 21.93 & 395.70  &       & 1.43 & 1415.23 & 25.42 & 14.93 & 15.54 & 4227.11  &$-13.13$  & 4285.29 & $-7.04$ & $-9.13$  \\[+1mm]
   P18+lensing+non-CMB (new) & 2342.43  & 21.94 & 395.77  & 9.12  & 1.71 & 1415.41 & 25.52 & 14.93 & 15.17 & 4242.01  & $-7.25$  & 4298.75 & $-2.45$ & $-3.25$  \\[+1mm]
\end{tabular}
\\[+1mm]
Note: $\Delta\chi^2$, $\Delta\textrm{DIC}$, and $\Delta\textrm{AIC}$ indicate the values relative to those of the tilted flat $\Lambda\textrm{CDM}$ model for the same combination of data sets. For the tilted flat $\Lambda$CDM model AIC$=2819.80$ (P18), $2828.71$ (P18+lensing), and $4303.26$ (P18+lensing+non-CMB (new)). All $\chi^2$ values are computed at the corresponding model best-fit cosmological parameter values. See section IV B of \cite{deCruzPerez:2022hfr} for detailed descriptions of individual CMB $\chi^2$'s.
\end{ruledtabular}
}
\label{tab:chi2_lcdm}
\end{table*}

\begin{table*}
\caption{Individual and total $\chi^2$ values for the best-fit flat and non-flat XCDM inflation models. Deviance information criterion (DIC) and Akaike information criterion (AIC) are also listed.  }
{\scriptsize
\begin{ruledtabular}
\begin{tabular}{lcccccccccccccc}
    Data sets   & $\chi_{\textrm{plik}}^2$  & $\chi_{\textrm{lowl}}^2$  & $\chi_{\textrm{simall}}^2$  & $\chi_{\textrm{lensing}}^2$ &  $\chi_{\textrm{prior}}^2$  &  $\chi_{\textrm{SN}}^2$  & $\chi_{\textrm{BAO}}^2$  &  $\chi_{H(z)}^2$   &  $\chi_{f\sigma_8}^2$ &  $\chi^2_{\textrm{total}}$      & $\Delta\chi^2$  & DIC & $\Delta\textrm{DIC}$  & $\Delta\textrm{AIC}$ \\[+0mm]
 \hline \\[-2mm]
    \multicolumn{15}{c}{Flat XCDM model}  \\
  \hline \\[-2mm]               
   Non-CMB (new)             &          &       &         &       &      & 1411.93 & 21.66 & 14.84 & 10.75 & 1459.18  &$-10.75$  & 1468.74 & $-9.37$ & $-8.75$  \\[+1mm]
   P18                       & 2341.65  & 22.46 & 395.79  &       & 1.51 &         &       &       &       & 2761.40  & $-4.40$  & 2815.67 & $-2.26$ & $-2.40$  \\[+1mm]
   P18+lensing               & 2342.37  & 22.14 & 395.65  & 8.61  & 1.81 &         &       &       &       & 2770.58  & $-4.13$  & 2824.21 & $-2.24$ & $-2.13$  \\[+1mm]
   P18+non-CMB (new)         & 2348.35  & 22.33 & 396.21  &       & 1.77 & 1413.17 & 25.60 & 14.97 & 17.44 & 4239.85  & $-0.39$  & 4294.20 &  +1.87  &  +1.61    \\[+1mm]
   P18+lensing+non-CMB (new) & 2347.50  & 22.54 & 396.48  & 9.01  & 1.57 & 1413.38 & 25.41 & 14.98 & 18.19 & 4249.05  & $-0.21$  & 4303.30 &  +2.10  &  +1.79    \\[+1mm]
 \hline\\[-2mm]                  
    \multicolumn{15}{c}{Flat XCDM+$A_L$ model}  \\
  \hline \\[-2mm]
   P18                       & 2337.36  & 21.64 & 395.60  &       & 1.29 &         &       &       &       & 2755.89  & $-9.91$  & 2813.08 & $-4.85$ & $-5.91$  \\[+1mm]
   P18+lensing               & 2341.84  & 21.98 & 395.62  & 9.04  & 1.94 &         &       &       &       & 2770.43  & $-4.28$  & 2825.81 & $-0.64$ & $-0.28$  \\[+1mm]
   P18+non-CMB (new)         & 2338.34  & 21.56 & 395.67  &       & 1.29 & 1412.51 & 26.83 & 14.90 & 13.88 & 4224.98  &$-15.26$  & 4283.50 & $-8.83$ &$-11.26$  \\[+1mm]
   P18+lensing+non-CMB (new) & 2343.23  & 21.75 & 395.69  & 9.77  & 2.03 & 1412.70 & 26.87 & 14.93 & 13.95 & 4240.92  & $-8.34$  & 4296.89 & $-4.31$ & $-4.34$  \\[+1mm]
 \hline\\[-2mm]                  
    \multicolumn{15}{c}{Non-flat XCDM model [Planck $P(q)$]}  \\
  \hline \\[-2mm]            
   Non-CMB (new)             &          &       &         &       &      & 1412.51 & 22.67 & 14.77 & 10.85 & 1460.80  &$-13.29$  & 1468.14 & $-9.97$ & $-9.29$  \\[+1mm]
   P18                       & 2336.63  & 21.31 & 395.63  &       & 1.34 &         &       &       &       & 2754.91  &$-10.89$  & 2810.86 & $-7.07$ & $-6.89$  \\[+1mm]
   P18+lensing               & 2341.99  & 21.93 & 395.68  & 9.29  & 1.51 &         &       &       &       & 2770.40  & $-4.31$  & 2827.00 &  +0.55  & $-0.31$  \\[+1mm]
   P18+non-CMB (new)         & 2346.82  & 22.99 & 396.59  &       & 1.81 & 1412.60 & 25.56 & 14.79 & 17.50 & 4238.67  & $-1.57$  & 4294.75 &  +2.42  &  +2.43    \\[+1mm]
   P18+lensing+non-CMB (new) & 2346.33  & 23.13 & 396.00  & 9.28  & 2.09 & 1412.98 & 25.43 & 14.85 & 18.18 & 4248.26  & $-1.00$  & 4303.54 &  +2.34  &  +3.00    \\[+1mm]
 \hline\\[-2mm]                  
    \multicolumn{15}{c}{Non-flat XCDM+$A_L$ model [Planck $P(q)$]}  \\
  \hline \\[-2mm]
   P18 ($A_L>0.8$)           & 2336.44  & 21.20 & 395.52  &       & 1.30 &         &       &       &       & 2754.46  &$-11.34$  & 2811.61 & $-6.32$ & $-5.34$  \\[+1mm]
   P18+lensing               & 2341.46  & 22.28 & 395.66  & 9.04  & 1.83 &         &       &       &       & 2770.28  & $-4.43$  & 2829.13 &  +2.68   & +1.57    \\[+1mm]
   P18+non-CMB (new)         & 2337.83  & 21.81 & 395.74  &       & 1.37 & 1412.23 & 27.55 & 14.81 & 13.50 & 4224.83  &$-15.41$  & 4285.15 & $-7.18$ & $-9.41$  \\[+1mm]
   P18+lensing+non-CMB (new) & 2341.91  & 21.97 & 395.81  &10.06  & 1.55 & 1412.20 & 27.24 & 14.79 & 14.17 & 4239.70  & $-9.56$  & 4298.54 & $-2.66$ & $-3.56$  \\[+1mm]
    \hline\\[-2mm]                  
    \multicolumn{15}{c}{Non-flat XCDM model [new $P(q)$]}  \\
 \hline \\[-2mm]                
   Non-CMB (new)             &          &       &         &       &      & 1411.93 & 21.98 & 14.88 & 10.71 & 1459.51   &$-13.27$ & 1468.73 & $-9.38$ & $-9.27$  \\[+1mm]
   P18                       & 2338.49  & 21.43 & 396.14  &       & 1.81 &         &       &       &       & 2757.86   & $-7.94$ & 2811.78 & $-6.15$ & $-3.94$  \\[+1mm]
   P18+lensing               & 2341.99  & 21.73 & 395.82  & 8.96  & 2.07 &         &       &       &       & 2770.57   & $-4.14$ & 2826.60 & +0.15   & $-0.14$  \\[+1mm]
   P18+non-CMB (new)         & 2347.31  & 22.91 & 396.21  &       & 1.83 & 1412.59 & 25.67 & 14.81 & 17.24 & 4238.57   & $-1.67$ & 4294.90 & +2.57   & +2.33    \\[+1mm]
   P18+lensing+non-CMB (new) & 2345.80  & 23.23 & 396.59  & 9.00  & 1.82 & 1412.99 & 25.12 & 14.81 & 18.60 & 4247.96   & $-1.30$ & 4304.26 & +3.06   & +2.70    \\[+1mm]
 \hline\\[-2mm]                  
    \multicolumn{15}{c}{Non-flat XCDM+$A_L$ model [new $P(q)$]}  \\
 \hline \\[-2mm]     
   P18 ($A_L > 0.8$)         & 2336.38  & 21.16 & 395.52  &       & 1.34 &         &       &       &       & 2754.40   &$-11.40$ & 2811.71 & $-6.22$ & $-5.40$  \\[+1mm]
   P18+lensing               & 2341.66  & 21.74 & 395.68  & 9.20  & 1.98 &         &       &       &       & 2770.27   & $-4.44$ & 2828.10 &  +1.65   &  +1.56   \\[+1mm]
   P18+non-CMB (new)         & 2337.48  & 21.79 & 395.75  &       & 1.35 & 1412.30 & 27.39 & 14.80 & 13.66 & 4224.52   &$-15.72$ & 4284.84 & $-7.49$ & $-9.72$  \\[+1mm]
   P18+lensing+non-CMB (new) & 2341.80  & 22.10 & 395.78  & 9.71  & 1.85 & 1412.28 & 27.47 & 14.79 & 13.97 & 4239.76   & $-9.50$ & 4298.27 & $-2.93$ & $-3.50$  \\[+1mm]
\end{tabular}
\\[+1mm]
Note: $\Delta\chi^2$, $\Delta\textrm{DIC}$, and $\Delta\textrm{AIC}$ indicate the values relative to those of the tilted flat $\Lambda\textrm{CDM}$ model for the same combination of data sets. For the tilted flat $\Lambda$CDM model AIC$=2819.80$ (P18), $2828.71$ (P18+lensing), and $4303.26$ (P18+lensing+non-CMB (new)). All $\chi^2$ values are computed at the corresponding model best-fit cosmological parameter values. See section IV B of \cite{deCruzPerez:2022hfr} for detailed descriptions of individual CMB $\chi^2$'s.
\end{ruledtabular}
}
\label{tab:chi2_xcdm}
\end{table*}

\section{Discussion}\label{sec:Discussion}

In this work we have employed P18 data, (P18) CMB lensing data, and non-CMB data to place constraints on the cosmological parameters of twelve cosmological models: flat $\Lambda$CDM (+$A_L$), non-flat $\Lambda$CDM Planck $P(q)$ (+$A_L$), non-flat $\Lambda$CDM new $P(q)$ (+$A_L$), flat XCDM (+$A_L$), non-flat XCDM Planck $P(q)$ (+$A_L$), and non-flat XCDM new $P(q)$ ($+A_L$). In order to compare how well each of these models fit observational data, we use two statistical criteria: the AIC (Eq.\ \eqref{eq:AIC}) and the DIC (Eq.\ \eqref{eq:DIC}). Furthermore, with the help of two estimators (Eqs.\ \eqref{eq:Tension_estimator_1} and \eqref{eq:Tension_estimator_2}), we determine the mutual (in)consistency between results obtained from two different data sets when analyzed in a given model. 

According to the statistical tools utilized, three of the twelve cosmological models studied are rejected at 3$\sigma$ significance due to incompatibilities between results obtained from different data sets, provided that these data are correct and free from unaccounted systematic errors. From the P18 vs.\ non-CMB (new) and P18+lensing vs.\ non-CMB (new) data comparisons, the three cosmological models ruled out at confidence levels exceeding $3\sigma$ are the flat XCDM, the non-flat XCDM Planck $P(q)$, and the non-flat XCDM new $P(q)$ models. Interestingly, when non-CMB (old) data is replaced by non-CMB (new) data, the non-flat $\Lambda$CDM Planck $P(q)$ model is no longer rejected at more than 3$\sigma$, which is what we found in \cite{deCruzPerez:2022hfr} where we used non-CMB (old) data. 

When we compare the cosmological parameter constraints obtained with P18 data and non-CMB (new) data in the flat XCDM model, we observe that the values of all primary parameters disagree at more than $2\sigma$. In particular, $\Omega_{b}h^2$, $\Omega_{c}h^2$, 100$\theta_{\text{MC}}$, and $\ln(10^{10}A_s)$ differ at $-2.14\sigma$, 2.34$\sigma$, 2.09$\sigma$, and $-2.63\sigma$ respectively, with the difference in the values of the dark energy equation of state parameter $w$ ($-4.80\sigma$) being the largest one. Including the lensing data in the analysis and considering the P18+lensing data vs.\ non-CMB (new) data comparison barely changes the level of disagreement. In this case, the difference between the two sets of cosmological parameter values for $\Omega_{b}h^2$, $\Omega_{c}h^2$, 100$\theta_{\text{MC}}$, and $\ln(10^{10}A_s)$ are $-2.13\sigma$, 2.27$\sigma$, 2.10$\sigma$, and $-2.65\sigma$, respectively, while the $w$ difference is $-4.27\sigma$. We also use two statistical estimators to measure the degree of concordance/discordance between the results obtained with two data sets in a given model. When comparing the P18 and non-CMB (new) results, we have $\log_{10}\mathcal{I} = -2.125$ and $p = 0.056\%$ ($\sigma = 3.448$) with both statistical estimators indicating a decisive degree of disagreement. As for the results obtained when we study the case of P18+lensing data vs.\ non-CMB data we obtain $\log_{10}\mathcal{I} = -2.247$ and $p = 0.039\%$ ($\sigma = 3.555$), which is also decisive.  Therefore, in light of these results we conclude that the flat XCDM model is ruled out at more than 3$\sigma$ significance. 

The other two cosmological models rejected by these data at 3$\sigma$ significance are the non-flat XCDM Planck $P(q)$ and the non-flat XCDM new $P(q)$ models. For both models, when we compare the results obtained with P18 and non-CMB (new) data we find smaller differences compared to those for the flat XCDM model. In the Planck $P(q)$ model, for the primary parameters common to the flat $\Lambda$CDM model the largest difference is found for $100\theta_{\text{MC}}$ ($-2.10\sigma$) whereas for the curvature parameter $\Omega_k$ and the dark energy equation of state parameter $w$ we find differences between the two values at $1.98\sigma$ and $-0.50\sigma$, respectively. In regard to the incompatibilities found between P18 data and non-CMB (new) data results we obtain $\log_{10}\mathcal{I} = -3.421$ and $p = 0.003\%$ ($\sigma = 4.294$) which means that the level of discordance exceeds 3$\sigma$ and thus argues against the joint analysis of these two data sets in the context of the non-flat XCDM Planck $P(q)$ model. When the lensing data are included in the mix the level of disagreement between the results from the two data sets is reduced, but is still high enough to reject this model at more than 3$\sigma$. The primary parameters $100\theta_{\text{MC}}$, $\Omega_k$, and $w$ values disagree at $-2.10\sigma$, $2.59\sigma$, and $-1.20\sigma$ respectively whereas for the tension estimators we obtain $\log_{10}\mathcal{I} = -1.824$ for the first one and $p = 0.069\%$ ($\sigma = 3.396$) for the second one. Similar results are obtained for the non-flat XCDM new $P(q)$ model but at a less severe level of disagreement. When we compare P18 and non-CMB (new) data results, the largest differences between the two sets of results are for $100\theta_{\text{MC}}$, $\Omega_k$, and $w$, which differ by $-1.79\sigma$, $1.82\sigma$, and $-0.61\sigma$ respectively. As expected, the statistical estimators reconfirm the tension observed in the differences in parameter values, with the first one being $\log_{10}\mathcal{I} = -3.125$ while for the second one we have $p = 0.007\%$ ($\sigma = 3.960$), with both indicating a decisive level of discordance between the results. No significant changes are observed when the P18+lensing data vs.\ non-CMB (new) data case is studied. The primary parameter $100\theta_{\text{MC}}$, $\Omega_k$, and $w$ values differ at $-1.79\sigma$, $2.14\sigma$, and $-1.21\sigma$ and for the statistical estimators of the level of tension we find $\log_{10}\mathcal{I} = -1.942$ and $p = 0.155\%$ ($\sigma = 3.164$). Given these results, both non-flat XCDM models with $A_L =1$ are ruled out at $3\sigma$ confidence level. 

Allowing the lensing consistency parameter $A_L$ to vary makes P18 data and P18+lensing data results compatible with non-CMB (new) data results in the XCDM models. In the flat XCDM model when P18 data are considered we find $A_L = 1.180^{+0.062}_{-0.10}$ which favors $A_L>1$ over $A_L =1$ by $1.80\sigma$, and $w=-1.23^{+0.31}_{-0.59}$ indicating a preference for phantom-behavior at $0.74\sigma$ significance, while for P18+non-CMB (new) data we get $A_L = 1.222\pm 0.063$ ($3.52\sigma$) and $w=-0.964\pm 0.024$ ($1.50\sigma$ preference for quintessence-like behavior), and finally when P18+lensing+non-CMB (new) data are analyzed we obtain $A_L = 1.101\pm 0.037$ ($2.73\sigma$) and $w=-0.968\pm 0.024$ ($1.33\sigma$ preference for quintessence-like behavior). When $A_L$ is allowed to vary, the tensions between cosmological parameter constraints obtained from the two different data sets are alleviated with respect to the case when $A_L=1$. When we look at the P18 vs.\ non-CMB (new) data case we find $\log_{10}\mathcal{I} = -0.364$ and $p = 3.619\%$ ($\sigma = 2.095$) whereas for the P18+lensing vs.\ non-CMB (new) data case we get $\log_{10}\mathcal{I} = -0.506$ and $p = 1.742\%$ ($\sigma = 2.378$), therefore, according to the statistical estimators the flat XCDM+$A_L$ model is not ruled out at 3$\sigma$ significance by these data. Regarding the performance of the flat XCDM+$A_L$ model with respect to the flat $\Lambda$CDM model, when the P18+lensing+non-CMB (new) data set is considered we obtain $\Delta\text{DIC} = -4.31$ which means that the first model is positively favored over the flat $\Lambda$CDM model. 

Very similar results are found for the non-flat XCDM Planck $P(q)+A_L$ and the non-flat XCDM new $P(q)+A_L$ cosmological models. In particular, when P18 data are analyzed we get for the XCDM Planck [new] $P(q)+A_L$ model $\Omega_k=-0.073^{+0.065}_{-0.029}$ [$-0.072^{+0.065}_{-0.030}$], $w = -1.36^{+1.1}_{-0.53}$ [$-1.39^{+1.1}_{-0.54}$], and $A_L < 1.20$ [$< 1.19$] which indicates a preference for a closed Universe of $1.12\sigma$ [$1.11\sigma$] and a preference of $0.33\sigma$ [$0.35\sigma$] in favor of phantom-like behavior. Jointly analyzing P18 and non-CMB (new) data, we obtain $\Omega_k=0.0011\pm 0.0019$ [$0.0011\pm 0.0019$] for the curvature parameter, favoring an open Universe by $0.58\sigma$ [$0.58\sigma$], while for the dark energy equation of state parameter we get $w =-0.958\pm 0.027$ [$-0.959\pm 0.026$] indicating a preference for the quintessence-like behavior by $1.56\sigma$ [$1.58\sigma$]. As for the lensing parameter, we find $A_L = 1.217\pm 0.064$ [$1.213\pm 0.064$] showing a preference for $A_L > 1$ at $3.39\sigma$ [$3.33\sigma$]. Finally when we analyze the largest data set considered in this work, namely P18+lensing+non-CMB (new), we find for the three non-standard parameters $\Omega_k = 0.0015\pm 0.0019$ [$0.0015\pm 0.0019$], $w= -0.958\pm 0.026$ [$-0.959\pm 0.027$], and $A_L = 1.102\pm 0.037$ [$1.101\pm 0.038$]. According to these results there is a $0.79\sigma$ [$0.79\sigma$] preference for an open Universe, a preference for quintessence-like behavior of $1.62\sigma$ [$1.52\sigma$], and the option $A_L > 1$ is preferred at $2.76\sigma$ [$2.66\sigma$]. It is important to note that while P18 data show a preference for a closed Universe ($\Omega_k<0$) and phantom-like behaviour ($w<-1$), when non-CMB (new) data are included in the mix either in the P18+non-CMB (new) case or in the P18+lensing+non-CMB (new) case, the opposite is true: an open Universe ($\Omega_k>0$) and quintessence-like behavior ($w>-1$) are favored. In regard to the statistical estimators, for the non-flat XCDM Planck [new] $P(q)+A_L$ model, in the case of P18 vs.\ non-CMB (new) data we get $\log_{10}\mathcal{I} = -1.173$ [$-0.957$] and $p = 0.902\%$ [$0.778\%$] ($\sigma = 2.611$ [$2.662$]) and for P18+lensing vs.\ non-CMB (new) data we have $\log_{10}\mathcal{I} = -0.275$ [$-0.312$] and $p = 3.026\%$ [$2.409\%$] ($\sigma = 2.167$ [$2.256$]). As noted previously, it is the inclusion of a varying $A_L$ parameter that brings concordance to the cosmological parameter constraints obtained from the two different data sets. As for the performance of the non-flat XCDM+$A_L$ models we obtain $\Delta\text{DIC}=-2.66 [-2.93]$ pointing to a positive preference for the non-flat models, with a varying equation of state parameter and lensing parameter, over the flat $\Lambda$CDM model with $A_L = 1$. 

It is interesting and important to note that for the largest data set we study, the P18+lensing+non-CMB (new) one, all models with a varying $A_L$ parameter are positively favored over the flat $\Lambda$CDM one. This seems to indicate that consideration of this phenomenological parameter might be necessary to get a better fit to these cosmological data. On the other hand, we should take into account that the flat $\Lambda$CDM model is the simplest observationally-consistent cosmological model among all the models studied in this paper. It passes all consistency tests we have subjected it to and it does so with a constant $\Lambda$ (not an evolving dark energy component) and with flat hypersurfaces ($\Omega_k =0$) and with the lensing consistency parameter set to unity ($A_L =1$). It is also interesting and important to note that analyses of updated PR4 Planck data in the $\Lambda$CDM+$A_L$ model \citep{Rosenberg:2022sdy, Tristram:2023haj} find $A_L$ values less inconsistent with $A_L = 1$, and that this change is partly due to the data set update and partly due to the different likelihoods used.

There appears to be a mismatch between the observed smoothing of some parts of the P18 CMB power spectra and the predicted smoothing due to weak gravitational lensing in the six-parameter flat $\Lambda$CDM model that best fits these data. 
One possible alternative that could resolve this issue is considering negative values of the curvature parameter $\Omega_k<0$ (closed geometry), which in turn allows for a larger non-relativistic matter density parameter $\Omega_m$ values if $\Omega_\Lambda$ is held constant, compared to the case with $\Omega_k=0$, thus incrementing the amount of lensing. Another alternative, which is phenomenological rather than physical, is to re-scale the gravitational potential power spectrum with the phenomenological lensing consistency parameter $A_L$, which automatically increases the amount of lensing for $A_L>1$. The second option allows $\Omega_m$ to remain in the low-value region (compared to the case either with $\Omega_k=0$ or with $\Omega_k<0$), so that the amount of structure formation is reduced, affecting the value of $\sigma_8$, which is lower for all $A_L$-varying models than for models with $A_L=1$. Consequently, the better performance of $A_L$-varying models highlighted above may be related to the simultaneous alleviation of the lensing anomaly and the $\sigma_8$ tension, when CMB and non-CMB (new) data are jointly analyzed, that affect the flat $\Lambda$CDM model.

Consistent with what we found earlier for the $\Lambda$CDM models \cite{deCruzPerez:2022hfr}, when only the CMB data (either P18 or P18+lensing data) are used it is not possible to obtain nearly model-independent cosmological parameter constraints. To obtain model-independent constraints it is necessary to include non-CMB data in the mix. Actually, even considering P18+lensing+non-CMB (new) data, which is the most restrictive data set we use, it is only possible to get almost model-independent constraints when comparing the models with $A_L =1$ or when comparing the models that allow for the variation of $A_L$. Among models with $A_L =1$ the differences between the primary and derived cosmological parameter values always remain below $1\sigma$. No particular parameter has more model-dependent values uniformly across models; parameter value differences vary, depending on the two models that are under comparison. Therefore, it is not unreasonable to claim that P18+lensing+non-CMB (new) data are able to set nearly model-independent cosmological parameter constraints among the models with $A_L=1$. On the other hand, when comparing the constraints obtained for the models with $A_L=1$ and those obtained for the $A_L$-varying models we observe differences of close to $2\sigma$ for $\sigma_8$ and close to $1\sigma$ for the second most affected parameter, $\ln(10^{10}A_s)$.

\section{Conclusions}\label{sec:Conclusions}

Under the condition that the data sets we have employed in this work are correct and free from unaccounted systematics, the statistical estimators we have used show that three of the twelve cosmological models we have studied are rejected at 3$\sigma$ or more because they cannot simultaneously accommodate either P18 and non-CMB (new) data or P18+lensing and non-CMB (new) data. These models are the flat XCDM model and the non-flat XCDM Planck and new $P(q)$ models. 

As we showed in \cite{deCruzPerez:2022hfr}, in the $\Lambda$CDM models neither P18 nor P18+lensing data are able to completely break the geometrical degeneracy between the  $\Omega_m - H_0 - \Omega_k - A_L$ parameters and, as expected, this remains true here when we additionally consider a varying dark energy equation of state parameter $w$. According to the results obtained in this work, there are three options for the flat $\Lambda$CDM model to deal with the lensing anomaly, if we consider only one additional parameter at a time: $\Omega_k<0$, $w>-1$, or $A_L>1$. In the case of the non-flat $\Lambda$CDM models with $A_L=1$, P18 data favor a more negative value of the curvature parameter, which is compensated by a larger value of the matter density parameter $\Omega_m$ and a lower value of the Hubble constant $H_0$. On the other hand, the results obtained for the flat XCDM model show that a more negative dark energy equation of state parameter $w$ value is favored, compensated by a smaller value of $\Omega_m$ and a higher value of $H_0$, contrary to the case of the non-flat models. Finally in the case of the flat $\Lambda$CDM+$A_L$ model, due to the weak correlations of the $A_L$ parameter with the other free parameters, we do not observe significant changes in the values of $\Omega_m$ and $H_0$ with respect to the flat $\Lambda$CDM model. According to the DIC values, among these three options the most favored one when it comes to fitting P18 data turns out to be $\Omega_k < 0$: for the two non-flat $\Lambda$CDM models we get $\Delta\textrm{DIC} = -7.34$ [Planck $P(q)$] and $\Delta\textrm{DIC} = -6.39$ [new $P(q)$], respectively, which indicates that both models are strongly favored over the flat $\Lambda$CDM model. On the other hand, for the flat $\Lambda$CDM+$A_L$ model we get $\Delta\textrm{DIC} = -5.52$ whereas for the flat XCDM model $\Delta\textrm{DIC} = -2.26$, meaning that both the $\Lambda$CDM+$A_L$ model and the flat XCDM model are positively favored over the standard flat $\Lambda$CDM model, with the flat $\Lambda$CDM+$A_L$ model favored over the flat XCDM model.

Although CMB lensing data are not as restrictive as P18 data, in some cases we find non-negligible changes in the cosmological parameter constraints when P18+lensing data are used instead of P18 data alone. For the non-flat $\Lambda$CDM models, we find that $\Omega_k$ is still negative but its absolute value decreases, moving closer to flat spatial geometry. In addition, $\Omega_m$ decreases whereas $H_0$ increases with respect to the values from the P18 data analysis. When we move from P18 data to P18+lensing data, the evidence in favor of $A_L>1$ decreases for the flat $\Lambda$CDM+$A_L$ model, but the results are barely changed  for the flat XCDM model, indicating that the XCDM model is not very sensitive to CMB lensing data.

Non-CMB (new) data by themselves do not support the results that we have just summarized. For the non-flat $\Lambda$CDM models, with non-CMB (new) data the evidence in favor of $\Omega_k<0$ (closed geometry) subsides and instead positive values of the curvature parameter ($\Omega_k > 0$, open geometry) are found, contrary to what happens when analyzing non-CMB (old) data, \cite{deCruzPerez:2022hfr}. Also, with non-CMB (new) data smaller values of $\Omega_m$ and higher values of $H_0$ are found with respect to the results obtained with P18 data. As for the flat XCDM model, non-CMB (new) data favor larger values of $\Omega_m$ and lower values of $H_0$ compared to those from P18 data. From the DIC values, we see that with non-CMB (new) data the non-flat $\Lambda$CDM Planck $P(q)$ and $\Lambda$CDM new $P(q)$ models have $\Delta\textrm{DIC} = +1.41$ and $\Delta\textrm{DIC} = +2.05$, respectively. These positive values indicate that the two models are not favored over the flat $\Lambda$CDM model. However, the flat XCDM model, which has $\Delta\textrm{DIC}=-9.37$ for non-CMB (new) data, is strongly preferred when compared to the standard flat $\Lambda$CDM model. 

The significant discrepancies found between the results obtained with P18/P18+lensing data and non-CMB (new) data led us to make use of statistical estimators to more properly quantify the tensions. As reported above, these tests reveal that the three XCDM models with $A_L=1$ are ruled out at $3\sigma$ or more. For the non-flat $\Lambda$CDM models, we observe that the level of tension in the cases of P18 vs.\ non-CMB (new) data and P18+lensing vs.\ non-CMB (new) data are reduced compared to the cases where non-CMB (old) data were used instead of non-CMB (new) data, with the main consequence being that here these models are not ruled out at the $3\sigma$ threshold. Regarding the XCDM models with a varying $A_L$ parameter we note that while the levels of tension are still high, they are less than $3\sigma$ and not enough to disallow the joint analysis of P18, lensing, and non-CMB (new) data. 

Overall, considering only the nine models not ruled out by discordances between parameter values determined from different data sets, for the P18+lensing+non-CMB (new) data set, we find little deviation from a flat geometry and moderate deviation from a cosmological constant, with the biggest deviations being $\Omega_k = 0.0015 \pm 0.0019$ in the XCDM Planck $P(q)+A_L$ and XCDM new $P(q)+A_L$ models, which favor open geometry and are 0.79$\sigma$ from flat geometry, and $w = -0.958\pm 0.026$ in the XCDM Planck $P(q)+A_L$ model, which favors quintessence-like dynamical dark energy and is 1.62$\sigma$ from a  cosmological constant. Interestingly, in all six non-flat models that are not ruled out at 3$\sigma$ or more, open geometry is mildly favored, and in all three XCDM+$A_L$ models (that are not ruled out at 3$\sigma$ or more), quintessence-like dynamical dark energy is moderately favored. In the $A_L = 1$ non-flat $\Lambda$CDM cases, we find for the P18+lensing+non-CMB (new) data set $\Omega_k = 0.0009 \pm 0.0017$ [$0.0008 \pm 0.0017$] for the Planck [new] $P(q)$ model, favoring open geometry at 0.53$\sigma$ [0.47$\sigma$].

Our cosmological parameter constraints obtained for the flat $\Lambda$CDM model, when P18+lensing+non-CMB (new) data are considered, are the most restrictive results to date. In particular, for the six primary parameters we get $\Omega_b h^2 = 0.02249\pm 0.00013$, $\Omega_c h^2 = 0.11849 \pm 0.00084$, $100\theta_{\text{MC}}=1.04109\pm 0.00028$, $\tau = 0.0569\pm 0.0071$, $n_s = 0.9685\pm 0.0036$, and $\ln(10^{10}A_s)=3.046\pm 0.014$. Additionally, for the derived parameters, we find $H_0=68.05\pm 0.38$ km s$^{-1}$ Mpc$^{-1}$, $\Omega_m = 0.3059\pm 0.0050$, and $\sigma_8 = 0.8077\pm 0.0057$. Among models with $A_L = 1$, these values show almost model-independent consistency, with differences always below $1\sigma$. However, when we compare these cosmological parameter values with those obtained for the $A_L$-varying models, we observe larger differences. In particular, for the six varying $A_L$ models relative to the flat $\Lambda$CDM model we find the following maximum differences for each parameter. For models with $w=-1$, we find $-0.35\sigma$ for $\Omega_b h^2$, $0.53\sigma$ for $\Omega_c h^2$, $-0.12\sigma$ for $100\theta_{\text{MC}}$, $0.81\sigma$ for $\tau$, $-0.41\sigma$ for $n_s$, $0.95\sigma$ for $\ln(10^{10}A_s)$, $0.71\sigma$ for $H_0$, $0.75\sigma$ for $\Omega_m$, and $1.08\sigma$ for $\sigma_8$, whereas comparing $w$-varying model results we get $-0.55\sigma$ for $\Omega_b h^2$, $0.81\sigma$ for $\Omega_c h^2$, $-0.27\sigma$ for $100\theta_{\text{MC}}$, $0.69\sigma$ for $\tau$, $-0.66\sigma$ for $n_s$, $0.88\sigma$ for $\ln(10^{10}A_s)$, $0.21\sigma$ for $H_0$, $0.13\sigma$ for $\Omega_m$, and $1.80\sigma$ for $\sigma_8$. 

When P18+lensing+non-CMB (new) data are analyzed, interesting trends related to the lensing consistency parameter $A_L$ are observed. For models with fixed $A_L=1$, we have on average evidence for open spatial geometry with positive curvature parameter $\Omega_k>0$ ($0.63\sigma$) and a quintessence-like dark energy equation of state parameter $w>-1$ ($0.63\sigma$). For models with varying $A_L$ parameter, we get $\Omega_k>0$ ($0.52\sigma$), $w>-1$ ($1.49\sigma$), and $A_L>1$ ($2.59\sigma$). Therefore, among the various non-standard parameters explored in this work, the lensing consistency parameter is the one that most deviates from the standard flat $\Lambda$CDM model $A_L = 1$ value. This conclusion is supported by the findings from the DIC values which indicate that, when P18+lensing+non-CMB (new) data are used, all $A_L$-varying models are positively favored over the flat $\Lambda$CDM model.  

The conclusions we have drawn in this work depend on the data sets we have employed. The recent analysis of the updated PR4 Planck data set \cite{Tristram:2023haj} results in updated values for the curvature parameter ($\Omega_k=-0.012\pm 0.010$) and the lensing consistency parameter ($A_L=1.039\pm 0.052$). These new measurements show less evidence in favor of non-flat hypersurfaces and $A_L>1$ compared to the results obtained from P18 data, partly as a consequence of the updated PR4 Planck data set and partly as a consequence of the different likelihoods used in the analyses. (One might view the second source of difference as a systematic and it might then be appropriate to account for it as an additional systematic error on cosmological parameter values.) 

More and better cosmological data are required in order to clarify whether the tensions highlighted in our work, affecting the standard flat $\Lambda$CDM model of cosmology, can really be interpreted as hints of new physics or if they arise from some unaccounted systematics in the cosmological data we have used. Given the circumstances, the flat $\Lambda$CDM model remains the simplest (largely) observationally consistent cosmological model.

\acknowledgements
J.d.C.P.\ was supported by the Margarita Salas fellowship funded by the European Union (NextGenerationEU). C.-G.P.\ was supported by a National Research Foundation of Korea (NRF) grant funded by the Korea government (MSIT) No.\ RS-2023-00246367.


\providecommand{\noopsort}[1]{}\providecommand{\singleletter}[1]{#1}%

\end{document}